\documentclass[11pt,oneside,a4paper]{article}
\usepackage{amssymb}
\usepackage{amsmath}
\usepackage[titletoc,title]{appendix}
\usepackage[T1]{fontenc}
\usepackage{graphicx}
\usepackage{float}
\usepackage[bookmarks=true]{hyperref}
\usepackage{multirow}
\usepackage{newtxtext,newtxmath}
\usepackage{threeparttable}
\usepackage{url}
\usepackage{hyperref}
\usepackage{subfig}
\usepackage{upgreek}
\usepackage{xcolor}

\makeatletter
\g@addto@macro\bfseries{\boldmath}
\makeatother

\definecolor{Blu}{rgb}{0.,0.,1.}
\definecolor{oucrimsonred}{rgb}{0.6, 0.0, 0.0}
\definecolor{persianblue}{rgb}{0.11, 0.22, 0.73}
\definecolor{forestgreen}{rgb}{0.13,0.35,0.13}
\definecolor{red}{rgb}{1.,0.,0.}

\newcommand{\Fig}[1]{Figure~\ref{#1}}
\newcommand{\Sec}[1]{Section~\ref{#1}}

\newcommand{\Tab}[1]{Table~\ref{#1}}

\newcommand{\eprk}{\ensuremath{\epsilon'_K}}
\newcommand{\Reps}{\ensuremath{{\rm Re}\,\epsilon'_K/\epsilon_K}}

\newcommand{\kl}{K_L}
\newcommand{\klpionn}{K_L \to \pi^0 \nu \overline{\nu}}

\newcommand{\kpien}{K_L \to \pi^\pm e^\mp \nu}

\newcommand{\klgg}{K_L \to 2\gamma}
\newcommand{\klpiopio}{K_L \to \pi^0 \pi^0}
\newcommand{\klpiopiopio}{K_L \to \pi^0 \pi^0 \pi^0}
\newcommand{\klppm}{K_L \to \pi^+ \pi^- \pi^0}

\newcommand{\pt}{p_{\mathrm{T}}}
\newcommand{\zvtx}{z_{\mathrm{vtx}}}

\newcommand{\ie}{\textit{i}.\textit{e}.}

\makeatletter
\newcommand\subsubsubsection{\@startsection{paragraph}{4}{\z@}%
                                     {-3.25ex\@plus -1ex \@minus -.2ex}%
                                     {1.5ex \@plus .2ex}%
                                     {\normalfont\normalsize\bfseries}}
\makeatother

\hypersetup{colorlinks, citecolor=oucrimsonred, linkcolor=persianblue, urlcolor=oucrimsonred}

\begin{document}

\begin{center}

\Large{\bf Searches for new physics with high-intensity kaon beams}

\vspace*{3mm}
\large{\bf Contributed paper to Snowmass 2021}\\

\vspace{10mm}
\Large{The KOTO\footnote{Contact persons: Hajime Nanjo (nanjo@champ.hep.sci.osaka-u.ac.jp), Tadashi Nomura (tadashi.nomura@kek.jp), Y.W.~Wah(ywah@uchicago.edu)},
LHCb\footnote{Contact persons: Francesco Dettori (francesco.dettori@cern.ch), Diego Martinez Santos (diego.martinez.santos@cern.ch)}, and 
NA62/KLEVER\footnote{Contact persons: Cristina Lazzeroni (cristina.lazzeroni@cern.ch), Matthew Moulson (moulson@lnf.infn.it)} Collaborations\\
and the US Kaon Interest Group\footnote{W.~Altmannshofer, L.~Bellantoni, E.~Blucher, G.J.~Bock, N.H.~Christ, D.~Denisov, Y.~Grossman, S.H.~Kettell, P.~Laycock, J.D.~Lewis, H.~Nguyen, R.~Ray, J.L.~Ritchie, P.~Rubin, R.~Tschirhart, Y.~Wah, E.~Worcester, E.D.~Zimmerman; Contact person: Elizabeth Worcester (etw@bnl.gov)}}

\vspace*{4mm}
\Large{28 April 2022}
\end{center}

\vspace*{10mm}

\begin{abstract}
The availability of high-intensity kaon beams at the J-PARC Hadron Experimental Facility and the CERN SPS North Area, together with the abundant forward production of kaons at the LHC, gives rise to unique possibilities for sensitive tests of the Standard Model in the quark flavor sector. 
Precise measurements of the branching ratios for the flavor-changing neutral current decays $K\to\pi\nu\bar{\nu}$ can provide unique constraints on CKM unitarity and, potentially, evidence for new physics. 
Building on the success of the current generation of fixed-target experiments, initiatives are taking shape in both Europe and Japan to measure the branching ratio for $K^+\to\pi^+\nu\bar\nu$ to $\sim$5\% and for $K_L\to\pi^0\nu\bar\nu$ to $\sim$20\% precision.
These planned experiments would also carry out lepton flavor universality tests, lepton number and flavor conservation tests, and perform other precision measurements in the kaon sector, as well as searches for exotic particles in kaon decays.
Meanwhile, the LHCb experiment is ready to restart data taking with a trigger upgrade that will vastly increase its sensitivity for rare $K_S$ decays and complementary hyperon decays. We overview the initiatives for next-generation experiments in kaon physics in Europe and Japan, identifying potential contributions from the US high-energy physics community.
\end{abstract}

\newpage

\newpage
\setcounter{tocdepth}{3}
\tableofcontents

\newpage
\setcounter{page}{1}
\pagestyle{plain}

\section{Introduction}
\subsection{A very brief history}
In 1947, with a cloud chamber observing cosmic rays, George Rochester and Clifford Butler discovered the neutral kaon~\cite{Rochester:1947}, dubbed the `strange' particle at the time. In 1949, Rosemary Brown and Cecil Powell observed the three-pion decay of the charged kaon~\cite{Brown:1949}, dubbed the `tau' particle. 
In the early fifties, Murray Gell-Mann recalled~\cite{Cronin:2013} that ``...Although the presence of Enrico (Fermi) in the back row was terrifying, nothing went seriously wrong. One day, in the course of explaining the strangeness idea, I discussed the neutral meson $K^0$ and $\overline{K^0}$, which are antiparticle of each other. Enrico objected that nothing new was involved, since one could take, instead of the $K^0$ and $\overline{K^0}$ fields, their sum and difference, which would then describe two neutral spin zero particle, each its own antiparticle ... I was able to answer Enrico by saying that in certain decays of this neutral strange particle, the sum and difference would indeed be the relevant fields, but that in the production process a single $K^0$ could accompany a lambda hyperon, which a $\overline{K^0}$ could not. Later (with A. Pais, 1955~\cite{Pais:1955}), when I wrote up the neutral K particle situation, I thanked Fermi for his question...''. $K_1$ and $K_2$ are the admixtures of $K^0$ and $\overline{K^0}$ with even and odd respectively, under the CP transformation. With the assumption of CP invariance, the $K_2$ was forbidden to decay to two pions and it lives a lot longer than the $K_1$. 

However, the person who really introduced this particle to the public was Leon M. Lederman and his collaborators. Their work on the long-lived neutral $K$ mesons was published in 1956~\cite{Lederman:1956}. The paper contained a description of the long-lived V-particle found in the Brookhaven Cosmotron. They established its lifetime, and observed some of its decays including semi-leptonic (so called $K_{e3}$ and $K_{\mu3}$) and three-pion final states. In late 1956, parity violation was experimentally observed by C.S.~Wu~\cite{Wu:1957}, and it shook the physics world that parity is violated in the weak interaction. 

In 1964, CP violation was discovered by James Christenson, James Cronin, Val Fitch and Ren\'e Turlay with about 50 samples of $K_2$ to two pion decays~\cite{Christenson:1964}. The experiment was designed to study the regeneration of the neutral kaons. When asked, Cronin said the two pions final state study was the very last item to be studied on a list of more than twenty physics analyses. The neutral kaon system, via its quantum mechanical interference (mass mixing), provides the precision measurements of the mass difference and decay width. 
The question of the origin of CP violation beyond mixing (the so-called $\epsilon'$ direct CP violating effect) was settled some 35 years later.
The NA31 experiment at CERN found the first evidence of direct CP violation in 1988 with three standard deviations from zero. However shortly after, the E731 experiment at Fermilab reported a measurement consistent with zero. A better precision was needed and a new generation of detectors was built.
Direct CP violation was finally established by the NA48 experiment at CERN and the KTeV experiment at Fermilab in the late 1990s.
The history of kaon physics parallels particle physics, and there may be further surprises as discussed.

For over 70 years, experimental studies of kaons have played a singular role in propelling the development of the Standard Model (SM). As in other branches of flavor physics, the continuing experimental interest in the kaon sector derives from the possibility of conducting precision measurements, particularly of suppressed or rare processes, that may reveal the effects of new physics with mass-scale sensitivity exceeding that which can be explored directly, e.g., at the LHC, or even at a next-generation hadron collider. Because of the relatively small number of kaon decay modes and the relatively simple final states, combined with the relative ease of producing intense kaon beams, kaon decay experiments are in many ways the quintessential intensity-frontier experiments.

\subsection{The golden kaon decay modes}
\label{sec:golden}

\begin{figure}[htbp]
\centering
\includegraphics[width=0.4\textwidth]{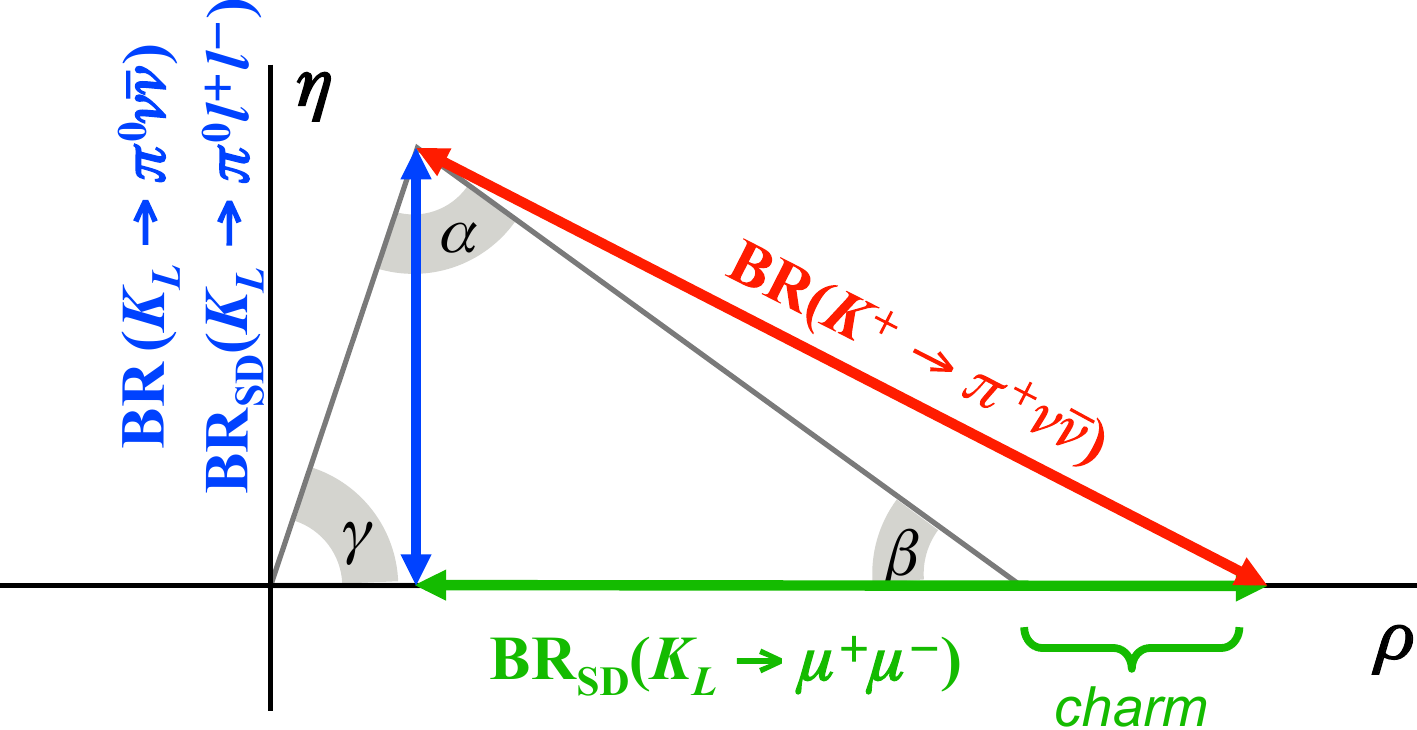}
\vspace{-2mm}
\caption{Determination of the unitary triangle with rare kaon decays.} 
\label{fig:ut}
\end{figure}

Rare kaon decays provide information on the unitary triangle, as illustrated in \Fig{fig:ut}. These are flavor-changing neutral current processes (FCNC) that probe the $s\to d\nu\bar\nu$ or $s\to d\ell^+\ell^-$ transitions. They are highly GIM-suppressed and their SM rates are very small. Complications from long-distance (LD) physics affect the modes unevenly. As seen from \Tab{tab:fcnc}, the $K\to\pi\nu\bar\nu$ decays are the least affected. The branching ratios (BRs) for the $K\to\pi\nu\bar\nu$ decays are among the observable quantities in the quark-flavor sector most sensitive to new physics.

\begin{table}[hbtp]
\centering
\begin{tabular}{lcccc}\hline\hline
Decay & $\Gamma_{\rm SD}$ & 
Theory error & SM BR $\times 10^{11}$ & Experimental BR $\times 10^{11}$ \\ \hline
$K_L\to\mu^+\mu^-$ & 10\% & 30\% & $79\pm12$ (SD) & $684\pm11$ \\
$K_L\to\pi^0 e^+ e^-$ & 40\% & 10\% & $3.2\pm1.0$ & $<28$ \\
$K_L\to\pi^0\mu^+\mu^-$ & 30\% & 15\% & $1.5\pm0.3$ & $<38$ \\
$K^+\to\pi^+\nu\bar{\nu}$ & 90\% & 4\% & $8.4\pm1.0$ & $10.6^{+4.1}_{-3.5}$ \\
$K_L\to\pi^0\nu\bar{\nu}$ & >99\% & 2\% & $3.4\pm0.6$ & $<300$ \\
\hline\hline
\end{tabular}
\caption{FCNC kaon decays with high sensitivity to short-distance (SD) physics. Theory errors refer to approximate uncertainty on LD-subtracted rate, excluding parametric contributions. Experimental upper limits on decay BRs are quoted at 90\% CL.}
\label{tab:fcnc}
\end{table}

\subsubsection{$K\to\pi\nu\nu$ in the Standard Model}

\begin{figure}[htbp]
\centering
\includegraphics[width=0.6\textwidth]{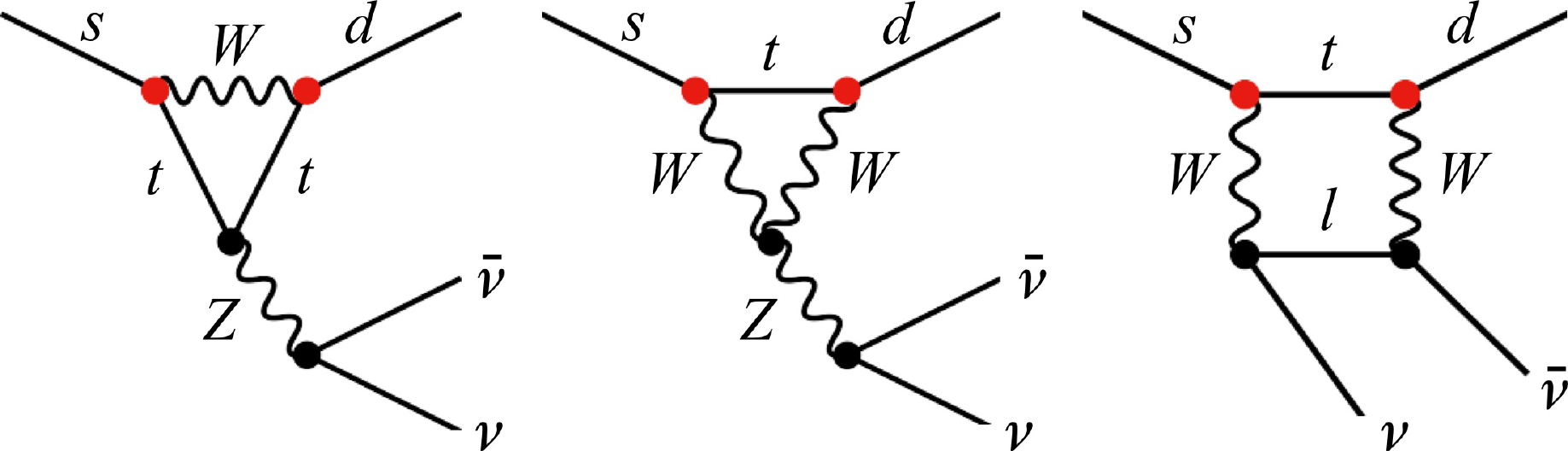}
\caption{Diagrams contributing to the process $K\to\pi\nu\bar{\nu}$.} 
\label{fig:fcnc}
\end{figure}

The $K\to\pi\nu\bar{\nu}$ decays probe the $s\to d\nu\bar{\nu}$ transition  via the $Z$-penguin and box diagrams shown in \Fig{fig:fcnc}. For several reasons, the SM calculation for their BRs is particularly clean~\cite{Cirigliano:2011ny}.
\begin{itemize}
\item The loop amplitudes are dominated by the top-quark contributions. The $K_L$ decay violates $CP$; its amplitude involves the top-quark 
contribution only. In fact, ${\rm BR}(K_L\to\pi^0\nu\bar{\nu})$ gives 
a direct measurement of the height (and therefore the area, $J$) 
of the unitary triangle. Small corrections to the amplitudes from the lighter quarks come into play for the charged channel. 
\item The hadronic matrix elements for these decays can be obtained from the
precise experimental measurements of the $K\to\pi e\nu$ ($K_{e3}$) decay rates.
\item There are no long-distance (LD) contributions from processes with
intermediate photons.
This point in particular distinguishes the
$K\to\pi\nu\bar{\nu}$ decays from the other modes appearing in \Fig{fig:ut}.
\end{itemize}
In the SM, the uncertainties on the $K\to\pi\nu\bar{\nu}$ rates are entirely dominated by the uncertainties on the CKM matrix elements $|V_{ub}|$ and $|V_{cb}|$ and the angle $\gamma$.
Using values for these parameters from the analysis of tree-level
observables, Buras et al. obtain~\cite{Buras:2015qea}
\begin{equation}
  \begin{split}   
    {\rm BR}(K_L\to\pi^0\nu\bar{\nu}) & = (3.4 \pm 0.6)\times10^{-11},\\
    {\rm BR}(K^+\to\pi^+\nu\bar{\nu}) & = (8.4 \pm 1.0)\times10^{-11},
  \end{split}
  \label{eq:buras}
\end{equation}
where the uncertainties are dominated by the external contributions from $V_{cb}$ and $V_{ub}$.
These uncertainties can be reduced by adding information on the CKM
parameters from the analysis of loop-level observables~\cite{Buras:2015qea}. The CKM analysis has been revisited several times, with values of the relevant CKM parameters taken from the PDG's global CKM fit~\cite{Brod:2021hsj}, or from approaches optimized to optimize the uncertainty on the $K\to\pi\nu\bar{\nu}$ BRs~\cite{Buras:2021nns}.
The results of these analyses for the BRs vary within the bracket expected from the CKM uncertainties.
Assuming no new-physics effects in $\epsilon_K$ and $S_{\psi K}$, the BRs can be determined independently of $|V_{cb} |$ as ${\rm BR}(K^0\to\pi^0\nu\bar{\nu}) = (2.94 \pm 0.15)\times10^{-11}$ and ${\rm BR}(K^+\to\pi^+\nu\bar{\nu}) =  (8.60 \pm 0.42)\times10^{-11}$~\cite{Buras:2021nns}.
In order to avoid the potential for bias to creep in from new physics sources contributing to loop-level observables, we continue to use the evaluation of Eqs.~(\ref{eq:buras}) as our reference value for the SM BRs.  
As a demonstration of the size of the intrinsic theoretical uncertainties,
if the CKM parameters are taken to be exact, the SM BRs of Eqs.~(\ref{eq:buras}) become
$(3.36\pm0.05)\times10^{-11}$ and $(8.39\pm0.30)\times10^{-11}$, respectively.
Because of the corrections from lighter-quark contributions, these
uncertainties are larger for the $K^+$ case.

\subsubsection{$K\to\pi\nu\bar\nu$ and the search for new physics}

Because the $K\to\pi\nu\bar\nu$ decays are strongly suppressed and calculated very precisely in the SM, their BRs are potentially sensitive to mass scales of hundreds of TeV, surpassing the sensitivity of $B$ decays in most SM extensions~\cite{Buras:2014zga}.


Because the amplitude for $K^+\to\pi^+\nu\bar{\nu}$
has both real and imaginary parts, while the amplitude for
$K_L\to\pi^0\nu\bar\nu$ is
purely imaginary, the decays have different sensitivity to new sources of CP violation.
Measurements of both BRs would therefore be extremely useful not only
to uncover evidence of new physics in the quark-flavor sector, but also
to distinguish between new physics models.
\begin{figure}[htb]
\centering
\includegraphics[width=0.6\textwidth]{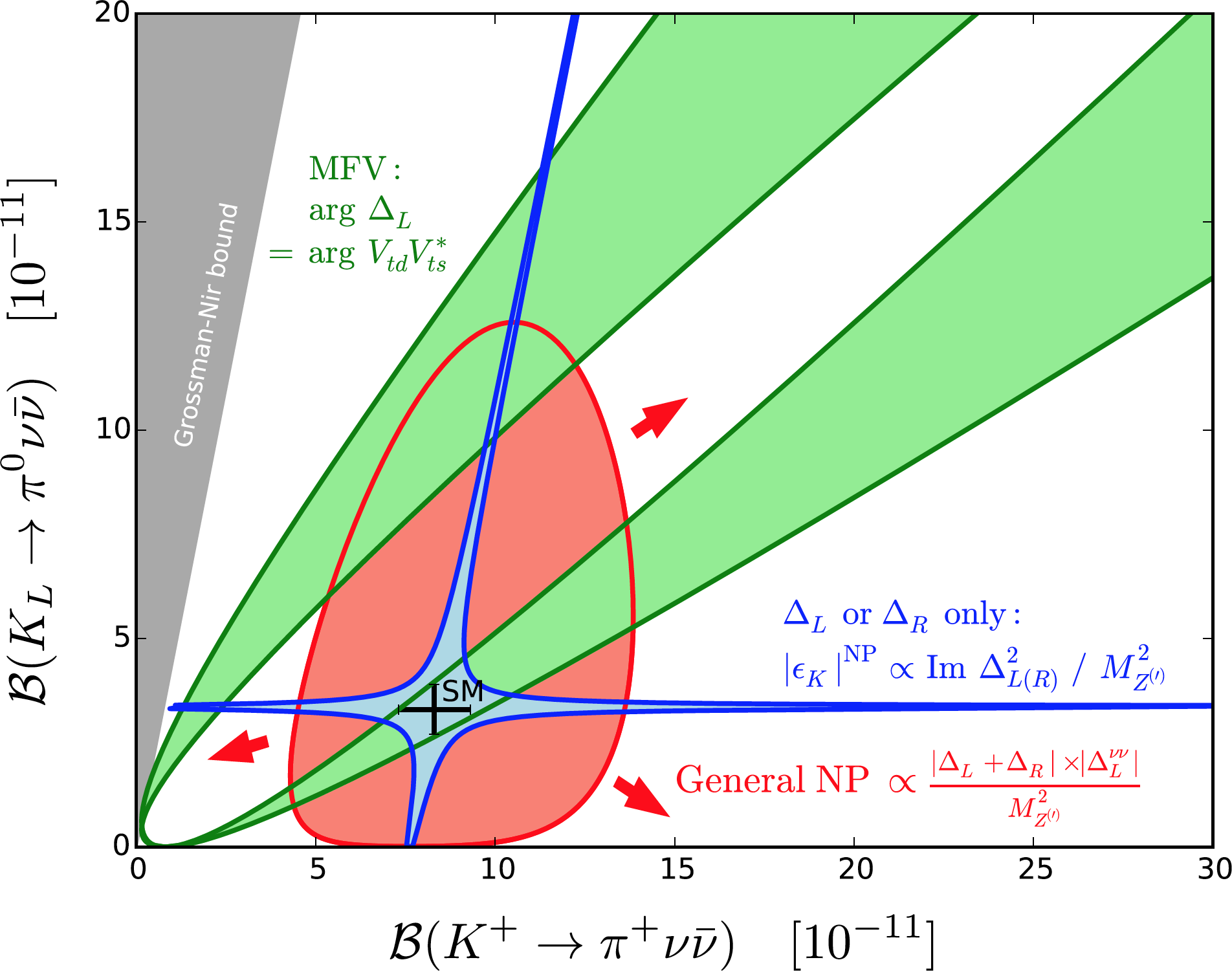}
\caption{Scheme for BSM modifications of $K\to\pi\nu\bar{\nu}$ BRs.} 
\label{fig:KpnnBSM}
\end{figure}
\Fig{fig:KpnnBSM}, reproduced from Ref.~\cite{Buras:2015yca}, illustrates a general scheme for the expected
correlation between the $K^+$ and $K_L$ decays under various scenarios.
If the new physics has a CKM-like structure of flavor interactions, with only couplings to left-handed quark currents, the values for ${\rm BR}(K^+\to\pi^+\nu\bar{\nu})$ and ${\rm BR}(K_L\to\pi^0\nu\bar{\nu})$
will lie along the green band in the figure. Specifically, for new physics models with minimal flavor violation, constraints from other flavor observables limit the expected size of the deviations of the BRs to within about 10\% of their SM values~\cite{Isidori:2006qy}.
If the new physics contains only left-handed or right-handed couplings
to the quark currents, it can affect $K\bar{K}$ mixing ($\Delta F = 2$) as well as $K\to\pi\nu\bar{\nu}$ decays ($\Delta F = 1$). Then, to respect constraints from the observed value of the $K\bar{K}$ mixing parameter $\epsilon_K$, the BRs for the $K^+$ and $K_L$ decays must lie on one of the two branches of the blue cross-shaped region~\cite{Blanke:2009pq}. This is characteristic of littlest-Higgs models with $T$ parity~\cite{Blanke:2009am}, as well as models with flavor-changing $Z$ or $Z'$ bosons and pure right-handed or left-handed couplings~\cite{Buras:2014zga,Buras:2012jb}.
    In the most general case, if the new physics has an arbitrary flavor
    structure and both left-handed and right-handed couplings, the
    constraint from  $\epsilon_K$ is evaded and there is little correlation,
    as illustrated by the red region.
    This is the case for the general MSSM framework (without minimal flavor
    violation) \cite{Isidori:2006qy} and in Randall-Sundrum models
    \cite{Blanke:2008yr}.

As discussed in \Sec{sec:kp_pnn_stat}, ${\rm BR}(K^+\to\pi^+\nu\bar\nu)$ has recently been measured to 40\% precision by NA62, following upon the collection of a few candidate events in the late 1990s and early 2000s by the Brookhaven E787 and E949 experiments. The BR for the neutral decay $K_L\to\pi^0\nu\bar\nu$ has not yet been measured.
Up to small corrections, considerations of isospin symmetry lead to the
model-independent bound
$\Gamma(K_L\to\pi^0\nu\bar{\nu})/\Gamma(K^+\to\pi^+\nu\bar{\nu}) < 1$~\cite{Grossman:1997sk}, known as the Grossman-Nir bound.
Thus the limit on ${\rm BR}(K^+\to\pi^+\nu\bar{\nu})$
from NA62 discussed in \Sec{sec:kp_pnn_stat} implies ${\rm BR}(K_L\to\pi^0\nu\bar{\nu}) < 0.81\times10^{-9}$.

The observation of three candidate events in the KOTO 2016--2018 data set, although ultimately found to be statistically compatible with the expected background, initially gave rise to some theoretical speculation about mechanisms for evading the Grossman-Nir bound, mainly involving the apparent enhancement of ${\rm BR}(K_L\to\pi^0\nu\bar{\nu})$ by the contribution from $K_L$ decays into final states containing a $\pi^0$ (or two photons) and one or more new, light particles with very weak SM couplings, the most studied case being $K_L\to\pi^0 S$, with $S$ a light scalar.
Since the Grossman-Nir relation is based on general isospin symmetry arguments that relate the decay amplitudes of charged and neutral kaons, it would be expected to hold also for $K_L\to\pi^0 S$ and $K^+\to\pi^+ S$. Indeed, in most of these studies, the Grossman-Nir bound is evaded for experimental reasons: NA62 might not detect $K^+\to\pi^+ S$ events with $m_S \approx m_{\pi^0}$ because of cuts to eliminate $K^+\to\pi^+\pi^0$ background, for example~\cite{Fuyuto:2014cya,Kitahara:2019lws,Egana-Ugrinovic:2019wzj}, or because of issues related to the different acceptances of KOTO and NA62 for different values of the lifetime of $S$~\cite{Kitahara:2019lws}. New physics scenarios have also been proposed in which the Grossman-Nir bound is violated at the level of the decay amplitudes~\cite{Ziegler:2020ize}.
Although, as discussed in \Sec{sec:klpnn_status}, there is no evidence for the enhancement of ${\rm BR}(K_L\to\pi^0\nu\bar{\nu})$ at this time, with the increased sensitivity expected from improvements to KOTO, and especially, from the next generation of $K_L\to\pi^0\nu\bar{\nu}$ searches, some of these scenarios may regain relevance.  

Like ${\rm BR}(K_L\to\pi^0\nu\bar{\nu})$, the parameter \eprk\ also gives a direct measurement of the height of the unitary triangle. Experimentally, \Reps\ has been measured by NA48~\cite{Batley:2002gn} and KTeV~\cite{Abouzaid:2010ny}, leading to the average $\Reps = (16.6\pm2.3)\times10^{-4}$~\cite{ParticleDataGroup:2020ssz}. In principle, the experimental value of \Reps\ significantly constrains
the value of ${\rm BR}(K_L\to\pi^0\nu\bar{\nu})$ to be expected in
any given new-physics scenario.
However, because of the delicate balance between the amplitudes
for the hadronic matrix elements of different operators, it is 
difficult to perform a reliable calculation of \Reps\ in the SM.
In recent years, there has been significant progress in evaluating the hadronic matrix elements on the lattice. In 2015, the RBC-UKQCD Collaboration obtained the
result $\Reps = (1.38\pm5.15\pm4.59)\times10^{-4}$
from a lattice calculation~\cite{Bai:2015nea}, 2.1$\sigma$ less than the experimental value. 
However, the detailed systematic analysis of the components of this result, in particular, the evaluation of the $\Delta I = 1/2$ $K\to\pi\pi$ amplitude $A_0$, led to significant improvements in the calculation. The updated result, 
$\Reps = (21.7\pm2.6\pm6.2\pm5.0)\times10^{-4}$, where the third uncertainty is from neglected isospin-breaking corrections, is much more consistent with the SM expectation~\cite{RBC:2020kdj}. An estimate based on the same calculation of the hadronic matrix elements, with evaluation of the isospin-breaking corrections from chiral-perturbation theory as well as NNLO QCD contributions to electroweak and QCD penguins, gives $\Reps = (13.9\pm5.2)\times10^{-4}$~\cite{Aebischer:2020jto}, in agreement with a recent estimate from chiral-perturbation theory including isospin-breaking corrections $\Reps = (14\pm5)\times10^{-4}$~\cite{Cirigliano:2019cpi}. Although at the moment the experimental result is in agreement with SM expectations, the uncertainties on the expected value remain larger than those of the measurement by about a factor of two, leaving significant room for contributions from new physics to the value of \Reps. As an example, the effects on the BRs for the $K\to\pi\nu\bar{\nu}$ decays from an O(10-TeV) $Z'$ boson with flavor-violating quark couplings or from modified $Z$ couplings from renormalization-group mixing with the $Z'$ have been explored with an effective field theory framework, 
with present constraints from $\epsilon_K$, \Reps, and $\Delta m_K$~\cite{Aebischer:2020mkv,Aebischer:2022vky}. These studies show that enhancements of 100\% or more are possible for ${\rm BR}(K_L\to\pi^0\nu\bar{\nu})$, with smaller enhancements ($<50\%$) for ${\rm BR}(K^+\to\pi^+\nu\bar{\nu})$, depending on the details of the scenario assumed.

As discussed in \Sec{sec:strange_lf}, observations of lepton-flavor-universality-violating phenomena are mounting in the $B$ sector.
Many explanations for such phenomena
predict strong third-generation couplings and thus significant changes
to the $K\to\pi\nu\bar{\nu}$ BRs through couplings to final states with
$\tau$ neutrinos~\cite{Bordone:2017lsy}.
Measurements of the $K\to\pi\nu\bar\nu$ BRs are therefore critical to interpreting the data from rare $B$ decays, and may demonstrate that these effects are a manifestation of new degrees of freedom such as
leptoquarks~\cite{Buttazzo:2017ixm,Fajfer:2018bfj,Marzocca:2021miv}.
See \Sec{sec:strange_lf} for more general considerations about the interplay of strange physics and lepton flavor anomalies.


\subsubsection{Experimental status for $K^+\to\pi^+\nu\bar{\nu}$}
\label{sec:kp_pnn_stat}

During the period from the mid 1990s to the mid 2000s, the American physics
community demonstrated a particularly keen interest in measuring the BRs for the $K\to\pi\nu\bar{\nu}$ decays. 
Although many of the planned experiments were never realized, a considerable amount of R\&D was conducted, resulting in important inputs for the 
conceptual design of any new $K\to\pi\nu\bar{\nu}$ experiment.

From 1995 to 2002, the Brookhaven experiments E787 and E949 made the first 
measurement of ${\rm BR}(K^+\to\pi^+\nu\bar{\nu})$ using low-energy kaons
from the AGS stopped in an active target surrounded by a hermetic array of detectors for energy and range measurements to identify the $\pi^+$ for candidate signal events. An innovative readout system was used for full  waveform digitization of the signals from the detectors, providing additional constraints. The first-phase experiment, E787, took data from 1995 to 1998 and observed two signal candidates~\cite{Adler:2001xv}; in 2002, the upgraded experiment, E949, was terminated 12 weeks into a scheduled 60-week AGS run after having collected an additional five candidate events~\cite{Artamonov:2009sz}. In late 2011, ORKA, a stopped-kaon experiment based on the substantially upgraded E949 apparatus relocated to the Fermilab Main Injector, was proposed~\cite{Comfort:2011zz}. ORKA aimed to collect as many as 1000 $K^+\to\pi^+\nu\bar\nu$ events in a five-year run, but development was terminated by the US DOE's P5 Panel in May 2014.

The CKM experiment, a decay-in-flight experiment to measure  ${\rm BR}(K^+\to\pi^+\nu\bar\nu)$ at Fermilab with a 22-GeV, RF-separated beam, was proposed in 2001~\cite{Frank:2001aa}. CKM, with its 100-m length and 25-m-long fiducial volume, Cherenkov particle identification for both beam and secondary particles, straw-chamber tracking in vacuum, and hermetic photon vetoes, was in many ways a precursor to the NA62 experiment at CERN. Unfortunately, the P5 Panel could not identify funding for the experiment in 2003, and an attempt to reformulate the proposal to reduce costs, largely by abandoning the idea of the RF-separated beam~\cite{Cooper:2005zz}, was not successful.

\begin{figure}[tb]
\centering
\includegraphics[width=\textwidth]{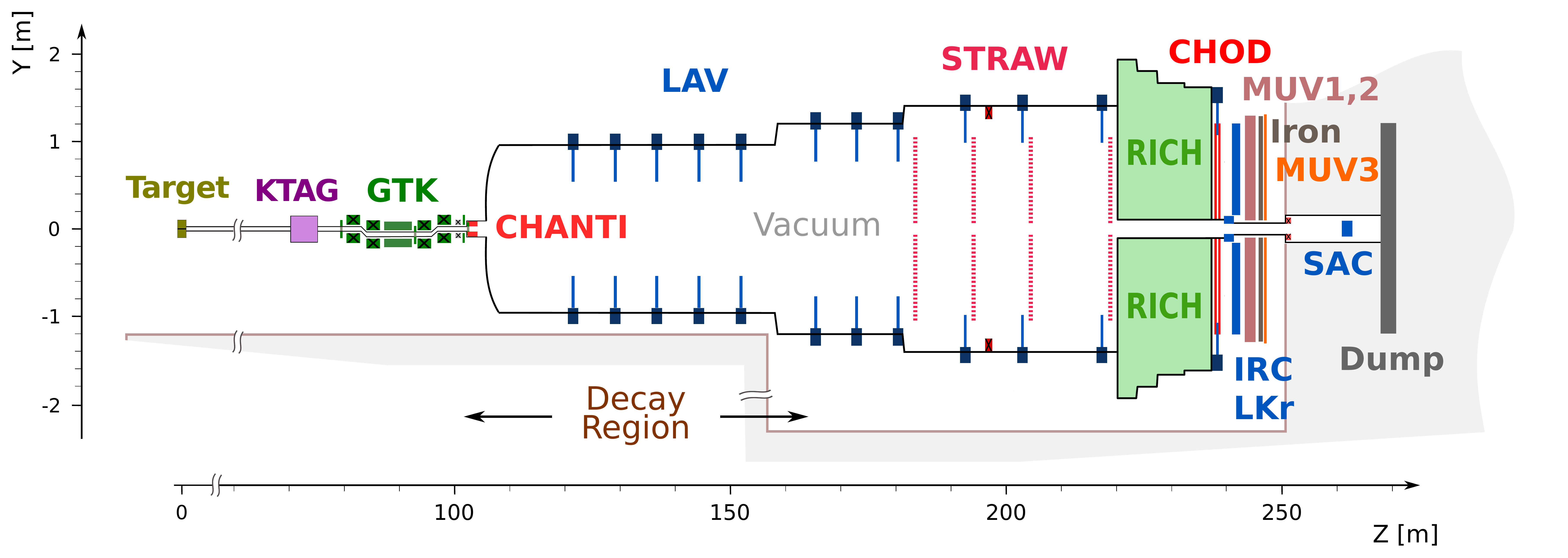}
\caption{Schematic diagram (side view) of the NA62 setup.}
\label{fig:NA62}
\vspace{-3mm}
\end{figure}

Following on the successes of NA31 and NA48, the NA62 experiment at the CERN SPS was proposed~\cite{Anelli:2005xxx} in 2005 to measure ${\rm BR}(K^+\to\pi^+\nu\bar\nu)$ with a precision of about 10\%. Like its predecessors, NA62~\cite{NA62:2010xx,NA62:2017rwk} is installed in the ECN3 experimental hall in the SPS North Area. The experimental setup is schematically illustrated in \Fig{fig:NA62}. $K^+\to\pi^+\nu\bar\nu$ decays are observed in flight. The experimental signature is an upstream $K^+$ coming into the experiment and decaying to a $\pi^+$, with no other particles present. Precise reconstruction of the primary and secondary tracks allows the abundant two-body decays to be rejected on the basis of the missing mass. The remainder of the experiment's rejection power (in particular, for decays without closed kinematics) comes from redundant particle identification systems and highly-efficient, hermetic photon veto detectors. 

A 400-GeV primary proton beam from the SPS at a nominal intensity of
$3.3\times10^{12}$ protons per pulse (ppp) is collided on a beryllium target to produce the 75-GeV unseparated positive secondary beam, which has a total
rate of 750~MHz and consists of about 6\% $K^+$. The beamline opens into the 
vacuum tank about 100~m downstream of the target. The vacuum tank is 
about 110~m long;
the fiducial volume (FV) 
occupies the first 60~m. About 10\% of the $K^+$s entering the experiment
decay in the FV, for a rate of 4.5~MHz.

Both the beam particle and the decay secondary are identified and tracked. 
A differential Cherenkov counter (KTAG) identifies the kaons in the beam; the beam spectrometer (Gigatracker, or GTK), consisting of a series of hybrid silicon pixel tracking detectors installed in an achromat in the beam line, measures the track parameters. The magnetic spectrometer for the secondary particles consists of four straw chambers operated inside the vacuum tank. $\pi/\mu$ separation for secondary particles is provided by the three downstream hadronic calorimeters/muon vetoes (MUV) and an 18-m-long ring-imaging Cherenkov counter (RICH) at the
end of the vacuum tank.

Rejection of photons from $\pi^0\to\gamma\gamma$ decays is important for the elimination of 
many background channels, in particular $K^+\to\pi^+\pi^0$. 
For these decays, requiring the secondary $\pi^+$ to have $p < 35$~GeV guarantees that the two photons from the $\pi^0$ have a total energy of at least 40~GeV. The missing-mass cut provides a rejection power of $10^4$, so the probability for the photon vetoes to miss both photons must be less than $10^{-8}$. Ring-shaped large-angle photon vetoes (LAVs) are placed at 12~stations along the vacuum volume. Downstream of the vacuum volume, the NA48 liquid-krypton calorimeter (LKr) vetoes 
forward, high-energy photons.
Smaller calorimeters around and at the downstream end of the beam pipe
complete the coverage for photons at very small angles.

After commissioning periods in 2014 and 2015, NA62 took physics data in
2016~\cite{NA62:2018ctf}, 
2017~\cite{NA62:2020fhy}, and 
2018~\cite{NA62:2021zjw}, collecting $2.2\times10^{18}$ protons on target (POT) corresponding to $6\times10^{12}$ $K^+$ decays in the fiducial decay volume, as summarized in \Tab{tab:na62_run2}.

\begin{table}
\centering
\begin{tabular}{lccc}\hline\hline
& 2016 & 2017 & 2018 \\ \hline
Days of running & 45 & 160 & 217 \\
Typical intensity, \% of nominal & 40 & 55 & 65 \\
Kaon decays in FV & $1.2\times10^{11}$ & $1.5\times10^{12}$ & $2.7\times10^{12}$ \\
Sensitivity ($10^{-10}$) & $3.15\pm0.24$ & $0.389\pm0.024$ & $0.111\pm0.007$ \\
Expected signal (SM) & $0.267\pm0.037$ & $2.16\pm0.28$ & $7.6\pm1.0$ \\
Expected background & $0.15^{+0.09}_{-0.04}$ & $1.46\pm0.30$ & $5.4^{+1.0}_{-0.8}$ \\
Candidates observed & 1 & 2 & 17 \\ \hline\hline
\end{tabular}
\caption{Summary of NA62 data taking in 2016~\cite{NA62:2018ctf}, 
2017~\cite{NA62:2020fhy}, and 
2018~\cite{NA62:2021zjw}.}
\label{tab:na62_run2}
\end{table}

The trigger for the signal-event (PNN) sample was designed to recognize the 1-track + missing energy topology: a hardware trigger at level 0 used the signal from the RICH as a time reference with additional conditions on multiplicities in the veto hodoscopes, the number of clusters and energy in the LKr and the number of hits in MUV3, while a software trigger at level 1 performed fast online event reconstruction to select events with a single, positive track, a $K^+$ signal from the 
KTAG, and conditions on the LAVs. The normalization and control sample was collected with a simple trigger with loose multiplicity cuts in the trigger hodoscopes. The principal steps for offline event selection were the reconstruction of the vertex between beam and secondary tracks, kinematic selection,
pion identification and muon rejection, and vetoes on any 
additional activity in the LAVs, LKr, small-angle vetoes, trigger hodoscopes, and tracking detectors. 

\begin{figure}[htbp]
\centering
\includegraphics[width=0.45\textwidth]{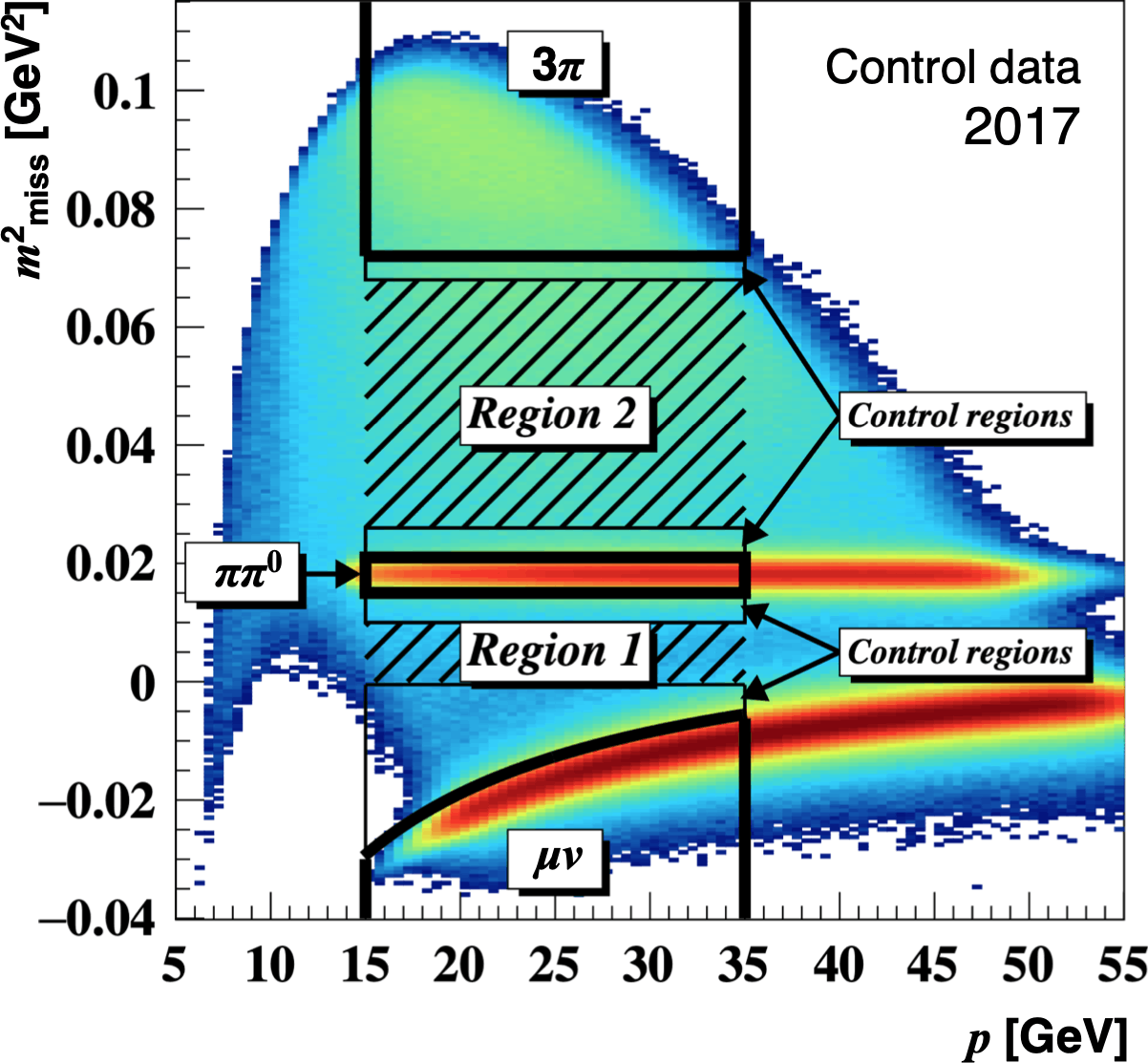}
\caption{Signal, background, and control regions in the $p_{\pi^+}$ vs.\ $m^2_{\rm miss}$ plane used for the NA62 analysis of $K^+\to\pi^+\nu\bar{\nu}$ with 2016--2017 data (for 2018, region 2 was extended to $p_{\pi^+} = 45$~GeV).}
\label{fig:na62_reg}
\end{figure}

The single-track topology was reconstructed starting from the beam track in the GTK in time with a signal from the KTAG. Good events were required to have a single positive track in the straw chambers and spectrometer. Upstream-downstream track matching was based on timing criteria ($\sigma_t \sim 100$~ps) and geometrical compatibility (distance of closest approach between upstream and downstream tracks less than 4 mm, with the decay vertex in the fiducial volume, $110 < z < 165$~m).
The missing mass in the $\pi^+$ track-identification hypothesis was calculated with a resolution $\sigma(m^2_{\rm miss})$ of about 0.001~GeV$^2$, allowing the definition of blinded signal and control regions as illustrated in \Fig{fig:na62_reg}:
\begin{eqnarray*}
{\rm Region~1:} & & 0 < m^2_{\rm miss} < 0.10~{\rm GeV}^2,\:\: 15 < p_{\pi^+} < 35~{\rm GeV};\\
{\rm Region~2:} & & 0.26 < m^2_{\rm miss} < 0.68~{\rm GeV}^2,\:\: 15 < p_{\pi^+} < 35~{\rm GeV}\:\: (45~{\rm GeV~for~2018}).
\end{eqnarray*}

The RICH was used to identify pions with $>75\%$ efficiency and to suppress muons at the level of 99.5\% over the fiducial momentum interval. Information on the energy sharing and cluster shape in the LKr and the hadron calorimeters MUV1--2, in combination with the veto on activity in MUV3, provided several orders of magnitude of additional muon rejection ($6\times10^{-6}$ at $p_{\pi^+} = 15~{\rm GeV}$ and improving at higher momentum), with $\sim$80\% efficiency for pion tracks.
The photon veto detectors (LAVs, LKR, IRC, and SAC), together with conditions on extra activity in the straws and trigger hodoscopes, were used to reject $K^+\to\pi^+\pi^0(\gamma)$ events with $p_{\pi^+}$ in the fiducial range down to the level of $2.7\times10^{-8}$, as verified by tag-and-probe measurements of the single-photon detection efficiencies folded with 
$K^+\to\pi^+\pi^0(\gamma)$ kinematics by Monte Carlo simulation
\cite{NA62:2020pwi}. The veto efficiency for events with extra activity allowed a separate limit to be placed on the invisible branching ratio of the $\pi^0$: $<4.4\times 10^{-9}$ at 90\% CL.

The remaining backgrounds from $K^+\to\mu^+\nu$, $K^+\to\pi^+\pi^0$, and $K^+\to\pi^+\pi^+\pi^-$ decays were estimated from the fractions of the missing-mass distributions in the tails extending into the signal regions measured with control samples. For example, the control sample of $K^+\to\mu^+\nu$ decays was obtained by applying the $\pi\nu\bar{\nu}$ selection criteria to the data collected with the control trigger, but with the $\pi^+/\mu^+$ separation criteria inverted. The control sample of $K^+\to\pi^+\pi^0$ was obtained by applying the single-track and $\pi^+$ identification criteria to the control sample, but without applying the photon veto criteria, instead reconstructing the $\pi^0$ in the LKr and using the missing mass calculated from the $\pi^0$ momentum and the nominal $K^+$ momentum for event selection. The $K^+\to\pi^+\pi^0$ control sample does not include $K^+\to\pi^+\pi^0\gamma$ events with the energy of the radiative photon high enough to shift the value of the missing mass into signal region 2. Such events are much more likely to be rejected than non-radiative events due to the presence of the additional photon, and the very small corresponding background contribution is estimated with Monte Carlo studies folding the single-photon detection efficiency with the $K^+\to\pi^+\pi^0\gamma$ kinematics.   

\begin{table}
    \centering
    \begin{tabular}{lcc}\hline\hline
    Process & Expected events & Expected events \\
    & 2017 data & 2018 data \\ \hline
    $K^+\to\pi^+\nu\bar{\nu}$ & $2.16\pm0.13\pm0.26_{\rm ext}$ & $7.58\pm0.40\pm0.90_{\rm ext}$\\
    $K^+\to\pi^+\pi^0(\gamma)$ & $0.29\pm0.04$ & $0.75\pm0.05$ \\
    $K^+\to\mu^+\nu(\gamma)$ & $0.15\pm0.04$ & $0.64\pm0.08$ \\
    $K^+\to\pi^+\pi^- e^+\nu$ & $0.12\pm0.08$ & $0.51\pm0.10$ \\
    $K^+\to\pi^+\pi^+\pi^-$ & $0.008\pm0.008$ & $0.22\pm0.08$ \\
    $K^+\to\pi^+\gamma\gamma$ & $0.005\pm0.005$ & $<0.01$ \\
    $K^+\to\pi^0\ell^+\nu$ & $<0.001$ & $<0.001$ \\
    Upstream background & $0.89\pm0.31$ & $3.30^{+0.98}_{-0.73}$ \\
    Total background & $1.46\pm0.33$ & $5.42^{+0.99}_{-0.75}$ \\ \hline \hline 
    \end{tabular}
    \caption{Expected numbers of signal and background events in NA62 data in the 2017~\cite{NA62:2020fhy} 
    and 2018~\cite{NA62:2021zjw} datasets.}
    \label{tab:na62_bkg}
\end{table}
\begin{figure}[htbp]
\centering
\includegraphics[width=0.7\textwidth]{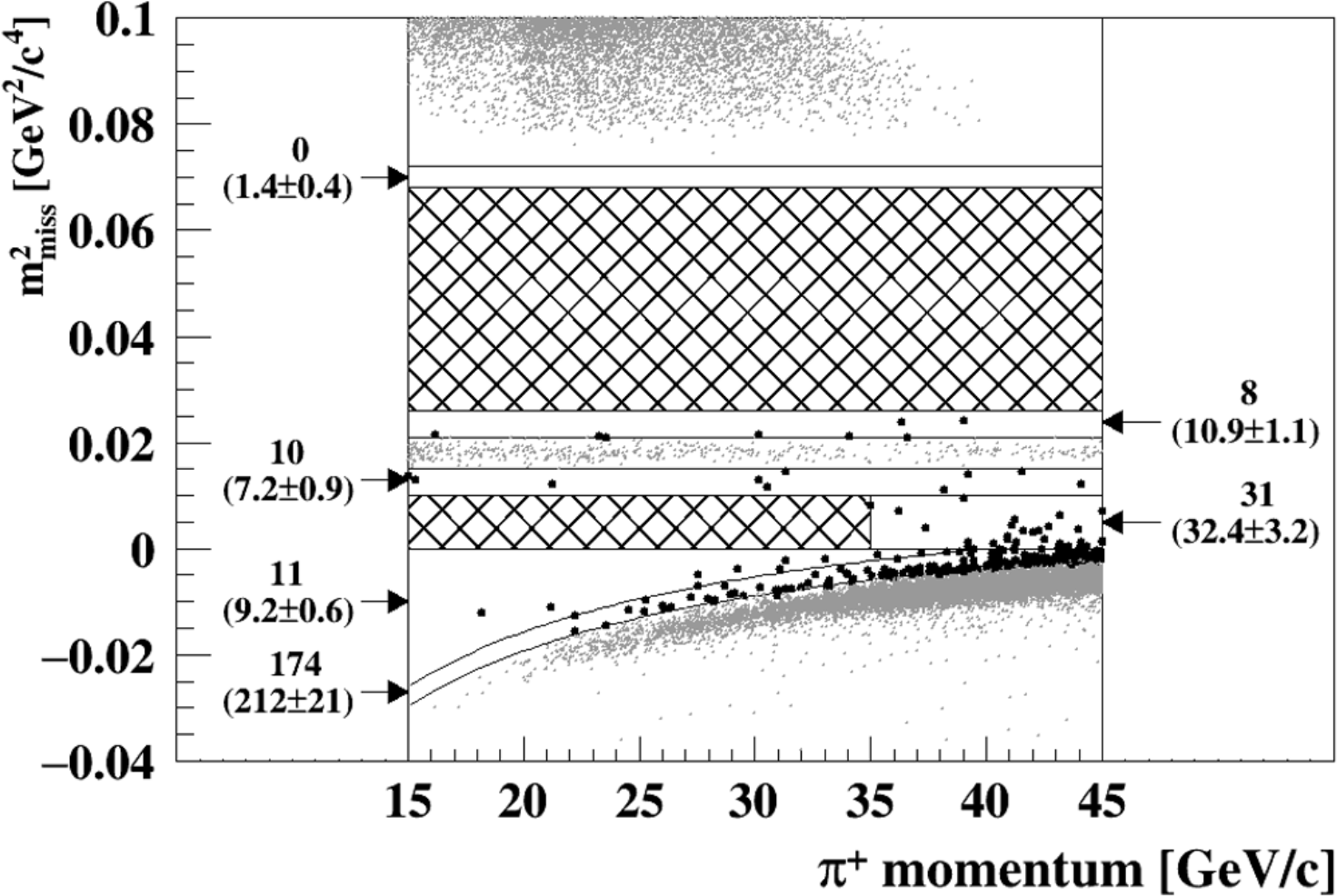}
\caption{Distribution in the $(p_{\pi^+},m^2_{\rm miss})$ plane of events in 2018 data passing all selection cuts, before unblinding of the signal regions. The observed and expected numbers of background events are indicated for each control region~\cite{NA62:2021zjw}.}
\label{fig:na62_blind}
\end{figure}
\begin{figure}[htbp]
\centering
\includegraphics[width=0.6\textwidth]{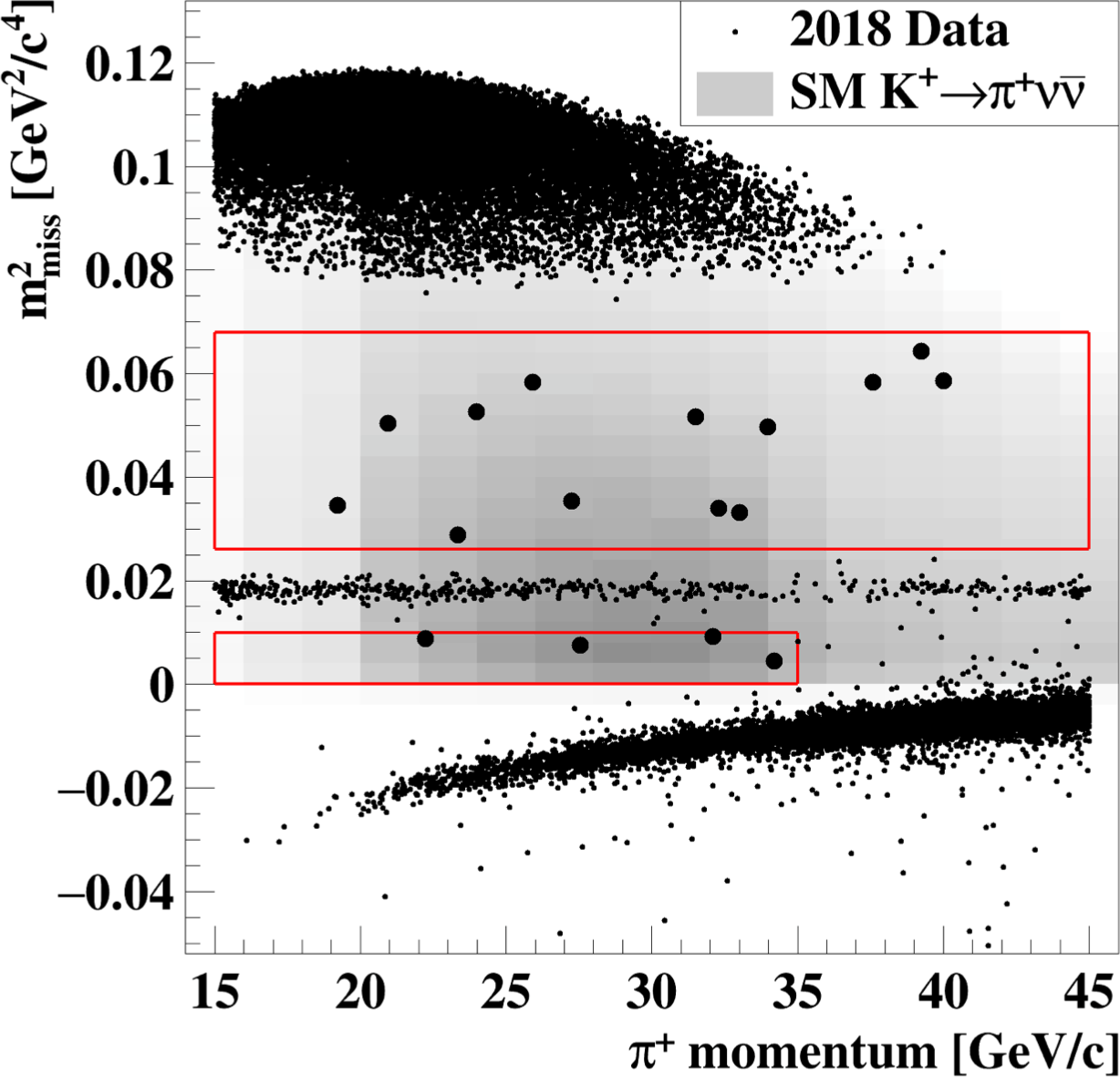}
\caption{Distribution in the $(p_{\pi^+},m^2_{\rm miss})$ plane of events in 2018 data passing all selection cuts, showing 17 events in the signal regions after unblinding~\cite{NA62:2021zjw}.}
\label{fig:na62_unblind}
\end{figure}
Accidental matching of a beam track and a $\pi^+$ track in the spectrometer can occur when a $K^+$ decays or interacts upstream of the last GTK station. For this reason, events are discarded in which the $\pi^+$ track extrapolates upstream to the exit from the final collimator (the ``box cut''). However, some of these ``upstream background'' events may not be eliminated. Consider the following scenario:
\begin{itemize}
    \item A $K^+$ identified as a kaon in the KTAG decays to $\pi^+\pi^0$ inside the GTK, and the photons from the $\pi^0$ are not observed. The $\pi^+$ from the decay makes it through or around the final collimator and is seen in the spectrometer, but a beam track is not reconstructed or is reconstructed poorly in the GTK.
    \item A $\pi^+$ in accidental coincidence with the $K^+$ leaves a track in the GTK but does not decay and does not emerge from the beam pipe, and so is not observed in the spectrometer.
    \item The $\pi^+$ from the $K^+$ decay undergoes multiple scattering in the first tracking station, so that its track is reconstructed after the scattering. Offline, it is extrapolated upstream and matched to the 
    track from the accidental $\pi^+$, forming a vertex in the fiducial volume. Because of the scattering, it does not extrapolate to the exit of the final collimator and is not rejected. 
\end{itemize}
This background was quantified by developing a Monte Carlo model for the upstream background and validating it against control data samples on the basis of distributions of such quantities as the distance of closest approach between beam and spectrometer tracks. Then, the final estimates of the upstream background were validated by comparison with subsamples defined by different inversions of background suppression cuts, such as the box cut, the missing-mass cuts, a veto on extra GTK activity, and a veto on the activity in the CHANTI (a charged-particle veto around the beamline located just downstream of the GTK). 
Through the start of the 2018 run, a large region around the exit of the final collimator was used for upstream background rejection, keeping the background level acceptable, but incurring a significant loss in signal acceptance. A new final collimator was installed just after the start of 2018 run, allowing the use of a much smaller rejection region. Table~\ref{tab:na62_bkg} compares the number of expected events for the signal and most important background channels in 2017 and 2018, demonstrating the effect of the new collimator and other improvements to the analysis, such as the extension of the upper limit of the fiducial momentum region from 35~GeV to 45~GeV for signal selection region 2, as well as of moderate increases in running time and intensity.
The distribution in the $(p_{\pi^+},m^2_{\rm miss})$ plane of the events in 2018 data passing all selection cuts, before unblinding of the signal regions, is presented in \Fig{fig:na62_blind}. The observed and expected numbers of background events in each control region are indicated and demonstrate satisfactory agreement. The same distribution with the signal regions unblinded,
showing the 17 $K^+\to\pi^+\nu\bar{\nu}$ candidates in 2018 data, is presented in \Fig{fig:na62_unblind}.

In all, from the 20 candidate events observed in the 2016--2018 data (see \Tab{tab:na62_run2}), NA62 obtains
\begin{equation}
{\rm BR}(K^+\to\pi^+\nu\bar\nu) = \left(10.6\,^{+4.0}_{-3.4}\left|_{\rm stat}\pm 0.9_{\rm syst}\right.\right)\times 10^{-11}.
\end{equation}
This result, the first measurement of ${\rm BR}(K^+\to\pi^+\nu\bar{\nu})$,
is in agreement with the SM prediction, and provides evidence for the existence of the decay with $3.4\sigma$ significance.
In data-taking during LHC Run III, which started in July 2021 and will continue until the start of Long Shutdown III, NA62 will reduce the uncertainty on this result, aiming to collect enough events to reach order 10\% precision%
~\cite{NA62:2019xxx,NA62:2021xxx}, 
as discussed in \Sec{sec:na62_ls3}.


\subsubsection{Experimental status for $K_L\to\pi^0\nu\bar{\nu}$}
\label{sec:klpnn_status}

\begin{figure}[htbp]
\centering
\includegraphics[width=0.8\textwidth]{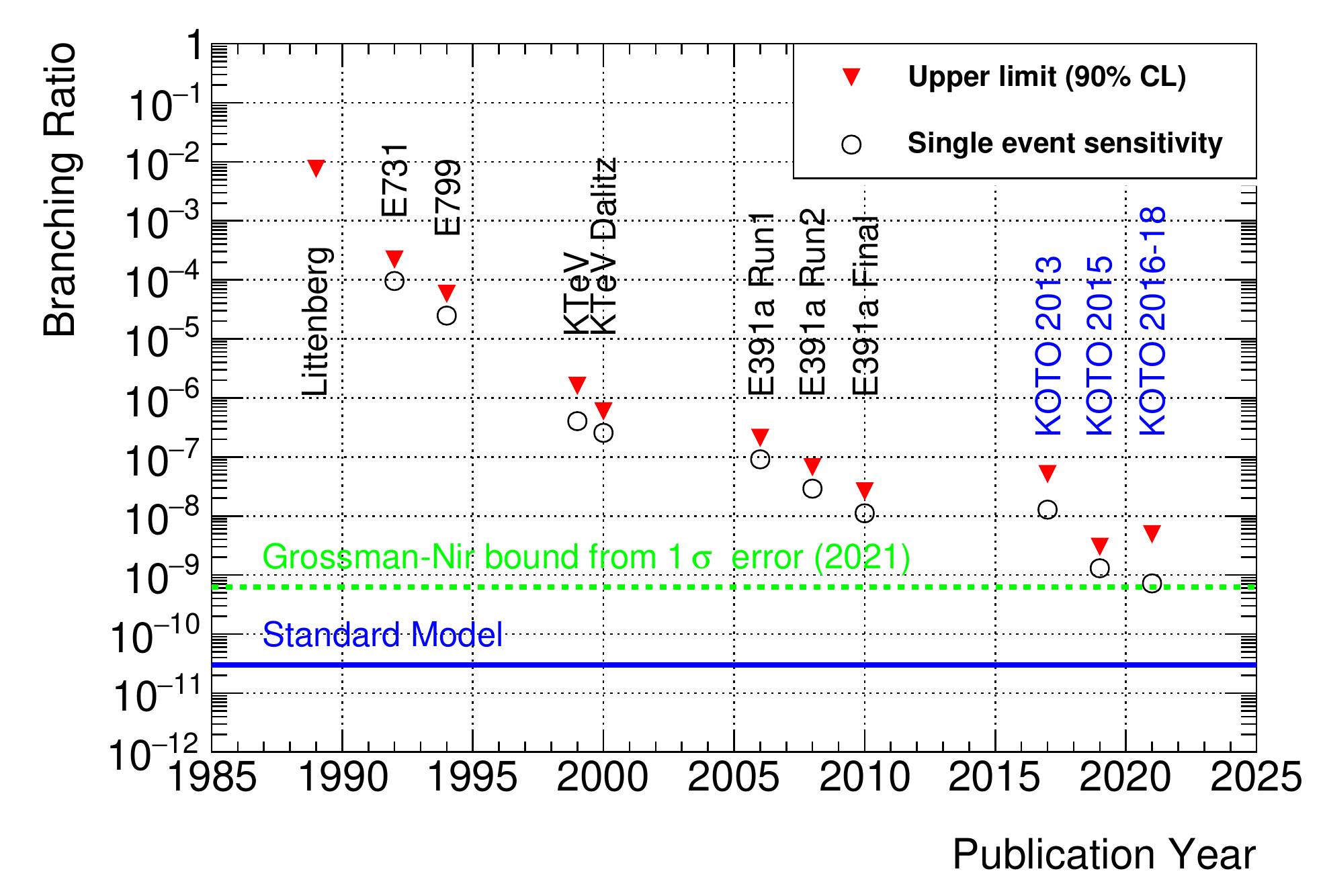}
\caption{History of $K_L\to\pi^0\nu\bar\nu$ searches~\cite{Littenberg:1989ix,Graham:1992pk,E779:1994amx,
KTeV:1998taf,E799-IIKTeV:1999iym,E391a:2006fxm, Ahn:2007cd,E391a:2009jdb,KOTO:2016vwr,Ahn:2018mvc,KOTO:2020prk}.}
\label{fig:evolutionKL}
\end{figure}

The history of searches for the $\klpionn$ decay is shown in Fig.~\ref{fig:evolutionKL}
with the upper limits and the single event sensitivities.
The decay was firstly studied by interpreting $\klpiopio$ data~\cite{Littenberg:1989ix}.
The E731, the E799, the KTeV experiments at Fermilab and the KEK E391a experiment updated the upper limit of the branching ratio~\cite{Graham:1992pk,E779:1994amx,KTeV:1998taf, E799-IIKTeV:1999iym,E391a:2006fxm, Ahn:2007cd, E391a:2009jdb}. Currently, the KOTO experiment, a successor of the KEK E391a experiment, is the only running experiment. KOTO has been improving the sensitivity~\cite{KOTO:2016vwr,Ahn:2018mvc,KOTO:2020prk}.

As was the case for the $K^+$ decay, there were complementary proposals
for measurements of the neutral decay at Brookhaven and at Fermilab during
the late 1990s and early 2000s. 
As in the case for the charged decay, 
neither experiment was ultimately realized, but substantial R\&D work was carried out, of considerable
interest for the design of new experiments.

The KOPIO experiment at Brookhaven \cite{Comfort:2015xx} was designed to exploit the capability of the AGS to provide microbunched slow-extracted proton beam to create a low-momentum
neutral beam with a time structure allowing the use 
of time-of-flight constraints for signal identification. During the extraction flat top, the primary beam was to be delivered in 200~ps pulses every 40~ns, with an extinction of $10^{-3}$ between pulses. The neutral secondary beam, of peak momentum 0.7~GeV, was tightly collimated in the vertical plane only, to maximize the beam solid angle (and therefore flux) while preserving useful geometrical constraints. A calorimeter at the downstream end of the decay
volume was used to precisely measure the arrival times and energies of
photons from candidate signal decays; a preradiator in front of the
calorimeter was used to obtain the angles of incidence of these photons
as well. This information, together with the a priori knowlegde of the
time structure of the beam, allowed for event-by-event reconstruction of
the $K_L$ momentum by time-of-flight, providing redundant constraints for
signal identification and background rejection. The highly innovative
KOPIO proposal was supported by a decade of planning and R\&D;
ultimately, however, funding for rare decay physics at Brookhaven was
terminated in 2005.

The proposed high-energy experiment at Fermilab, KAMI~\cite{KAMI:2001cyd}, took
a more conventional approach. The 120-GeV primary proton beam from the
Main Injector was to be used to obtain a tightly collimated neutral
secondary beam at an angle of 12--20 mrad, with a mean momentum of
about 20 GeV. The total length of the beamline and detector was about
200~m, including an evacuated volume hermetically enclosed by photon vetoes
of about 90~m. The fiducial volume spanned approximately the first 65~m of
the vacuum tank. A unique feature of KAMI was the presence of a
charged-particle spectrometer consisting of five stations of scintillating
fiber planes straddling a large-gap dipole magnet, making possible a
complete physics program in rare $K_L$ decays, as well as providing
additional constraints in the reconstruction of control samples of
abundant $K_L$ decays to keep measurement systematics under control.
The KAMI proposal was rejected by the Fermilab PAC in 2001 because
of concerns that the target sensitivity (100 events) might not be
reached due to insufficient redundancy to guarantee rejection of
background from unanticipated sources.  

The E391a experiment at KEK 12-GeV Proton Synchrotron was the first dedicated experiment to search for the $K_L\to\pi^0\nu\overline{\nu}$ decay.
With the data taken in 2005, 
the single event sensitivity of $1.1\times 10^{-8}$ was achieved, and the upper limit of the branching ratio was set to be  $2.6\times 10^{-8}$ at 90\% CL~\cite{E391a:2009jdb}.
A narrow $K_L$ beam (``pencil beam'') was prepared with 10-m-long beam line.
At the downstream of the beam line, the E391a detector with a decay volume was prepared. Most of the sub-detectors and the decay volume were in the 4-m-diameter 9-m-long evacuated vacuum tank.
The observed particles are required to be only two photons from a $\pi^0$ to identify the signal.
Two photons from a $\pi^0$ were detected with an endcap calorimeter downstream of the decay volume.
A hermetic veto system surrounding the decay volume ensured no other particles detected.
The $\pi^0$ decay vertex was reconstructed assuming the two photons from a $\pi^0$ decay.
The decay vertex could be assumed on the beam axis owing to the "pencil beam".
The four vectors of the two photons and the $\pi^0$ were fully reconstructed.
The transverse momentum ($p_T$) of the $\pi^0$ can be larger due to missing two neutrinos.
Therefore, the $p_T$ can be used to identify the signal.
This ``pencil beam'' method was demonstrated in the E391a experiment, 
which gave a basis of the next high sensitivity experiments.

\begin{figure}[htbp]
\centering
\includegraphics[width=\textwidth]{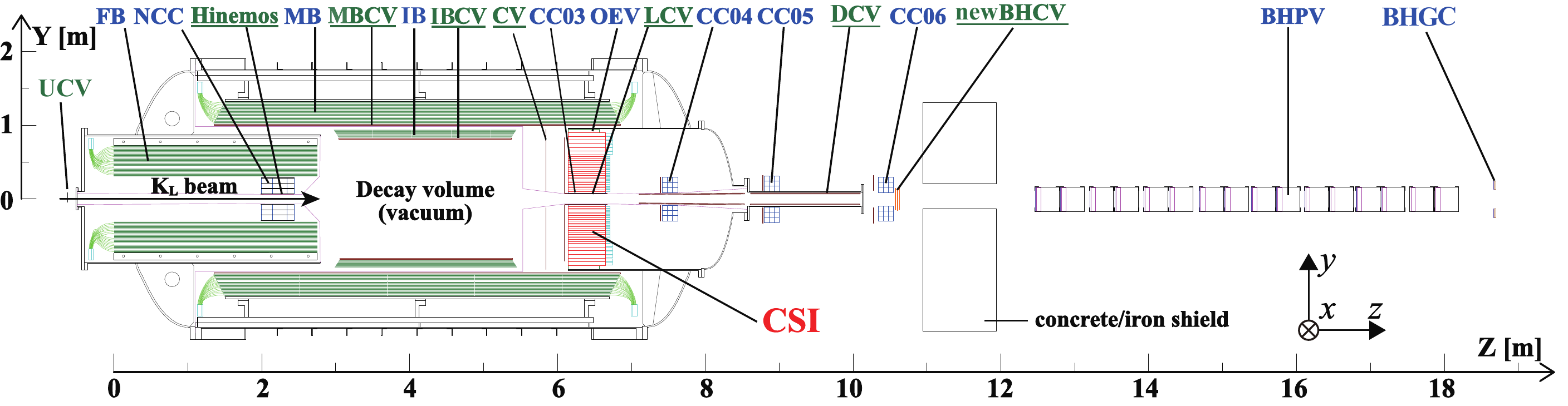}
\caption{Schematic diagram of the KOTO detector.}
\label{fig:KOTO}
\end{figure}
The KOTO experiment at J-PARC was designed with the ``pencil beam'' method.
The salient features of the experiment are the use of a highly collimated,
neutral beam and very high performance
hermetic calorimetry and photon vetoing.  
The 30-GeV primary proton beam is interacted on a gold target,
and secondary particles are generated.
Neutral beam is prepared from the secondaries with 20-m-long beam line.
Two collimators in the beam line shape the beam in a solid angle of 7.8~$\mu$sr at an
extraction angle of $16^\circ$, which provides $K_L$ beam with 
a peak momentum of 1.4~$\mathrm{GeV}/c$ in a $8\times 8~\mathrm{cm}^2$
cross-section at the beam line exit. 
Contamination of short-lived and charged particles in the beam
is suppressed with the long beam line and a sweeping magnet, while the photon flux is reduced using a 7-cm long lead absorber. 
At the downstream of the beam line, 
the KOTO detector with the decay volume is located (Fig.~\ref{fig:KOTO}).
Most of the sub-detectors and the decay volume are in the 4-m diameter 9-m long evaculated chamber.
Two photons from a $\pi^0$ are detected with a calotimeter (CsI) composed of
undoped CsI crystals. The $\pi^0$ vertex and the $p_T$ are reconstructed with the ``pencile beam'' method.
The reconstructed vertex position and $p_T$ of the $\pi^0$ are used to define
the signal region. 
The $\klgg$ decay is suppressed by requiring higher $p_T$ of the
reconstructed $\pi^0$. Other backgrounds from $K_L$ decays involving charged particles or additional photons
are reduced using the hermetic veto system surrounding the decay volume.

The J-PARC has achieved 64 kW beam power in the 30-GeV-proton slow extraction, which corresponds to $7\times 10^{13}$ protons per 5.2-s cycle. Shorter cycle of 4.2 s and more protons in a spill are expected after an on-going accelerator upgrade in 2021-2022,
and 100-kW beam power is expected to be achieved.

The KOTO experiment currently set an upper limit of the branching ratio at $3\times 10^{-9}$ (90\% CL) with no candidate events observed from the analysis of data collected in 2015~\cite{Ahn:2018mvc}. 
From the analysis of data collected in 2016-2018, KOTO observed three candidate events
at the single event sensitivity of $7.2\times 10^{-10}$, which is consistent to the expected number of background 1.2~\cite{KOTO:2020prk}.
\begin{figure}[htbp]
\centering
\includegraphics[width=0.6\textwidth]{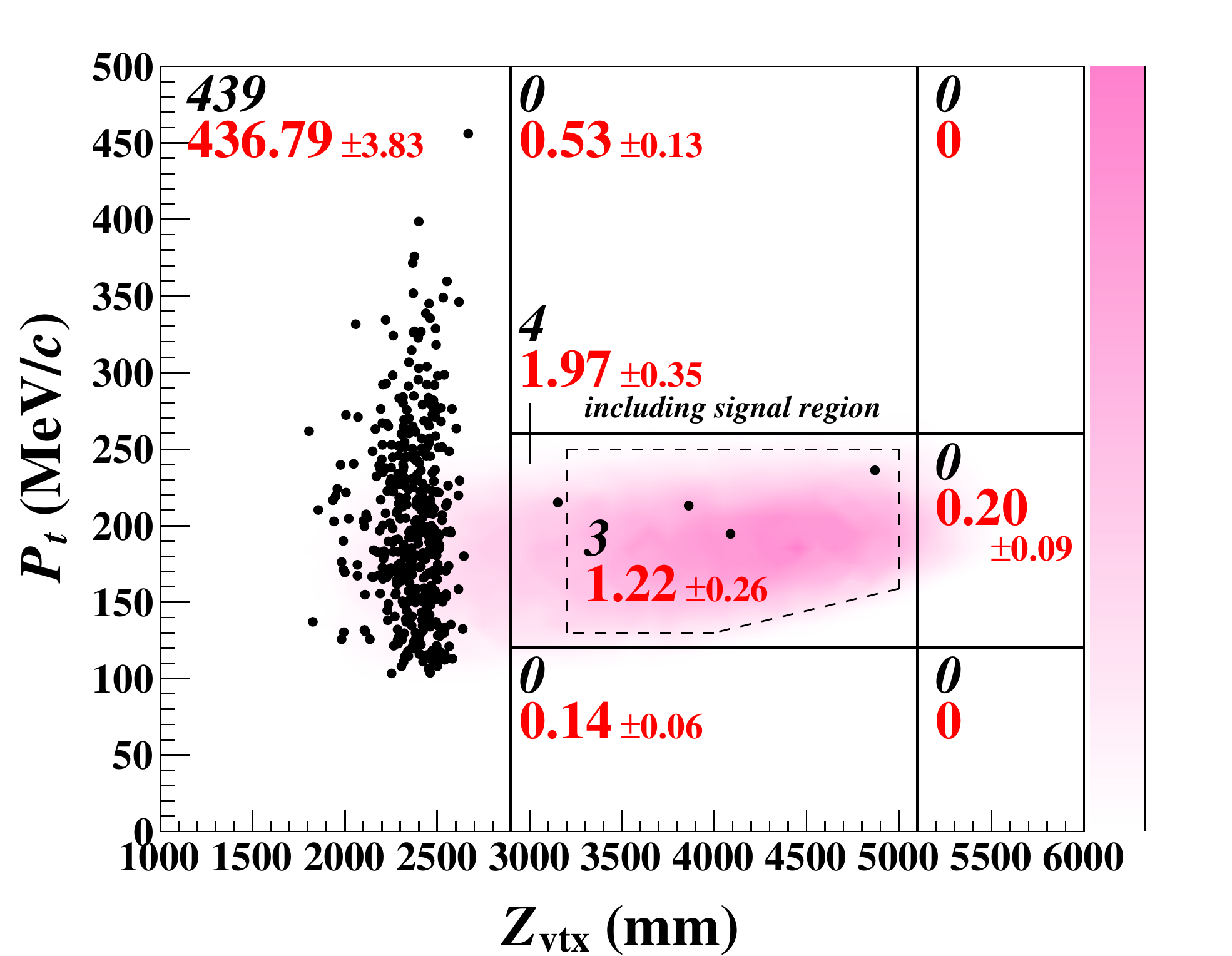}
\caption{Distribution of events in reconstructed $p_T$ and $z$ vertex plain after imposing the event selection~\cite{KOTO:2020prk}. The region surrounded with dotted lines is the signal region. Black dots corresponds to the observed events. The shaded contour shows the distribution of $\klpionn$ MC. The number of observed (expected) events in each region is shown with black (red) number.}
\label{fig:KOTOPtZ}
\end{figure}
\begin{table}
        \caption{Summary of the numbers of background events with a central value estimate~\cite{KOTO:2020prk}}
        \label{tab:KOTO2018_BGSummary}
        \centering
        \begin{threeparttable}[h]
        \begin{tabular}{llc}
                \hline \hline
                source & & Number of events\\
                \hline
                $K_L$                   & $\klpiopiopio$                                        & 0.01 $\pm$ 0.01 \\
                                                & $\klgg$       (beam halo)                             & 0.26 $\pm$ 0.07 \tnote{a}\\
                                                & Other $K_L$ decays                            & 0.005 $\pm$ 0.005 \\
                $K^{\pm}$               &                                                               & 0.87 $\pm$ 0.25 \tnote{a}\\
                Neutron                 & Hadron cluster                                        & 0.017 $\pm$ 0.002\\
                                                & CV $\eta$                                             & 0.03 $\pm$ 0.01\\
                                                & Upstream $\pi^0$                                      & 0.03 $\pm$ 0.03\\
                \hline
                total                           &                                                               &1.22 $\pm$ 0.26\\
                \hline \hline
        \end{tabular}
        \end{threeparttable}
\end{table}
 The contribution of each background source is shown in Table~\ref{tab:KOTO2018_BGSummary}.  The $K^\pm$ background contributes dominantly.
In this background, KOTO found charged kaon contaminates the neutral beam slightly,
and the decay $K^\pm\to\pi^0 e^\pm \nu$ mainly fakes the signal with $e^\pm$ missed due to detector inefficiency.  This charged kaon background is reduced with a low-mass in-beam charged particle counter at the entrance of the detector.  A (prototype) counter composed of plastic scintillating fibers was already installed and used in (2020) 2021.
Currently it reduces the background by one order or more. In 2022, a new counter with 0.2-mm-thick scintilltor film will be installed with possible reduction factor of 100.
The second largest one is $K_L\to 2\gamma$ decay in the beam halo region, which gives fake $p_T$. Multivariate analysis using cluster energies, positions, and shapes can be used to reduce this background, because actual vertex position of $\pi^0$ is far from the beam axis. Currently the background can be reduced by a factor of 30 or more.
KOTO will reach the single event sensitivity of $O(10^{-11})$ 
in mid 2020s
with possible 10 times larger data owing to the accelerator upgrade.
To go beyond this sensitivity, 
a next-generation experiment is needed in order to actually
measure the branching ratio.


\subsubsection{$K_{L(S)}\to \mu^+\mu^-$}
The $K_{L(S)}\to \mu^+ \mu^-$ decays are suppressed in the SM, which predicts~\cite{Ecker:1991ru, Isidori:2003ts, DAmbrosio:2017klp, KLMuMu_theory}:
\begin{equation}
{\rm BR}(K_{L} \to \mu^+ \mu^-) = 
\begin{cases} 
&(6.85 \pm 0.80_{\textrm{LD}} \pm 0.06_{\textrm{SD}}) \times 10^{-9}\\
& (8.11 \pm 1.49_{\textrm{LD}} \pm 0.1_{\textrm{SD}}) \times 10^{-9}
\end{cases}
\end{equation}
and
\begin{equation}
{\rm BR}(K_{S} \to \mu^+ \mu^-) =(5.18\pm 1.50_{\textrm{LD}} \pm 0.02_{\textrm{SD}})\times 10^{-12},
\end{equation}
where SD stands for Short-Distance, LD for Long-Distance, and the two ${\rm BR}(K_L\to\mu^+ \mu^-)$ possibilities correspond to the two values of the unknown sign of the LD amplitude $A_L^{\gamma\gamma}$.
These decays are very sensitive to dynamics beyond the standard model, such as scalar or vector Leptoquarks~\cite{Dorsner:2011ai,Mandal:2019gff,Bobeth:2017ecx} or the Minimal Supersymmetric Standard Model (MSSM)~\cite{Chobanova:2017rkj}. 
The current experimental measurement~\cite{E871:2000wvm,Akagi:1994bb,E791:1994xxb,ParticleDataGroup:2020ssz}, 
\begin{equation}
\textrm{BR}_{\rm exp}(K_L\to\mu^+\mu^-) = \left(6.84 \pm  0.11 \right) \times 10^{-9},
\end{equation}
agrees with one of the two predictions of the SM within the theoretical uncertainties, leaving no room for a BSM discovery in this decay alone unless the theory significantly improves. On the other hand, the current experimental upper bound set by LHCb~\cite{LHCb:2020ycd},
\begin{equation}
{\rm BR} (K_S\to\mu^+\mu^-) < 2.1\times 10^{-10},
\end{equation}
\noindent leaves still an almost two orders-of-magnitude uncharted territory.
In addition, a positive signal could allow a measurement of the $K_{S}/K_{L}$ interference term by means of a tagged analysis~\cite{DAmbrosio:2017klp}. This can be done in LHCb by tagging the $K^0$ flavor using accompanying particles in reactions like $pp\rightarrow K^0K^-X$ or  $pp\rightarrow K^0\Lambda^0X$, where the experiment would measure an effective yield:
\begin{eqnarray}
\textrm{BR}(K_S \to \mu^+ \mu^-)_{\rm eff}=\tau_S \left( \int^{t_{{\rm max}}}_{t_{{\rm min}}} d t e^{- \Gamma_S t} \varepsilon (t)\right)^{-1}
 \Biggl[ \int^{t_{{\rm max}}}_{t_{{\rm min}}} d t \Biggl\{  \Gamma (K_{S} \to \mu^+ \mu^- )  e^{- \Gamma_S t}  \nonumber \\ 
 ~~+   \frac{ D f_K^2 M_K^3 \beta_{\mu}}{ {8} \pi} \textrm{Re}\left[  i \left( A_S A_L - \beta_{\mu}^2 {B_S^{\ast}} B_L \right) e^{ - i \Delta M_K t}  \right] e^{- \frac{ \Gamma_S + \Gamma_L}{2} t } \Biggr\} \varepsilon (t) \Biggr], 
\label{eq:effBR}
\end{eqnarray}
\noindent as a function of the dilution factor:
\begin{align}
 D= \frac{ K^0 - \bar{K^0} }  { K^0 + \bar{K^0} }.
 \label{eq:DE}
\end{align}

At the LHC the $K^0$ and $\bar{K^0}$ are produced approximately in equal amounts due to the high energy, but the dilution factor can be controlled experimentally if flavor tagging is provided by channels like the ones mentioned above. A measurement of the interference also allows determining the scalar amplitude of the decay,
which is theoretically extremely clean; moreover it is suggested that the interference 
can also be measured using $K_S$ regeneration in high intensity $K_L$ beams~\cite{Dery:2021mct}.


\subsubsection{$K_{L(S)}\to \pi^0 \ell^+ \ell^-$}
\label{sec:k0-pi0-ell-ell}

The FCNC transitions $K \to\pi \ell^+\ell^-$ ($\ell = e, \mu$) are dominated by single virtual-photon exchange $K \to \pi \gamma^* \to \pi \ell^+ \ell^-$ if allowed by CP invariance as in $K_S$ and $K^\pm$ decays. However, this contribution is CP violating for $K_L \to\pi^0 \ell^+\ell^-$ and consequently this process has become a point of reference for studying the CP-violating sector of the SM~\cite{Cirigliano:2011ny}.

Like $K_L \to\pi^0\nu\bar{\nu}$, the short-distance contribution to the $K_L \to\pi^0\ell^+\ell^-$ decays depends on $|V_{ts}^*V_{td}|$ and thus measures the height of the unitarity triangle shown in Fig. \ref{fig:ut}. To the extent that there is little constraint on the CP-violating phase of the $s\to d\ell^+\ell^-$ transition, measurement of the $K_L \to\pi^0\ell^+\ell^-$ BRs may reveal the effects of new physics \cite{Smith:2014mla}. However, unlike in the case of $K_L \to\pi^0\nu\bar{\nu}$, there are long-distance contributions to the decay amplitude that are of the same order of magnitude as the short-distance component. First, there is an indirect CP-violating contribution of the type $K_L\to K_S\to\pi^0\ell^+\ell^-$ the magnitude of which can be obtained from measurements of the $K_S$ decays. Second, for $K_L\to\pi^0\mu^+\mu^-$, there is a CP-conserving long-distance contribution mediated by $K_L\to\pi^0\gamma\gamma$ (this is helicity suppressed for the $e^+e^-$ mode). In the SM, ${\rm BR}(K_L\to\pi^0e^+e^-) = 3.2\left(^{+8}_{-9}\right)\times10^{-11}$ and ${\rm BR}(K_L\to\pi^0\mu^+\mu^-) = 1.29(24)\times10^{-11}$~\cite{Mertens:2011ts}. The best experimental limits are from KTeV: ${\rm BR}(K_L\to\pi^0e^+e^-) < 28 \times10^{-11}$~\cite{AlaviHarati:2003mr} and ${\rm BR}(K_L\to\pi^0\mu^+\mu^-) < 38 \times10^{-11}$~\cite{AlaviHarati:2000hs} at 90\% CL.

Branching ratios of the $K_S\to\pi^0\ell^+\ell^-$ decays were measured by NA48/1~\cite{Batley:2003mu,Batley:2004wg}:
\begin{equation}
{\rm BR}(K_S\to\pi^0e^+e^-) =
\left(3.0^{+1.5}_{-1.2}\right)\times 10^{-9}, \quad
{\rm BR}(K_S\to\pi^0\mu^+\mu^-) = \left(2.9^{+1.5}_{-1.2}\right)\times 10^{-9}.
\end{equation}
In the $e^+e^-$ case, there are complications from photon-conversion background from $K_S\to\pi^0\pi^0$, but the level of precision achieved is similar to that for $\mu^+\mu^-$. Significant progress on the measurements of these decays is required to be able to provide a more precise determination of the indirect CP-violating contribution to the $K_L \to\pi^0\ell^+\ell^-$ decay rates.

\subsection{Broader physics programme}
\label{sec:broader_programme}

The large datasets of ${\cal O}(10^{14})$ kaon decays to be collected by the future experiments aiming at $K\to\pi\nu\bar\nu$ and $K_L\to(\pi^0)\ell^+\ell^-$ measurements, as well as the state-of-the-art detectors involving precision tracking, redundant particle identification, high performance calorimetry and hermetic photon vetoes, lead to a broad physics program in addition to the ``golden modes'' discussed in Section~\ref{sec:golden}. A review of rare kaon decays is provided in Ref.~\cite{Cirigliano:2011ny}, while the relevant hidden-sector scenarios are reviewed in Ref.~\cite{Goudzovski:2022vbt}. Moreover, future $K^+$ decay experiments may serve as a proof of concept for a tagged neutrino facility~\cite{Perrin-Terrin:2021jtl}.


\subsubsection{Interplay of kaon physics with lepton flavour anomalies}
\label{sec:strange_lf}

Significant hints for violation of lepton universality have been seen at various experiments, usually referred to as `anomalies'. These include the $(g-2)_{\mu}$ anomaly~\cite{Muong-2:2021ojo}, the $B^+$ anomalies in $b\to c\ell\nu$ reported by LHCb, Belle, and BaBar, and the $B^0$ anomalies observed by $b\to s\ell\ell$ at LHCb. The latter gets further strengthened by small (but consistent) departures of $b\to s\mu\mu$ exclusive branching fractions with respect to the SM prediction, including differential branching fractions in $B^0\to K^*\mu\mu$ decays. The combined global significance for BSM dynamics in $b\to s\ell\ell$ transitions is at present at the 4.3--5.5$\sigma$ level~\cite{Hurth:2021nsi,Isidori:2021vtc}. A broad set of models are being proposed by the theory community to address those, most notably leptoquark models, but also other possibilities like $Z'$ models or RPV SUSY models. 
The strangeness decays are needed in this scenario in order to determine the dynamics behind those anomalies: the  $s\to d\ell\ell$ transition is the equivalent to $b\to s\ell\ell$ for the second generation of quarks, and, similarly, $s\to u\ell\nu$ is the equivalent of $b\to c\ell\nu$. Thus, strange decays allow to understand how the BSM dynamics couples to each quark generation or, in other words, its flavor structure. 
Some recent examples for the interplay between these anomalies and the strangeness decays can be found in ~\cite{Mandal:2019gff,Marzocca:2021miv,Heeck:2022znj}. Lepton flavor universality violation is related to lepton flavor violation, and from that perspective, bounds of lepton flavor violation in strange decays are also required to constrain possible models for the $B$ anomalies~\cite{Borsato:2018tcz}.


\subsubsection{Rare kaon decay measurements}
\label{sec:rare_kaon_decays}

\subsubsubsection{Lepton flavour universality (LFU) tests}
\label{sec:lfu_tests}
Leptonic kaon decays represent a very sensitive LFU probe. 
In the SM, these decays are mediated by a $W$ boson with universal coupling to leptons.
They are helicity suppressed, due to the $V-A$ structure of the charged current coupling, and the ratio of their amplitudes
\begin{equation}
R_{K} = \frac{\Gamma(K^+\to e^+\nu)}{\Gamma(K^+\to\mu^+\nu)} = \frac{m_e^2}{m_{\mu}^2}\cdot {\left(\frac{m_K^2-m_e^2}{m_K^2-m_{\mu}^2}\right)}^2\cdot \left(1+\delta^K_{\rm EM}\right)
\end{equation}
is very clean theoretically, as the strong interaction dynamics cancels out at the first order, and the hadronic structure dependence appears only through the electroweak corrections~$\delta^K_{\rm EM}$. In particular, the SM prediction for $R_{K}$ is currently known at the sub-permille level~\cite{Cirigliano:2007xi}:
\begin{equation}
R_K =  (2.477 \pm 0.001)\times10^{-5}.  \label{eq:rk_sm}
\end{equation}
Lepton universality can be violated in many extensions of the SM with non-trivial flavor structure.
Models with mass-dependent couplings, for instance, such as models with an extended Higgs sector~\cite{ PhysRevD.74.011701}, or leptoquarks~\cite{BUCHMULLER1987442, Davidson:1993qk} can be constrained by precise tests of LFU.
Moreover, $R_K$ is sensitive to the neutrino mixing parameters within SM extensions involving a fourth generation of quarks and leptons~\cite{lacker2010simultaneous} or sterile neutrinos~\cite{Abada:2012mc}.

The first measurements of $R_K$ were performed in the 1970s~\cite{PhysRevLett.29.1274,HEARD1975327,HEINTZE1976302}. The current PDG world average~\cite{ParticleDataGroup:2020ssz},
\begin{equation}
R_K^{\rm exp} = (2.488 \pm 0.009) \times 10^{-5},
\end{equation}
is dominated by a result from an early phase of the NA62 experiment~\cite{NA62:2012lny} and is 
%
consistent with the earlier measurements and with the SM expectation. The experimental precision on $R_K$ is still an order of magnitude larger than the uncertainty in the SM prediction, which motivates further measurements with improved accuracy.

Compared to the leptonic modes, the semi-leptonic decays are polluted by the hadronic structure dependence, typically described by a form factor with free parameters. However, the flavour-changing neutral current mediated $K\to\pi\ell\ell$ processes allow probing lepton universality at the one-loop level. If LFU holds, the form factor parameters have to be equal in both the electron and the muon channels~\cite{Crivellin:2016vjc}.

The physics of $K^+\to\pi^+\ell^+\ell^-$ ($\ell = e, \mu$) decays~\cite{Ecker:1987, DAmbrosio:1998gur, DAmbrosio:2018ytt} is similar to that of $K_S\to\pi^0\ell^+\ell^-$ discussed in Section~\ref{sec:k0-pi0-ell-ell}. In the $K^+$ decay case, interference with the long-distance amplitude from $K^+\to\pi^+\gamma^*$ leads to a CP violating charge asymmetry for the $e^+e^-$ channel, however the rate asymmetry in the SM is too small to be
observed~\cite{DAmbrosio:1998gur}. The $K^+\to\pi^+\ell^+\ell^-$ amplitudes
include the common weak form factor
$W(z; a_+, b_+) = G_F m_K^2 (a_+ + b_+ z) + W^{\pi\pi}(z)$, where
$z = m_{\mu\mu}^2/m_K^2$, $a_+$ and $b_+$ are free real parameters, and
$W^{\pi\pi}(z)$ is a complex function describing the contribution from the
three-pion intermediate state with a $\pi^+\pi^-\to\gamma^*$ transition. The
most precise form factor and BR measurements to date has been
performed by the E865 and NA48/2 experiments~\cite{E865:1999ker, NA482:2009pfe, NA482:2010zrc}:
\begin{align}
\text{E865: } {\rm BR}(K^+\to\pi^+e^+e^-)&= 2.94(15) \times 10^{-7}, \,a_+ = -0.587(10), \,b_+ = -0.655(44),\\
\text{NA48/2: }{\rm BR}(K^+\to\pi^+e^+e^-) &= 3.11(13) \times 10^{-7}, \,a_+ = -0.578(16), \,b_+ = -0.779(66),\\
\text{NA48/2: }{\rm BR}(K^+\to\pi^+\mu^+\mu^-) &= 9.62(25) \times 10^{-7}, \,a_+ = -0.575(39), \,b_+ = -0.813(145).
\end{align}
These results are compatible with LFU within uncertainties. Further measurements are in progress with the NA62 Run~1 dataset.


\subsubsubsection{Lepton flavor and lepton number conservation tests}
\label{sec:lfv-lnv}

Discovery of lepton number (LN) or lepton flavor number (LF) violation would be a clear indication of new physics. The minimal Type-I seesaw model~\cite{Mohapatra:1979ia} provides a source of LN violation through the exchange of Majorana neutrinos, while processes violating LF conservation can occur via the exchange of leptoquarks or $Z'$ boson, as well as in SM extensions with light pseudoscalar bosons. Using the dataset collected in 2016--18 using prescaled di-lepton trigger chains, the NA62 experiment has published ${\cal O}(10^{-11})$ upper limits on the branching fractions of LF and LN violating decays $K^+\to\pi^-(\pi^0)\ell^+\ell^+$, $K^+\to\pi^\pm\mu^\mp e^+$ and $\pi^0\to\mu^-e^+$~\cite{NA62:2019eax,NA62:2021zxl,NA62:2022tte}. Most of these searches are not limited by backgrounds, and the corresponding sensitivities are expected to improve linearly with the integrated kaon flux. This, along with the possible improvements to the trigger scheme, would allow probing these and other LN/LF violating processes at the ${\cal O}(10^{-12})$ or lower level at future $K^+$ experiments.

LF violating observables in the $K_L$ sector include the branching fractions of the $K_L\to(\pi^0)(\pi^0)\mu^\pm e^\mp$ and $K_L\to e^\pm e^\pm \mu^\mp\mu^\mp$ decays. These could be accessed at the ${\cal O}(10^{-12})$ sensitivity level at a high-intensity $K_L$ experiment with a magnetic spectrometer (\Sec{sec:intermediateKL}), improving on the earlier limits reported by the BNL-E871 and KTeV experiments.


\subsubsubsection{Chiral perturbation theory tests}

The non-perturbative nature of the strong interaction in kaon physics presents a major theoretical challenge for describing most of the kaon decays. Chiral perturbation theory (ChPT) provides a universal framework for treating leptonic, semileptonic and nonleptonic decays, including rare and radiative modes. Hadronic uncertainties are parametrized in ChPT by a number of parameters, the so-called low-energy constants~\cite{Cirigliano:2011ny}. Precision measurements of kaon decay rates and differential distributions represent important tests of the ChPT predictions, and provide crucial inputs to the theoretical development of low-energy hadron physics.

The rare and radiative decays that could be studied at future kaon experiments include, among others: $K^+\to e^+\nu\gamma$, $K^+\to\pi^0 e^+\nu\gamma$, $K_L\to\pi^- e^+\nu\gamma$~\cite{Bijnens:1992en}, $K\to\pi\gamma\gamma$~\cite{Ecker:1987hd}, $K\to\pi\gamma \ell^+\ell^-$~\cite{Donoghue:1997rr, Gabbiani:1998tj}, $K\to\pi\pi\ell\nu$~\cite{Bijnens:1989mr}, $K^+\to\ell_1^+\nu\ell_2^+\ell_2^-$~\cite{Bijnens:1992en}, and $K\to\pi\pi\pi\gamma$~\cite{DAmbrosio:1996jmq}.


\subsubsection{Neutral pion decays}
\label{sec:neutral_pion_decays}

High intensity rare kaon decay experiments also represent $\pi^{0}$ factories, since neutral pions are produced in about 30\% of both $K^{+}$ and $K_L$ decays. Moreover, the $\pi^0$ production can be tagged by other (charged) particles originating from the kaon decay vertex. The $\pi^0$ decays proceed via the $\pi^0\to\gamma^*\gamma^*$ vertex, described by the transition form factor, which also enters the computation of the hadronic light-by-light scattering contributing to the muon anomalous magnetic moment~\cite{Aoyama:2020ynm}. Future kaon experiments with dedicated di-electron trigger chains would be able to improve the precision of the measurements of the rare decays $\pi^0 \to e^+e^-e^+e^-$~\cite{KTeV:2008pev} and $\pi^0\to e^+e^-$~\cite{KTeV:2006pwx}, and thus provide an important input to the theoretical modeling of the form factor. In addition, a new precise measurement of the branching ratio for the latter decay would resolve an existing $2\sigma$ tension with the SM prediction~\cite{Husek:2014tna}.


\subsubsection{Searches for dark sectors in kaon decays}

Measurements of rare kaon decays, including dedicated studies of differential distributions, represent crucial probes of light dark sectors. The sensitivity provided by the kaon sector owes to both the suppression of the total kaon decay width, and the availability of the very large dataset. Dark sector searches in kaon decays have attracted much interest recently. The status of current searches over a broad range of hidden sector scenarios, the impact of current experiments on the parameter space of these scenarios, and the future projections are discussed in a recent comprehensive review~\cite{Goudzovski:2022vbt}.

The NA62 collaboration is pursuing a dedicated program of dark-sector searches using the Run~1 dataset collected in 2016--18. The results published so far include the world's most stringent upper limits on the dark scalar ($\varphi$) mixing parameter obtained from searches for the $K^+\to\pi^+\varphi$ decay~\cite{NA62:2021zjw,NA62:2020pwi,NA62:2020xlg}, heavy neutral lepton ($N$) mixing parameters $|U_{\ell 4}|^2$ ($\ell=e;\mu$) obtained from searches for the $K^+\to\ell^+N$ decays~\cite{NA62:2017qcd,NA62:2020mcv,NA62:2021bji}, and the dark photon ($A'$) mixing parameter obtained from a search for the $\pi^0\to\gamma A'$ decay~\cite{NA62:2019meo}. The sensitivity of these searches is expected to improve in future at least as the square root of the integrated $K^+$ flux. Possible improvements to the data collection and analysis (e.g. software trigger development) would bring further improvements in sensitivity.

Similarly, the KOTO experiment has published a search for the $K_L\to\pi^0\varphi$ decay, and is  searching for other hidden-sector scenarios with photons in the final state ($K_L\to\gamma\gamma X_{\rm inv}$, $K_L\to\pi^0\pi^0 X_{\rm inv}$, $K_L\to\gamma A'$). Next-generation $K_L\to\pi^0\nu\bar\nu$ experiments would improve significantly the sensitivity of these searches, while a $K_L$ experiment with a magnetic spectrometer (\Sec{sec:intermediateKL}) would widen the physics reach by addressing additional dark-sector scenarios.


\subsubsection{Hyperon decays}

Hyperons physics has been always complementary to kaon physics, thanks to baryon number and spin properties which change the sensitivity to SM and new interactions. While CP violation in kaon decays was discovered early, that in hyperons is yet to be established. Important efforts in this direction were done by the HyperCP dedicated experiment~\cite{HyperCP:2004kbv} at Fermilab. In fact, the recent sudden change in the average of the $\alpha$ decay parameter value of the $\Lambda \to p \pi^-$ decays, thanks to the BESIII and CLAS experiment measurements~\cite{BESIII:2018cnd,Ireland:2019uja}, has renewed the interest in possible measurements of CP violation in hyperon decays.



The $\Sigma^+\to p\mu^+\mu^-$ decay represents an $s\to d\mu^+\mu^-$ transition and provides complementary information to the equivalents from kaon decays~\cite{Geng:2021fog} and attracted theoreticians attention since 1962~\cite{Lyagin:1962} as probe of neutral currents. An evidence of this decay was first obtained at HyperCP~\cite{HyperCP:2005mvo}. While the branching fraction measured by HyperCP is consistent with the SM prediction, the dimuon mass spectrum showed hints of a resonance near the dimuon mass threshold. More recently, LHCb also found an evidence for this decay, obtaining the result~\cite{LHCb:2017rdd}:
\begin{equation}
{\rm BR}(\Sigma^+\to p\mu^+\mu^-) = \left(2.2_{-1.3}^{+1.8}\right)\times 10^{-8},
\end{equation}
although with no sign of any new resonance, excluding the central value of the HyperCP branching fraction with intermediate particle hypothesis.

\section{Experiments at the J-PARC Hadron Experimental Facility}

\subsection{The KOTO experiment until 2021}

The KOTO collaboration is searching for the $\klpionn$ decay at the J-PARC KOTO experiment.
For the branching ratio of the decay,
we set an upper limit of $3.0\times 10^{-9}$ at 90\% CL with the data collected in 2015, and reached a single event sensitivity of $7.2\times 10^{-10}$ with the data collected in 2016--18. We took 2 times more data in 2019--21 than in 2016--18 with detector upgrades, and we are analyzing the data.

\subsection{The KOTO experiment from 2022}
The J-PARC accelerator achieved  the beam power 65 kW ($7\times 10^{13}$ protons per 5.2~s cycle)
in 2021 in the slow extraction.
With an upgrade of J-PARC accelerator in 2021--22, shorter repetition cycle (4.2~s) and smoother beam-spill will be expected. The beam power of 80~kW is expected in 2022, and is expected to be increased to 100~kW in the following 2--3 years. For the effect of the instantaneous beam intensity on the accidental loss of $\klpionn$ acceptance,
we expect that the 100-kW beam will give better situation compared to the 65-kW beam in 2021.
We expect to reach the single event sensitivity at $O(10^{-11})$ around 2025
with the following DAQ and detector upgrades.

\subsubsection{Upgrade of trigger and DAQ system}
\label{sect:kotodaqupgrade}
KOTO will upgrade the trigger and DAQ system in 2022
in order to handle higher data rate form higher beam power with the accelerator upgrade.

Waveforms for all the detector are recorded with 
125-MHz or 500-MHz ADCs.
In the current trigger, KOTO used a level-1 trigger  
from the energy and timing information of each detector component,
and used a cluster-counting trigger for the number of calorimeter clusters. 
With the trigger,  the waveform data were sent
to a level-2 module (currently without any triggering function), 
and then
to a level-3 PC cluster at J-PARC.
Event building was performed at the PC cluster, 
and the data were sent to a computing center at KEK. 

For the trigger and DAQ upgrade, 
the level-2 module with a bottle-neck was replaced 
with a FPGA-based event builder.
The data after event building will be send to a new PC cluster
with GPUs. The GPU will perform waveform-based data compression,
and the CPU power can be used for complicated event selections.
With those efforts, KOTO plan to keep the same data rate from J-PARC to the KEK computing center even with twice higher input data rate.

\subsubsection{Upgrade of upstream charged particle veto counter}

In the analysis of data taken in 2016--18, KOTO found a background source from charged kaon in the beam. Charged kaons are produced from the interaction of $K_L$ or pions in the beam line materials.
Charged kaons produced in the beamline downstream of a sweeping magnet can enter the KOTO detector. The flux of the charged kaon was $O(10^{-5})$ compared to the $K_L$, but $K^\pm\to \pi^\pm e^\mp \nu$
becomes a background when $e^\pm$ is not detected. This was a main background in the analysis of 2016-18 data. In 2020, we installed a charged particle counter called UCV in the beam to detect charged kaons in the beam, and veto the event. 
In 2021, we installed a new version of UCV with a sheet of 
0.5-mm square plastic scintillating fibers.
The efficiency for the charged particle is 90\% or more.
Beam particles scatters in the UCV and increase the hit rate of detectors near the UCV, 
which increases the accidental loss of the signal acceptance.
$K_L$s or neutrons scatters in the UCV into the beam halo region, 
which increases halo-$K_L$ or halo-neutron background.
In order to reduce these effects, 
KOTO plans to install another UCV with 160-mm-square 
0.2-mm-thick plastic scintillator film. 
The scintillation photons escaping out from the scintillator surface is collected with aluminized-mylar film, and PMTs are used to detect the photons.
The charged particle efficiency of 99\% was achieved with a 
prototype counter.

\subsubsection{Installation of second sweeping magnet}
KOTO plans to install a new sweeping magnet
at the downstream-end of the beam line in order to reduce the charged kaons. This will reduce the charged-kaon flux by a factor of ten or more, and the charged-kaon background can be negligible
with both the new UCV counter and the second magnet.

\subsection{KOTO step-2 experiment}
\subsubsection{Concept of KOTO step-2}
Although KOTO will reach the sensitivity at $O(10^{-11})$ around 2025,
it would take longer time toward the sensitivity predicted by the Standard Model,  $3\times 10^{-11}$,
considering the operation plan of the Main Ring accelerator and the expected running time in future.
We thus should have a new experiment that can discover and observe 
a large number 
of $\klpionn$ events and measure its branching ratio.

An idea of such a next-step experiment, called KOTO step-2, was already mentioned in the KOTO proposal in 2006~\cite{KOTOproposal}.
Now we have 
gained
experiences in the KOTO experiment and are ready to consider and design the next-generation experiment more realistically.
We eagerly consider
the realization of KOTO step-2 in the early phase of the extension of the Hadron Experimental Facility.

\begin{figure}[ht]
\begin{minipage}{0.5\linewidth}
\includegraphics[width=\linewidth]{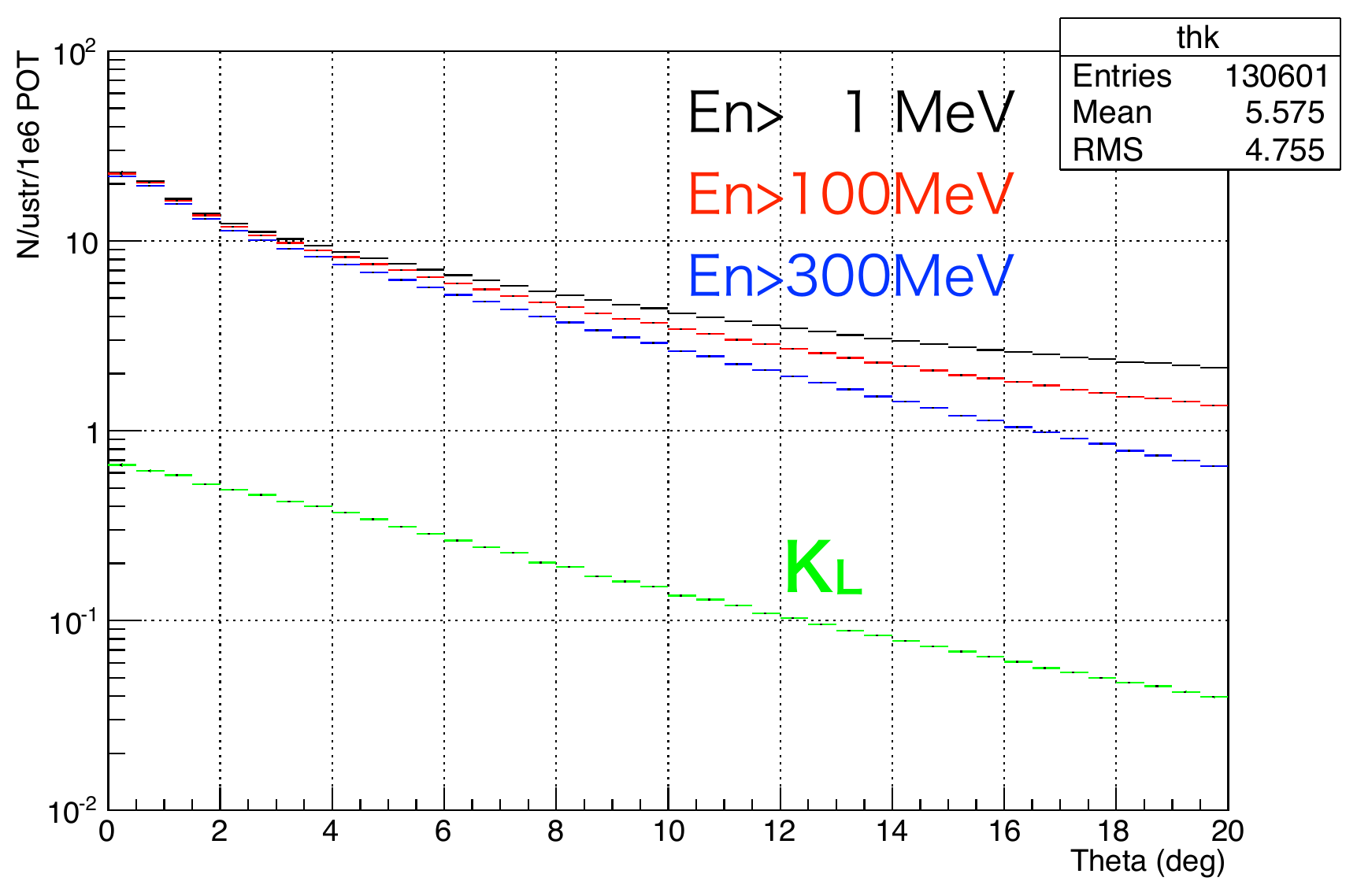}
\end{minipage}
\begin{minipage}{0.5\linewidth}
\includegraphics[width=\linewidth]{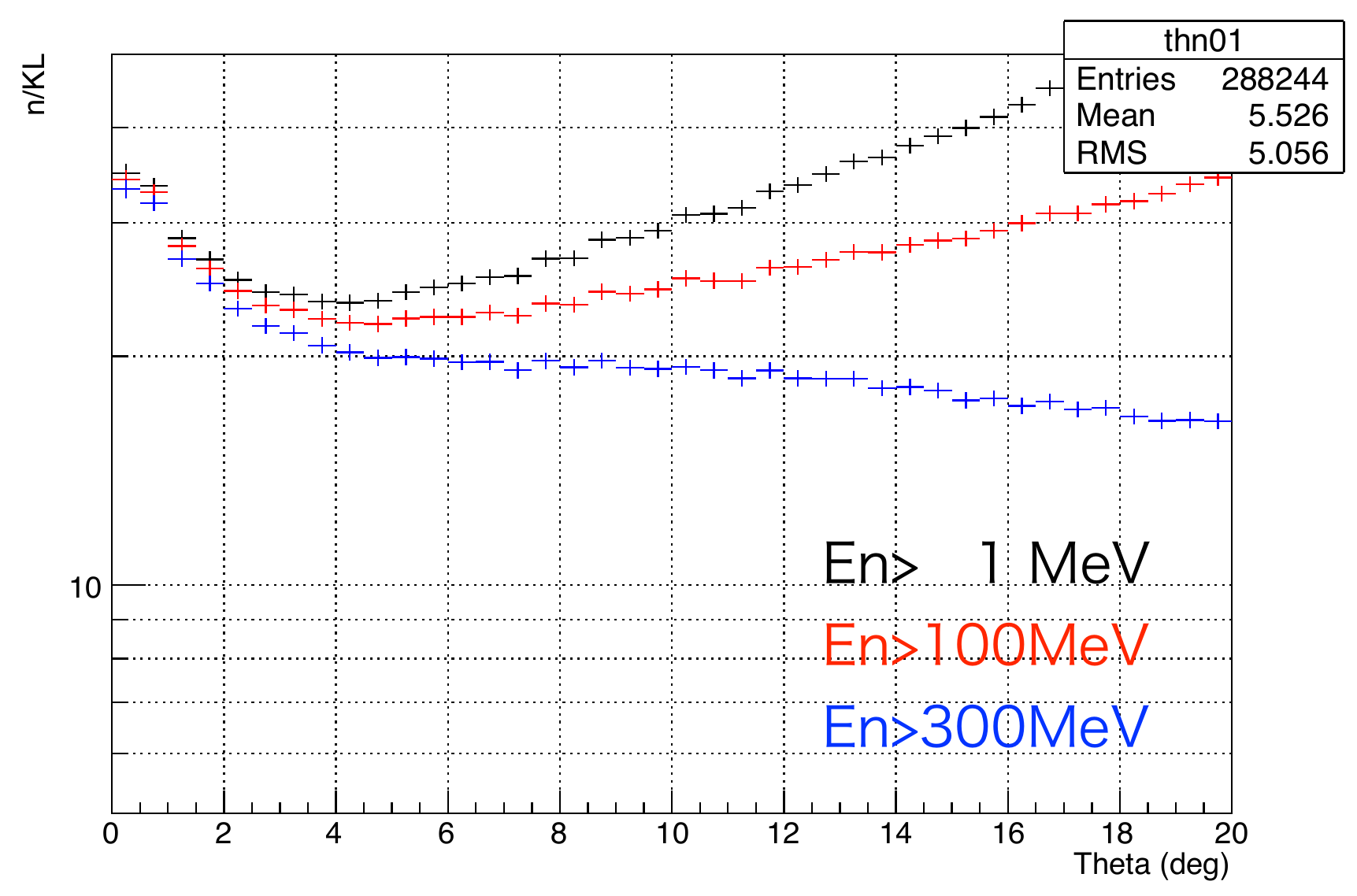}
\end{minipage} 
\caption{\label{fig:vsangle}Simulated $K_L$ and neutron yields (left) and their ratio (right) as functions of the production angle~\cite{ref:kaon2019}. 
The yields were evaluated at 1~m downstream of the target, 
normalized by the solid angle ($\mu$str).
Black, red, and blue points indicate the results when selecting neutrons with their energies of more than 1, 100, and 300~MeV, respectively.}
\end{figure}

\begin{figure}[ht]
\centering
\includegraphics[width=0.8\linewidth]{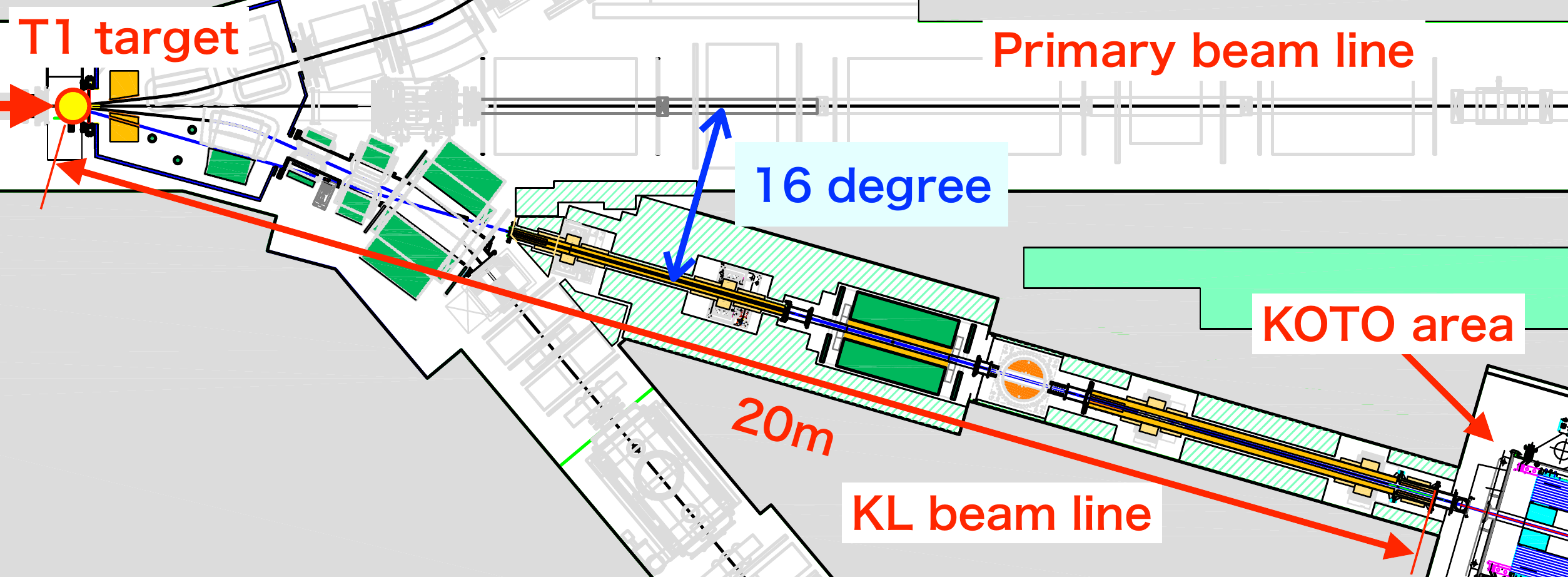}
\caption{\label{fig:kotobl}Schematic drawing of the KL beam line for the KOTO experiment in the current Hadron Experimental Facility.}
\end{figure}

\begin{figure}[ht]
\centering
\includegraphics[width=\linewidth]{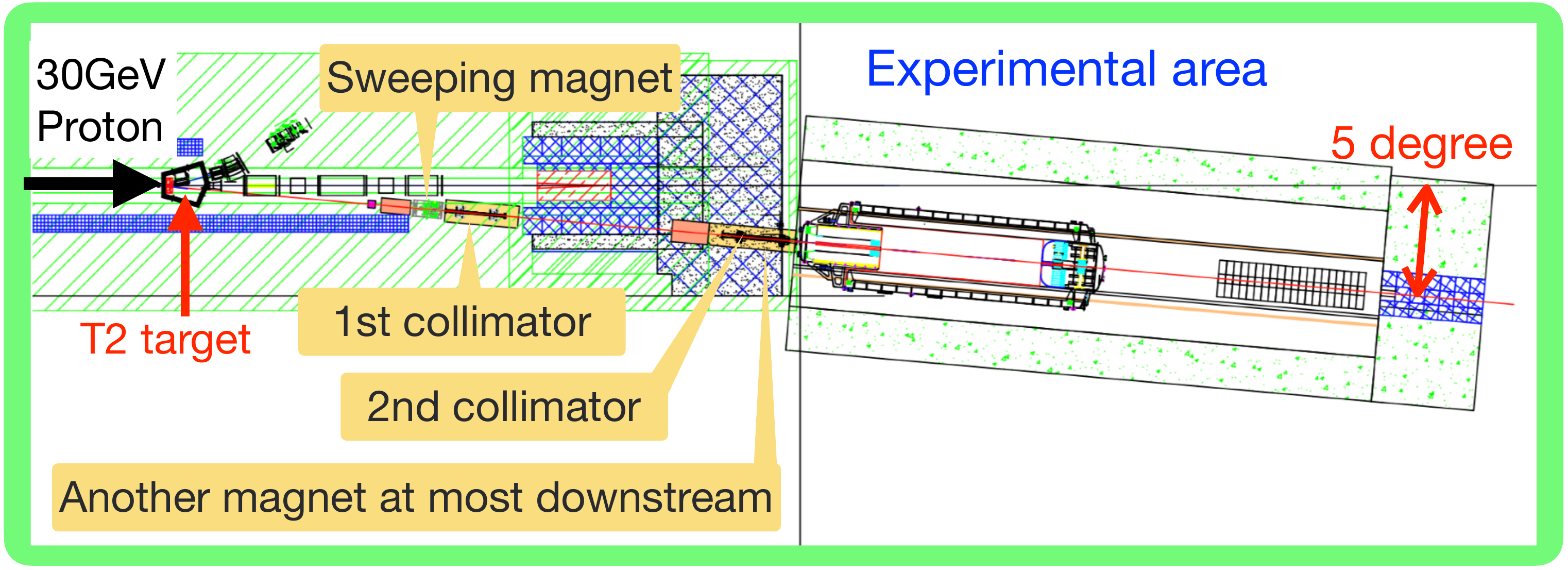}
\caption{\label{fig:koto2bl}Schematic drawing of the KOTO step-2 beam line in the Extended Hadron Experimental Facility.
The experimental area is located behind the beam dump. Distance between the T2 target and the experimental area is assumed to be 43~m in the present studies.
A modeled cylindrical detector with 20~m in length and 3~m in diameter is described in the experimental area as a reference.}
\end{figure}

To achieve a higher experimental sensitivity for the $\klpionn$ measurement, 
 we must consider to maximize the $\kl$ flux, the detection acceptance of the signal, and the signal-to-background ratio.
The $\kl$ flux is determined by the production angle and the solid angle of the secondary neutral beam.
Other parameters are the achievable intensity of the primary proton beam and the target properties (material, thickness, etc.).
The production angle is defined as the angle between the primary proton beam and secondary neutral beam directions. 
Figure~\ref{fig:vsangle} shows the $\kl$ and neutron yields and the neutron-to-$\kl$ flux ratio as functions of the production angle when a 102-mm-long gold target is used.
KOTO step-2 chooses the production angle of 5~degrees as an optimum point for a higher $\kl$ flux with a smaller neutron fraction in the beam. 
In case of the KOTO experiment, the production angle is 16-degree, as shown in Fig.~\ref{fig:kotobl}, 
which was chosen to utilize the T1 target with the experimental area away from the primary beam.
In order to realize the 5-degree production while keeping the solid angle of the neutral beam as large as possible,
$\ie$ with the shortest beam line, a new experimental area behind the primary beam dump and a new target station close to the dump are necessary.
Figure~\ref{fig:koto2bl} shows a possible configuration in the Extended Hadron Experimental Facility, utilizing the second target (T2).

Here we introduce the KOTO step-2 detector base design.
The acceptance of the signal is primarily determined by the detector geometry.
The basic detector configuration is same as the current KOTO experiment, a cylindrical detector system with an electromagnetic calorimeter to detect two photons from the $\pi^0$ decay at the end cap. 
A longer $\kl$ decay region and a larger diameter of the calorimeter are being considered to obtain a larger acceptance.
The 5-degree production also provides a benefit in view of the signal acceptance; a harder $\kl$ spectrum than KOTO is expected, and two photons are boosted more in the forward direction, and thus the acceptance gain by a longer decay region can be utilized.

The ability of the signal discovery and the precision of the branching ratio measurement depend on the signal-to-background ratio, as well as the expected number of observed events.
The background level is affected by many factors such as the beam size at the detector, the flux of beam particles (neutrons, $\kl$) leaking outside the beam (beam-halo),
 charged kaons in the neutral beam, and detector performances.


\subsubsection{KL2 beamline}
Table~\ref{tab:beampar} summarizes the beam parameters for KOTO step-2
and those in the current KOTO experiment. 
\begin{table}[ht]
\caption{\label{tab:beampar}Beam parameters for KOTO step-2 and the current KOTO experiment.}
\begin{center}
\begin{threeparttable}
\begin{tabular}{lll}
\hline
&KOTO step-2\tnote{*}&KOTO\\
\hline
Beam power&100~kW&64~kW (100~kW in future)\\
Target&102-mm-long gold&60-mm-long gold\\
Production angle&5$^{\circ}$&16$^{\circ}$\\
Beam line length&43~m&20~m\\
Solid angle&4.8~$\mu$sr&7.8~$\mu$sr\\
\hline
\end{tabular}
\begin{tablenotes}\footnotesize
\item[*] 
Note the parameters for step-2 are tentative for this study.
\end{tablenotes}
\end{threeparttable}
\end{center}
\end{table}

\begin{figure}[ht]
\centering
\begin{minipage}{0.45\linewidth}
\includegraphics[width=\linewidth]{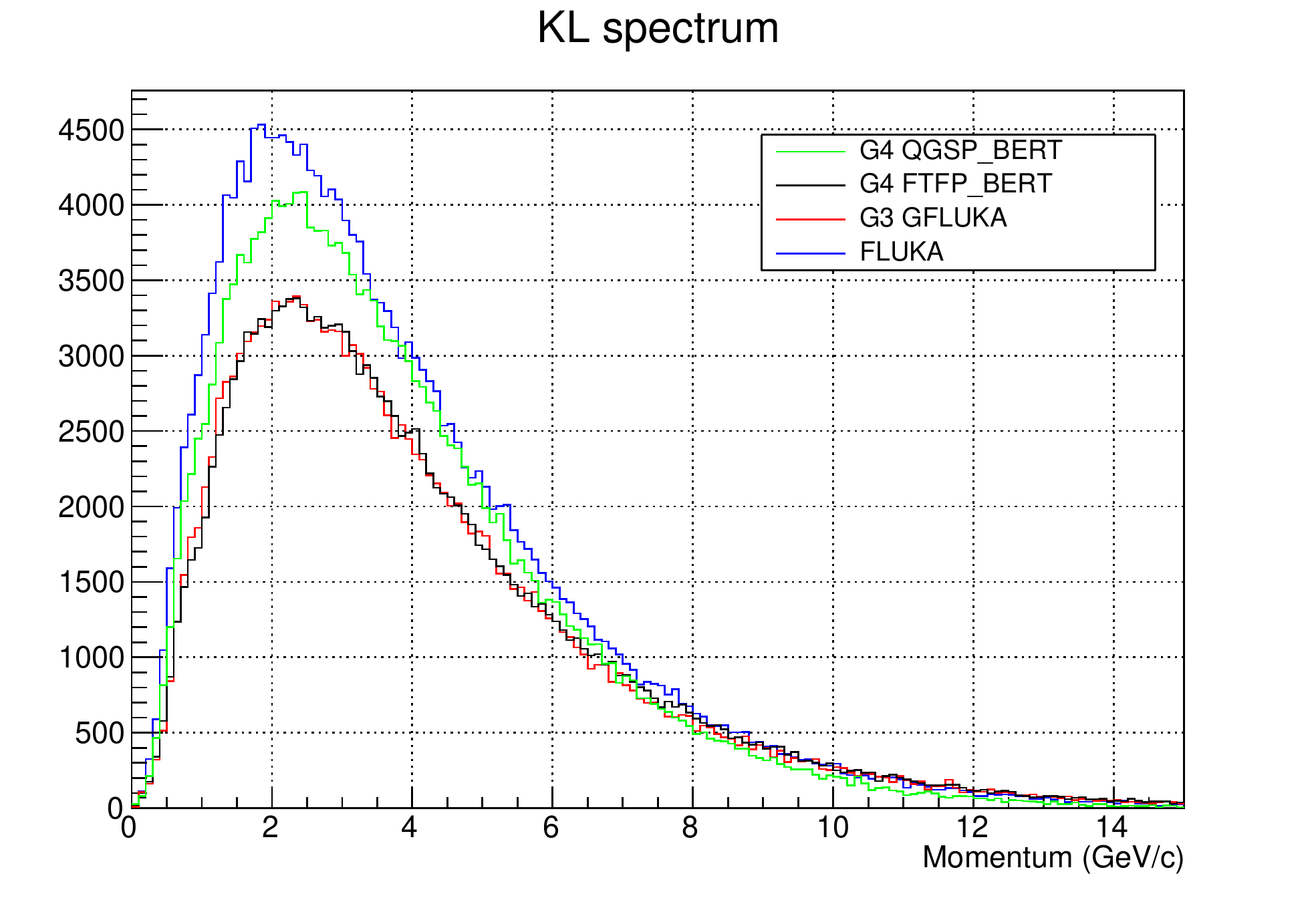}
\end{minipage}
\begin{minipage}{0.45\linewidth}
\includegraphics[width=\linewidth]{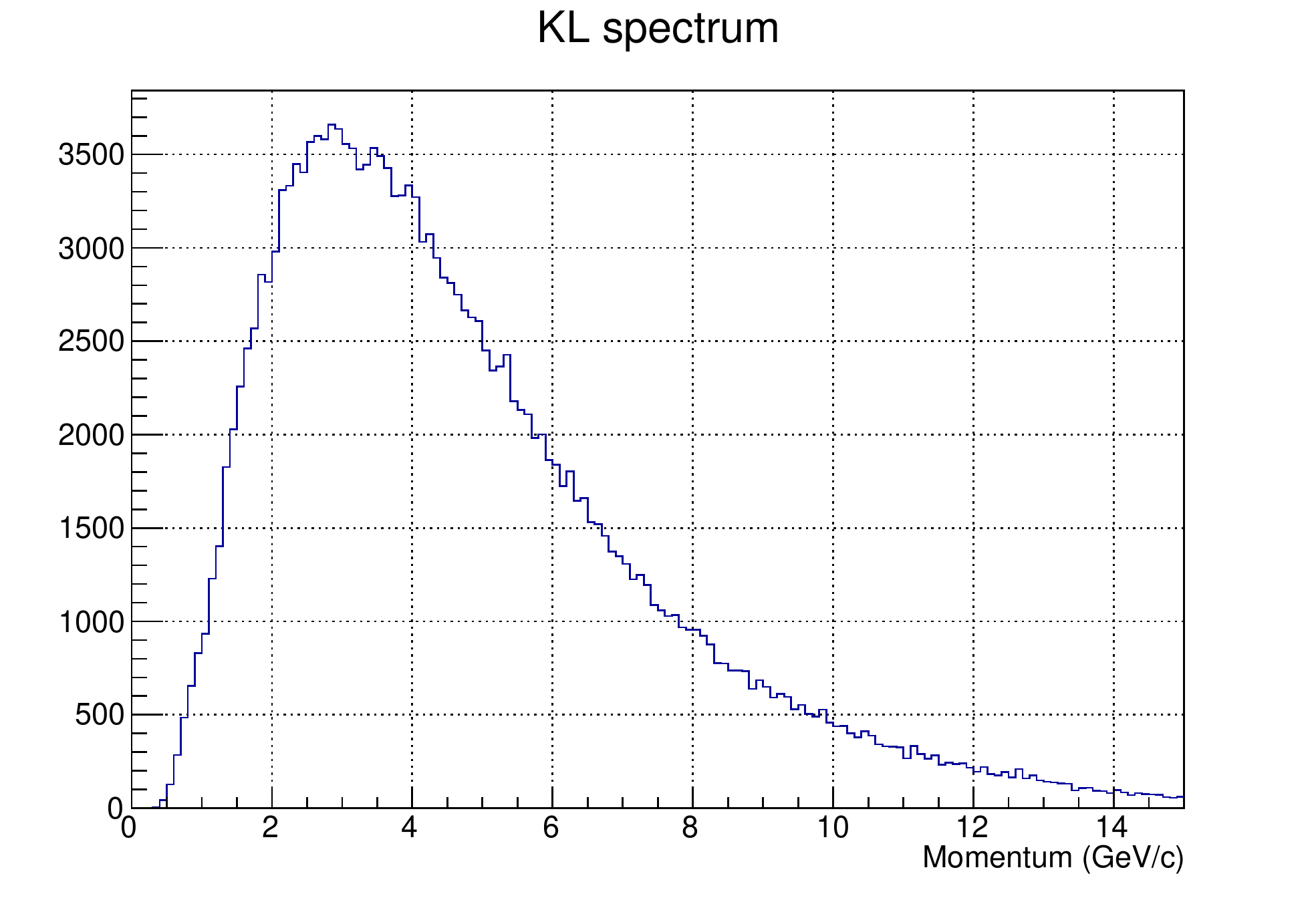}
\end{minipage}
\caption{\label{fig:klmom}
$\kl$ spectra at 1~m from the T2 target by the target simulation (left) and at the exit of the KL2 beam line at 43~m from the T2 target by the beam line simulation (right). In the left plot, the results by using various simulation packages are also shown, as well as the result by the GEANT3-based simulation (labeled ``G3 GFLUKA'') which is our default in this study.}
\end{figure}

A T2 target used in the KOTO step-2 study was 
a gold rod with its diameter of 10~mm and length of 102~mm, which corresponds to 1$\lambda_{I}$ (interaction length).
Spectra and relative yields of $\kl$ at 1 m from the target were obtained from simulations as shown  in Fig.~\ref{fig:klmom} (left),
where GEANT3, GEANT4 (10.5.1 with a physics list of QGSP\_BERT or FTFP\_BERT), and FLUKA (2020.0.3) results are shown for comparison.
The resultant $\kl$ fluxes were found to agree with each other within 30\%. GEANT3 provided 
the smallest $\kl$ yield
and thus is considered to be a conservative choice in the discussion of the sensitivity.

\begin{figure}[ht]
\centering
\begin{minipage}{0.45\linewidth}
\includegraphics[width=\linewidth]{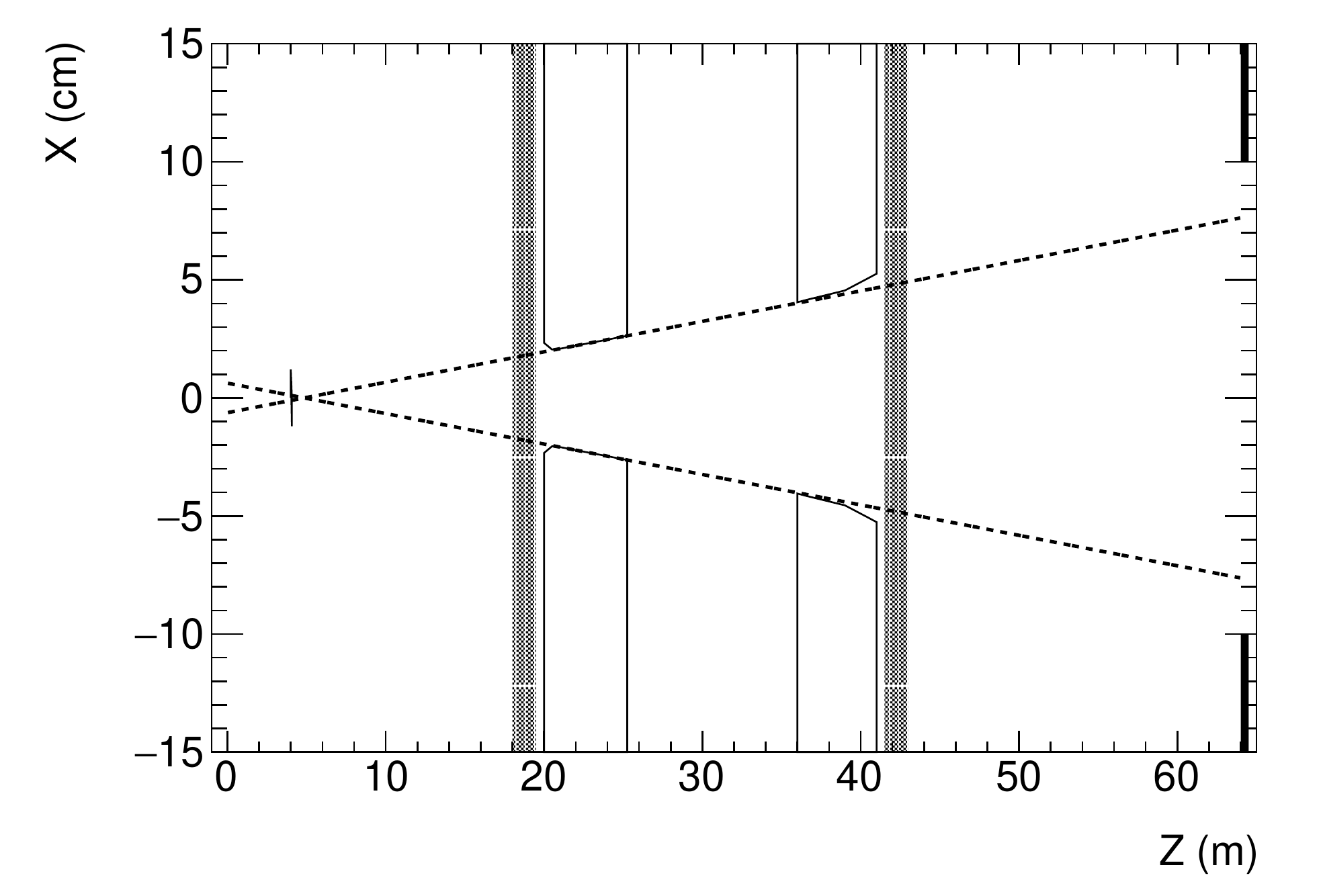}
\end{minipage}
\begin{minipage}{0.45\linewidth}
\includegraphics[width=\linewidth]{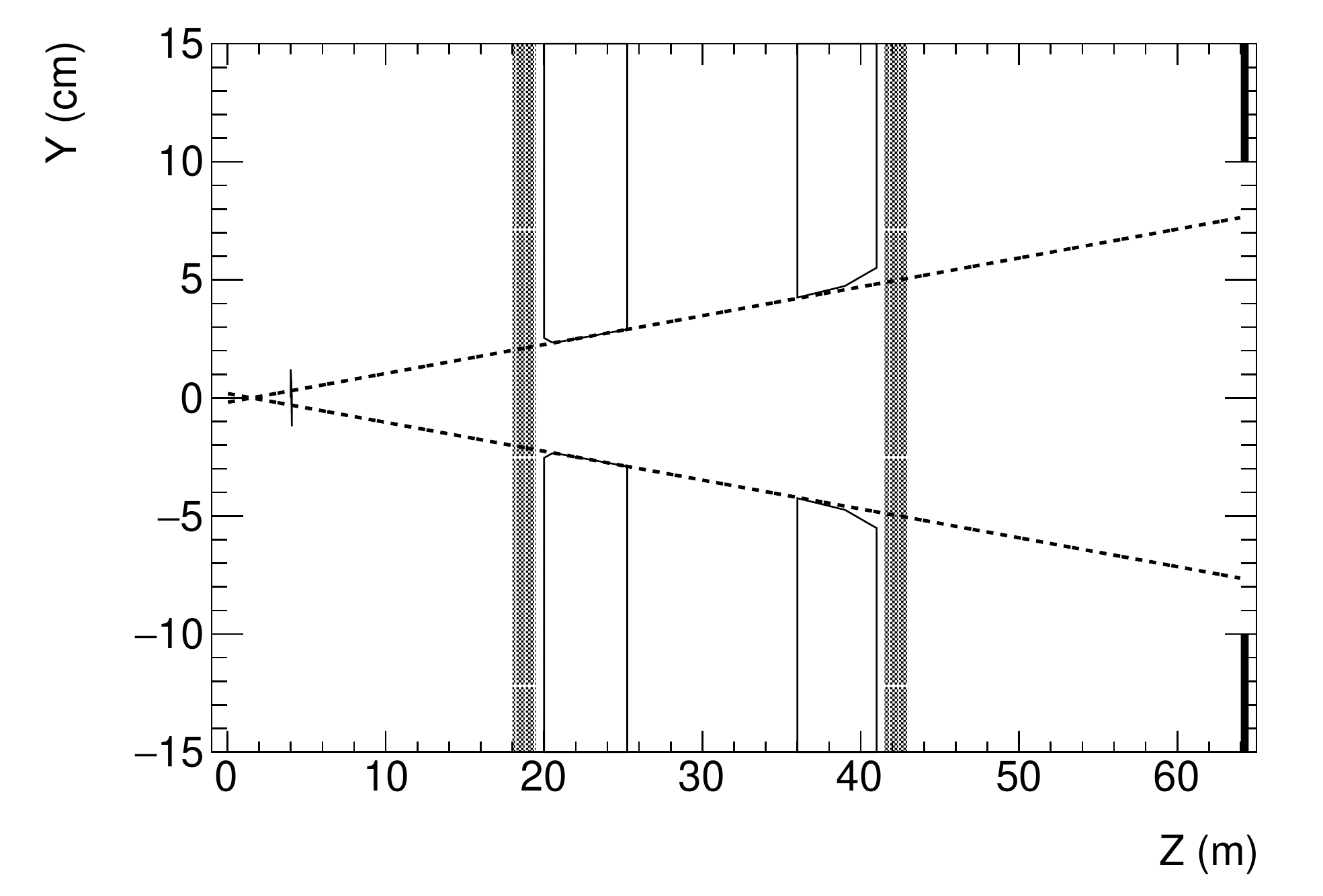}
\end{minipage}
\caption{\label{fig:collXY}Design of the KL2 beam line in x-z (left) and y-z (right) views.
The short vertical lines at the upstream show the photon absorber,
the solid lines in the middle show the two collimators,
the shaded regions show the two sweeping magnets,
the dahsed lines show the 4.8-$\mu$str 
collimation lines defined with the first collimator, and
the black filled regions at the downstream
show inner region of the calorimeter of the KOTO step-2 detector,
where the beam size is 15-cm square.
 }
\end{figure}

The beam line of KOTO step-2 (KL2 beam line) (Fig.~\ref{fig:collXY})
consists of two stages of 5-m-long collimators, a photon absorber in the beam,
and two sweeping magnets to sweep out charged particles from the target or produced in the beam line.
The photon absorber, made of 7-cm-long lead, is located at 7~m downstream of the target.
The first collimator, starting from 20~m from the target, defines the solid angle and shape of the neutral beam.
The solid angle is set to be 4.8~$\mu$str.
The second collimator, starting from 38~m from the target, cut the particles coming from the interactions at the photon absorber and the beam-defining edge of the first collimator. The bore shape of the second collimator is designed not to be seen from the target so that particles coming directly from the target do not hit the inner surface and thus do not generate particles leaking 
outside the beam.
The first sweeping magnet is located upstream of the first collimator,
and the second sweeping magnet is located downstream of the second collimator
(The effect of the second sweeping magnet was evaluated independently from the beam line without the second sweeping magnet).

\begin{figure}[ht]
\centering
\begin{minipage}{0.45\linewidth}
\includegraphics[width=\linewidth]{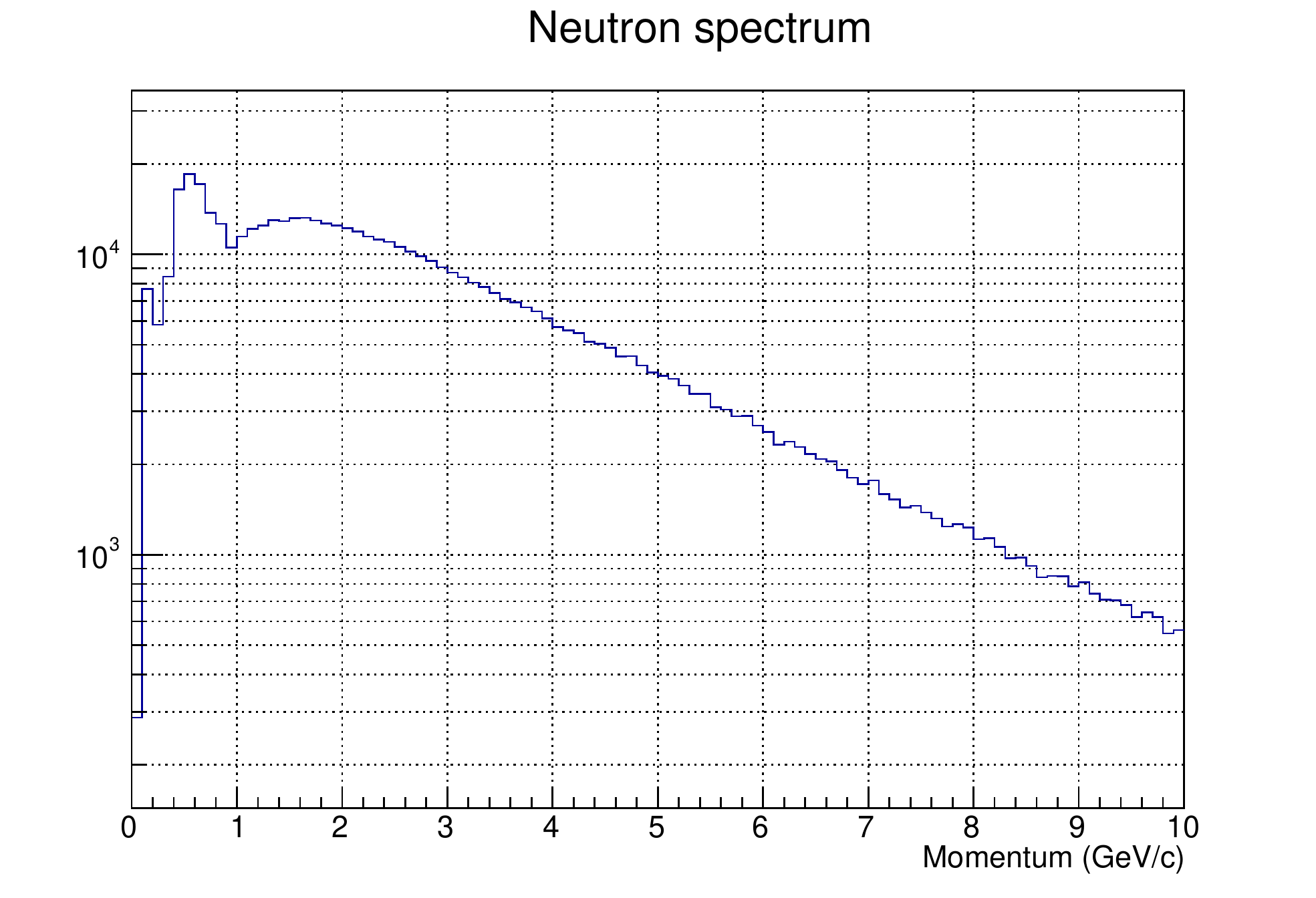}
\end{minipage}
\begin{minipage}{0.45\linewidth}
\includegraphics[width=\linewidth]{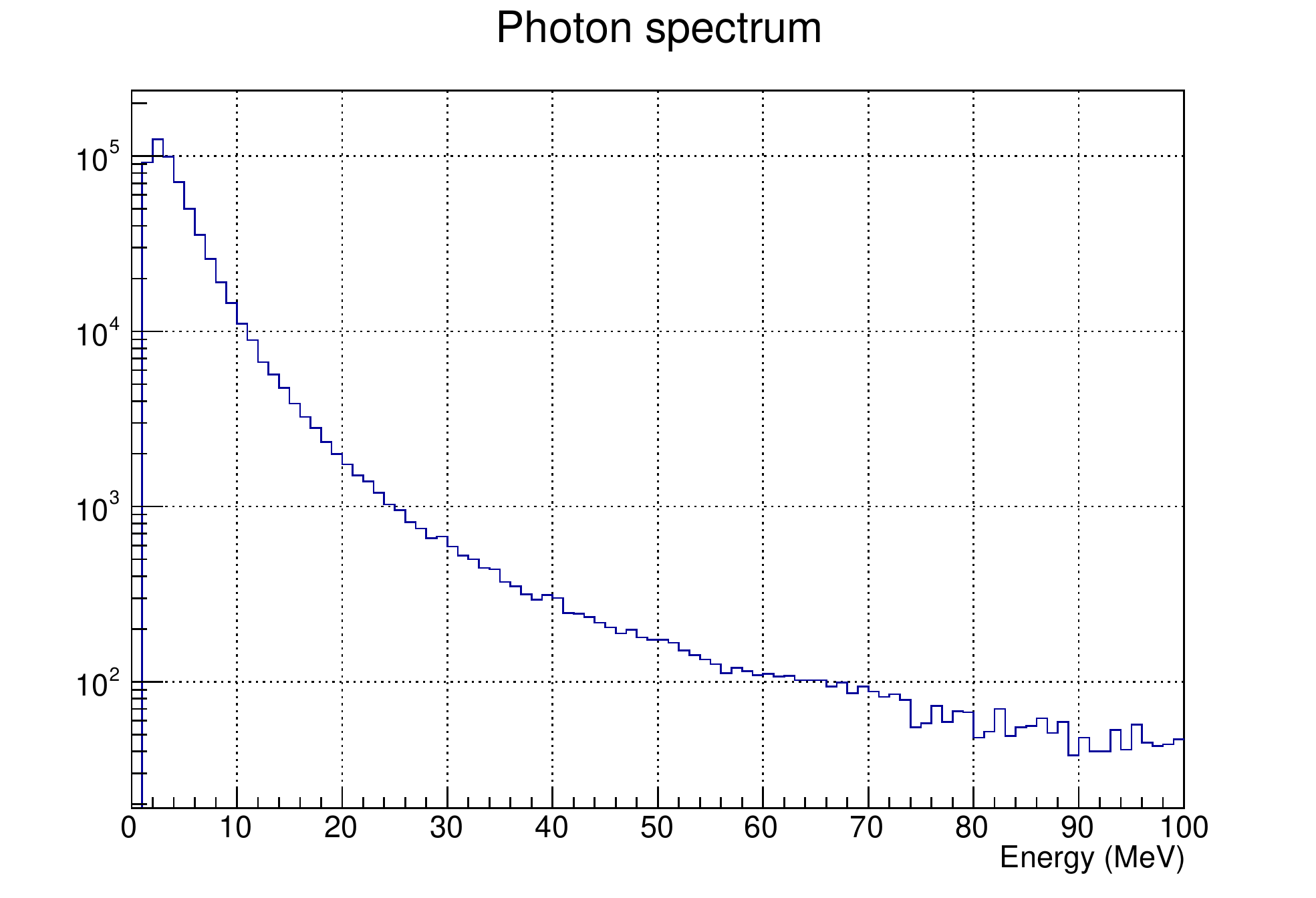}
\end{minipage}
\caption{\label{fig:ngspectra}Simulated neutron (left) and photon (right) spectra at the exit of the beam line.}
\end{figure}

Figures~\ref{fig:klmom} (right) and \ref{fig:ngspectra} show the simulated spectra of $\kl$, neutrons, and photons at the exit of the KL2 beam line, respectively.
The $\kl$ yield was evaluated to be $1.1\times 10^7$ per $2\times 10^{13}$ protons on the target (POT).
Note that the beam power of 100~kW corresponds to $2\times 10^{13}$~POT per second with 30~GeV protons.
The resultant $\kl$ flux per POT is 2.6 times higher than that of the current KOTO experiment.
The $\kl$ spectrum peaks at 3~GeV/c, while it is 1.4~GeV/c in the current KOTO experiment.
The simulated particle fluxes are summarized in Table~\ref{tab:yield}.
\begin{table}[ht]
\caption{\label{tab:yield}Expected particle yields estimated by the simulations.}
\begin{center}
\begin{threeparttable}
\begin{tabular}{cccc}
\hline
\multirow{2}{*}{Particle}&\multirow{2}{*}{Energy range}&Yield & On-spill rate\\
&&(per $2\times 10^{13}$~POT) & (MHz)\\
\hline
$\kl$&&$1.1\times 10^7$&24\\
\hline
\multirow{2}{*}{Photon}&$>$10~MeV&$5.3\times10^7$&110\\
&$>$100~MeV&$1.2\times10^7$&24\\
\hline
\multirow{2}{*}{Neutron}&$>$0.1~GeV&$3.1\times10^8$&660\\
&$>$1~GeV&$2.1\times10^8$&450\\
\hline
\end{tabular}
\begin{tablenotes}\footnotesize
\item[]
The beam power of 100~kW corresponds to $2\times 10^{13}$~POT/s with 30~GeV protons.
The on-spill rate means the instantaneous rate during the beam spill, assuming 2-second beam spill every 4.2 seconds. 
\end{tablenotes}
\end{threeparttable}
\end{center}
\end{table}

\begin{figure}[ht]
\centering
\centering
\begin{minipage}{0.45\linewidth}
\includegraphics[width=\linewidth, height=\linewidth]{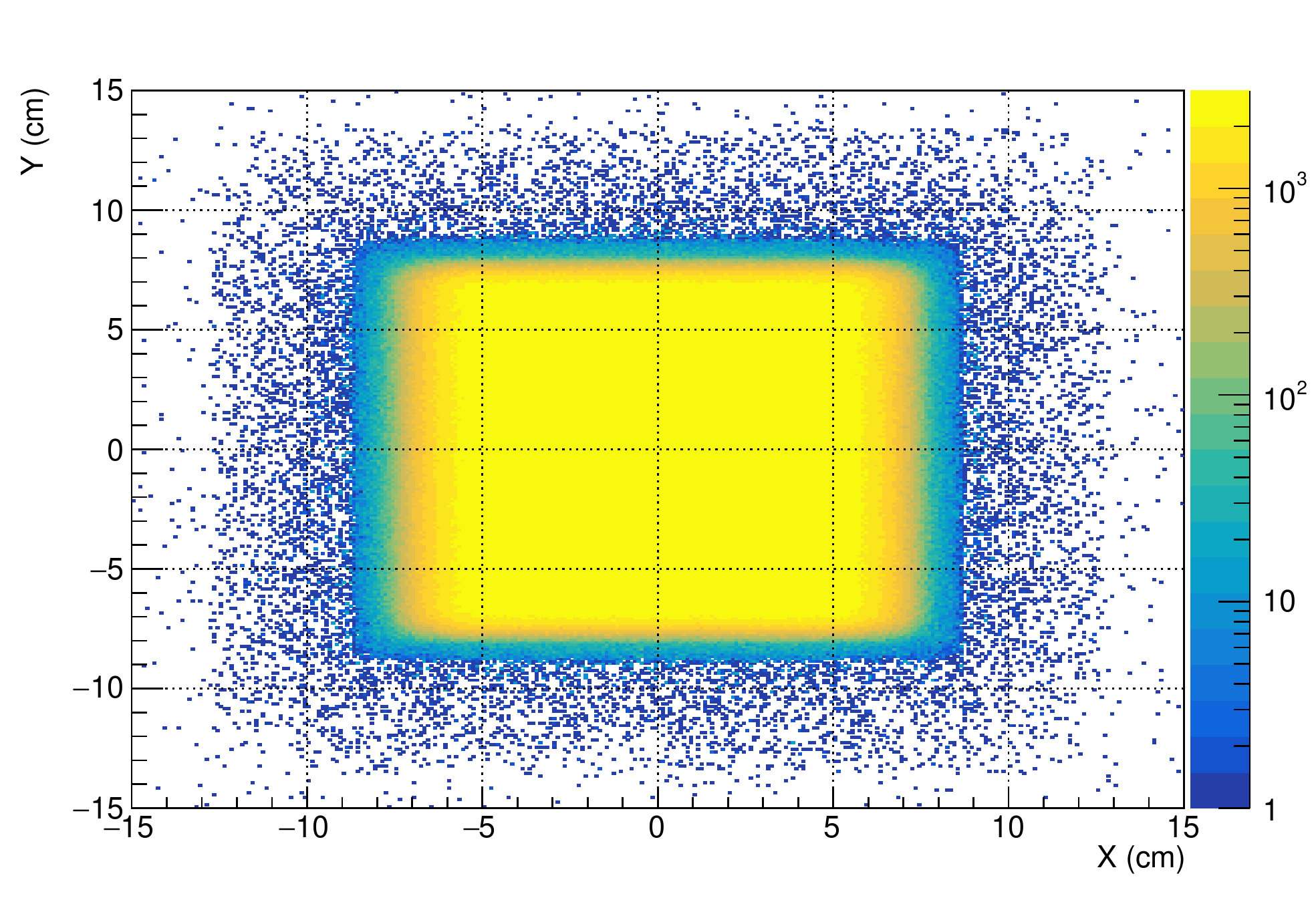}
\end{minipage}
\begin{minipage}{0.45\linewidth}
\includegraphics[width=\linewidth]{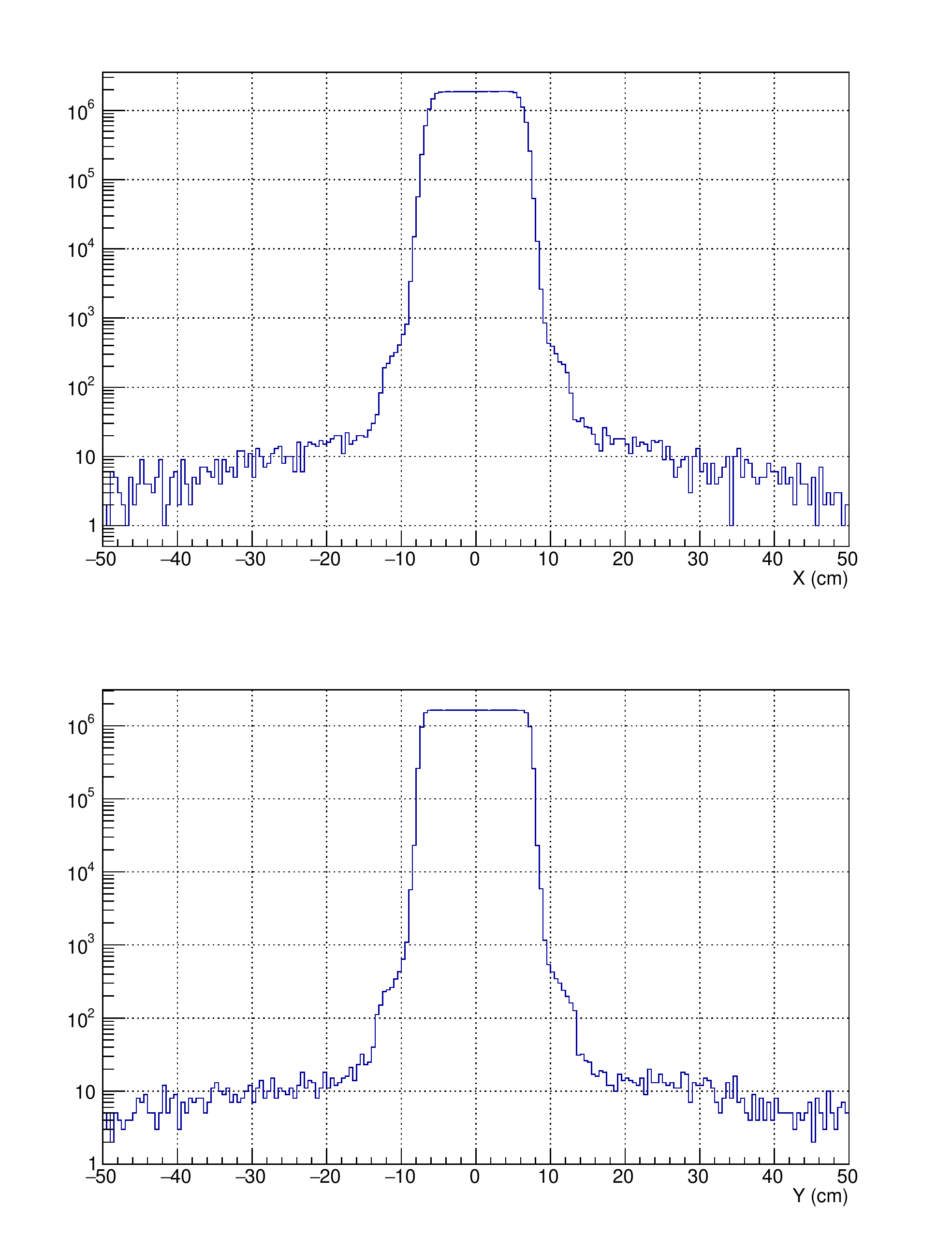}
\end{minipage}
\caption{\label{fig:nprof}Beam shape at the end-cap plane, represented by the neutron profile.
The left figure shows the distribution in $\pm 15$~cm of the beam center, and the right top (bottom) histogram indicates the horizontal (vertical) distribution in $\pm 50$~cm of the beam center.}
\end{figure}

\subsubsubsection{Halo neutrons}
Figure~\ref{fig:nprof} shows the neutron profile at the assumed calorimeter location, 64~m from the T2 target. As shown in the figure, the neutral beam is shaped so as to be a square at the calorimeter location.
Evaluation of neutrons spreading to the beam halo region, called ``halo neutron'', is important since they are potential sources of backgrounds due to their interaction with the detector materials.
Here we define the core and halo neutrons as those inside and outside the $\pm$10~cm region at the calorimeter, respectively.
The ratio of the halo neutron yield to the core yield was found to be $1.8 \times 10^{-4}$.

\subsubsubsection{Charged kaons in the neutral beam}
The contamination of the charged kaons in the neutral beam line is harmful, since $K^\pm$ 
decays such as $K^\pm \to \pi^0 e^\pm \nu$
 in the detector region can mimic the signal, which were pointed out in the analysis of the KOTO experiment~\cite{KOTO:2020prk}.
The major production point of charged kaons is the second collimator. 
Neutral particles ($\kl$ and neutrons) hit the inner surface of the collimator and the produced charged kaon in the interactions can reach the end of the beam line. Charged pions from $\kl$ decays hitting the collimator also can produce charged kaons.
According to the beam line simulation, 
the flux ratio of the charged kaon and $\kl$ entering the decay region is $R(K^\pm/\kl)=4.1\times 10^{-6}$
without the second sweeping magnet.
To evaluate the reduction by the second sweeping magnet, 
we conducted another beam line simulation with a sweeping magnet that provides a magnetic field of 2~Tesla in 1.5~m~long at the end of the beam line.
We confirmed that it can reduce the ratio to $R<1.1\times 10^{-6}$, which is limited by the simulation statistics.

%

\subsubsection{Activities in the experimental area behind the beam dump}
\begin{figure}[ht]
\centering
\includegraphics[width=\linewidth]{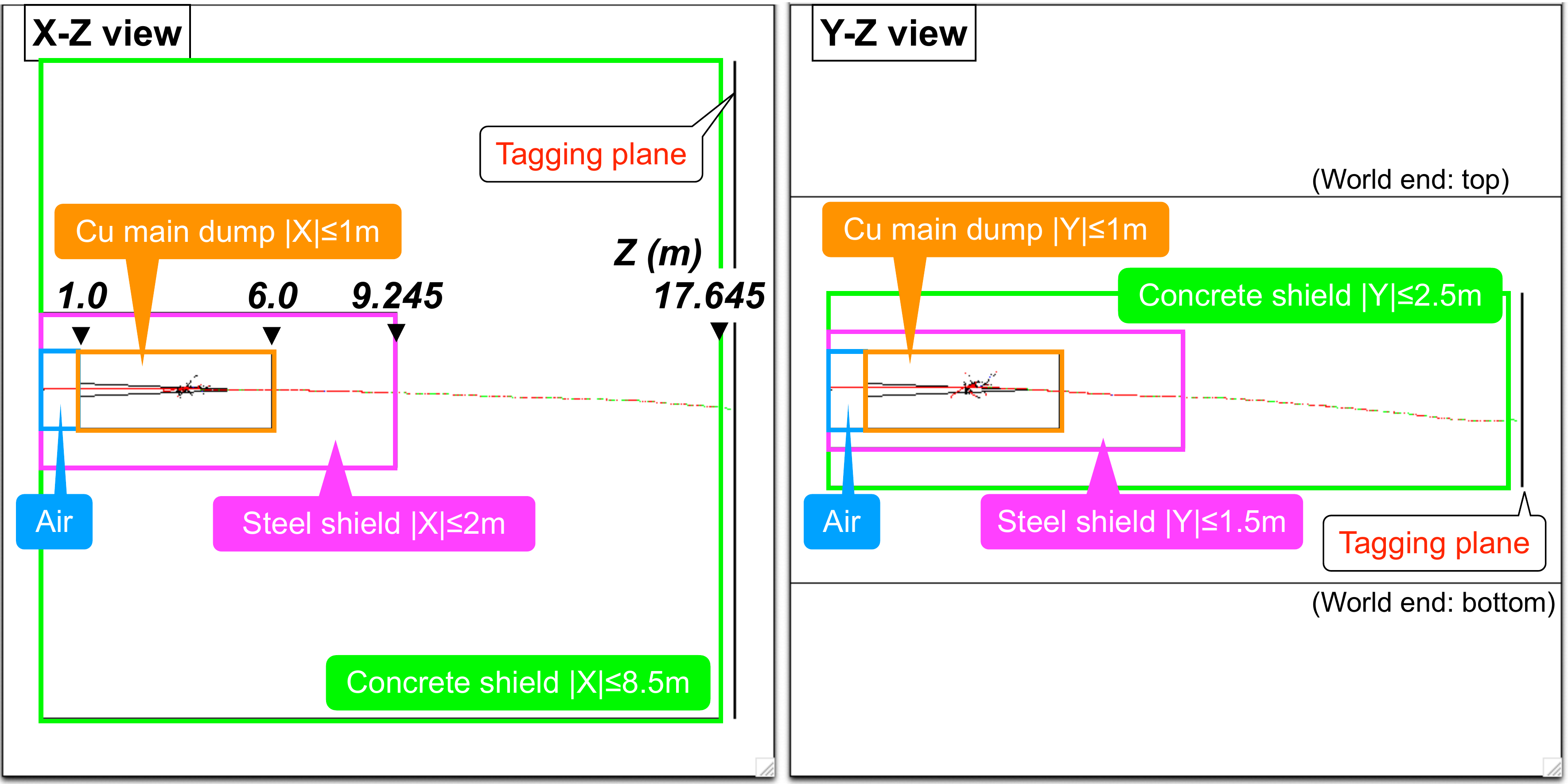}
\caption{\label{fig:bd}Model of the current primary beam dump used in the simulation study
in horizontal (left) and vertical cutaway views.}
\end{figure}

To realize the production angle of 5 degree, the experimental area must be located behind the primary beam dump.
There is a concern that many particles from interactions in the beam dump penetrate the shield and reach the experimental area, which cause a high rate of accidental hits in the detector.
To evaluate the flux in the experimental area, a GEANT3-based simulation of the beam dump was conducted.
Figure~\ref{fig:bd} illustrates the model of the current beam dump.
The main body of the dump is made of copper, on which a tapered hole exists to spread the proton hit position along the beam direction and thus distribute the heat dissipation. Steel and concrete shields follow behind the copper dump.
In this simulation, the proton beam with the size ($\sigma$) of 3~cm is injected into the dump, which roughly indicate the parameter in the current operation.

\begin{figure}[ht]
\centering
 \begin{minipage}{0.45\linewidth}
  \includegraphics[width=0.9\linewidth]{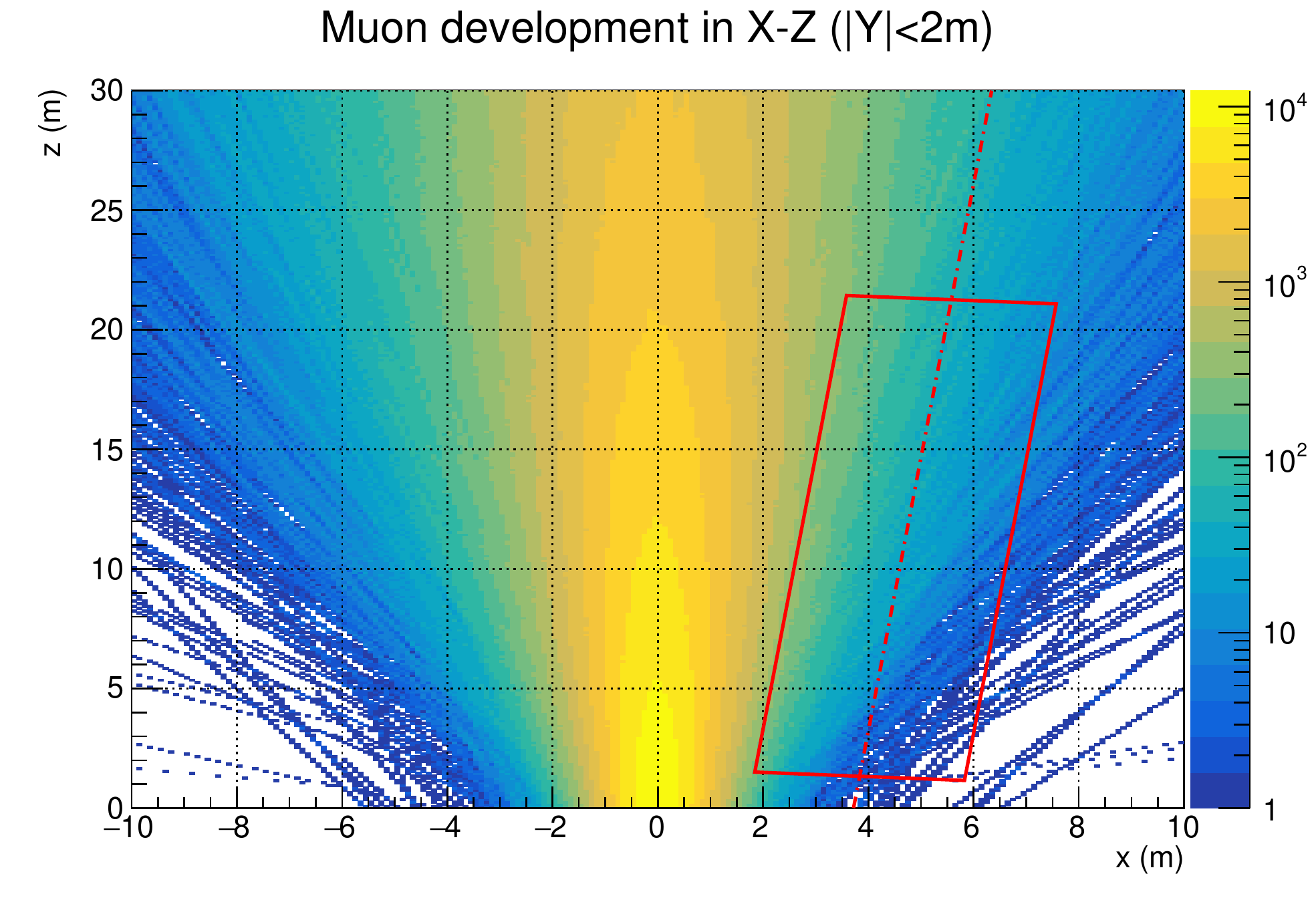}
  \caption{\label{fig:bd-mu-develop}Simulated development of the muon tracks behind the current beam dump, distributed in the horizontal (x) and the beam direction (z). The red box indicates the assumed location of the KOTO step-2 detector, and the red dashed line indicates the center of the KL2 beam line.}
 \end{minipage}
 \hspace{0.05\linewidth}
 \begin{minipage}{0.45\linewidth}
  \includegraphics[width=0.9\linewidth]{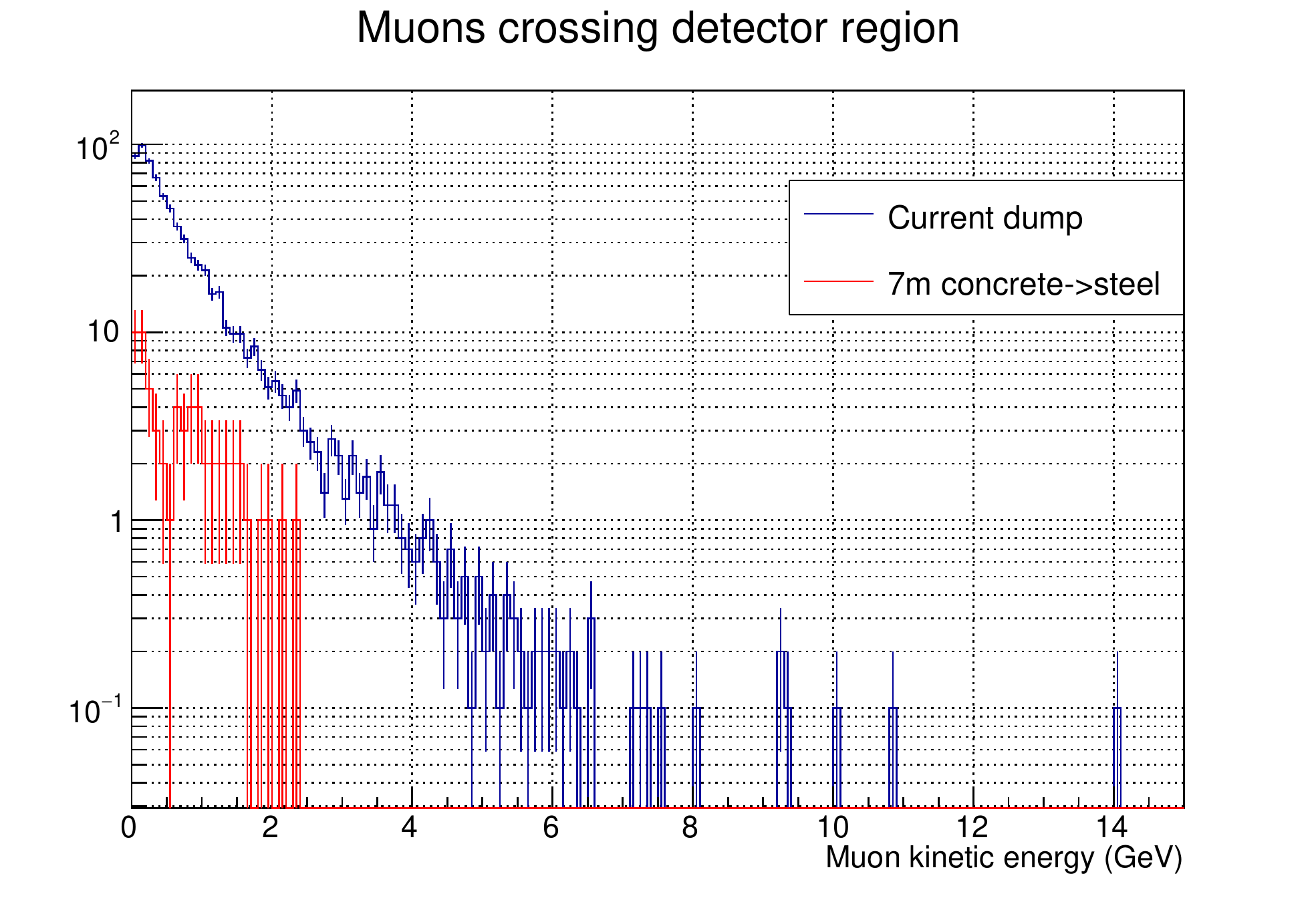}
  \caption{\label{fig:bd-emu}Energy distribution of muons whose extrapolated trajectory crosses the KOTO step-2 detector region in case of the current dump (blue) and the modified dump (red).}  
 \end{minipage}
\end{figure}

Almost all the particles entering the area were muons. 
As shown in Fig.~\ref{fig:bd-mu-develop}, the KOTO step-2 detector is not in the most intense region but still in the region where the flux is high.
The on-spill counting rate due to the muons which cross the detector region was estimated to be 15~MHz when the 100~kW beam is fully injected into the dump.
The contribution of other particles such as neutrons were found to be small, about 25~kHz even integrated over the whole region behind the dump.

To reduce the muon flux, a part of the concrete shield must be replaced with steel shield.
Replacing the material of the 4~m (horizontal) $\times$ 3~m (vertical) $\times$ 7~m (beam direction) volume from concrete to steel reduces the flux by an order of magnitude to 1.3~MHz.
Figure~\ref{fig:bd-emu} shows the energy distribution of the muons crossing the detector region before and after the replacement.

\subsubsection{KOTO step-2 detector}
\begin{figure}[ht]
 \includegraphics[width=\textwidth]{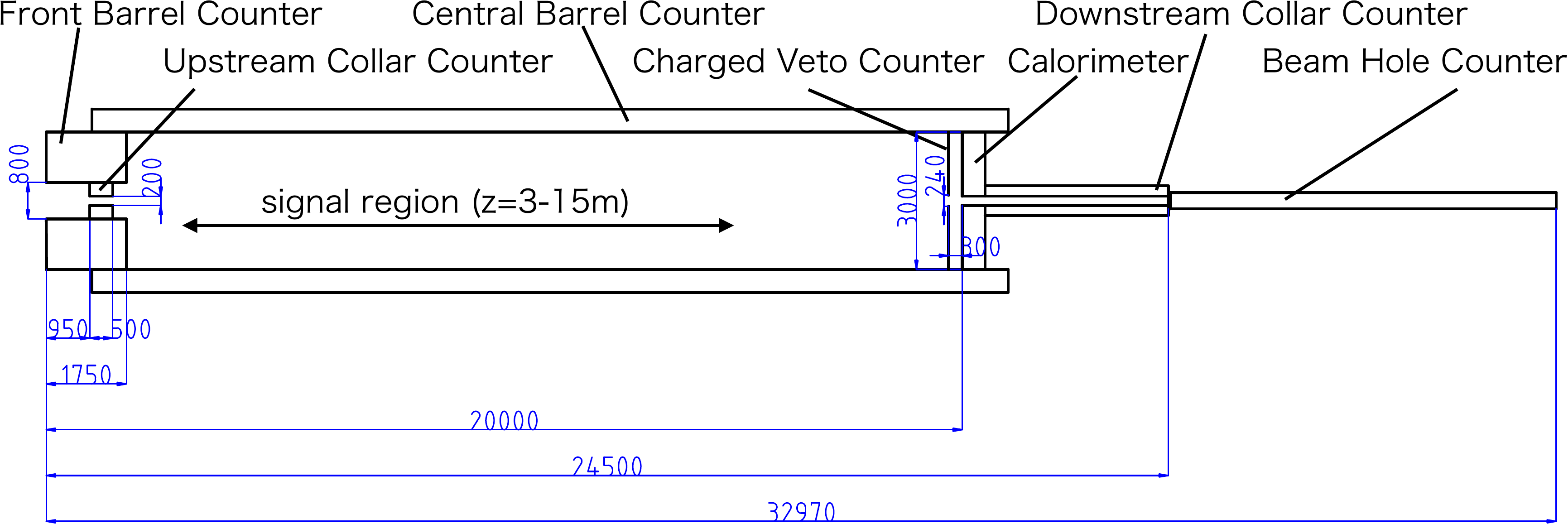}
 \caption{Conceptual KOTO step-2 detector.
 The upstream edge of the Front Barrel Counter is 44~m from T2 target.}
 \label{fig:conceptualDetector}
\end{figure}

A conceptual detector used in the base design is shown in Fig.~\ref{fig:conceptualDetector}.
The $z$ axis is defined on the beam axis pointing downstream
with the origin at the upstream surface of the Front Barrel Counter,
which is 44~m from the T2 target (43-m long beam line and 1-m long space).
The following conceptual detector is designed;
the diameter of the calorimeter is 3~m to gain the signal acceptance.
The beam hole in the calorimeter is 20~cm $\times$ 20~cm to accept
the 15~cm $\times$ 15~cm beam size.
The $z$ position of the calorimeter is 20~m with
a larger decay volume to enhance the $K_L$ decay.
The larger decay volume is effectvie in the KOTO step-2,
because the higher $K_L$ momentum keeps the signal acceptance.

The Charged Veto Counter
is a charged-particle veto counter at 30~cm upstream of the calorimeter
to veto $K_L\to \pi^{\pm} e^{\mp}\nu$, $K_L\to \pi^{\pm} \mu^{\mp}\nu$,
or $K_L\to \pi^+\pi^-\pi^0$.
The beam hole at the Charged Veto Counter is 24~cm $\times$ 24~cm to avoid
$\pi^0$ or $\eta$ produced by
the interaction with neutrons.
The Front Barrel Counter is 1.75-m long and
the Upstream Collar Counter is 0.5-m long
to veto $K_L\to3\pi^0$ decay at upstream of and inside the Front Barrel Counter.
The Central Barrel Counter is 20-m long mainly to veto $K_L\to 2\pi^0$
by detecting extra photons from the decay.
The Downstream Collar Counter is 4-m long to veto particles passing
through the beam hole in the calorimeter but going
outside the beam region.
The Beam Hole Counter covers in-beam region starting from 24.5~m
to veto particles escaping through the calorimeter beam hole.
The Front Barrel Counter, the Upstream Collar Counter,
the Central Barrel Counter,
and the Downstream Collar Counter,
act as either photon and charged-particle veto
for the conceptual design.
For the Beam Hole Counter,
we introduce two separate counters, 
a beam-hole charged-veto counter at the upstream and 
a beam-hole photon-veto counter.
The signal region is defined in a 12-m region from 3~m to 15~m in $z$.

The modeling of the calorimeter was based on the 50-cm long CsI-crystal calorimeter in the current KOTO detector.
Those of the front and central barrel counters, and the upstream and downstream collar counters
were based on lead/plastic-scintillator sandwich counters.
Those of the beam hole charged / photon veto counters were
based on the current KOTO in-beam detectors.

\subsubsection{Beam Condition and Event Selection}
\subsubsubsection{Beam conditions}
The beam condition and running time as shown in Table~\ref{tab:beam}
were assumed in the study.
\begin{table}[ht]
 \centering
 \caption{Assumed beam and running time.}\label{tab:beam}
 \begin{tabular}{lll}\hline
  Beam power & 100~kW &(at 1-interaction-length T2 target) \\
  && ($1.1\times 10^7 K_L/2\times 10^{13}~\mathrm{POT}$)\\
  Repetition cycle &4.2~s  & \\
  Spill length &2~s  & \\
  Running time &$3\times 10^7$~s  & \\\hline
 \end{tabular}
\end{table}

\subsubsubsection{Signal Reconstruction}
We evaluated yields of the signal and backgrounds
with Monte Carlo simulations.
The calorimeter response was simulated either with model responses
or with shower simulations in the calorimeter.
We assumed 50-cm-long CsI crystals for the calorimeter material in the shower simulations.

A shower is generated by a particle incident on the calorimeter.
A cluster is formed based on energy deposits
in the calorimeter segmented in x-y directions.
Assuming the incident particle to be a photon,
the energy and position are reconstructed.
We treat it as a photon in the later analysis
regardless of the original particle species.

\begin{figure}[ht]
 \centering
 \includegraphics[width=0.5\textwidth]{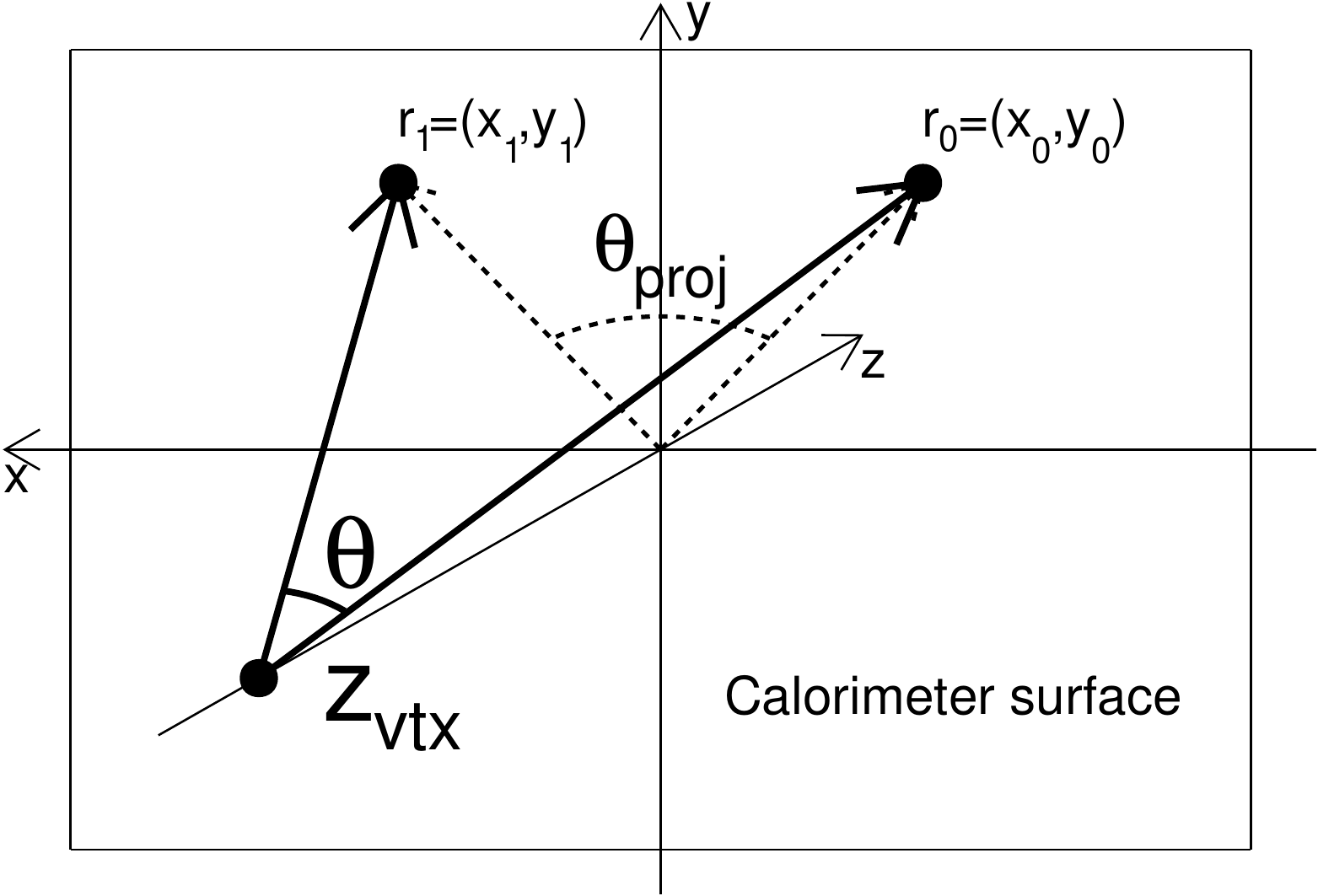}
 \caption{Geometrical relation in the vertex reconstruction.}\label{fig:zvert}
\end{figure}

A $\pi^0$ is reconstructed from the two photons on the calorimeter;
The opening angle of two photon-momentum directions ($\theta$)
can be evaluated with the energies of the two photons ($E_0, E_1$)
from 4-momentum conservation:
\begin{align*}
 {p}_{\pi^0}=&{p}_0+{p}_1,\\
 m_{\pi^0}^2=& 2 E_0 E_1 (1-\cos\theta),
\end{align*}
where ${p}_{\pi^0}$ is four-momentum of the $\pi^0$,
${p}_0$ and ${p}_1$ are four-momenta
of two photons, and $m_{\pi^0}$ is the nominal $\pi^0$ mass.
The vertex position of the $\pi^0$ is assumed to be on the $z$ axis
owing to the narrow beam, and
the $z$ vertex position ($\zvtx$) is calculated from the
geometrical relation among $\theta$ and hit positions
${\bf r}_0=(x_0, y_0)$, ${\bf r_1}=(x_1, y_1)$ on the calorimeter
as shown in Fig.~\ref{fig:zvert}.
The $\zvtx$ gives the momenta of two photons,
and the sum of the momenta gives momentum of $\pi^0$.
Accordingly, the $\pi^0$ transverse momentum ($\pt$) is obtained.

\begin{figure}[ht]
 \centering
 \includegraphics[width=0.5\textwidth]{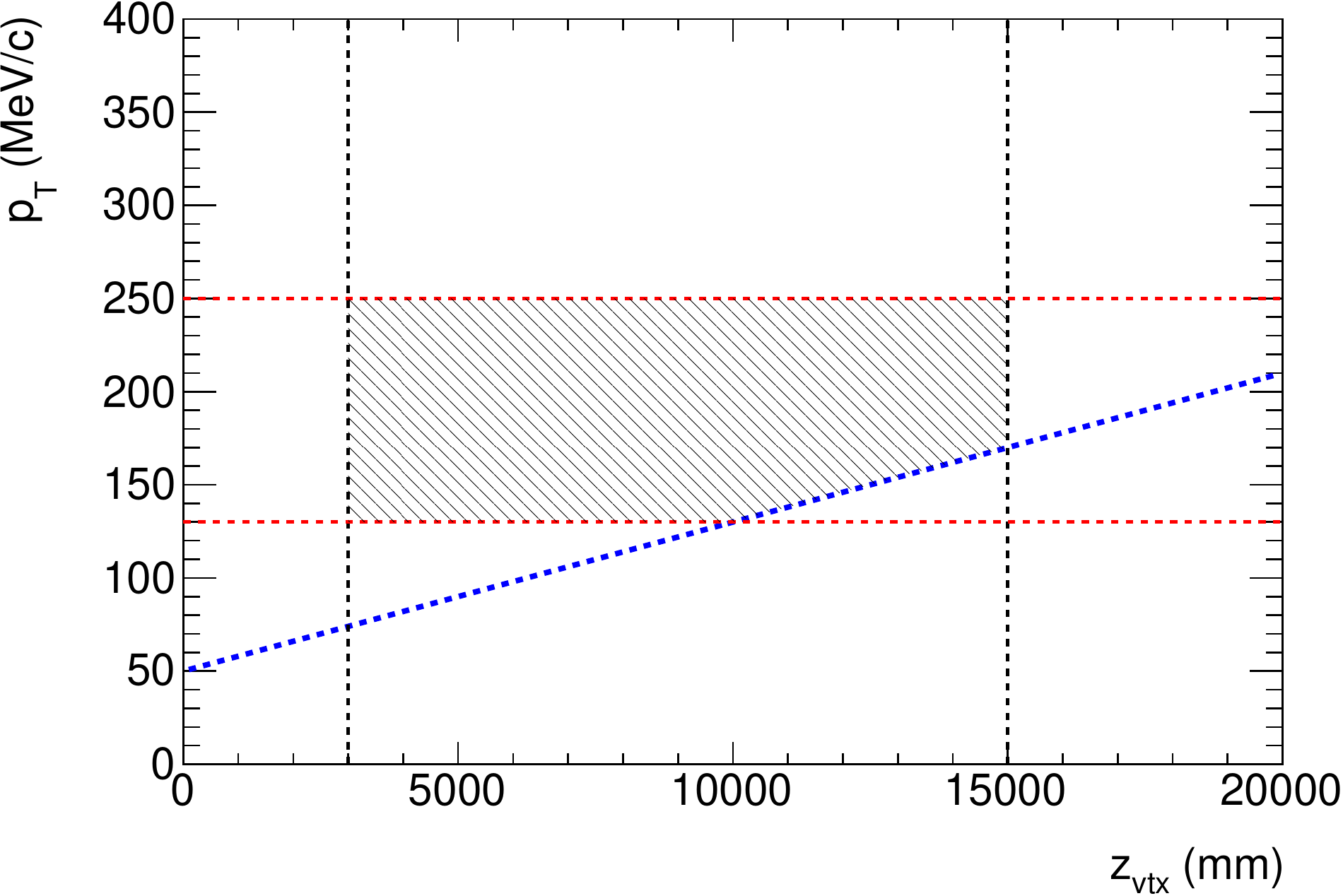}
\caption{The shaded area shows the $\pt$ criteria in the $\zvtx$-$\pt$ plane. The blue dotted line shows the tighter $\pt$ criteria in the downstream region.}
\label{fig:hexiagonal}
\end{figure}

\subsubsubsection{Event selection}\label{sec:cuts}
We use the following event selections
for the events with two clusters in the calorimeter.
\begin{enumerate}
 \item Sum of two photon energies : $E_0+E_1>500~\mathrm{MeV}$.
 \item Calorimeter fiducial area :
       $\sqrt{x_0^2+y_0^2}<1350~\mathrm{mm}$,
       $\sqrt{x_1^2+y_1^2}<1350~\mathrm{mm}$.
 \item Calorimeter fiducial area :
       $\mathrm{max}(|x_0|, |y_0|)>175~\mathrm{mm}$,
       $\mathrm{max}(|x_1|, |y_1|)>175~\mathrm{mm}$.
 \item Photon energy :
       $E_0>100~\mathrm{MeV}$,
       $E_1>100~\mathrm{MeV}$.
 \item Distance between two photons : $|{\bf r}_1-{\bf r}_0|>300~\mathrm{mm}$.
 \item Projection angle ($\theta_{\mathrm{proj}}$ as shown in Fig.~\ref{fig:zvert}) : $\theta_{\mathrm{proj}}\equiv \mathrm{acos}\left( \frac{{\bf r}_0\cdot{\bf r}_1}{|{\bf r}_0||{\bf r}_1|}\right)<150^{\circ}$.
 \item $\pi^0$ decay vertex : $3~\mathrm{m}<\zvtx <15~\mathrm{m}$.
 \item $\pi^0$ transverse momentum : $130~\mathrm{MeV}/c<\pt <250~\mathrm{MeV}/c$.
 \item Tighter $\pi^0$ $\pt$ criteria in the downstream (Fig.~\ref{fig:hexiagonal}):
       $\frac{\pt}{(\mathrm{MeV}/c)}>\frac{\zvtx}{(\mathrm{mm})}\times 0.008+50$.
 \item Selection to reduce hadron cluster background\\
       In order to reduce neutron clusters,
       cluster shape, pulse shape, and 
       depth information of the hits in the calorimeter
       are used as in the analysis of the KOTO step-1.
       The signal selection efficiency of $0.9^3=0.73$ is assumed.
 \item Selection to reduce halo $K_L\to 2\gamma$  background\\
       The photon incident-angle information is used to reduce the halo $K_L\to 2\gamma$  background
       as in the KOTO step-1. The signal selection efficiency of $0.9$ is assumed.
\end{enumerate}
The first 5 selections ensure the quality of the photon cluster;
The sum of the photon energies is useful to reduce a trigger bias,
because we plan to use the sum of the calorimeter energy for the trigger.
The edge region of the calorimeter is avoided to reduce the energy leak outside the calorimeter.
Higher energy photons give good resolution.
Large distance between the two photons
reduces the overlap of two clusters.

The next four are kinematic selections.
The projection angle selection requires
no back-to-back configuration of the two photons to reduce $K_L\to2\gamma$.
Larger $\pi^0$ $\pt$ is required
to match the kinematics of the signal.
The tighter $\pt$ selection is required in the downstream region,
because the reconstructed $\pt$ tends to be larger due to 
worse $\pt$ resolution for the decay near the calorimeter.

The last two are the identification criteria with the calorimeter
to discriminate photon and neutron clusters, or
to discriminate correct and incorrect photon-incident angles.

\subsubsection{Signal sensitivity estimation}

\subsubsubsection{Decay probability and geometrical acceptance of two photons at the calorimeter}
The decay probability is found to be
\begin{align*}
P^{\mathrm{truth}}_{\mathrm{decay}}=&
\frac{\text{Number of $K_L$'s that decayed in $3~\mathrm{m}<z<15~\mathrm{m}$}}
{\text{Total number of $K_L$'s at $z=-1$~m}} = 9.9\%.
\end{align*}
The geometrical acceptance is evaluated to be 
\begin{align*}
A^{\mathrm{truth}}_{\mathrm{geom}}=&
\frac
{\text{Number of $K_L$'s with 2$\gamma$'s in the calorimeter that decayed in $3<z<15$~m}}
{\text{Number of $K_L$'s that decayed in $3~\mathrm{m}<z<15~\mathrm{m}$}} = 24\%.
\end{align*}

\subsubsubsection{Cut acceptance}
The overall acceptance for the conditions 1--6 and 8--9\footnote{The seventh condition, $\zvtx$ selection, is already treated in the previous section.}
listed in Section~\ref{sec:cuts}
is 40\%. The assumed acceptance for all the additional cuts
to reduce the hadron-cluster background
and halo $K_L\to 2\gamma$ is $0.9^4=66\%$. Including all the above, the cut acceptance is 26\%.

\subsubsubsection{Accidental loss}
In order to veto background events,
we set a timing window (veto window)
to detect extra particles 
with respect to the two-photon hit timing at the calorimeter.
The width of the veto window is set to 40 ns for the Central Barrel Counter,
30 ns for the beam-hole charged-veto counter,
6 ns for the beam-hole photon-veto counter,
and 20 ns for the other counters.

When the $\klpionn$ signal is detected with the calorimeter,
another $K_L$ might decay accidentally
and its daughter-particle may hit a counter at the same time.
Similarly, a photon or neutron in the beam might hit the Beam Hole Counter
at the same time.
These accidental hits will veto the signal if the hit timing is within the veto window.
We call this type of signal loss as ``accidental loss''.

The hit rates of counters other than Beam Hole Counters were
evaluated with $\kl$-decay simulations, where any hits in the counters
were included.

For the Beam-Hole Charged Veto Counter,
three layers of a MWPC-type wire chamber~\cite{Kamiji:2017deh} are assumed:
the thickness of the gas volume for each layer is 2.8 mm, and
the cathode plane is a 50-$\mu$m-thick graphite-coated polyimide film.
This design reduces the hit rate from neutral particles,
such as photons, neutrons, and $K_L$.
The layer-hit is
defined as
the energy deposit larger than
1/4 of the minimum-ionizing-particle peak.
The counter-hit is
defined as
two coincident layer-hits out of three layers,
which maintains the charged-particle efficiency
to be better than 99.5\%
with less contribution from neutral particles.

For the Beam-Hole Photon Veto Counter,
25 modules of lead-aerogel Cherenkov counters~\cite{Maeda:2014pga} are assumed.
A high-energy photon generates an $e^+e^-$ pair in the lead plate,
and these generate Cherenkov light in the aerogel radiator.
The Cherenkov light is guided by mirrors to a PMT.
The individual module-hit is defined with 
a 5.5-photoelectron (p.e.) threshold.
The counter-hit is defined with the 
consecutive three-module coincidence.

The hit rate ($r_i$), the veto width $(w_i)$, and
the individual loss $(1-\exp\left(- w_i r_i\right))$
for each detector are summarized
in Table~\ref{tab:rateWidth}.
The total accidental loss is evaluated:
\begin{align*}
 \text{Accidental loss}=&1-\exp\left(-\sum_i w_i r_i\right)\\
 =&39\%.
\end{align*}
\begin{table}[ht]
 \centering
 \caption{Summary of rate, veto width, and individual accidental loss.}\label{tab:rateWidth}
 \begin{tabular}{llll}
  Detector&Rate(MHz) & Veto width (ns) & Individual loss (\%)\\ \hline
  Front Barrel Counter&0.18 &20 & 0.4\\
  Upstream Collar Counter&0.80 &20 & 1.6\\
  Central Barrel Counter&2.21 &40 & 8.5\\
  Calorimeter&3.45 &20 & 6.7\\
  Downstream Collar Counter&0.97 &20 & 1.9\\\hline
  Beam-hole charged-veto&2.9 &30 & 8.3\\
  Beam-hole photon-veto&35.2 &6 & 19\\
  \hline
 \end{tabular}
\end{table}

\subsubsubsection{Shower-leakage loss}\label{sec:backsplash}
When two photons from the $\klpionn$ decay are detected
in the calorimeter,
the shower can leak both the downstream and upstream of the calorimeter,
and make hits on the other counters such as the Central Barrel Counter.
Such hits will veto the signal if the hit timing is within the veto window.
We call this signal loss ``shower-leakage loss''.
In particular,
we call the loss caused by
shower leakage toward the upstream
``backsplash loss''.

The Charged Veto Counter covers
the upstream side of the calorimeter, and would
suffer from the backsplash.
This backsplash loss was neglected assuming that 
the back-splash contribution can be distinguished
with the timing difference between the Charged Veot Counter and the calorimeter.

The large coverage of the Central Barrel Counter gives significant effect
on the backsplash loss. First, we introduce a timing definition
of the Central Barrel Counter ($t_{\mathrm{BarrelVeto}}$):
\begin{align*}
 t_{\mathrm{BarrelVeto}}=&
 t_{\mathrm{BarrelHit}}-
 \left[t_{\mathrm{CalorimeterHit}}
 -\frac{z_{\mathrm{Calorimeter}}-z_{\mathrm{BarrelHit}}}{c}
 \right].
\end{align*}

\begin{figure}[ht]
 \centering
 \includegraphics[page=2,width=0.5\textwidth]{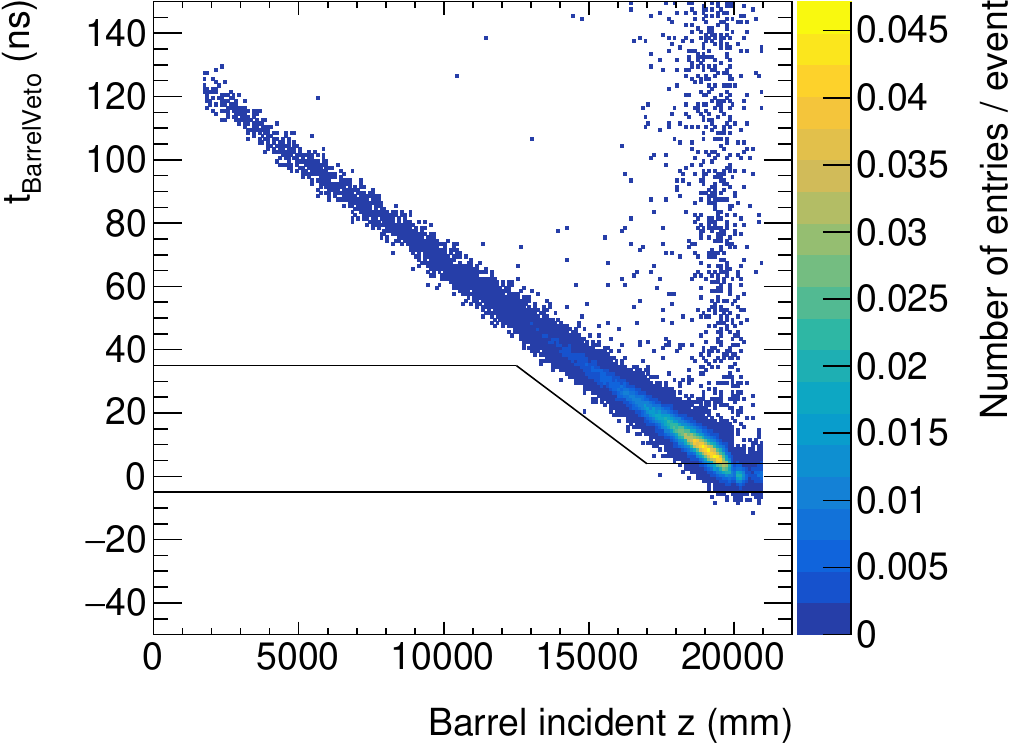}
 \caption{Incident timing of the backsplash particles
 on the Central Barrel Counter
 as a function of the incident $z$ position.
 The smearing with the timing resolution is applied
 depending on the incident energy.
 }\label{fig:BStiming}
\end{figure}

Figure~\ref{fig:BStiming} shows the incident timing of the
backsplash particles on the Central Barrel Counter
as a function of the incident $z$ position.
The timing-smearing is applied with the timing resolution
depending on the incident energy.
Events with larger $t_{\mathrm{BarrelVeto}}$ at $z\sim 20000~\mathrm{mm}$
are generated with neutrons from the electromagnetic shower.
Events with smaller $z$ tend to have larger $t_{\mathrm{BarrelVeto}}$
due to longer flight distance.
This clear correlation can be used to exclude the backsplash particles
from the veto to reduce the backsplash loss.
We loosen the veto criteria
at the downstream region:
for $z$ from 12.5~m to 17~m, the largest timing of the veto window
is changed from 35~ns to 4~ns linearly, and 
for $z>17~\mathrm{m}$, it is 4~ns.
The region within the two lines in the figure shows the 
new veto region, which gives the survival probability
of $91\%$, equivalently the backsplash loss of 9\%.

In addition, if we added $z$
smearing using a Gaussian with the $\sigma$ of 500~mm,
the same survival probability is obtained.

\begin{figure}[ht]
 \centering
 \includegraphics[page=8,width=0.5\textwidth]{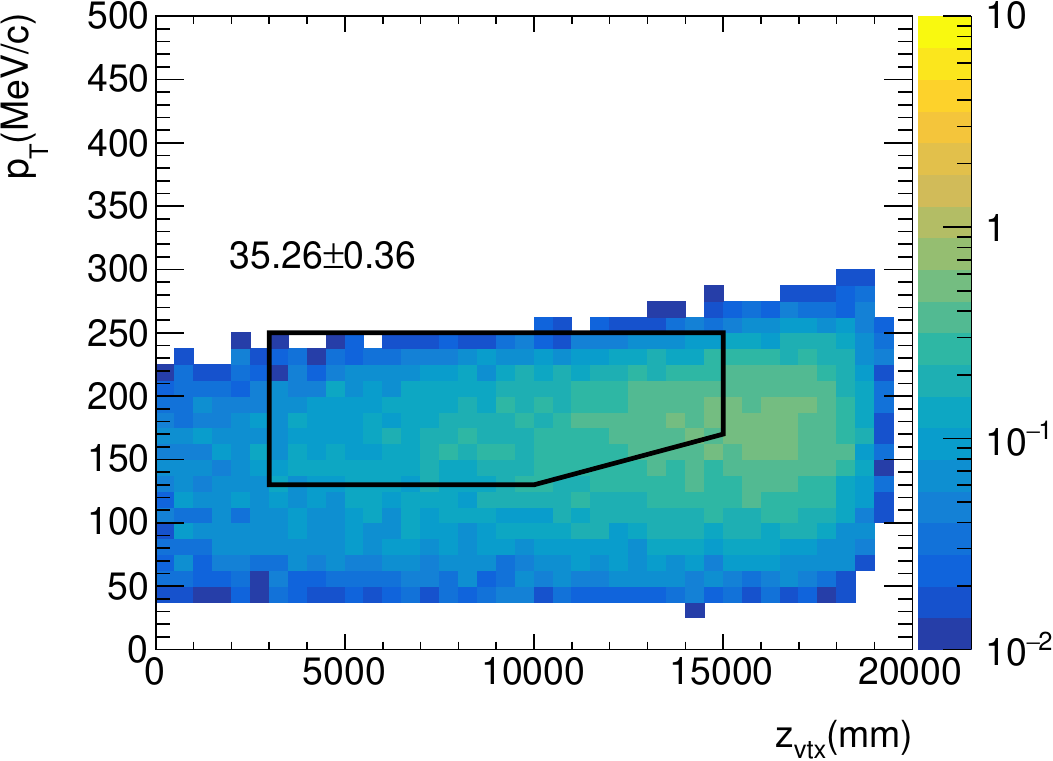}
 \caption{Distribution in the $\zvtx$-$\pt$ plane for $\klpionn$
 for the running time of $3\times 10^7$~s.
 All the cuts other than $\pt$ and $\zvtx$ cuts are applied.
 }
 \label{fig:ptz_pi0nn}
\end{figure}

\subsubsubsection{Signal yield}
We evaluate the number of the signal ($S$) in $3\times 10^7$ s running time
with $\mathrm{BR}_{\mathrm{\klpionn}}=3\times 10^{-11}$:
\begin{align*}
 S=&
 \frac{(\text{beam power})\times (\text{running time})}
 {(\text{beam energy})}
 \times (\text{number of }K_L\text{/POT})\\
 &\times P_{\mathrm{decay}}
 \times A_{\mathrm{geom}}
 \times A_{\mathrm{cut}}
 \times (\text{1-accidental loss})
 \times (\text{1-backsplash loss}) \times \mathcal{B}_{\klpionn}\\
 =&
 \frac{
 (100~\mathrm{kW})\times (3\times 10^7~\mathrm{s})
 }
 {
 (30~\mathrm{GeV})
 }
 \times
 \frac{(1.1\times 10^7 K_L)}
 {(2\times 10^{13}~\mathrm{POT})}\\
 &\times 9.9\% \times 24\% \times 26\% \times (1-39\%) \times 91\%
 \times (3\times 10^{-11})\\
 =&35.
\end{align*}
Here, $P_{\mathrm{decay}}$ is the decay probability,
$A_{\mathrm{geom}}$ is the geometrical acceptance for the two photons
entering the calorimeter,
$A_{\mathrm{cut}}$ is the cut acceptance, and
$\mathcal{B}_{\klpionn}$ is the branching fraction of $\klpionn$.
Distribution in the $\zvtx$-$\pt$ plane is shown in Fig.~\ref{fig:ptz_pi0nn}.

\subsubsection{Background estimation}

\begin{figure}[ht]
 \centering
 \includegraphics[page=12,width=0.5\textwidth]{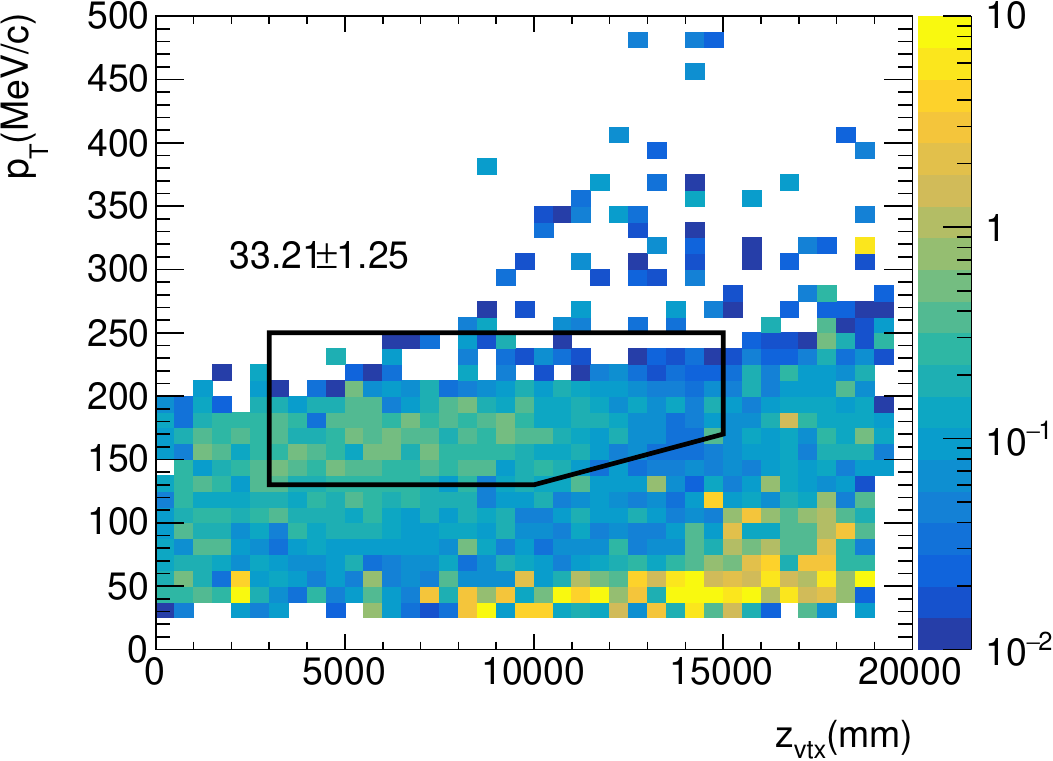}
 \caption{Distribution in the $\zvtx$-$\pt$ plane for $\klpiopio$
 background events
 for the running time of $3\times 10^7$~s.
 All the cuts other than $\pt$ and $\zvtx$ cuts are applied.
 }
 \label{fig:ptz_pi0pi0}
\end{figure}

\subsubsubsection{$\klpiopio$}
$\klpiopio$ becomes a background
when two clusters are formed at the calorimeter and
the other photons are missed in the following cases.
\begin{enumerate}
 \item Fusion background: Three photons enter the calorimeter,
       and two of them are fused into one cluster.
       The other one photon is missed due to the detector inefficiency.
 \item Even-pairing background: Two photons from a $\pi^0$-decay
       form two clusters in the calorimeter. Two photons from
       the other $\pi^0$ are missed due to the detector inefficiency.
 \item Odd-pairing background : One photon from a $\pi^0$ and
       one photon from the other $\pi^0$ form two clusters in the calorimeter.
       The other two photons are missed due to the detector inefficiency.
\end{enumerate}
The number of this background is evaluated to be 33.2,
in which 2 event come from the fusion, 27 events from the even-pairing,
4 events from the odd-pairing background.
Fig.~\ref{fig:ptz_pi0pi0} shows the distribution 
in the $\zvtx$-$\pt$ plane with all the cuts
other than the $\zvtx$ and $\pt$ selections.

Among the 33.2 background events,
both two photons are missed in the Central Barrel Counter
for 29 events, which gives the dominant contribution.

\begin{figure}[ht]
 \centering
 \includegraphics[page=20,width=0.5\textwidth]{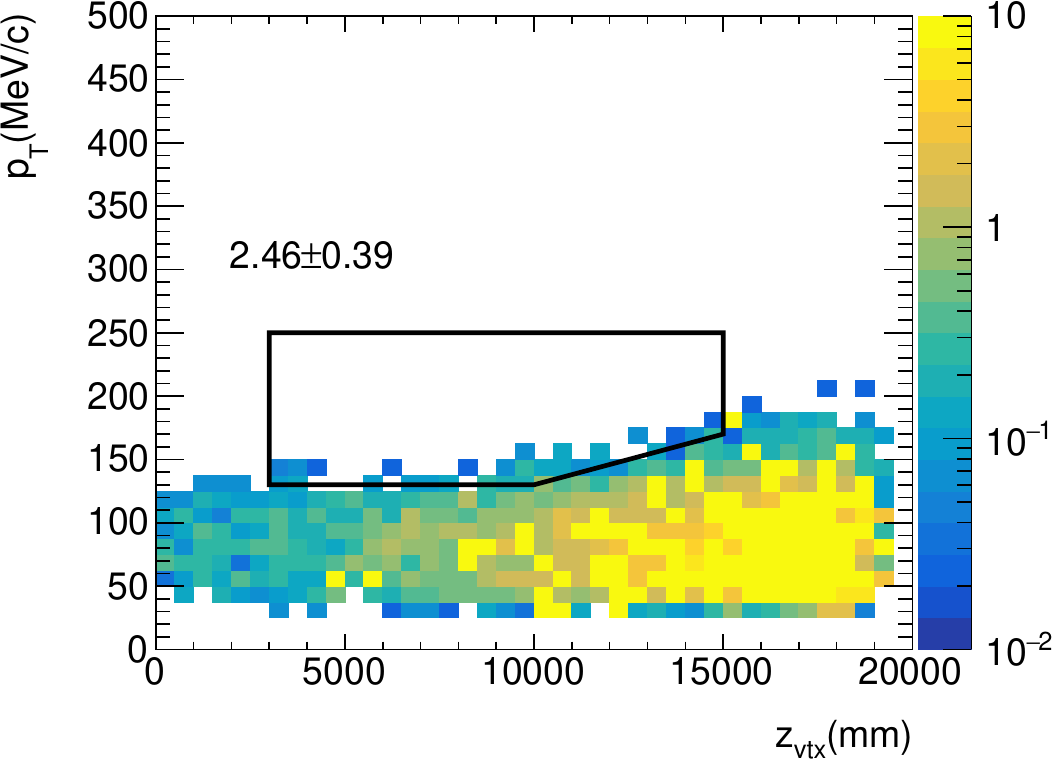}
 \caption{Distribution in the $\zvtx$-$\pt$ plane for $\klppm$
 background
 for the running time of $3\times 10^7$~s.
 All the cuts other than $\pt$ and $\zvtx$ cuts are applied.}
 \label{fig:ptz_pipipi0}
\end{figure}

\subsubsubsection{$\klppm$}
$\klppm$ becomes a background when $\pi^+$ and $\pi^-$ are not detected.
The number of this background events
is evaluated to be 2.5
as shown in Fig.~\ref{fig:ptz_pipipi0},
where one charged pion is lost in the Charged Veto Counter,
and the other is lost in the Beam Hole Counter.
The maximum $\pt$ of the reconstructed $\pi^0$ is
limited with the kinematics and the $\pt$ resolution.
The tighter $\pt$ selection in the downstream
makes the pentagonal cut in the $\zvtx$-$\pt$ plane
as shown in the figure.
This cut reduces the background
because the $\pt$ resolution is worse in the downstream.

\begin{figure}[ht]
 \centering
 \includegraphics[page=2,width=0.5\textwidth]{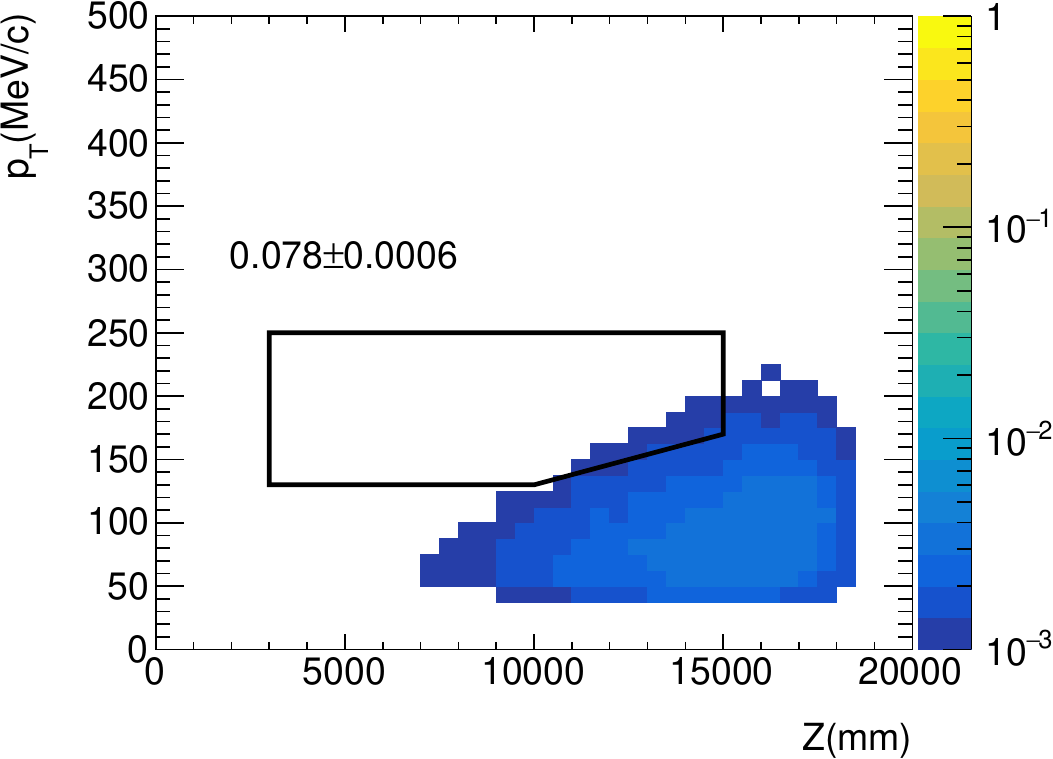}
 \caption{Distribution in the $\zvtx$-$\pt$ plane for Ke3
 background
 for the running time of $3\times 10^7$~s.
 All the cuts other than $\pt$ and $\zvtx$ cuts are applied.}
 \label{fig:ptz_ke3}
\end{figure}

\subsubsubsection{$\kpien$ (Ke3)}
The Ke3 background happens when the electron and the charged pion are not
identified with the Charged Veto Counter.
The number of background events is evaluated to be
0.08
with $10^{-12}$ reduction with the Charged Veto Counter 
as shown in Fig.~\ref{fig:ptz_ke3}.

\begin{figure}[ht]
 \centering
 \includegraphics[page=2,width=0.5\textwidth]{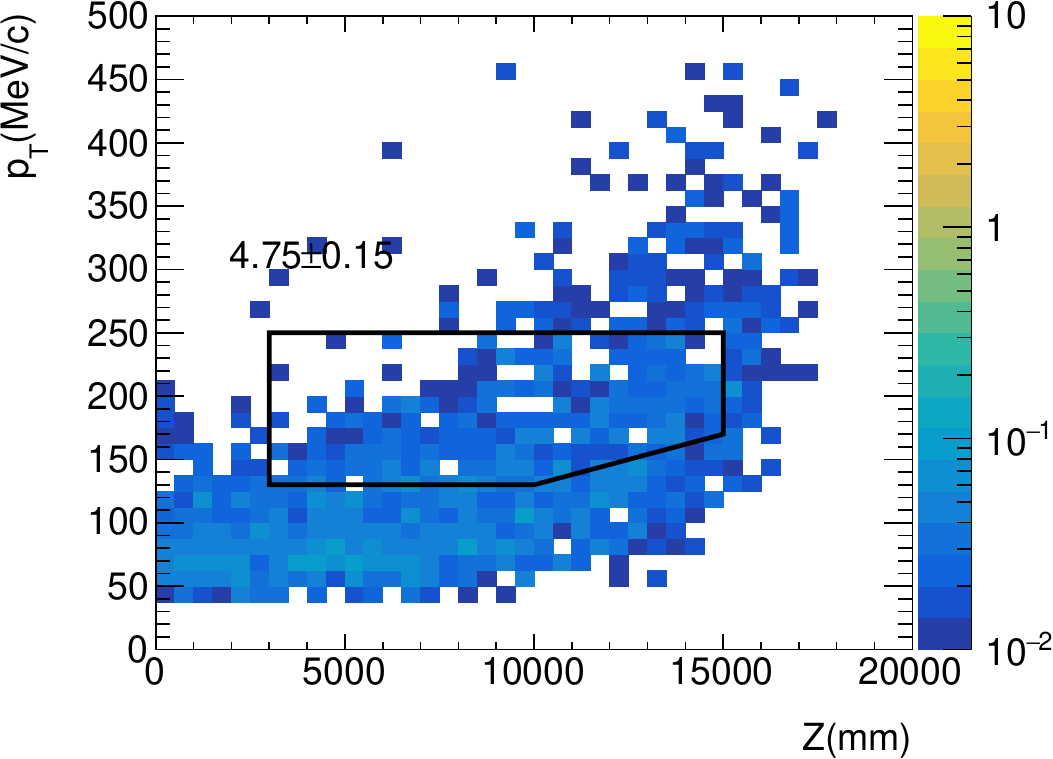}
 \caption{Distribution in the $\zvtx$-$\pt$ plane
 for halo $K_L\to2\gamma$ background
 for the running time of $3\times 10^7$~s.
 All the cuts other than $\pt$ and $\zvtx$ cuts are applied.}
 \label{fig:ptz_haloK}
\end{figure}

\subsubsubsection{$\klgg$ for halo $K_L$} \label{sec:haloKL}
$K_L$ in the beam scatters at the beam line components, and exists in the
beam halo region. When such a halo $K_L$ decays into two photons,
larger $\pt$ is possible
due to the assumption of the vertex on the $z$-axis.
The decay vertex is wrongly reconstructed
with the nominal pion mass assumption.
This fake vertex gives a wrong photon-incident angle.
Therefore, this halo $K_L\to2\gamma$ background can be reduced with
incident-angle information at the calorimeter.
We can reconstruct
another vertex with the nominal $K_L$ mass assumption,
which gives a correct photon-incident angle.
By comparing the observed cluster shape to 
those from the incorrect and correct photon-incident angles,
this background is reduced to be 10\% in the KOTO step-1, 
while keeping 90\% signal efficiency.
In this report, we assume that the background is reduced
to be 1\%,
because the higher energy photon in the step-2 will give
a better resolution in the photon-incident angle.
We will study it more in the future.
The number of this background is evaluated to be 4.8
as shown in Fig.~\ref{fig:ptz_haloK}.
For the halo $K_L$ generation in this simulation,
the core $K_L$ momentum spectrum and 
the halo neutron directions were used.
Systematic uncertainties on the flux and spectrum
are also one of the future studies.

\subsubsubsection{$K^\pm\to\pi^0 e^\pm\nu$}
$K^\pm$ is generated in the interaction of $K_L$,
neutron, or $\pi^\pm$
at the collimator in the beam line.
Here we assume that 
the second sweeping magnet
near the entrance of the detector
will reduce the contribution to be 10\%.

Higher momentum $K^\pm$ can survive
in the downstream of the second magnet,
and $K^\pm\to\pi^0 e^\pm\nu$
decay occurs in the detector.
This becomes a background if $e^\pm$ is undetected.
The kinematics of $\pi^0$ is similar to
$K_L\to\pi^0\nu\overline{\nu}$, therefore
this is one of the serious backgrounds.
Detection of $e^\pm$ is one of the keys to reduce the background.

\begin{figure}[ht]
 \centering
 \includegraphics[page=2,width=0.5\textwidth]{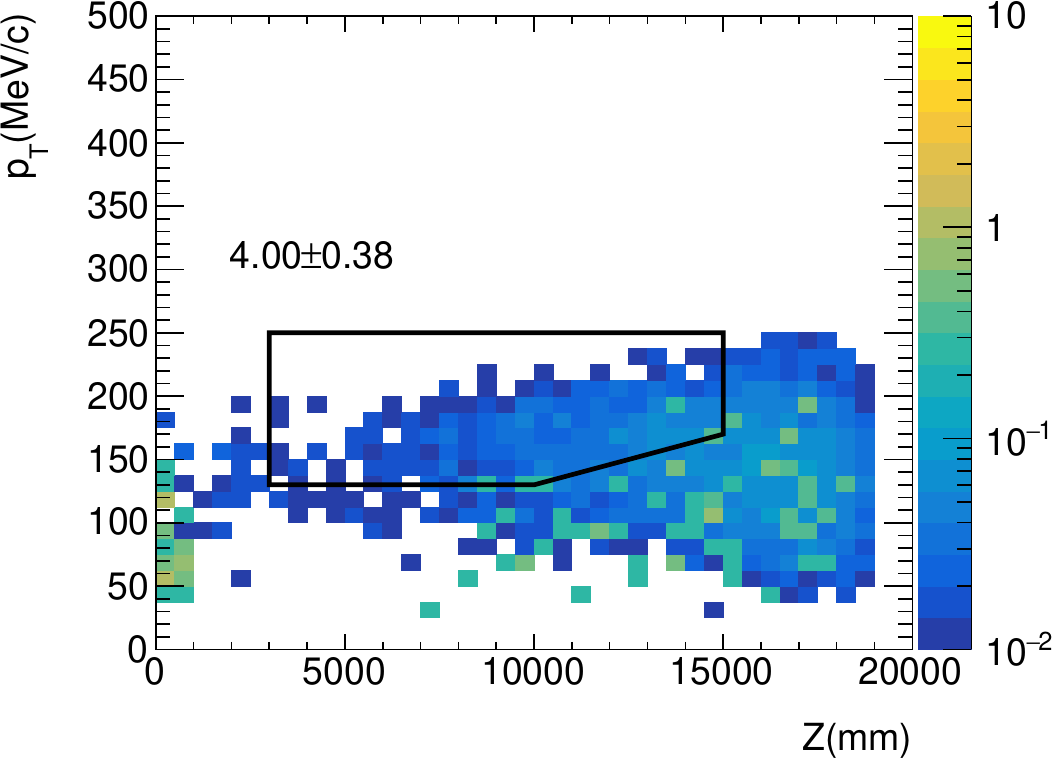}
 \caption{Distribution in the $\zvtx$-$\pt$ plane
 for $K^\pm\to\pi^0 e^\pm\nu$
 for the running time of $3\times 10^7$~s.
 All the cuts other than $\pt$ and $\zvtx$ cuts are applied.}
 \label{fig:ptz_Kplus}
\end{figure}

We evaluated the number of the background events to be 4.0 as shown
in Fig.~\ref{fig:ptz_Kplus}.
In the current beam line simulation, statistics is not large enough.
We use the $K_L$ momentum spectrum and directions
for the $K^\pm$ generation.
We will evaluate these with more statistics with the beam line simulation in the future.

\subsubsubsection{Hadron cluster}\label{sec:hadronCluster}
A halo neutron hits the calorimeter to produce a first hadronic shower, and another neutron in the shower travels inside the calorimeter, and produces a second hadronic shower apart from the first one. These two hadronic clusters mimic the signal.

We evaluated this background using halo neutrons prepared
with the beam line simulation,
and with a calorimeter composed of 50-cm-long CsI crystals.
A full-shower simulation was performed with those neutrons.

\begin{figure}[ht]
 \centering
 \includegraphics[page=2,width=0.5\textwidth]{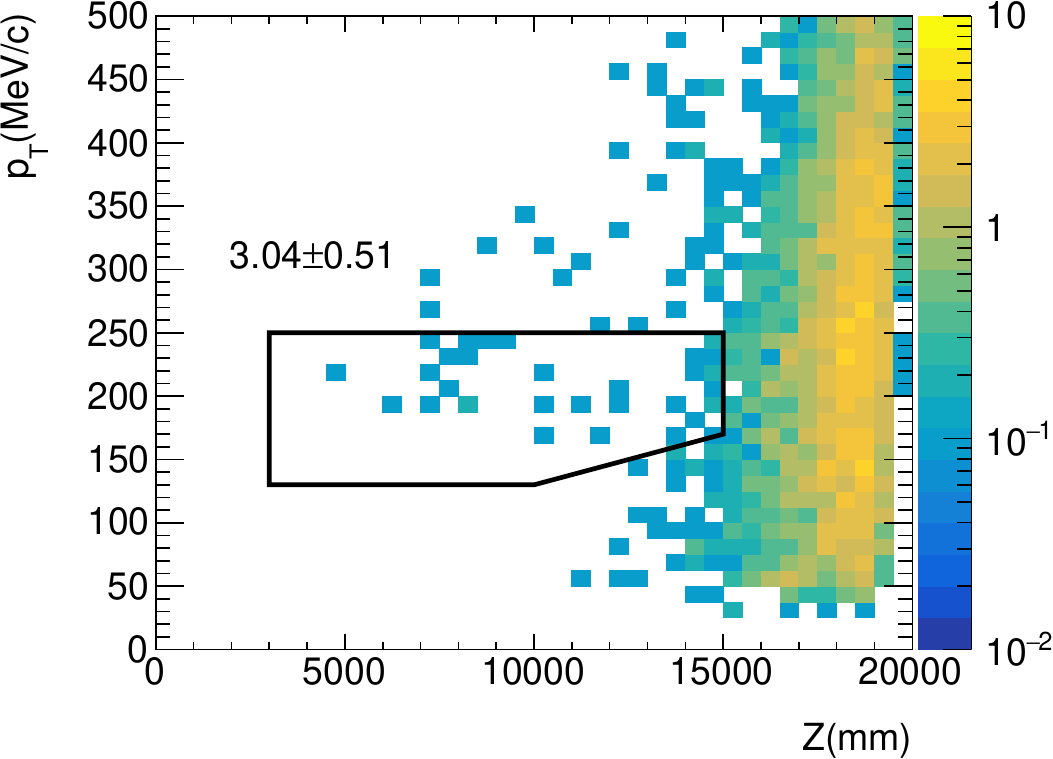}
 \caption{Distribution in the $\zvtx$-$\pt$ plane
 of the hadron cluster background
 for the running time of $3\times 10^7$~s.
 All the cuts other than $\pt$ and $\zvtx$ cuts are applied.}
 \label{fig:ptz_NN}
\end{figure}

We evaluated the number of background events to be 3.0
as shown in Fig.~\ref{fig:ptz_NN}.
Here we assume $10^{-7}$ reduction with
the cluster-shape information, pulse-shape information,
and shower depth information in the calorimeter.
In the KOTO step-1, we achieved to reduce the background to be
$\times ((2.5 \pm 0.01)\times 10^{-6})$
with 72\% signal efficiency
by using cluster and pulse shapes.
By using the shower depth information,
we also achieved to reduce it to be  $\times (2.1\times 10^{-2})$
with 90\% signal efficiency with small correlation to
the cluster and pulse shape cuts.
In total, $10^{7}$ reduction is feasible.
This reduction power is one of the requirements on the calorimeter design.

\subsubsubsection{$\pi^0$ production at the Upstream Collar Counter}
If a halo neutron hits the Upstream Collar Counter, and produces a $\pi^0$,
which decays into two photons, it mimics the signal.

\begin{figure}[ht]
 \centering
 \includegraphics[page=2,width=0.5\textwidth]{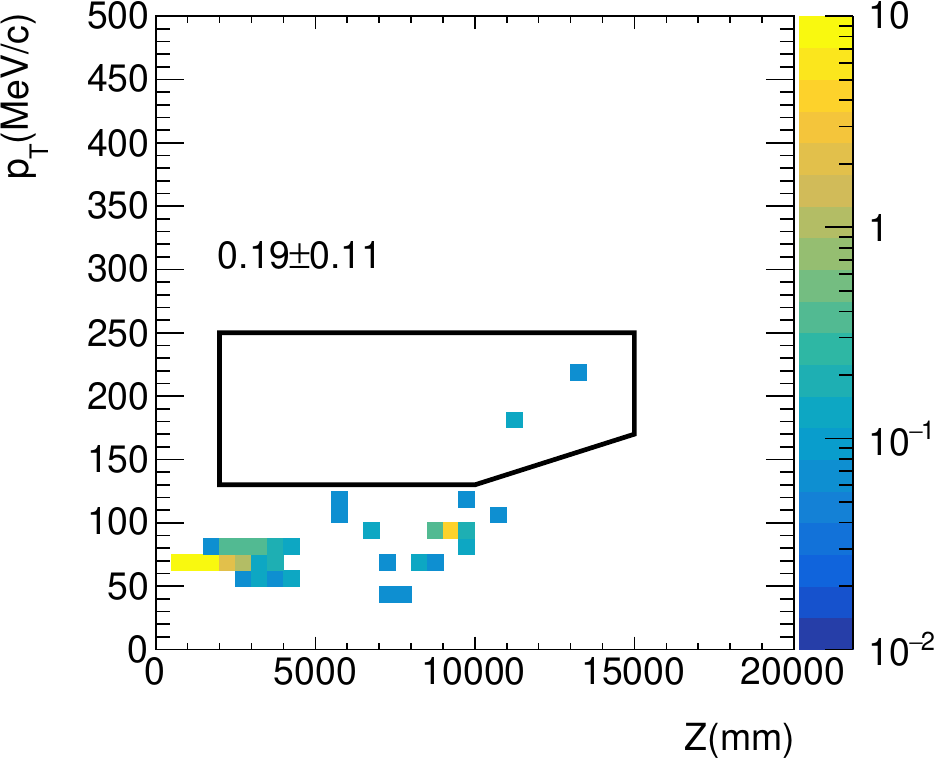}
 \caption{Distribution in the $\zvtx$-$\pt$ plane
 of the background from $\pi^0$ production at the
 Upstream Collar Counter
 for the running time of $3\times 10^7$~s.
 All the cuts other than $\pt$ and $\zvtx$ cuts are applied.}
 \label{fig:ptz_NCCpi0}
\end{figure}

Halo neutrons obtained from the beam line simulation are used
to simulate the $\pi^0$ production in the Upstream Collar Counter.
We assume fully-active CsI crystals for the detector.
Other particles produced in the $\pi^0$ production can deposit energy
in the detector, and can veto the event.
In the simulation, such events were discarded at first.
In the next step, only the $\pi^0$-decay was generated
in the Upstream Collar Counter.
Two photons from the $\pi^0$ can also interact 
the Upstream Collar Counter, and deposit energy in the counter.
Such events were also discarded.
The $\pi^0$ production near the downstream surface of the
Upstream Collar Counter mainly survives.
Finally when the two photons hit the calorimeter,
a full shower simulation was performed.
In this process, photon energy can be mis-measured due to
photo-nuclear interaction.
Accordingly the distribution of the events in the $\zvtx$-$\pt$ plane
was obtained as shown in Fig.~\ref{fig:ptz_NCCpi0}.
We evaluated the number of background events to be 0.19.

\subsubsubsection{$\eta$ production at the Charged Veto Counter}
A halo neutron hits the Charged Veto Counter, and produces a $\eta$.
The $\eta$ decays into two photons with the branching fraction of 39.4\%,
which can mimic the signal.
The decay vertex will be reconstructed at the upstream of the
Charged Veto Counter because the $\eta$ mass is
four times larger than the $\pi^0$ mass.

\begin{figure}[ht]
 \centering
 \includegraphics[page=2,width=0.5\textwidth]{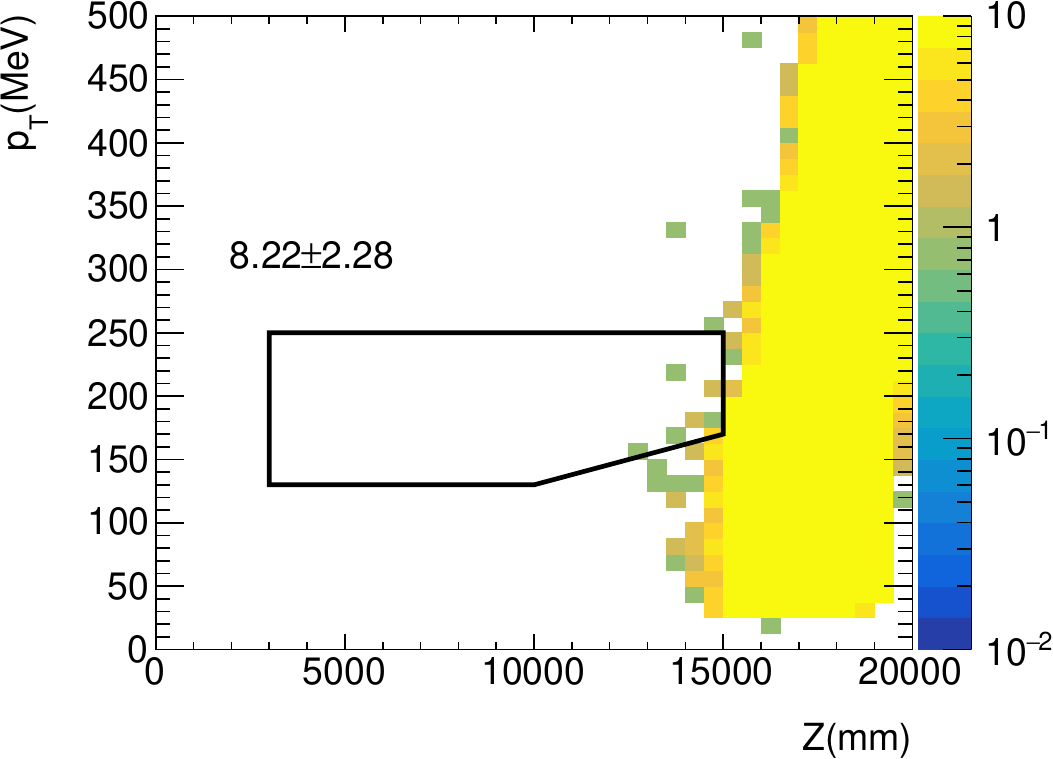}
 \caption{Distribution in the $\zvtx$-$\pt$ plane
 of the background from $\eta$ production at the Charged Veto Counter
 for the running time of $3\times 10^7$~s.
 All the cuts other than $\pt$ and $\zvtx$ cuts are applied.}
 \label{fig:ptz_CVeta}
\end{figure}

Halo neutrons obtained from the beam line simulation are used
to simulate the $\eta$ production in the Charged Veto Counter.
We assume a 3-mm-thick plastic scintillator at 30-cm upstream of
the calorimeter.
Other particles produced in the $\eta$ production can deposit energy in the
Charged Veto Counter, and can veto the event.
In the simulation, such events were discarded at first.
In the next step, only the $\eta$-decay was generated.
When two photons from the $\eta$ hit the calorimeter,
a full-shower simulation was performed.
Two clusters were formed, and 
the distribution of the events in the $\zvtx$-$\pt$ plane
was obtained as shown in Fig.~\ref{fig:ptz_CVeta}.
We evaluated the number of background events to be 8.2.

\subsubsubsection{Summary of the background estimation}
A summary of  the background estimations is shown in Table~\ref{tab:bg}.
The total number of background events is
$56.0\pm 2.8$, where $33.2\pm 1.3$ comes from
the $\klpiopio$ decay.
\begin{table}[ht]
 \centering
 \caption{Summary of background estimations.}\label{tab:bg}
 \begin{tabular}{lrr}\hline
  Background&Number&\\\hline
  $\klpiopio$&33.2 &$\pm$1.3 \\
  $\klppm$&2.5 &$\pm$0.4 \\
  $\kpien$& 0.08 & $\pm$0.0006\\
  halo $K_L\to 2\gamma$&4.8 &$\pm$0.2 \\
  $K^\pm\to\pi^0 e^\pm\nu$&4.0 &$\pm$0.4 \\
  hadron cluster&3.0 &$\pm$0.5 \\
  $\pi^0$ at upstream&0.2 &$\pm$0.1 \\
  $\eta$ at downstream&8.2 &$\pm$2.3 \\ \hline
  Total& 56.0 & $\pm$2.8 \\\hline
 \end{tabular}
\end{table}

\subsubsection{Outlook at J-PARC}
We assume $3\times 10^7$ s running time with
100 kW beam on a 1-interaction-length T2 target,
where the $K_L$ flux is
$1.1\times 10^7$ per $2\times 10^{13}$ protons on target.

The sensitivity and the impacts are calculated and 
summarized in Table~\ref{tab:sensitivity}.
Here we assume
that 
the statistical uncertainties in the numbers of events
are dominant.

The single event sensitivity is evaluated to be 
$8.5\times 10^{-13}$.
The expected number of background events is 56.
With the SM branching fraction of $3\times 10^{-11}$,
35 signal events are expected with a signal-to-background ratio (S/B) of 0.63.
 \begin{itemize}
  \item $4.7$-$\sigma$ observation is expected
	for the signal branching fraction
	$\sim 3\times 10^{-11}$.
  \item It indicates new physics at the 90\% confidence level (C.L.) if
	the new physics gives
	$44\%$ deviation on the BR from the SM prediction.
  \item It corresponds to $14\%$ measurement of the CP-violating CKM parameter $\eta$ in the SM (The branching fraction is proportional to $\eta^2$). 
 \end{itemize}

\begin{table}[ht]
 \centering
 \caption{Summary of the sensitivity and the impact.}
 \label{tab:sensitivity}
 \begin{threeparttable}
 \begin{tabular}{lcc}\hline
 & Formula & Value \\
 \hline
  Signal (branching fraction : $3\times 10^{-11}$) & $S$ & $35.3 \pm 0.4$\\
  Background & $B$ & $56.0 \pm 2.8$ \\
  \hline
  Single event sensitivity & $(3\times 10^{-11})/S$ & $8.5\times 10^{-13}$\\
  Signal-to-background ratio & $S/B$ & 0.63 \\
  Significance of the observation&  $S/\sqrt{B}$ & $4.7 \sigma$\\
  90\%-C.L. excess / deficit & $1.64\times \sqrt{S+B}$ & 16 events \\
  & $1.64\times \sqrt{S+B}/S $ & 44\% of SM\\
  Precision on branching fraction & $\sqrt{S+B}/S$ & 27\%\\
  Precision on CKM parameter $\eta$ &$0.5\times \sqrt{S+B}/S$ &14\%\\
  \hline
 \end{tabular}
 \begin{tablenotes}\footnotesize
 \item[*] Running time of $3\times 10^7$~s is assumed in the calculation.
 \end{tablenotes}
 \end{threeparttable}
\end{table}

\section{Experiments at the CERN SPS}

\subsection{NA62 until LS3}
\label{sec:na62_ls3}

As discussed in \Sec{sec:kp_pnn_stat}, NA62 has performed the first measurement of the $K^+\to\pi^+\nu\bar\nu$ branching ratio with $3.4\sigma$ significance, having collected 20~candidate events in 2016--2018 with an expected $S/B$ of about 1.4. As a result, ${\rm BR}(K^+\to\pi^+\nu\bar\nu)$ is now experimentally known to about 40\% precision. The original NA62 goal, however, was to measure the BR 
to within just over 10\%. To fully realize the potential of the experiment to achieve its original goal, in October 2019, NA62 submitted an addendum to its original proposal~\cite{NA62:2019xxx} to the CERN SPSC to extend the data taking throughout LHC Run III, until Long Shutdown 3 (LS3), currently foreseen at the end of 2025~\cite{LHCcomm:2022}. Additional data taking was initially approved for one year and began in July 2021. Finally, in 2021 
NA62 was approved until LS3. 

In advance of the 2021 run, numerous improvements to the experiment were made in order to allow data to be taken efficiently at the nominal beam intensity, including improvements to the efficiency and stability of the data-acquisition system, the introduction of additional detectors to provide increased rejection power against specific backgrounds, the reduction of losses from accidental coincidence vetoes, and the further optimization of the
$K^+\to\pi^+\nu\bar\nu$ analysis.

\begin{figure}[tb]
\centering
\includegraphics[width=0.7\textwidth]{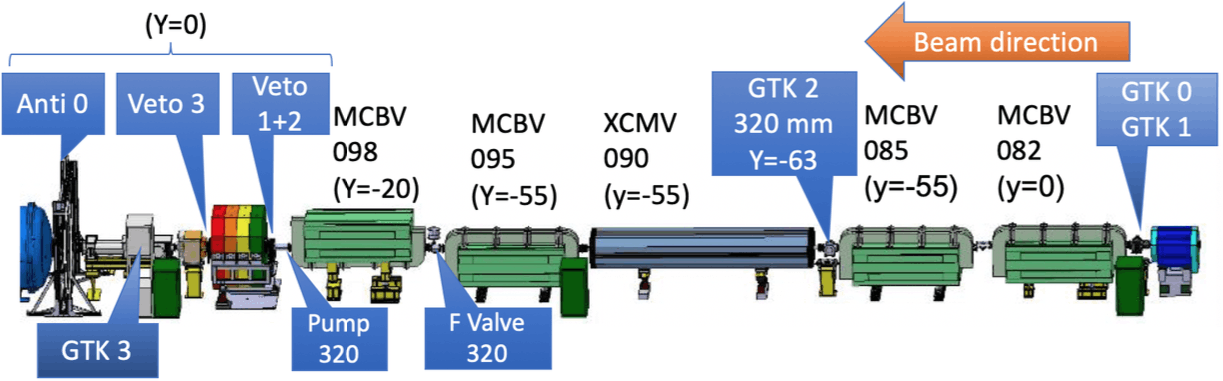}
\caption{Layout of the new GTK achromat and beamline elements for NA62 in 2021. The main changes include optimization of the achromat and the installation of additional detectors: a fourth GTK station (GTK0) placed next to GTK1, a new veto counter (Veto 1--3) around the beam pipe before and after the final collimator, and a new hodoscope (Anti 0) to veto muon halo background.}
\label{fig:na62_beam3}
\end{figure}

As noted in \Sec{sec:kp_pnn_stat}, the largest source of background in NA62 is from upstream $K^+$ decays and
interactions. To reduce this background, the beamline elements were rearranged around the GTK achromat, and a fourth station was added to the GTK beam tracker to reduce the influence of combinatorics in GTK track reconstruction (\Fig{fig:na62_beam3}).
An upstream charged-particle veto counter, consisting of three stations of 2-cm wide scintillator tiles mounted immediately above and below the beam pipe, was installed just upstream and downstream of the final collimator to detect photons and charged pions from upstream $K^+$ decays.
This veto counter is read out with a new digital front-end board data-acquisition chain consisting of a new, custom, on-detector FPGA TDC board that sends data over high-speed links to an off-detector front-end server equipped with FELIX data receiving PCIe cards. In addition, a new hodoscope (ANTI0) was installed at the entrance of the fiducial volume, which mainly serves to veto the muon halo background for new physics searches in beam dump mode. In standard running with the $K^+$ beam, the ANTI0 is expected to assist with the reduction of background trigger rates.

The above improvements allow the present selection criteria against the upstream background to be loosened, regaining signal acceptance. Steps to reduce losses from accidental coincidences in the veto detectors include improvements to the use of time-clustering information for LKr reconstruction and the adoption of multivariate analysis techniques to refine the veto from the LAV detectors and to make better used of the information from the small-angle photon vetoes and from the detectors used to reject events with charged products from photon interactions. These optimizations are currently being refined with the 2016--2018 dataset, for which the analysis has been completed.
\subsection{Kaon physics at the SPS after LS3}
\label{sec:sps_program}

Following upon the successful application at NA62 of the in-flight technique to measure ${\rm BR}(K^+\to\pi^+\nu\bar\nu)$, we envision a comprehensive program for the study of the rare decay modes of both $K^+$ and $K_L$ mesons, to be carried out with high-intensity kaon beams from the CERN SPS in multiple phases, including an experiment to measure ${\rm BR}(K^+\to\pi^+\nu\bar\nu)$ at the 5\% level and an experiment to measure ${\rm BR}(K_L\to\pi^0\nu\bar\nu)$ at the 20\% level. The detectors could also be reconfigured to allow measurements of
$K_L$ decays with charged particles, such as $K_L\to\pi^0\ell^+\ell^-$.

The success of this program depends on the delivery of a high-intensity, slow-extracted 400-GeV/$c$ proton beam from the SPS to the ECN3 experimental hall (where NA62 is located), which is ideally suited for
next-generation kaon experiments. Up to $10^{19}$ protons on target per year will be required. Studies indicate that, with duty-cycle optimization, sufficient proton fluxes can be delivered to allow a high-intensity kaon experiment to run as part of a robust fixed-target program in the CERN North Area~\cite{Bartosik:2018xxx}. The feasibility of upgrades to the primary beamline tunnel, target gallery, and experimental cavern to handle an intensity of up to six times that in NA62 has been studied
as part of the Physics Beyond Colliders initiative at CERN, and preliminary indications are positive~\cite{Banerjee:2018xxx,Gatignon:2018xxx}. 

To measure ${\rm BR}(K^+\to\pi^+\nu\bar\nu)$ to $5\%$ accuracy matching the irreducible theory precision, the experiment must be able to handle a beam intensity of at least four times that in NA62. The entire complement of NA62 detectors would require substantial upgrades in order to provide clean event reconstruction at the expected rates. Ideas for the new detectors
are inspired by R\&D efforts for HL-LHC.

The baseline design for the $K_L$ experiment is the KLEVER project~\cite{Ambrosino:2019qvz}, presented as part of the Physics Beyond Colliders
initiative~\cite{Beacham:2019nyx}. The KLEVER goal is to measure ${\rm BR}(K_L \to\pi^0\nu\bar\nu)$ to within 20\%.
The boost from the high-energy neutral beam facilitates the rejection of
background channels such as $K_L\to\pi^0\pi^0$ by detection of the
additional photons in the final state.
Background from $\Lambda \to n\pi^0$ decays in the fiducial volume must
be kept under control, and recent work has focused on the possibility
of a beamline extension and other adaptations of the experiment to ensure
sufficient rejection of this channel.  
The layout poses challenges for the design of the small-angle
vetoes, which must reject photons from $K_L$ decays escaping through the
beam exit amidst an intense background from soft photons and neutrons in the
beam. 

The $K^+$ and $K_L$ experiments would run consecutively, with the order to be determined, and use interchangeable detectors. The forward calorimetry and photon veto systems would be common to both experiments. Detector and beamline configuration changes between phases would be scheduled during LHC shutdown periods. As in NA62, the broad physics program of the high-intensity $K^+$ experiment includes studies of rare kaon decays, searches for lepton-flavor or lepton-number violating decays, and searches for hidden sectors in kaon and pion decays (Section~\ref{sec:broader_programme}). In the $K_L\to\pi^0\nu\bar\nu$ configuration, to optimize the measurement, the $K_L$ experiment would have no secondary tracking and limited PID capability. We foresee however the possibility of taking data with the downstream detectors configured for charged-particle measurements 
(secondary tracking and PID) but with a neutral beam. This will allow searches for radiative and lepton-flavor violating $K_L$ decays, $K_L$ decays to exotic particles, and particularly, measurements of the $K_L\to\pi^0\ell^+\ell^-$ decays, for which
only limits exist at present. These measurements would overconstrain the kaon unitarity triangle; to the extent that there is little constraint on the CP-violating phase of the $s\to d\ell^+\ell^-$ transition, the $K_L \to\pi^0\ell^+\ell^-$ measurements may reveal the effects of new physics~\cite{Smith:2014mla}.
\subsection{High-intensity beams}

Future experiments at the SPS for precision measurements of the BRs of the $K\to\pi\nu\bar{\nu}$ decays will require a very intense slow-extracted proton beam.

At a nominal intensity of $1.1\times10^{12}$ protons on target per effective second of spill, the current NA62 beam, with an angular acceptance of 12~$\mu$sr and a momentum acceptance of $\delta p/p \sim 1\%$, has a total rate of about 750~MHz and a kaon content of about 6\%. When going from charged to neutral beam, the momentum acceptance is defined by the fiducial volume and analysis cuts and is a considerable fraction of the total momentum distribution. However, the increase in flux is offset by the very tight collimation required for neutral beams (at most 0.5~$\mu$sr, in order to conserve $p_\perp$ constraints for background rejection, in the case of KLEVER). Then, because of the optimization of the production angle for the $K_L$ experiment, the smaller fiducial-volume acceptance for $K_L$ decays due to the longer lifetime, and the smaller standard-model BR for the $K_L\to\pi^0\nu\bar\nu$ decay, an increase of intensity by an order of magnitude is required to obtain the same rate of decays in the fiducial volume. Therefore, the statistical needs of the neutral-kaon experiment drive the intensity request.

The assumed proton beam intensity for KLEVER of $6.7\times10^{12}$ protons on target per effective second of spill ($2\times10^{13}$ ppp assuming a 4.8-s flat-top and a 3-s effective spill) is as high as considered reasonably achievable in an existing North Area installation, as described below.
For the KLEVER beam parameters, the total beam rate (excluding low-energy photons) is just under 700~MHz, and the total rate at which events must be vetoed while maintaining high signal efficiency is around 150~MHz.



\subsubsection{Primary proton beam}
\label{sec:proton_beam}

The proton beam for the North Area experiments is extracted from the SPS with electromagnetic (ZS) and magnetic (MST) septa in the LSS2 area and the beam is brought to the North Area targets via the TT20 transfer tunnel. Before arriving at the TCC2 target cave, the beam is split twice into three beams, which are brought to interact on targets T2, T4, and T6 inside TCC2.
Targets T2 and T4 are ``wobbling'' targets, at which multiple secondary beams can be produced.  The T4 target produces (in addition to the H6 and H8 beams for the EHN1 beamlines) the P42 proton beam, which is the primary beam for kaon experiments.
The P42 beam is transported over 800~m to the TCC8 target cave, which houses the T10 target used to produce the secondary beam for the ECN3 hall (where the NA62 experiment is currently installed).

We assume that the proposed $K\to\pi\nu\bar\nu$ experiments will be installed in ECN3. This will require 
upgrades to the P42 transport line, and
TCC8 target gallery.
These upgrades have been studied
by the Conventional Beams working group in the context of the 
Physics Beyond Colliders initiative~\cite{Banerjee:2018xxx,Gatignon:2018xxx}. Preliminary indications from these studies are described below.

\paragraph{Slow extraction to T4}
In 2018, the maximum intensity accelerated in the SPS under normal operation was $3.5\times10^{13}$ ppp, obtained using the multi-turn extraction scheme for the PS-to-SPS transfer~\cite{Bartosik:2018xxx}. This intensity is more than sufficient to allow both a high-intensity kaon experiment and 
a robust North Area fixed-target program to run concurrently.
In reality, however, the success of the experiment depends on the
total number of protons delivered to target during the running
period, as well as the quality of the spill.
In 2018, about $1.25\times10^{19}$ protons were delivered to the TCC2 targets. 
To observe 60 SM $K_L\to\pi^0\nu\bar{\nu}$ events, KLEVER will require a total
integrated intensity of $1\times10^{19}$ pot/year for five years. 

The SPS Losses and Activation Working Group and Physics Beyond Colliders Beam Dump Facility Working Group have made
general progress on issues related to the slow extraction
of high-intensity proton beams to the North 
Area. The emphasis of these studies has been on how to extract $4\times10^{19}$ pot onto the BDF target while delivering roughly the same number of protons ($1\times10^{19}$) per year to the TCC2 targets as in past years. 

The principal limitation on the sustained intensity on target (i.e., in pot/year) is from beam losses at the extraction septa, leading to machine activation, reduced component lifetime, and limitations on personnel access and maintenance. 
Several percent of the beam is lost in the extraction process. Primary proton scattering on the septum wires is the dominant source of the beam loss.
To keep the residual activation levels at around 5~mSv/h after 30 hours of cool-down while extracting around $5\times10^{19}$ pot per year will require a reduction in the activation per proton of approximately a factor four. The BDF study developed and tested methods to reduce the losses per proton at extraction by a factor of four.
A significant effort was made on several fronts to conceive, design, deploy and test where possible methods to reduce the extraction beam losses or to mitigate their effects through different material choice or handling. It is expected that, through a combination of concepts presented in this study, a factor of four reduction of the prompt extraction losses in LSS2 is within reach for operational scenarios.

The same extraction system in LSS2 and transfer line to TT20 are used for the BDF and the TCC2 targets, with the first TCC2 splitter (splitting the beam between the T2-T4 and the T6 targets) replaced to allow switching between the destinations BDF and North Area on a cycle-by-cycle basis.
Therefore, the factor of four in loss reduction should apply to both
BDF and North Area extraction. This substantially relaxes the limitation on the total number of protons extracted per year (to whatever target), to $5\times10^{19}$ pot.

For the TCC2 targets, the next limiting factor then becomes losses on the Lamberson splitter septa used to split the beam first between T2/T4 and T6, and then between T2 and T4. The fraction of beam lost on the splitters is at least 3\%, as estimated by intensity monitors in the NA.  Activation from splitter losses would then represent the limitation on the total number of protons delivered per year.
It may be possible to reduce the losses per proton during the splitting process to the North Area; this question certainly merits further study. 

\begin{figure}[htb]
\centering
\includegraphics[width=0.6\textwidth]{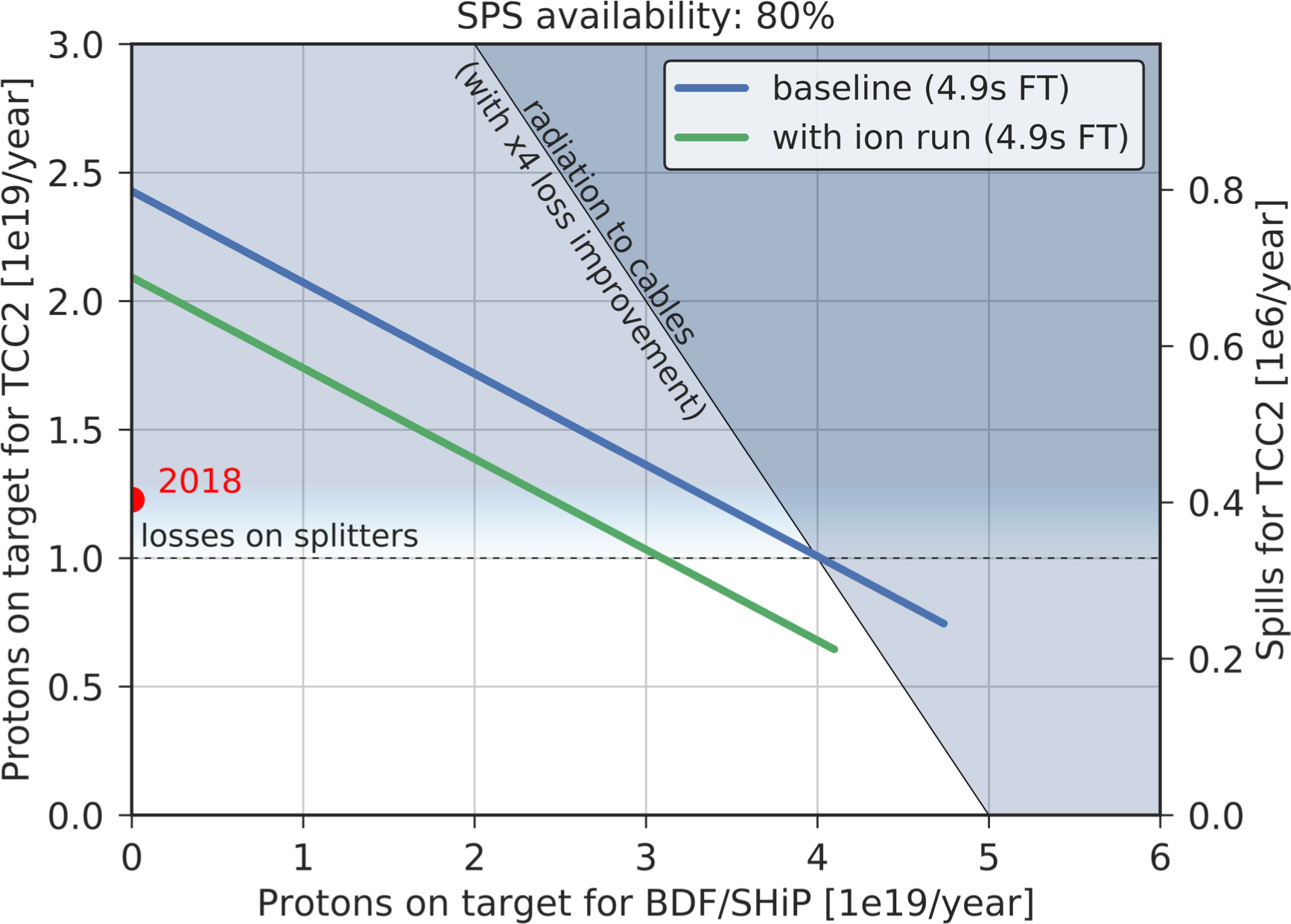}
\caption{Future proton sharing scenarios with (green) and without (blue) ion operation, with a 4.9-s flat top (FT) for the TCC2 experiments. Reproduced from Ref.~\cite{Bartosik:2018xxx}.}
\label{fig:sps_sx}
\end{figure}

Duty cycle optimization for the BDF and North Area has been studied under the assumptions that extraction losses can be reduced by a factor of four but that splitter losses limit the total number of protons per year delivered to the North Area targets to current levels \cite{Bartosik:2018xxx}. The studies were performed taking into account realistic supercycle compositions and respecting the SPS limits on power dissipation in the magnets. Two values of the flat-top length for the North Area targets were considered, 4.9 and 9.7 s, with the more favorable results being obtained for a 4.9-s flat top. The results are summarized in \Fig{fig:sps_sx}. 
As noted above, activation from splitter losses limit the North Area targets to 1--1.5$\times10^{19}$ pot/year. Limits from radiation resistance of cables in the extraction region are also illustrated. The blue and green lines indicate proton sharing between the TCC2 targets and BDF, with and without ion running, assuming 80\% SPS availability.  
These results indicate that with small compromises, BDF, K experiments, and the North Area test beam program might be able to run concurrently. It does not currently seem possible to run the BDF, K experiments, and a robust fixed-target program requiring significant intensity all at the same time without significant compromises.

It is important to note that this study was performed with the primary objective of establishing the feasibility of delivering $4\times10^{19}$~pot/year to the BDF. If splitter losses in TCC2 can be reduced even slightly, this would relax the limit on the number of protons delivered per year to the North Area targets, allowing for a more flexible allocation of the $5\times10^{19}$~pot/year extracted from the SPS among BDF, future kaon experiments, and other potential users of the North Area.

At the T4 target, the beam will be wide and parallel in the vertical
plane, so that it mostly misses the 2~mm thick T4 target plate. The small fraction that hits the target will be sufficient to produce the H6 and H8 beams without damaging the target head. On the other hand, most of the beam will not be attenuated by T4 and the transmission to T10 could be as high as 80\%. 

\paragraph{Target survival}
The present T10 target consists of a series of four beryllium rods of diameter 2 mm, each 100 mm long, placed end to end and suspended by 25-$\mu$m-thick aluminum foils clamped into aluminium flanges. The target is housed in an enclosure cooled by forced-air convection. 
Detailed studies have been carried out to determine whether this target can withstand an intensity of $2\times10^{13}$~ppp~\cite{Solieri:2020tar}, with FLUKA used to model the energy deposition in the target and ANSYS used for thermal and mechanical finite element analysis. The simulation was performed both for NA62 conditions ($3\times10^{12}$~ppp), as a baseline, and for KLEVER conditions ($2\times10^{13}$ ppp). For both modes a cycle time of 16.8~s was assumed, with a flat top duration of 4.8~s.
Appropriate beam parameters (production angle, beam spot size, divergence) were used in each case. In particular, for KLEVER, the diameter of the beryllium target rods was increased to 4~mm. This is useful not only for better target survival, but also to accommodate the larger production angle (8~mrad) to be used in KLEVER.
It was found that, at KLEVER intensities, creep-induced plastic deformation could progressively modify the shape of the target rods over the lifetime of the experiment. Additionally, such high temperatures in the rods are not compatible with the use of aluminum support foils.

A new target design to address these problems is relatively straightforward.
A more efficient cooling system can be implemented by using pressurized air to cool a contained volume surrounding the target (as opposed to cooling the whole target box), thereby strongly increasing the heat-transfer coefficient. Due to the relatively high surface-to-volume
ratio of the slender target rods, the steady state temperature of the rods is very sensitive to this parameter. To provide a precise and reliable mounting system under the increased temperatures and airflow, the aluminium foils will likely need to be substituted by thicker and more solid supports. The target mounting system developed for the CNGS experiment can be used as starting point for the new design. In the CNGS design~\cite{Bruno:2006xxx}, there is a better thermal contact between target rods and supports, and the compact and enclosed design of the vessel and stable supports allow pressurized air cooling to be used, thereby achieving much higher values of the heat transfer coefficient.

As noted in~\Sec{sec:neutral_beam}, a possible advantage of the use of a high-$Z$ target for production of a high-intensity neutral beam is that more of the photons produced in the primary interaction would shower in the target. This would reduce the required thickness of the photon converter placed in the dump collimator, and in turn, the fraction of neutral beam particles lost to inelastic interactions or scattered out 
of the beam. The energy-deposition studies described above for KLEVER conditions were carried out for graphite, CuCrZr, molybdenum, iridium, tungsten, and tantalum targets, each of thickness in nuclear interaction lengths equal to that for the beryllium target. For all of these materials except for graphite, the total energy deposited per primary is increased by a factor of 20 to 100, relative to that for beryllium. Because the higher-$Z$ targets are shorter, the increase in deposited energy density is even higher, and the heat exchange surface is smaller. These results suggest that the use of a high-$Z$ target material may not be compatible with an air cooling system. 
A design similar to that implemented for the hadron production target at J-PARC~\cite{Takahashi:2015xxx}, which features a gold target embedded in a water-cooled target holder made of CuCrZr, would be able to dissipate far more heat than a pressurized-air cooling system. In this case, it would be necessary to carefully evaluate the relative advantages of a thinner photon converter (high-$Z$ target with surrounding material) versus the better collimation made possible by the precise spatial constraints on the source of primary interactions (low-$Z$, suspended target). 

Despite the higher total energy deposition in graphite relative to beryllium, in operation, the peak temperature of a graphite target would actually be lower, due to the larger nuclear interaction length and higher emissivity of graphite. In addition, because of the toxicity of beryllium, graphite is an easier material to work with. For these reasons, a graphite target would be advantageous from an engineering point of view.

\paragraph{Dump collimator (TAX) survival}

The TAX collimators consist of two large modules that can be independently moved in the vertical direction, each featuring different collimator holes and beam dump positions. Different beam modes are made possible through the choice of different combinations of vertical positions and apertures of the two modules. In NA62, the T10 TAX is installed in TCC8, 24 meters downstream of the T10 target. A similar device, the T4 TAX, is used as the dump collimator for the P42 primary proton beam.
Each TAX module is made up of four 120~(H) $\times$ 80~(W) $\times$ 40~(D) cm$^3$ blocks, for a total depth of 160 cm. For the T10 TAX, the first two blocks of the first module are made of copper, while the remaining blocks are made of cast iron. The blocks rest on a water-cooled copper plate and are surrounded by iron shielding. They are moved vertically via screw jacks, with the TAX motors located 3.5~m away, behind the shielding. Energy deposition simulations with FLUKA and thermal/mechanical finite-element simulations with ANSYS have been performed~\cite{Solieri:2020tax} to assess the performance of the closed TAX for NA62 running in beam-dump mode and for KLEVER operation, as described above. In NA62 beam-dump mode, the T10 target is removed from the line and the full primary beam is dumped directly on the blocks. Scenarios for running at intensities of up to four times the present intensity were studied. 
KLEVER operation was simulated with the target in place, an 8~mrad targeting angle, and the TAX moved upstream to its position 5.5~m downstream of T10 in KLEVER, with corresponding reduction in the size of the beam spot. 

The response of the TAX to three successive pulses was simulated for intensities of 3, 6, 9, and $12\times10^{12}$~ppp (from one to four times the present value) to obtain maximum values of the temperature and cumulative plastic deformation (ratcheting) in the first TAX block.
Ratcheting is observed for the two highest intensities and is especially pronounced for the highest intensity, for which a plastic deformation of 1.7\% is reached in only three pulses. At the higher intensities, it is likely that the TAX would develop cracks around the collimator hole within the timeframe necessary to reach the steady state (less than 24 hours), while even for the case of the two lower intensity values, the presence of high amplitude stress cycles puts the material on the sides of the lower collimator hole at risk of fatigue failure, which could potentially be favoured by the irradiation damage that the material in that area is subjected to. Based on these results, cracks in the lower collimator hole are expected to form for prolonged operation in beam dump mode at intensities higher than $6\times10^{12}$~ppp.
For KLEVER, the specific locations of maximum stress change due to the differences in beam and beam-hole geometry, but the essential conclusion is similar: the area surrounding the collimator hole would be subjected to extensive plastic deformation after only one pulse.
In both the NA62 and KLEVER scenarios, the temperature increase due to the pulse is modest with respect to the steady state temperature. This indicates that the problems due to the high temperatures reached are not caused by the value of the peak energy deposition density in the components but, rather, by the inability of the cooling system to efficiently remove the deposited energy with each cycle.

One of the most effective ways to increase the cooling system efficiency is by placing the cooling elements as close as possible to the heating source. Embedding cooling tubes directly in the TAX blocks appears to be a natural solution. This concept has been successfully tested and implemented in several beam dump designs currently installed at CERN. 
To evaluate the thermal performance of such a solution, a conceptual design for a new TAX was developed to perform thermal calculations in ANSYS. 
Stainless steel tubes were embedded directly in the blocks at various distances from the central hole for the beam, and the performance was studied as a function of this distance and the water flow rate. 
As a proof of concept, these studies demonstrate that a straightforward design for the dump collimator should be able to handle the full KLEVER intensity.
For further studies, a more detailed design featuring multiple holes for use both in $K^+$ and $K_L$ mode at high intensity could be developed, with potential for optimization of the choice of materials and other aspects of the design.
The same arguments apply for the P42 TAX.

\paragraph{Other considerations}
There was initially a concern that the air containment in the TCC8 cavern would be insufficient for KLEVER operation with the existing ventilation approach. However, the air flow inside the TCC8 target cavern was measured to be very small, rendering an
expensive upgrade unnecessary.

The primary proton beam is incident on the T10 target at a downward angle of 8~mrad and is further deflected downward by an additional sweeping magnet following the target. FLUKA simulations for the prompt dose on the surface above ECN3 are under way; preliminary results indicate that an adequate solution can be found for shielding the target region. Muons in the forward direction can be dealt with by a mixed mitigation strategy involving additional upstream shielding and potentially a thicker earthen shield in the downstream region, possibly in combination with better fencing around the ECN3 area.


\subsubsection{Particle production at the target}
\label{sec:targ_prod}

To explore the parameter space available for designing a beamline in terms of the target material, length, and secondary production angle, good knowledge is required of particle production by 400-GeV protons on fixed target. Much of the available data is from the late 1970s and early 1980s, when specific measurements of charged-particle production by 400-GeV protons on beryllium targets as a function of $p$ and $p_\perp$ were carried out at CERN \cite{Atherton:1980vj}, and at Fermilab a neutral beamline was constructed in order to measure several of the neutral components \cite{Lundberg:1984pj,Beretvas:1986km,Jones:1980vp}.
Although these data are somewhat sparse, they can be used to benchmark current models of particle production in hadronic interactions, such that the models can be used for interpolation of the experimental data. These studies not only illustrate the main considerations in choosing the
parameters for the secondary beamlines, but also allow parameterizations to be extracted that can be used for the large scale event generation needed to design the experiments.  
In particular, the experimental data cited above has been used to benchmark results from FLUKA (version 1.0, March 2017) and Geant4 (version 4.10.03 and G4Beamline 3.04, with physics list FTFP\_BERT). The full results of these studies are available in Ref.~\cite{vanDijk:2018aaa}; below we highlight some of the essential aspects of the comparison of the existing data with the simulations.

\begin{figure}[htb]
  \centering
  \includegraphics[width=0.8\textwidth]{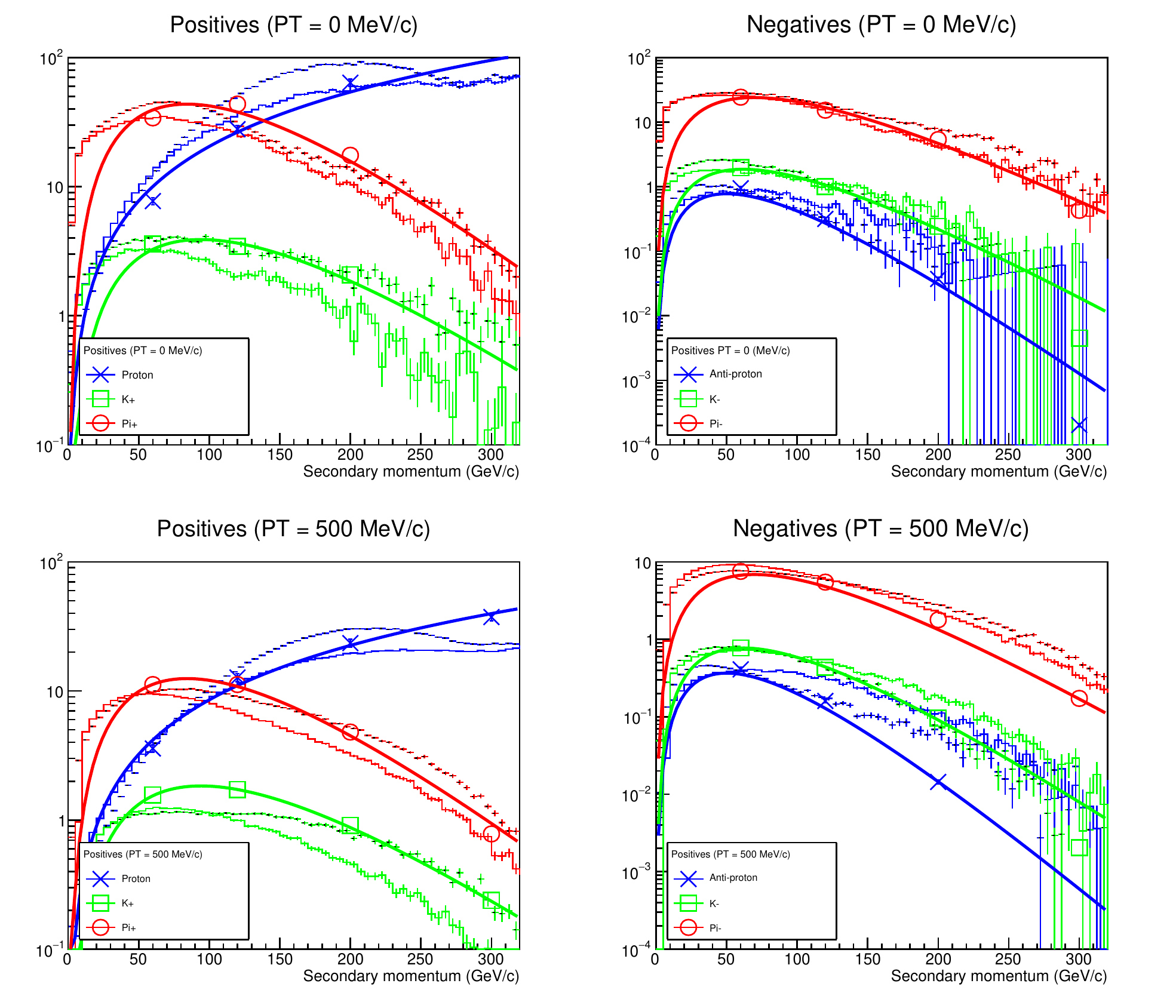}
  \caption{Benchmarking of inclusive charged particle production from 400 GeV protons on a thin beryllium target: protons (blue), pions (red), and kaons (green). The vertical axis is $d^2N/(dp\,d\Omega)$ (GeV$^{-1}$\,sr$^{-1}$), with $N$ the yield per interacting proton.
  Markers and solid curves show measurements from \cite{Atherton:1980vj} and parameterization therein; open and continuous histograms show results from FLUKA and Geant4, respectively.}
  \label{fig:Atherton_comp}
\end{figure}

Atherton et al.~\cite{Atherton:1980vj} measured absolute cross sections for charged particle production in proton-beryllium collisions at 400~GeV, for secondary momentum from 60 to 300~GeV, $p_\perp$ at 0 and 500~MeV, and various target lengths. 
These measurements have long served as a reference for particle production from beryllium targets in the North Area. In particular, they were used for the design of NA62.\footnote{The data on charged-particle production from \cite{Atherton:1980vj} were also used for the design of the NA31 and NA48 neutral beamlines: $K_L$ production rates were inferred from the $K^+$ and $K^-$ data from \cite{Atherton:1980vj} using a simple parton model.}
\Fig{fig:Atherton_comp} shows the benchmarking of the FLUKA and Geant4 results with the Atherton data for a 40-mm ($\sim$0.1~$\lambda_{\rm int}$) target. The Atherton measurements are shown with markers; the results of the FLUKA and Geant4 simulations are shown with open and continuous histograms, respectively. With some exceptions, the two simulation codes differ from the data and from each other by less than 50\%. FLUKA does a generally better job than Geant4 at matching the overall level of pion and kaon production in the data; both simulations show pion and kaon production peaking at much lower momentum than in data. Both codes dramatically underpredict proton production at high momentum; evidently, the description of the diffractive physics suffers from some limitations in either case, with Geant4 doing a better job than FLUKA for proton momenta up to about 150 GeV.
Unfortunately, the Atherton data do not constrain the momentum dependence of particle production at low momentum ($p<60$ GeV). Atherton et al.\ used an empirical form to parameterize the data, as shown by the solid curves \cite{Atherton:1980vj}; these parameterizations are merely for convenience and are not fit results.

\begin{figure}[htb]
  \centering
  \includegraphics[width=0.4\textwidth]{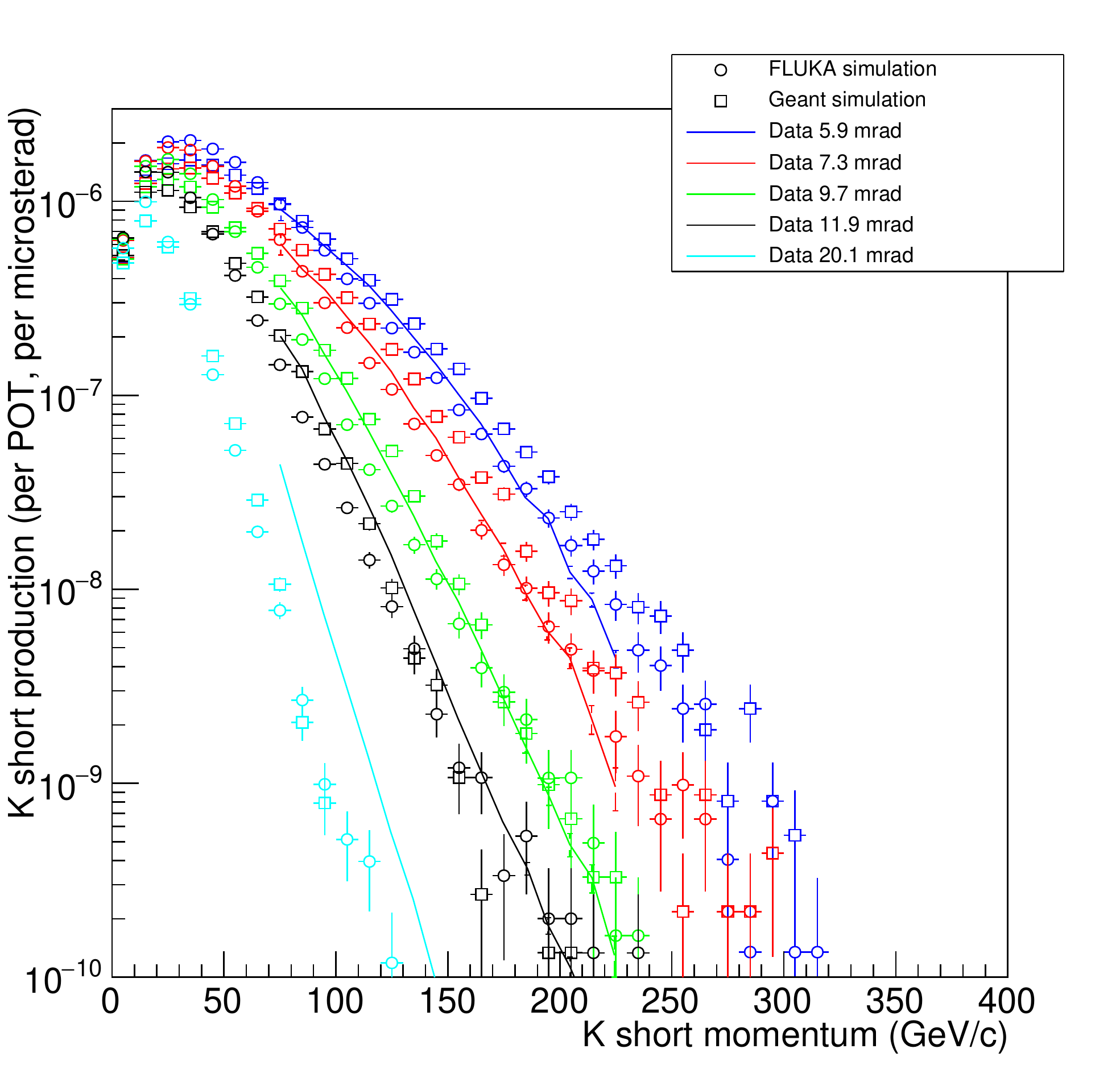}
  \includegraphics[width=0.4\textwidth]{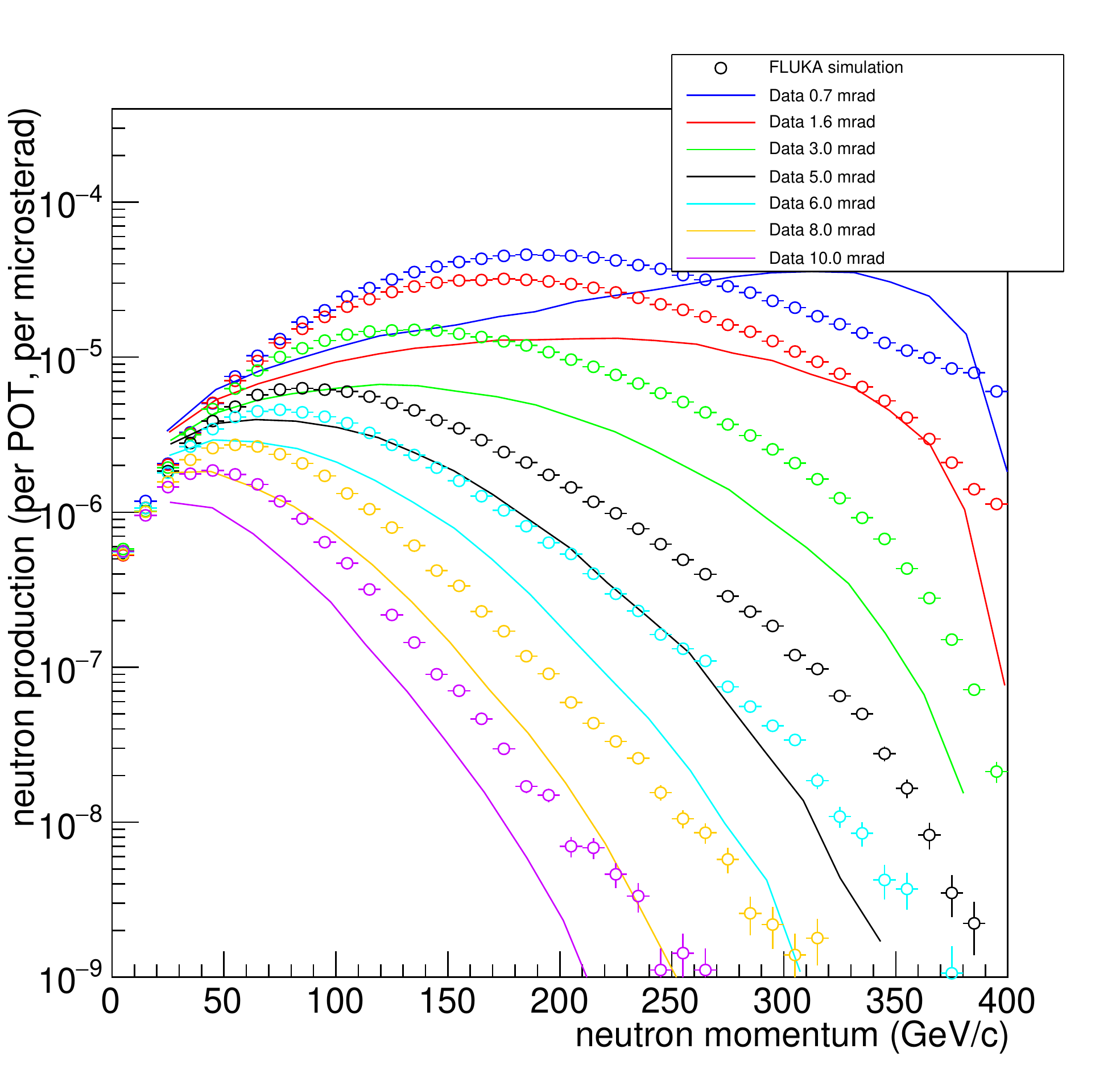}
  \\
  \includegraphics[width=0.4\textwidth]{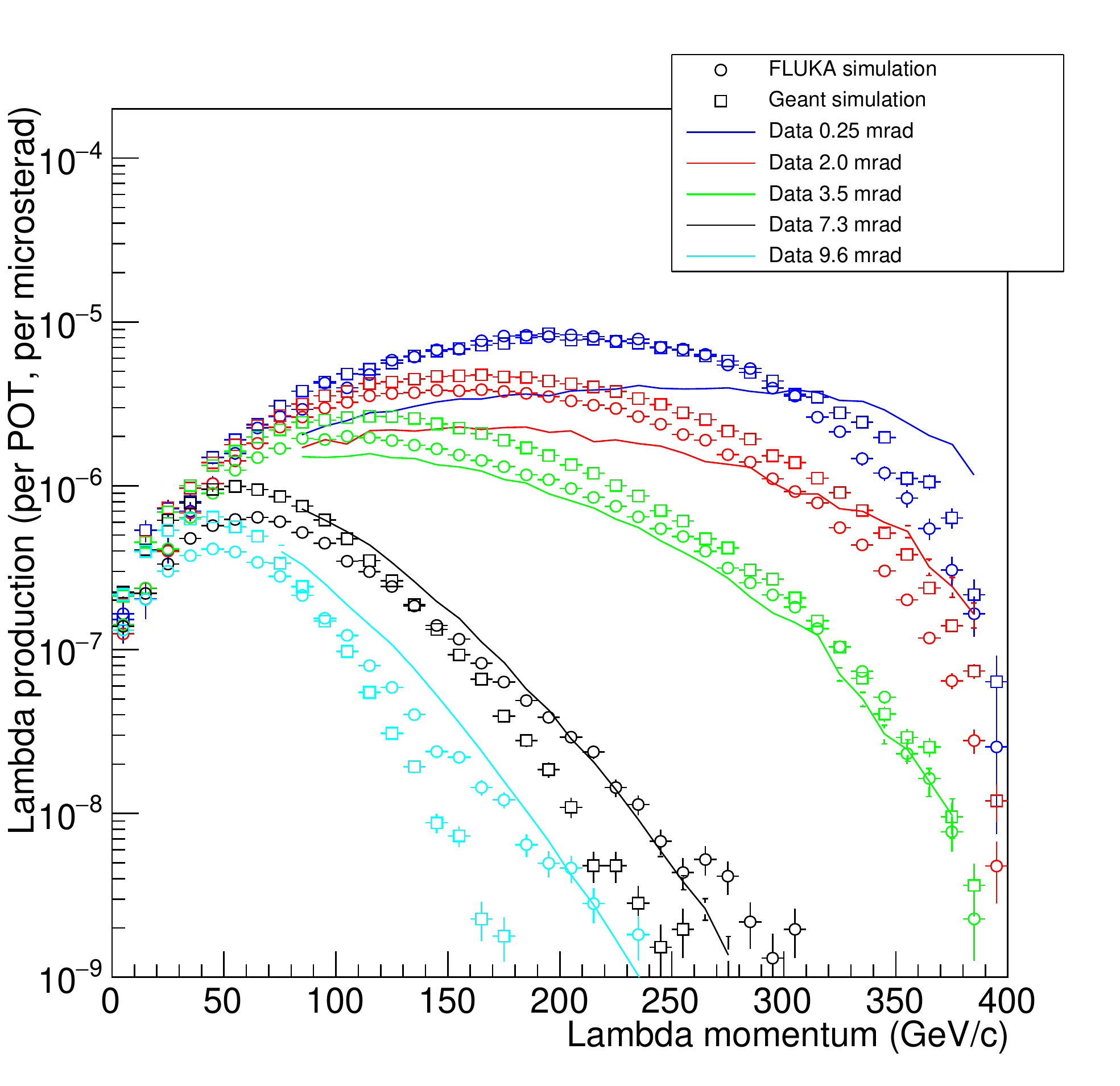}
  \includegraphics[width=0.4\textwidth]{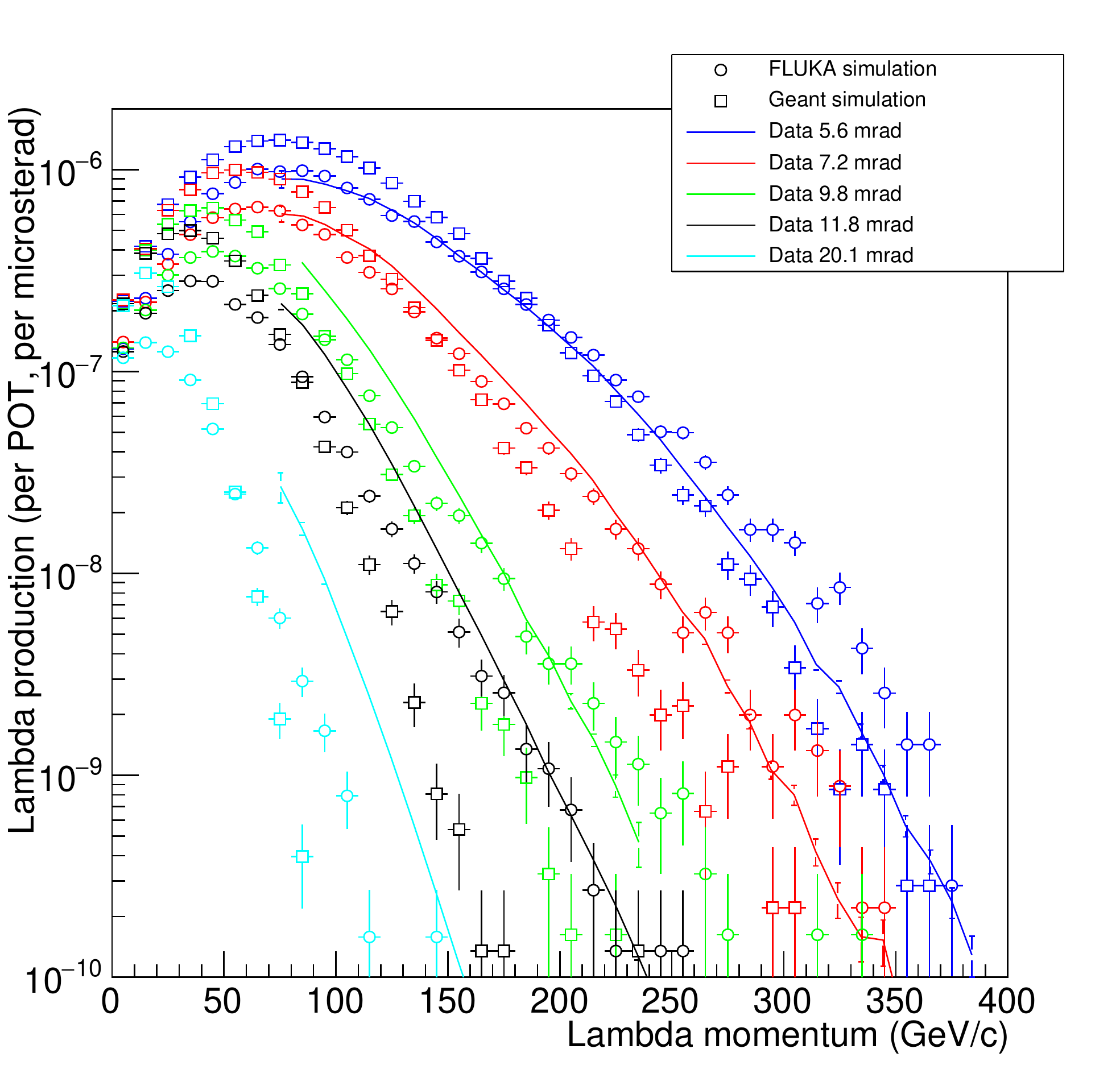}
  \caption{Benchmarking of inclusive neutral particle production from 400 GeV protons on a thin beryllium target, for $K_S$ mesons (upper left), neutrons (upper right), and $\Lambda$ baryons (bottom, two data sets). For each plot, the curves represent the experimental data \cite{Lundberg:1984pj,Beretvas:1986km,Jones:1980vp} for angular bins of $\pm0.5$~mrad about the value indicated in the legend, according to color; the markers illustrate corresponding results from the simulations (circles for FLUKA; squares for Geant4). The vertical axis is $d^2N/(dp\,d\Omega)$ (GeV$^{-1}$\,$\mu$sr$^{-1}$), with $N$ the yield per proton on target.}
  \label{fig:neutral_comp}
\end{figure}
For neutral particles, a series of experiments was performed at Fermilab to measure the production cross sections for $K_S$
\cite{Lundberg:1984pj,Beretvas:1986km}, $\Lambda$ \cite{Lundberg:1984pj,Beretvas:1986km} and neutrons \cite{Jones:1980vp}. The comparisons between data and simulation are shown in \Fig{fig:neutral_comp}. As above, the comparison is for 400~GeV primary protons interacting in a 40-mm beryllium target. For $K_S$, the agreement between the data and simulation results is mostly reasonable. FLUKA and Geant4 differ on the absolute production yields by generally less $50\%$, with FLUKA and Geant4 on average slightly underpredicting and overpredicting the data, respectively. There are small differences in the shapes of the momentum distributions, but, except for the point at $\theta = 20.1$~mrad, the main problem is the lack of data for $p < 100$~GeV, i.e., for much of the momentum range of interest. For neutrons, the agreement is less satisfactory, especially at high momentum and small angle, echoing the limitations of the simulations in describing forward particle production seen for the case of charged particles. Nevertheless, for production angles of $\theta = 3$~mrad and above, both simulations agree with data on the shape of the neutron momentum distributions and reproduce the absolute yields to within a factor of 2--3, with FLUKA generally overpredicting Geant4 and Geant4 comparing reasonably favorably with data for production angles around $\theta = 8$~mrad, which happens to coincide with the most favorable choices for production angle in a neutral beam experiment. It is important to note that the total neutron production in simulation is overestimated relative to the data.
For $\Lambda$ production, the comparison is in many ways similar to that for neutron production, as might be expected considering the similarity of the $\Lambda$ and neutron spectra and the influence of leading-baryon effects in inclusive $\Lambda$ production. For $\Lambda$s, the differences between data and simulation are most pronounced at small angles, with the momentum spectra observed in data significantly harder than in the simulations. At intermediate production angles (about 3.5 to 12 mrad), the simulated $\Lambda$ momentum spectra more closely resemble the data.

The neutral particle production data in \cite{Lundberg:1984pj,Beretvas:1986km,Jones:1980vp} also allow benchmarking of the simulations for thin ($\sim$0.1$\lambda_{\rm int}$) copper and lead targets, as done in \cite{vanDijk:2018aaa}, obtaining essentially the same conclusions as for beryllium as discussed above. 

\begin{figure}[htb]
  \centering
  \includegraphics[height=0.5\textheight]{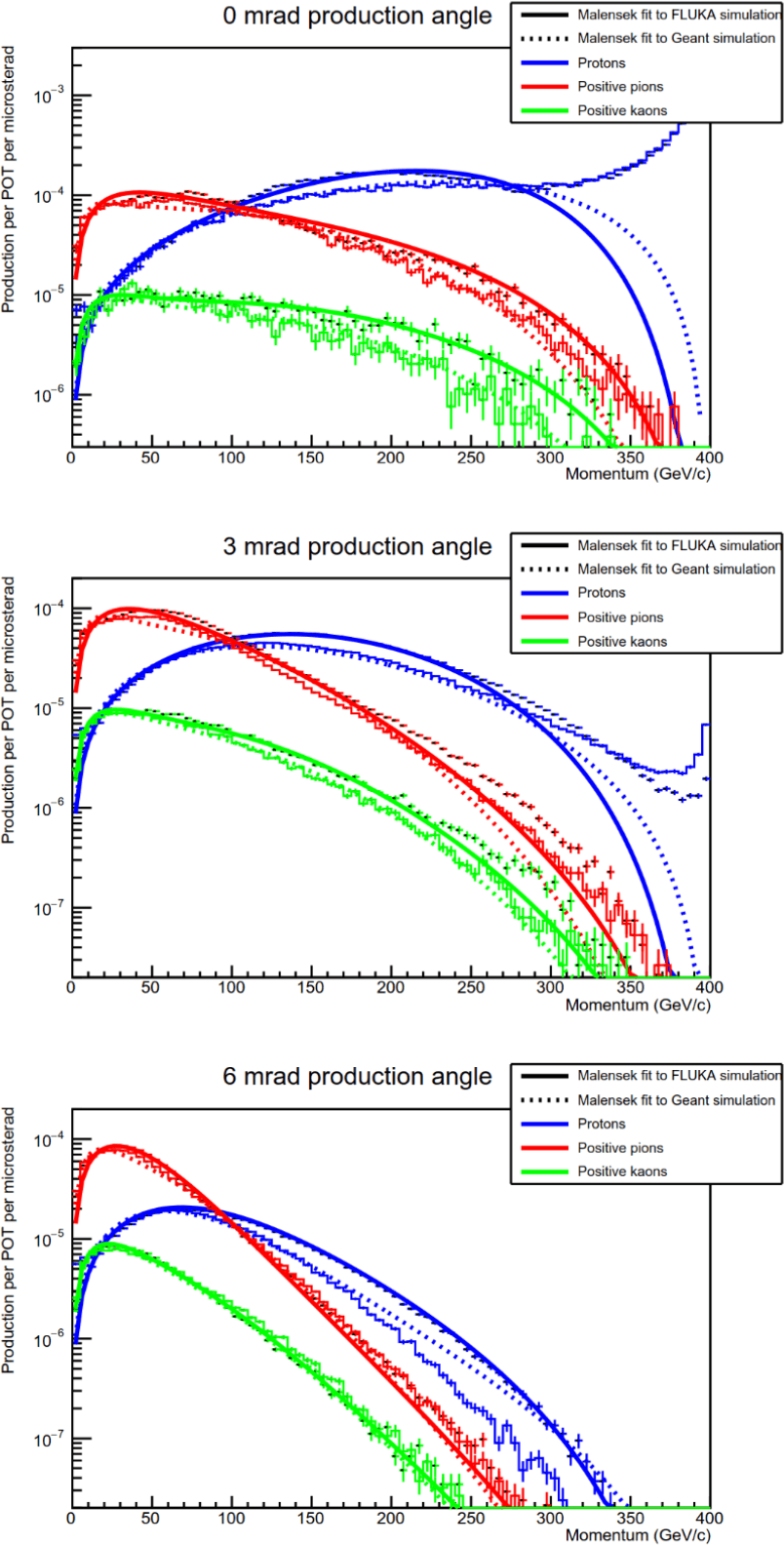}
  \hspace*{10mm}
  \includegraphics[height=0.5\textheight]{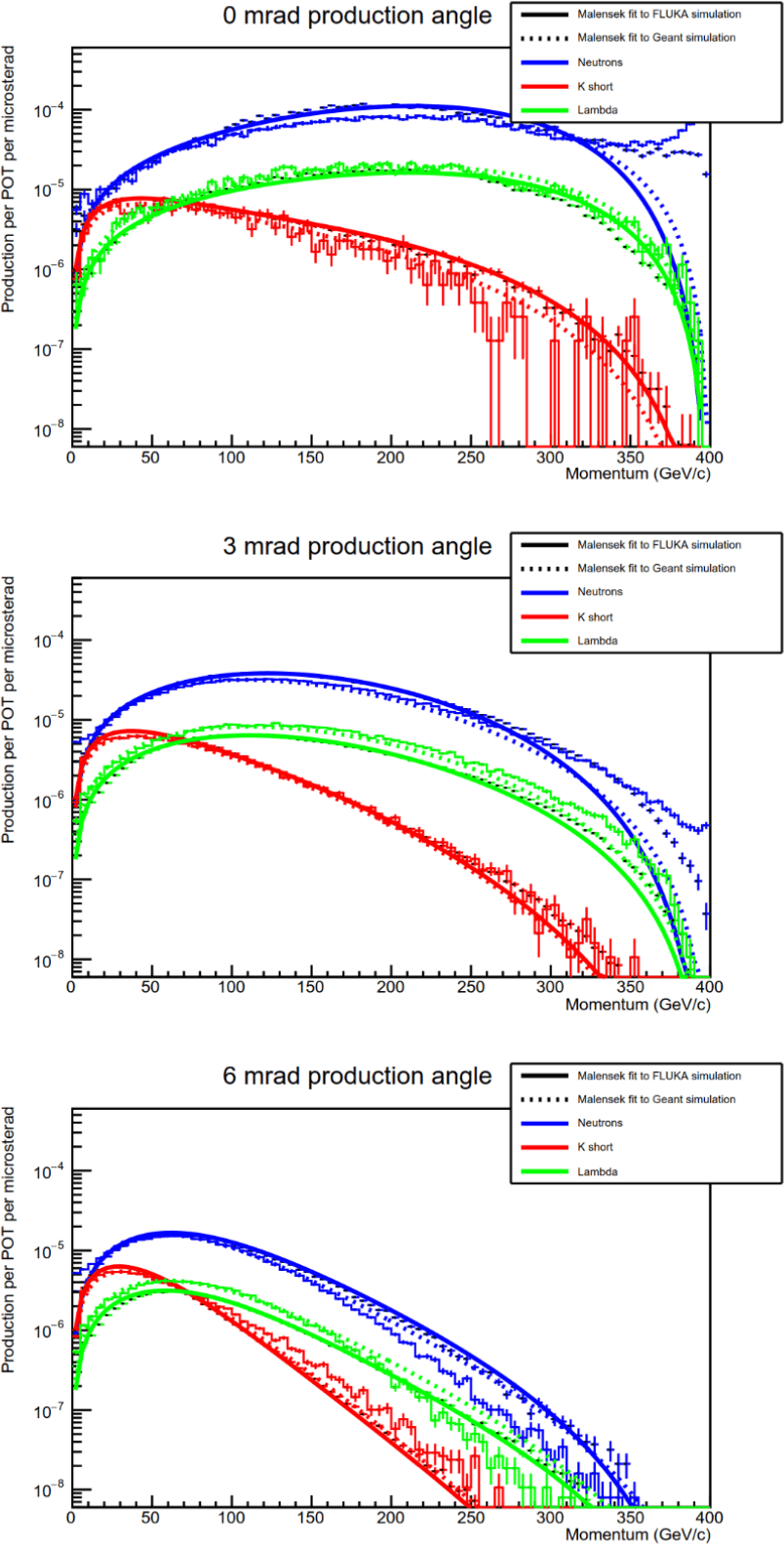}
  \caption{Parameterization of results on particle production from FLUKA and Geant4 simulations, for charged particles (left: $p$, blue; $\pi^+$, red; $K^+$, green) and neutral particles (right: $n$, blue; $K_S$, red; $\Lambda$, green). 
  The rows show the momentum spectra for different bins in the production angle $\theta$ (top to bottom: 0, 3, and 6~mrad).
  The solid (dashed) curves show fits with the Malensek parameterization to the FLUKA (Geant4) simulations results shown in the open (continuous) histograms. 
  The vertical axis is $d^2N/(dp\,d\Omega)$ (GeV$^{-1}$\,$\mu$sr$^{-1}$), with $N$ the yield per proton on target.
  }
  \label{fig:Malensek}
\end{figure}
The measurements described above and simulations for comparison were all obtained with targets thin enough to ensure that momentum loss and reinteraction of secondaries can be assumed to lead to negligible effects. For the full simulation of particle production, e.g., for the design of the KLEVER beamline, these effects are important. 
FLUKA and Geant4 simulated data sets for beryllium, copper, and lead targets of $\sim$1~$\lambda_{\rm int}$ were therefore generated and parameterized using an empirical parameterization of particle yields developed by Malensek~\cite{Malensek:1981em}. While the Atherton formula was intended only to describe the primary interaction for thin targets, the Malensek parameterization was intended to remain valid for thicker targets, for which reinteraction effects become important.
The Malensek parameterization, which was originally used to fit the charged-particle production measurements from Atherton, uses four free parameters to describe the momentum spectra $d^2N/(dp\,d\Omega)$;
the parameter values depend on target thickness and proton beam momentum.
The quality of the parameterization in reproducing the momentum spectra for simulated particles is illustrated in \Fig{fig:Malensek}.
The fitted Malensek parametrizations give, in most cases, a reasonable approximation of the simulated distributions. There are several clear deviations. The primary component of the protons is deliberately excluded from the fit. The fits for the Geant data are for most plots somewhat worse than for FLUKA, particularly at higher momenta.
Nevertheless, there is not usually more than a factor two between the results from FLUKA and Geant. This appears to be a reasonable index of the intrinsic uncertainty in the resulting parameterizations of particle production. The results of the fits with the Malensek parameterization to the FLUKA and Geant spectra for 400~GeV protons on 400-mm beryllium, copper, and lead targets are 
available in Ref.~\cite{vanDijk:2018aaa}. 
An important conclusion of this study is that for targets of identical thickness in terms of nuclear interaction length, no large differences were found in terms of hadronic production. The fits to the spectra for the beryllium target are the basis for the beam flux estimates assumed in the following sections.

\subsubsection{Charged kaon beam}
In NA62, the secondary beam is extracted at zero angle to maximize the total rate of $K^+$ decays; this increases the total beam rate considerably. With a four-fold increase (to $1.3\times10^{13}$~ppp, corresponding to two-thirds of the KLEVER intensity), 3~GHz of charged particles in the beam would need to be extracted into the experimental hall and tracked in the detector. At this intensity, the rate of $K^+\to\pi^+\nu\bar\nu$ decays in the fiducial volume would be twice the rate of $K_L\to\pi^0\nu\bar{\nu}$ decays in KLEVER at the full intensity of $2\times10^{13}$~ppp. The feasibility considerations relative to the high-intensity $K_L$ beam also apply to the $K^+$ beam, that is less intense.
The beam parameters other than intensity are assumed to be the same as for the current NA62 experiment; the main characteristics can be found in Refs.~\cite{NA62:2010xx,NA62:2017rwk}.


\subsubsection{Neutral kaon beam}
\label{sec:neutral_beam}

In designing a new neutral beamline for the KLEVER experiment, the most important parameters for the layout are the production angle and the solid angle of the beam aperture. 
The choice of production angle balances increased $K_L$ production and higher $K_L$ momentum at small angle against higher $K_L/n$ and $K_L/\Lambda$ ratios at larger angles. 
The choice of solid angle balances high beam flux against the need for tight collimation for increased $p_\perp$ constraints to reject the main background from $K_L\to\pi^0\pi^0$ with lost photons.
We initially assume that the 400 GeV proton beam from 
the SPS is incident on a 400-mm beryllium target, as in NA62, and consider other possibilities subsequently.

\begin{figure}[htb]
  \centering
  \includegraphics[width=0.45\textwidth]{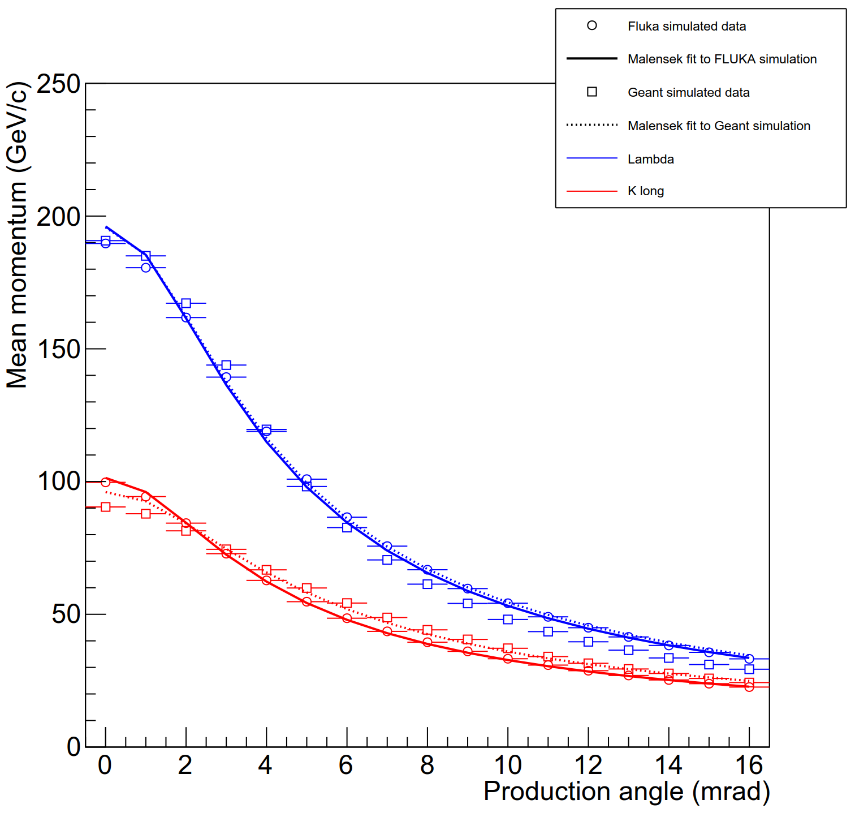}
  \includegraphics[width=0.45\textwidth]{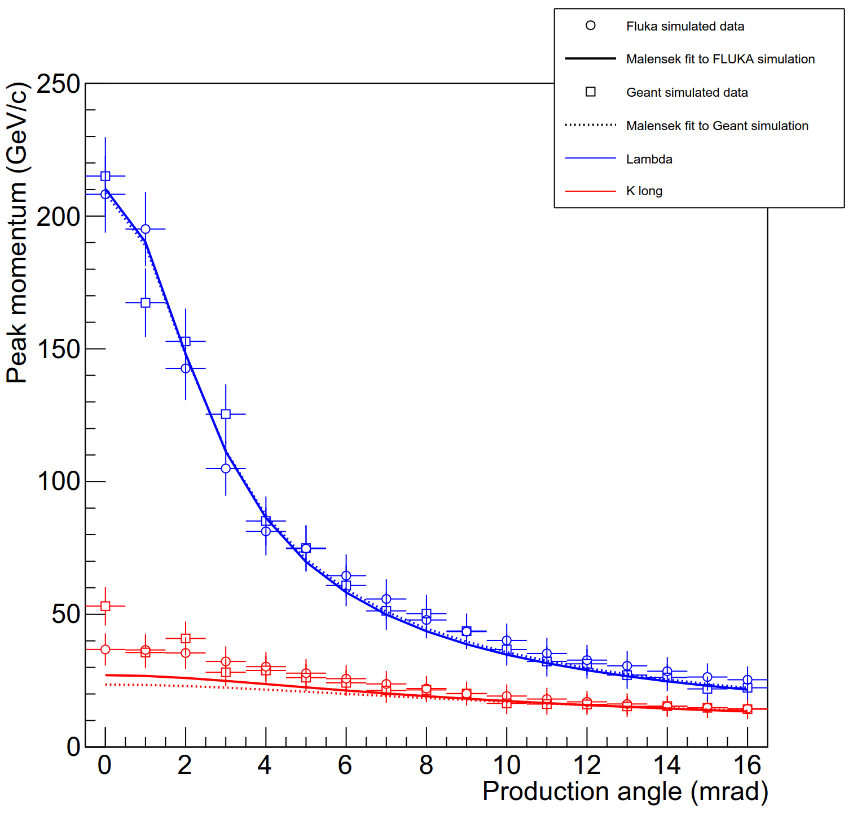}
  \caption{Mean (left) and peak (right) momentum for $K_L$ (red) and $\Lambda$ (blue) in the beam vs.\ beam production angle, from FLUKA (squares) and Geant4 (circles), with solid (dashed) curves from fits with the Malensek parameterization.}
  \label{fig:beam_mom_ang}
\end{figure}

\paragraph{Production angle and solid angle}
The $K_L$ and $\Lambda$ momenta in the beam are determined only by the production angle and collimation aperture and are broadly distributed, as seen in \Fig{fig:neutral_comp}.
\Fig{fig:beam_mom_ang} illustrates the evolution of the secondary $K_L$ and $\Lambda$ momentum distributions with production angle, as obtained with the FLUKA and Geant4 simulations described in \Sec{sec:targ_prod}.
As seen in \Fig{fig:neutral_comp}, there are some differences in the shapes of the momentum distributions from the two different simulations. 
These are most strongly reflected in the different values obtained for the peak momentum for small angles ($\theta < 2$~mrad). Nevertheless, consistent results are obtained for the mean momentum vs. angle, and the Malensek parameterization describes the dependence reasonably well, especially for $\theta > 4$~mrad. 

\begin{figure}[htb]
  \centering
  \includegraphics[width=0.45\textwidth]{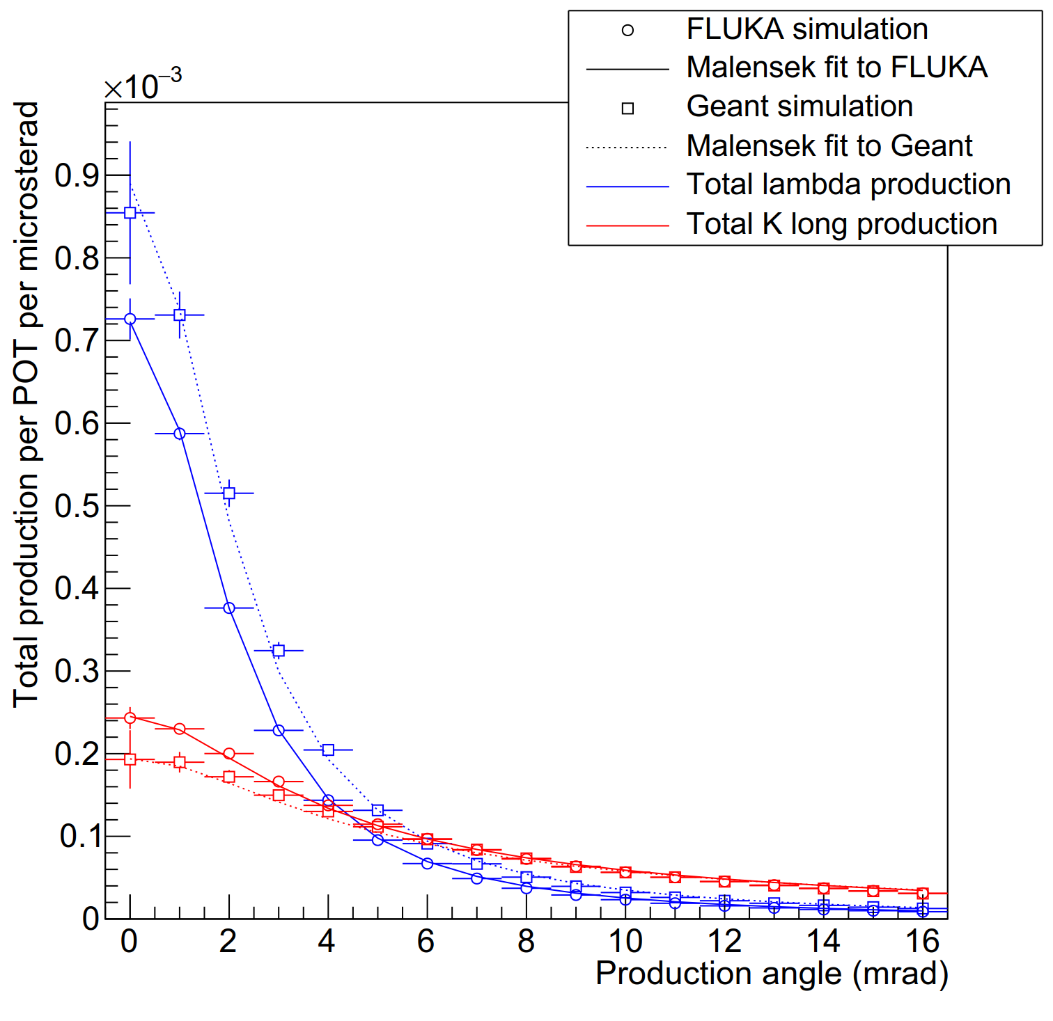}
  \includegraphics[width=0.45\textwidth]{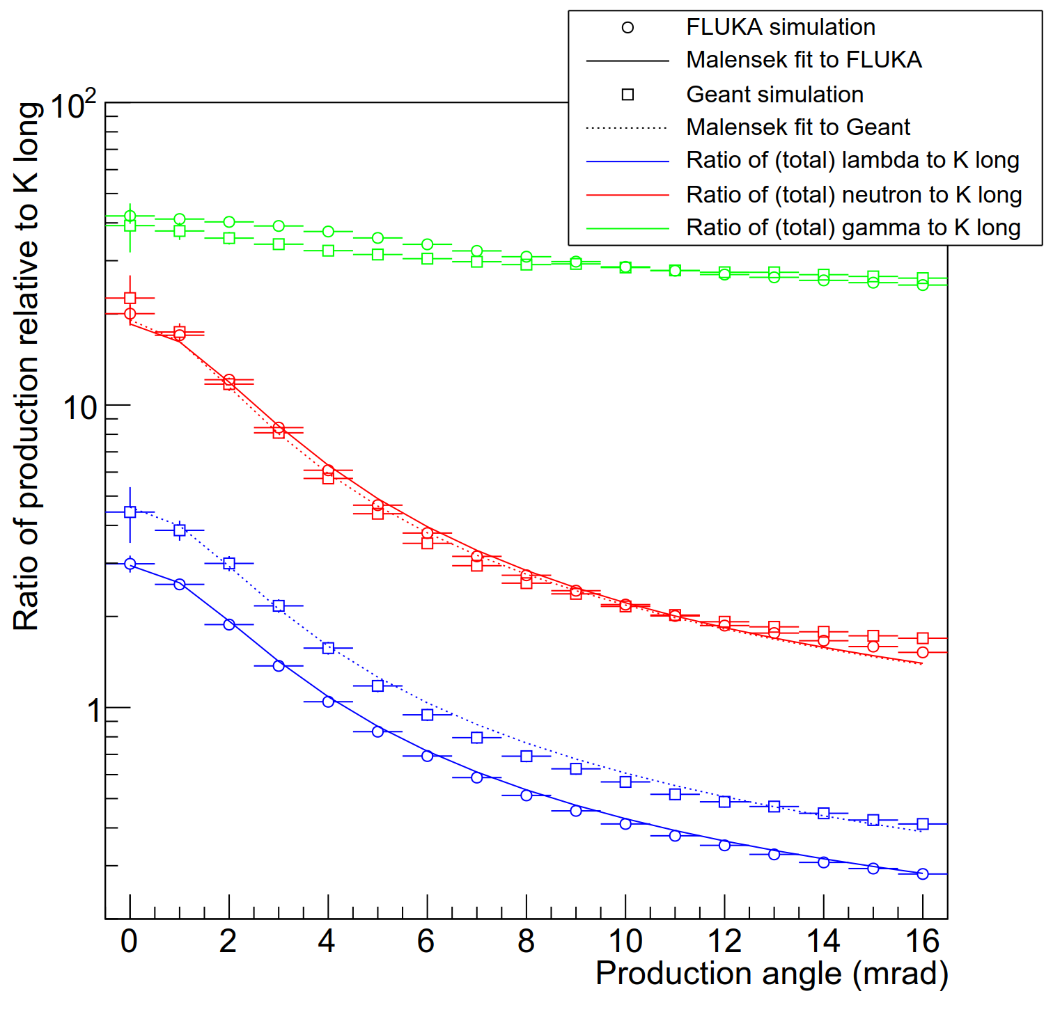}
  \caption{Left: Neutral beam fluxes per POT per $\mu$sr of beam acceptance for $K_L$ (blue) and $\Lambda$ (red) 
  vs.\ beam production angle.
  Right: Ratios of $\gamma$ (green), $n$ (red), and $\Lambda$ (blue) to $K_L$ fluxes vs.\ beam production angle. For both plots, FLUKA (Geant4) results are shown with squares (circles), with solid (dashed) curves from fits with the Malensek parameterization.}
  \label{fig:beam_prod_ang}
\end{figure}
\Fig{fig:beam_prod_ang}, left, shows the absolute beam $K_L$ and $\Lambda$ rates as functions of production angle. There is some disagreement between the rates predicted by FLUKA and Geant4, with FLUKA predicting slightly more $K_L$ and fewer $\Lambda$s than Geant does. However, both simulations show that the decrease in flux with increasing angle is stronger for $\Lambda$s than for $K_L$.

\begin{figure}[htb]
  \centering
  \includegraphics[width=0.45\textwidth]{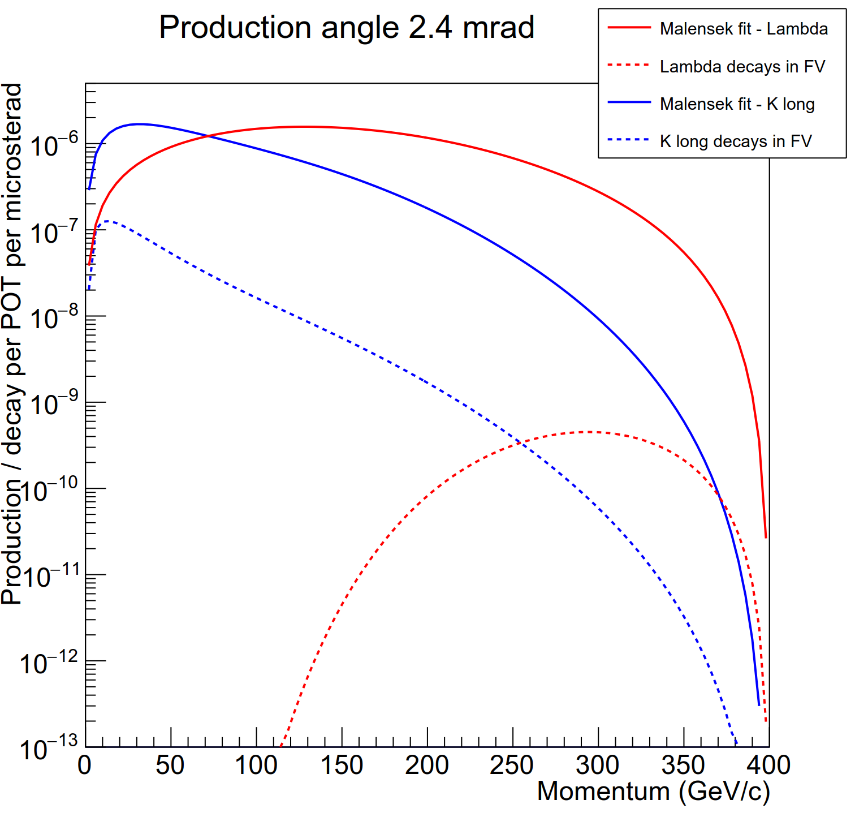}
  \includegraphics[width=0.45\textwidth]{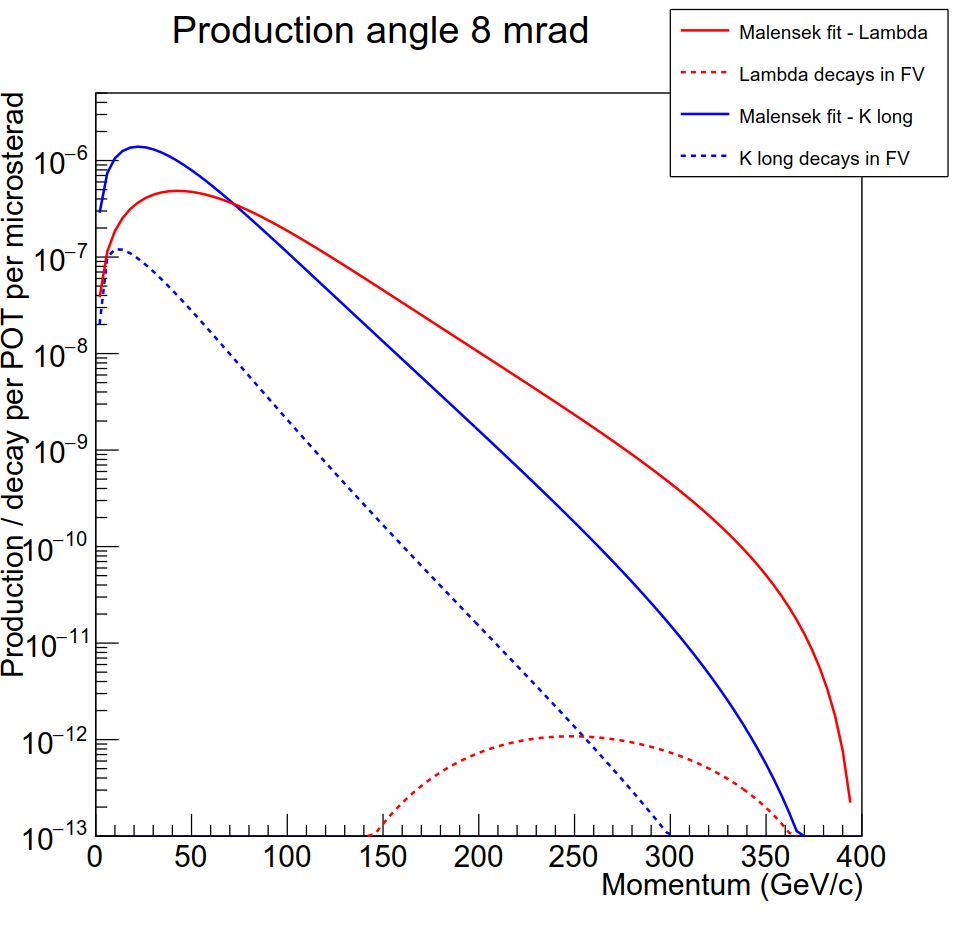}
  \caption{$K_L$ (blue) and $\Lambda$ (red) beam momentum spectra (number of particles in beam per POT per
  $\mu$sr of neutral beam acceptance) for production angles of 2.4~mrad (left) and 8~mrad (right). Solid curves show the Malensek parameterization of FLUKA results at production; dashed curves show the momentum spectra for particles decaying in the approximate fiducial volume of the experiment.}
  \label{fig:beam_mom}
\end{figure}
\begin{table}
    \centering
    \begin{tabular}{lcc}\hline\hline
          & 2.4 mrad & 8.0 mrad \\ \hline
         Mean $K_L$ momentum (GeV) & & \\
         \hspace{3ex}at production & 79 & 39 \\
         \hspace{3ex}in FV & 46 & 26.4 \\
         Mean $\Lambda$ momentum (GeV) & & \\
         \hspace{3ex}at production & 151 & 66 \\
         \hspace{3ex}in FV & 285 & 251 \\
         $K_L$ rate ($10^{-6}$/pot/$\mu$sr) & & \\
         \hspace{3ex}at production & 187 & 73 \\
         \hspace{3ex}in FV & 6.8 & 4.1  \\
         $\Lambda$ rate ($10^{-6}$/pot/$\mu$sr) & & \\
         \hspace{3ex}at production & 310 & 37 \\
         \hspace{3ex}in FV & 0.053 &  0.000132\\ \hline\hline
    \end{tabular}
    \caption{Parameters of $K_L$, $n$, and $\Lambda$ production with production angles of 2.4 and 8.0 mrad, from FLUKA. Mean momenta are for spectra as
    fitted with the Malensek parameterization.}
    \label{tab:beam_prod}
\end{table}

To further illustrate the trade-offs inherent in the choice of the beam angle, \Fig{fig:beam_mom} shows the $K_L$ and $\Lambda$ momentum distributions for two choices of production angle, 2.4 and 8~mrad, with absolute normalization. A production angle of 2.4~mrad was used for the $K_L$ beam in NA48, because this value maximizes the ratio $N^2(K_L)/N(n)$, with $N(K_L)$ the flux of kaons in the beam in a certain momentum interval ($70~{\rm GeV}<p_{K_L}<170~{\rm GeV}$) to the total flux of neutrons in the beam. This criteria was intended to balance the absolute $K_L$ production rate (which decreases with angle) against the ratio $N(K_L)/N(n)$
(which increases with angle). 
For a $K_L\to\pi^0\nu\bar{\nu}$ experiment, there are additional considerations. The extra boost provided by higher $K_L$ momentum facilitates the task of vetoing background channels with extra photons, such as $K_L\to\pi^0\pi^0$. On the other hand, with harder input momentum spectra, 
a smaller fraction of $K_L$s and a larger faction of $\Lambda$s decay inside the fiducial volume. $\Lambda\to n\pi^0$ decays constitute a potentially dangerous background, because the neutron will usually go unobserved.\footnote{The assumption for the design of the SAC is that the neutron will not interact in the detector 80\% of the time; see \Sec{sec:klever_sac}.}  
Although the $\Lambda$ decay length is only $c\tau_\Lambda = 7.89$~{\rm cm}, $\Lambda$s are produced at high momentum.
In \Fig{fig:beam_mom}, the solid curves show the momentum spectra, with absolute normalization for $K_L$s and $\Lambda$s at production; the dashed curves show the spectra for particles decaying in an approximate fiducial volume (in this case, 60~m long, starting 135~m downstream from the target).
For the 2.4~mrad production angle, production of $\Lambda$s with $p = 300$~GeV falls off by only a factor of $\sim$7 relative to at the peak of the momentum distribution around 120~GeV, while at $p = 300$~GeV, the effective decay length for $\Lambda$s is $\beta\gamma c\tau_\Lambda = 21.2$~m. As a result, with a production angle of 2.4 mrad, a considerable fraction of $\Lambda$s (0.17\%) reach the fiducial volume before decaying. Increasing the production angle 
ameliorates the problem considerably, principally by softening the $\Lambda$ spectrum at production, and secondarily by reducing the $\Lambda$ flux in the beam relative to the $K_L$ flux (see \Fig{fig:beam_prod_ang}).
The comparison between production angle settings is quantified in \Tab{tab:beam_prod}).
The move from 2.4~mrad to 8~mrad reduces the absolute rate of $K_L$ in the beam by nearly two thirds, but due 
to the increased acceptance of the fiducial volume for lower $K_L$
momentum, decays in the fiducial volume are only reduced by 40\%. Meanwhile, the $\Lambda$ acceptance is reduced by 
a factor of 240, such that the ratio of $\Lambda/K_L$ decaying in the fiducial volume (in this example) is on the order of $10^{-5}$, and the remaining $\Lambda$s have very high momentum. From \Fig{fig:beam_prod_ang}, the ratio of $n/K_L$ in the beam is seen to decrease from 10.9 to 2.6.

Use of a production angle greater than 8~mrad gives diminishing returns and causes additional complications in targeting the primary beam. At the relatively large production angle of 8~mrad, in order to maintain the sensitivity of the measurement given the limits on primary beam intensity, it
is necessary to make use of a larger solid-angle acceptance than that used in NA48 ($\Delta\theta = 0.15$~mrad, corresponding to $\Delta\Omega = \pi(\Delta\theta)^2 = 0.07$~$\mu$sr).
For $\Delta\Omega = 0.3$~mrad, the beam radius at the current position of the LKr calorimeter, 241.5~m from the target, is 73 mm; thus the beam just fits through the $r = 80$~mm central vacuum tube in the LKr. Use of a larger solid-angle bite requires replacement of the LKr, but this is foreseen for independent reasons (see \Sec{sec:klever_mec}). Ultimately, the choice of solid angle is limited by background from $K_L\to\pi^0\pi^0$ decays, as the $p_\perp$ distribution for reconstructed $\pi^0$s develops a tail towards higher values, undermining the effectiveness of the cut on $p_\perp$ used to identify signal events. 
The largest possible solid angle is $\Delta\theta = 0.4$~mrad, corresponding to $\Delta\Omega = 0.5$~$\mu$sr.

\paragraph{Layout}

\begin{figure}[htb]
  \centering
  \includegraphics[width=0.9\textwidth]{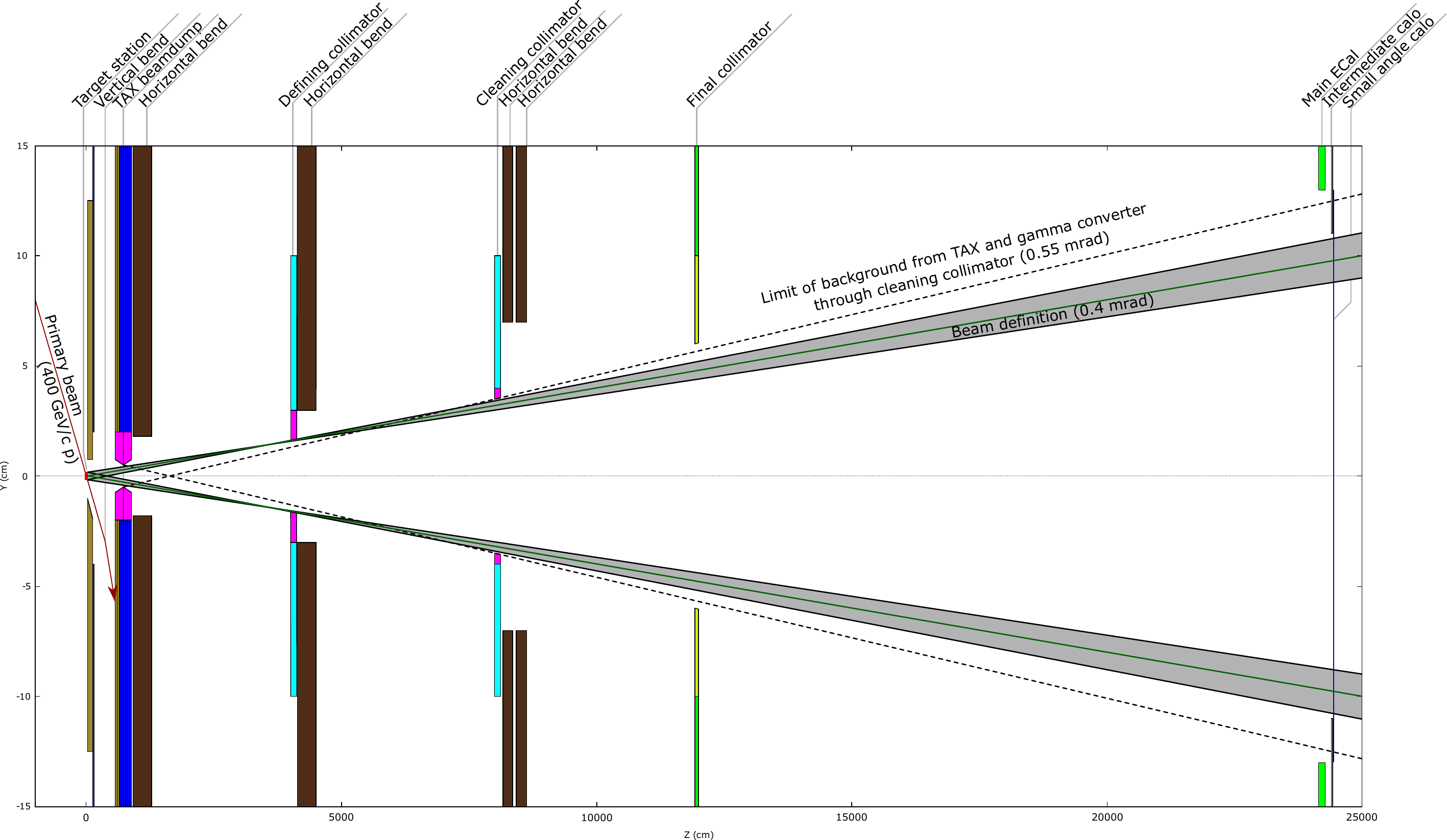}
  \caption{KLEVER neutral beamline layout as modeled in FLUKA,
    showing four collimation stages
    corresponding to dump, defining, cleaning, and final collimators.
    The final collimator is an active detector (AFC), built into the
    upstream veto. The aperture of the main calorimeter and coverage of the
    small-angle calorimeters is also shown.}
  \label{fig:beam}
\end{figure}
\begin{figure}[htb]
  \centering
  \includegraphics[width=0.5\textwidth]{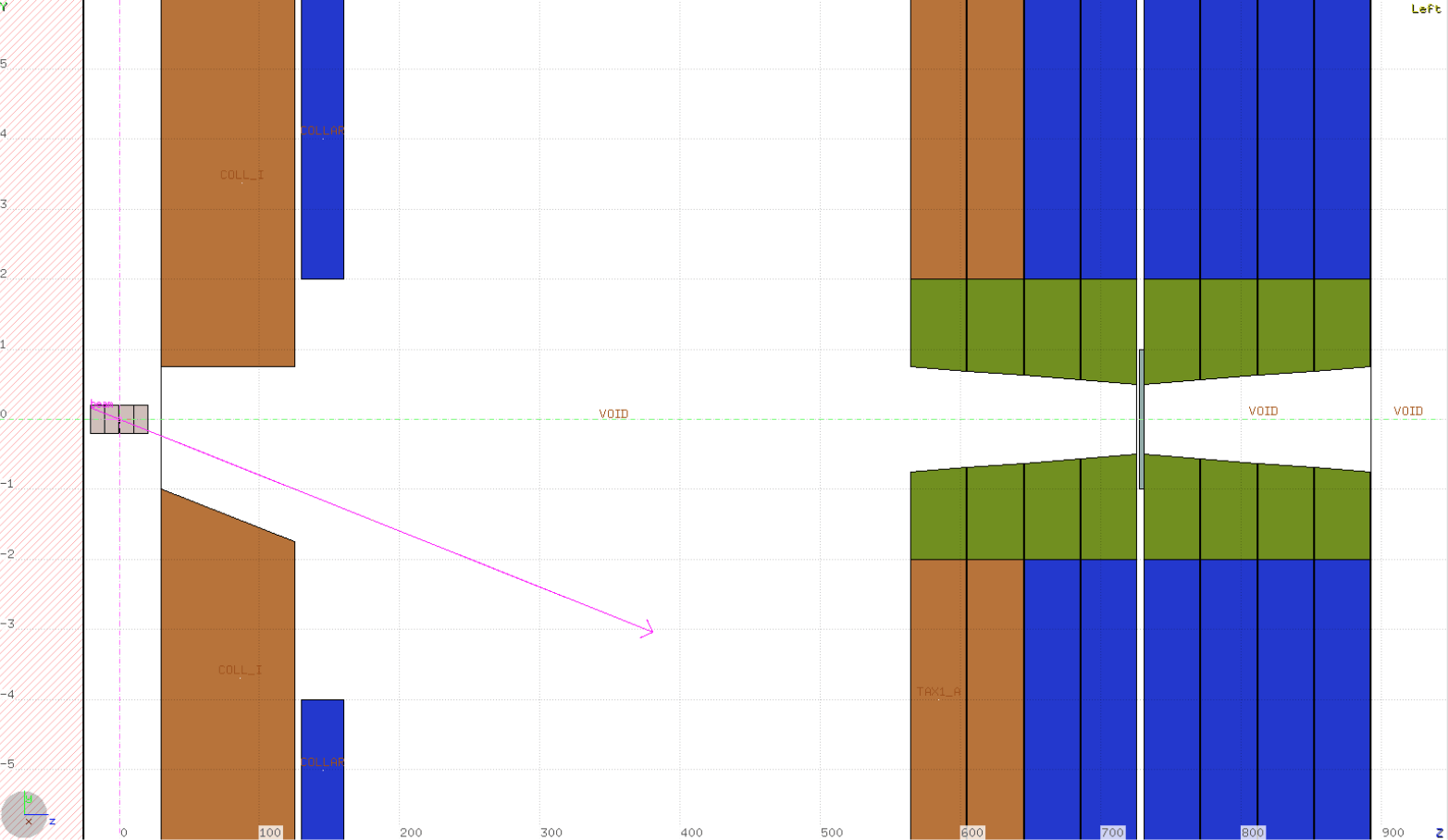}
  \caption{Detail from FLUKA showing the target region featuring the T10 target station internal collimator and steel collar, and the TAX, composed of copper and steel (blue) absorbers as described in \Sec{sec:proton_beam}. The photon absorber is inset in the middle of the tungsten TAX inserts (olive).}
  \label{fig:targ_area}
\end{figure}

A four-collimator neutral beamline layout for ECN3 has been developed \cite{VanDijk:2019oml}, as illustrated in Fig.~\ref{fig:beam}. The primary proton beam, sloping downwards at the desired production angle of 8 mrad, is incident on the T10 target. For the baseline design, the target is assumed to be a beryllium rod of 4~mm diameter. This is immediately followed by a vertical sweeping magnet (MTRV, 7.5~Tm), which deflects the protons downward by an additional 5.6 mrad.
The TAX dump collimator with a hole for the passage of the neutral beam is centered at a point 7.3~m downstream of the target. The TAX serves not only to dump the protons, but also to stop a large fraction of pions and kaons from the primary interaction before they can decay into muons, which contribute to background downstream. 
A photon converter consisting of about 9 $X_0$ of high-$Z$ material is positioned at the center of the TAX between the two modules, and reduces the flux of high-energy photons ($E>5$~GeV) in the neutral beam by about two orders of magnitude. 
A horizontal sweeping magnet downstream of the TAX (MTR, 7.5~Tm, 36~mm gap, $z=11$~m) ensures that the products of interactions on the edges of the collimator and photon converter (including the $e^+e^-$ pairs produced) are swept out of the acceptance of the subsequent collimation stages.
\Fig{fig:targ_area} shows a schematic of the area including the target and TAX as implemented in the FLUKA simulation of the beamline. 

A collimator at $z = 40$ m downstream of the target with $r = 16$~mm defines the beam solid angle, and a cleaning collimator at $z = 80$~m removes halo from particles interacting on the edges of the defining
collimator. Both are followed by horizontal sweeping magnets (MTN, 7.5~Tm, 60~mm gap, $z=43$~m for the defining collimator; $2\times$ MBPL, 3.8~Tm each, 110~mm gap, $z=83$~m for the cleaning collimator). The sweeping fields have been optimized to minimize muon backgrounds: at maximum field strength, the rates on the detectors are minimized with the polarities of the sweeping magnets for the dump, defining, and sweeping collimators in the left-right-right configuration. All of the sweeping magnets referred to here are elements of the existing K12 beamline for NA62 and can be reused.

The final stage of collimation is the active final collimator (AFC) incorporated into the upstream veto (UV) at $z=120$~m. The cleaning and final (AFC) collimators have apertures which are progressively larger than the beam acceptance, such that the former lies outside a cone from the (2~mm radius) target passing through the defining collimator, and the latter lies outside a cone from the TAX aperture passing through the cleaning collimator.

\paragraph{Particle fluxes and halo}
\begin{figure}
  \centering
  \includegraphics[width=0.45\textwidth]{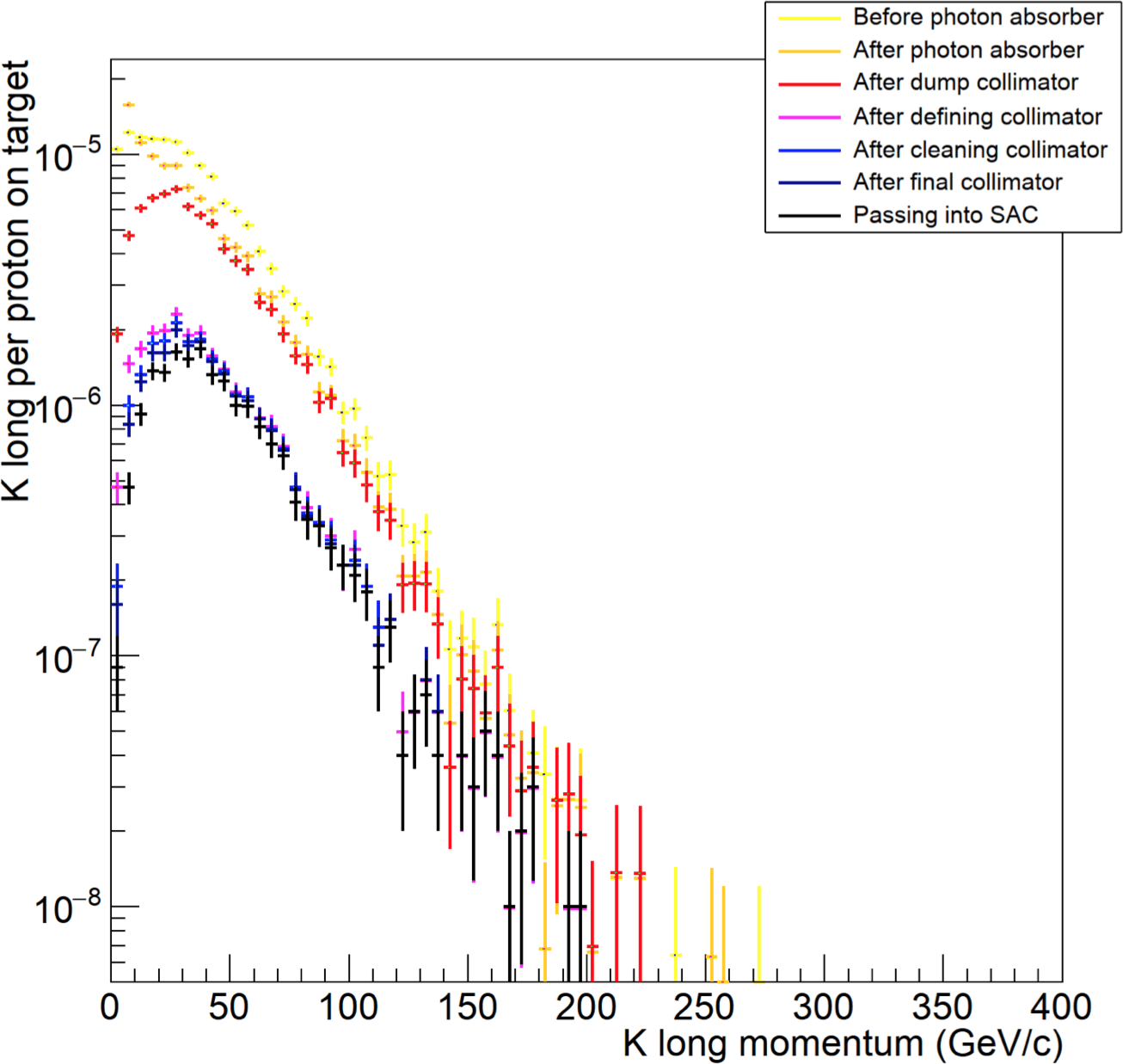}\\
  \includegraphics[width=0.45\textwidth]{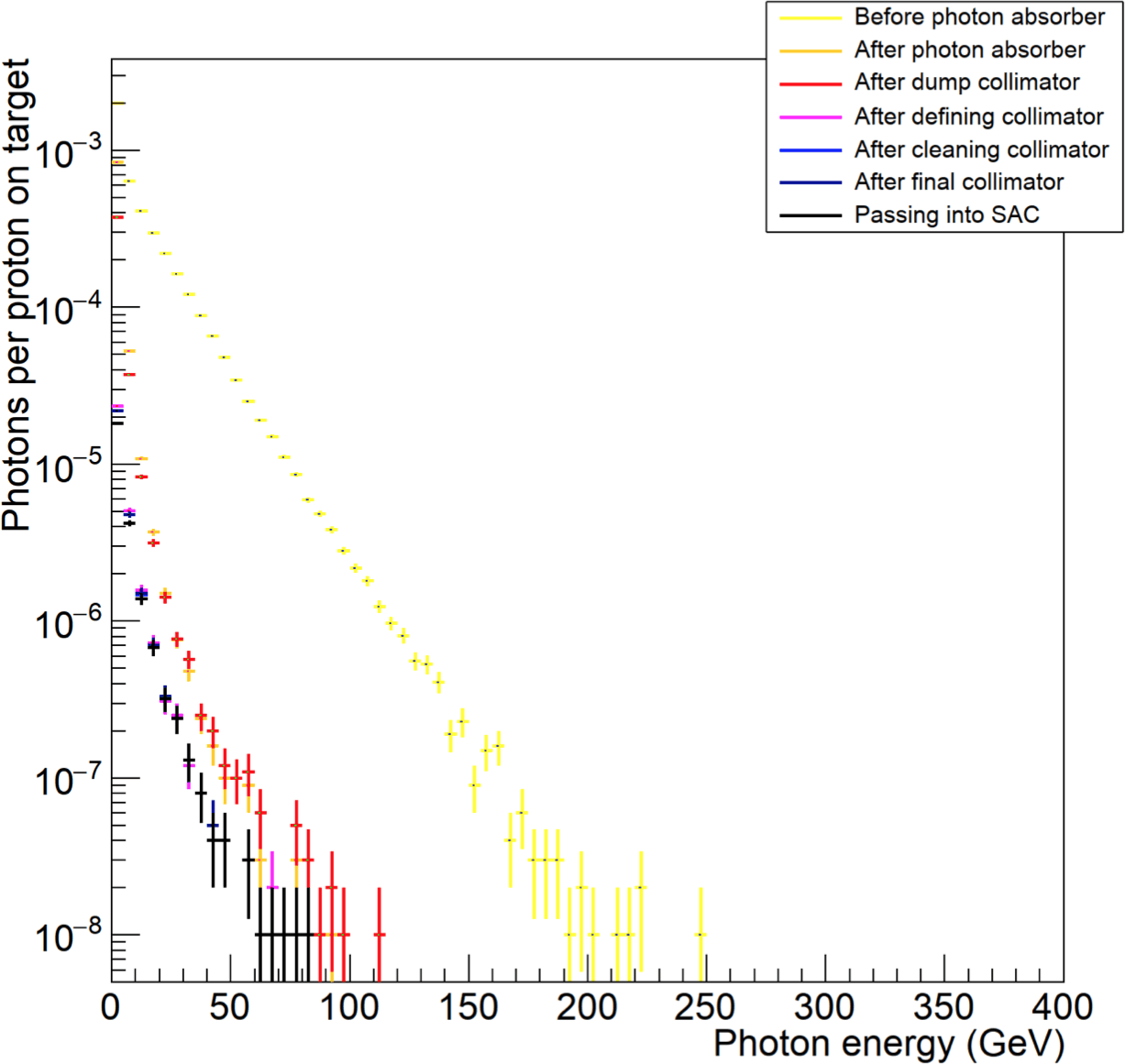}
  \includegraphics[width=0.45\textwidth]{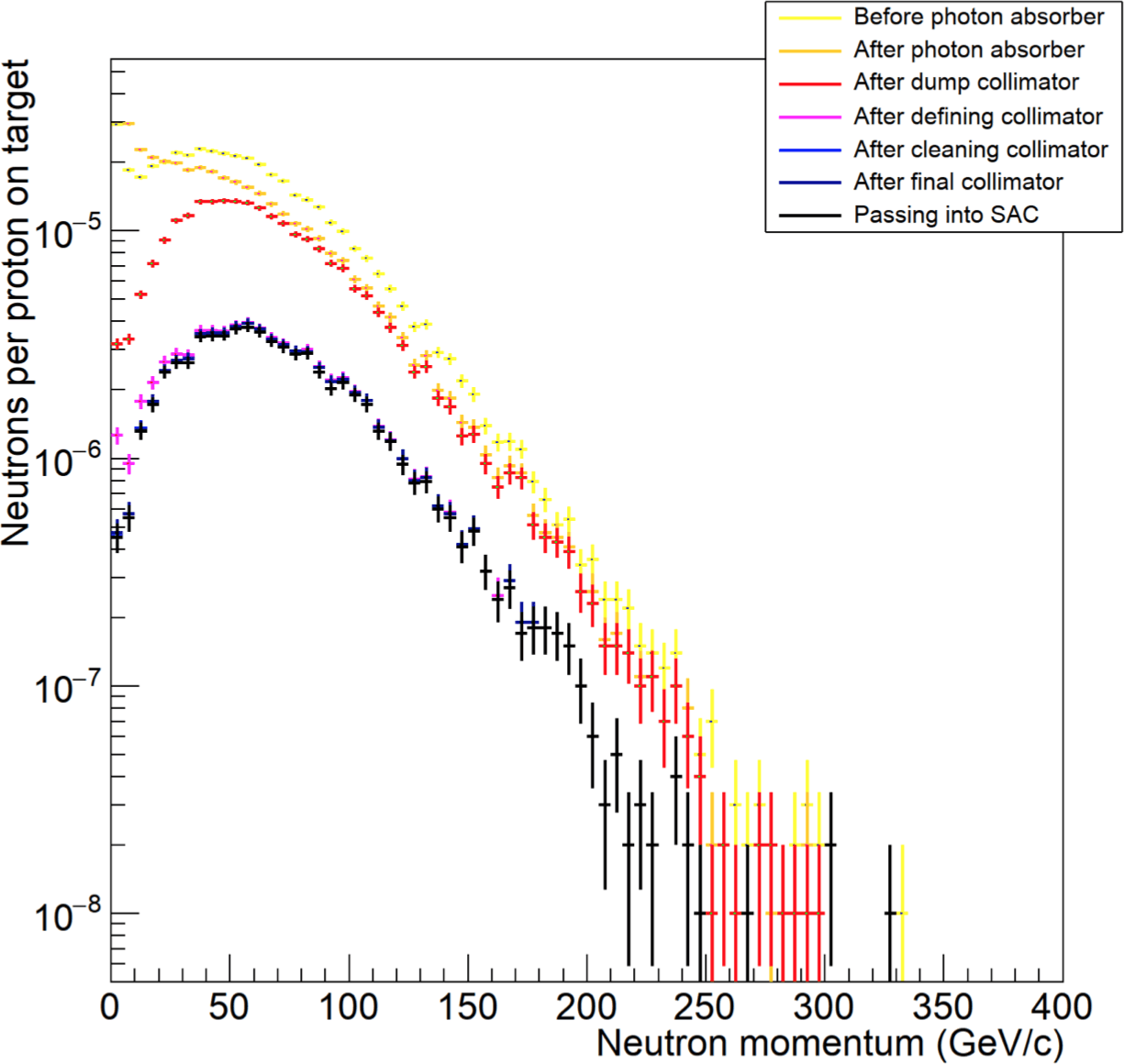}
  \caption{Momentum distributions for $K_L$s (top), photons (bottom left),
    and neutrons (bottom right) in the neutral beam at various points along
    the neutral beamline.}
  \label{fig:beam_flux}.
\end{figure}
\begin{table}
  \centering
  \small
  \begin{tabular}{lcccccc}\hline\hline
    \multicolumn{1}{c}{Beam component}  
    & \multicolumn{2}{c}{Before converter} 
    & \multicolumn{2}{c}{After converter} 
    & \multicolumn{2}{c}{After final collimator} \\
    & pot$^{-1}$ & GHz & pot$^{-1}$ & MHz & pot$^{-1}$ & MHz \\ \hline
    $\gamma$ & & & & & & \\
    \hphantom{2em} $E>1$~GeV & $4.2\times10^{-3}$ & 27.9
    & $9.1\times10^{-4}$ & 6100 & $2.97\times10^{-5}$ & 198 \\
    \hphantom{2em} $E>5$~GeV & $2.19\times10^{-3}$ & 14.6 
    & $7.1\times10^{-5}$ & 470 & $7.9\times10^{-6}$ & 53 \\
    \hphantom{2em} $E>10$~GeV & $1.55\times10^{-3}$ & 10.3 
    & $1.81\times10^{-5}$ & 121 & $3.15\times10^{-6}$ & 21 \\
    \hphantom{2em} $E>30$~GeV & $6.3\times10^{-4}$ & 4.2
    & $2.06\times10^{-6}$ & 13.7 & $6.2\times10^{-7}$ & 4.1 \\
    $n$ & & & & & & \\
    \hphantom{2em} $E>1$~GeV & $4.3\times10^{-4}$ & 2.88
    & $4.2\times10^{-4}$ & 2820 & $6.7\times10^{-5}$ & 440 \\
    $K_L$ & & & & & & \\
    \hphantom{2em} $E>1$~GeV & $1.37\times10^{-4}$ & 0.91
    & $1.29\times10^{-4}$ & 870 & $2.11\times10^{-5}$ & 140. \\
    \hline\hline
  \end{tabular}
  \caption{Particle fluxes in the neutral beam, as obtained from the FLUKA simulation of the beamline. The rates in MHz/GHz assume a primary beam intensity of $6.7\times10^{12}$ pot/s.}
\label{tab:beam_rates}
\end{table}

A detailed simulation of the entire beamline has been developed using the FLUKA package.
\Fig{fig:beam_flux} shows the momentum distributions for $K_L$s, photons,
and neutrons in the neutral beam at various stages of the beamline simulation:
at generation (i.e., at the exit from the target), after the converter and each of the three collimators, and at the end of the beamline, at the front face of the SAC. 
The normalization of these distributions provides the particle
fluxes in the neutral beam per incident proton. 
Of particular interest are the photon and neutron fluxes after the final
collimator.
The fluxes of photons, neutrons, and $K_L$s in the beam are summarized
in \Tab{tab:beam_rates}. 
For the purposes of the sensitivity estimates, $2.1\times10^{-5}$ $K_L$s enter the detector per proton incident on the target.
The beam rates (in GHz or MHz) in the table
assume a primary beam intensity of 
$2\times10^{13}$ ppp with an effective spill 
length of 3 seconds\footnote{The actual length 
is 4.8 s; the shorter effective spill length accounts for intensity fluctuations over the spill.}
In addition to the 140 MHz of $K_L$ mesons entering the detector, there are 440 MHz of neutrons: the $n/K_L$ ratio of about 3 observed for particle production at 8 mrad is not 
significantly changed during transport of the 
neutral beam. In addition to these particle 
fluxes, there are about 50 MHz of photons with $E>5$~GeV entering the detector, most of which are incident on the KLEVER small-angle calorimeter (SAC) at the downstream end of the detector (the
condition $E<5$~GeV condition corresponds to the SAC threshold).

The FLUKA simulation also contains an idealized representation of the KLEVER experimental setup, for the purposes of evaluating rates on the detectors from beam halo. These rates, with specific reference
to the KLEVER experimental geometry, are discussed in \Sec{sec:klever_rates}.

\paragraph{Target material and photon converter}

As seen from \Fig{fig:beam_flux} and \Tab{tab:beam_rates}, the photon converter in the TAX dramatically reduces the flux of photons in the beam, especially for high energy photons. This is critical to avoid blinding the SAC
at the downstream end of the beamline. 
In the baseline design, the converter is a tungsten prism of 32.9~mm thickness, corresponding to 9.4$X_0$, or 7.3 photon conversion lengths. This thickness is chosen to keep the rate of photons with $E > 5$~GeV below 40~MHz at the entrance to the SAC.
On the other hand, this thickness corresponds to 58\% of a nuclear collision length and 33\% of an interaction length, so that about 35\% of $K_L$ mesons and 40\% of neutrons interact in or are scattered out of the beam by the converter.\footnote{The sole effect of the converter on the $K_L$ and neutron flux in the beam is not readily obvious from \Fig{fig:beam_flux} and \Tab{tab:beam_rates} because particles scattered from the beam are eliminated by the defining collimator, which also limits the acceptance for particles from the target.} 

\begin{figure}
  \centering
  \includegraphics[width=0.45\textwidth]{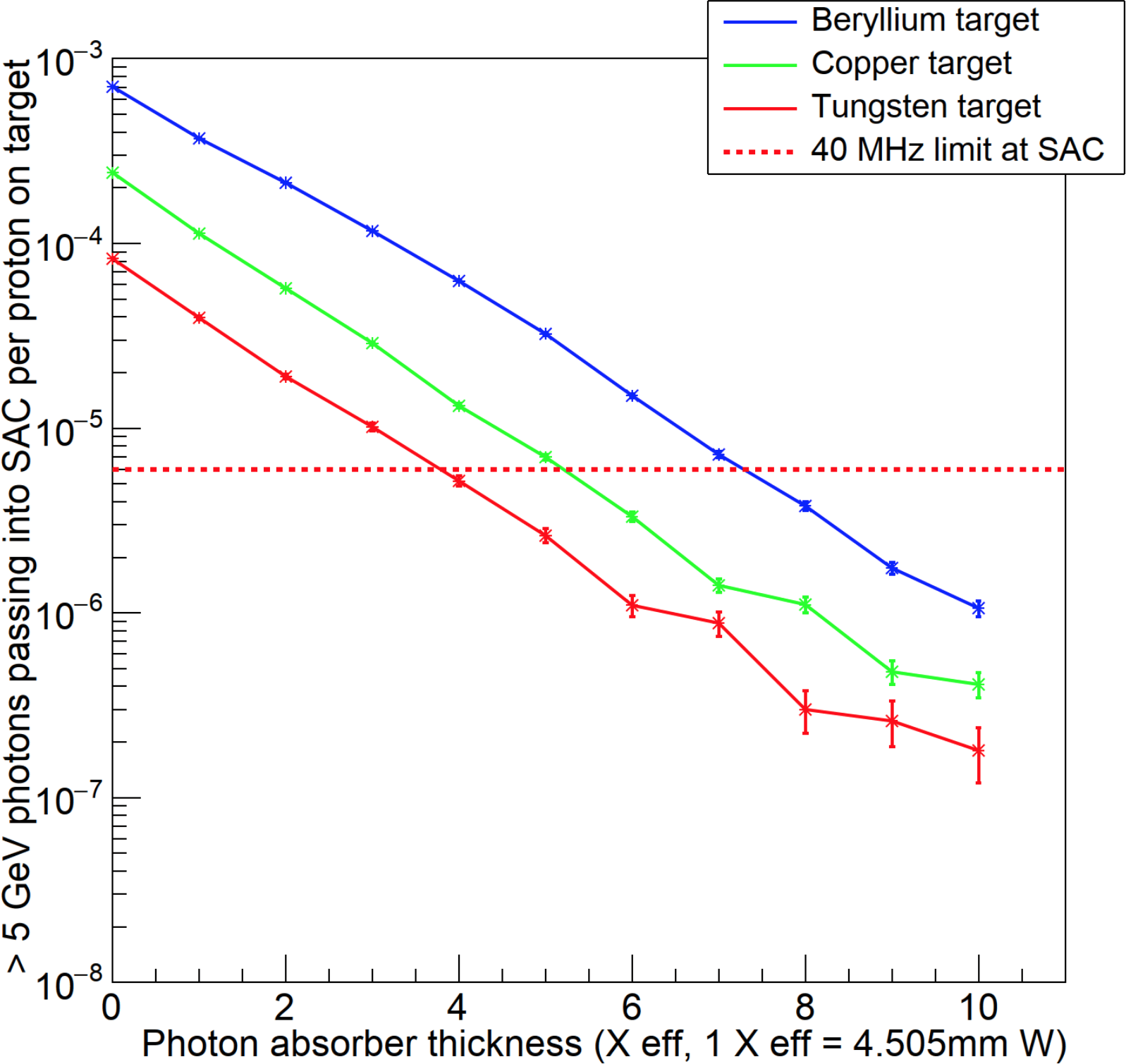}
    \caption{Beam fluxes for photons with $E>5$~GeV at the SAC as a function of photon converter thickness, for beryllium, copper, and lead targets. The dotted red line indicates the flux corresponding to a rate of 40 MHz, considered to be the maximum tolerable rate for high-energy photons from the beam on the SAC.}
  \label{fig:targ_photons}.
\end{figure}
The use of a higher-$Z$ material for the target is a potentially interesting alternative, as it would suppress the photon flux from the target without incurring such a significant loss of $K_L$ flux in the converter. Most of the prompt photons come from $\pi^0$ decays within the target, and the use of a high-$Z$ material causes many of these photons to be converted immediately, while the overall thickness of the target in nuclear interaction lengths is unchanged. (Recall that as described in \cite{vanDijk:2018aaa}, no large differences were found in terms of hadronic production for targets of identical thickness in nuclear interaction lengths.)
The rate of photons at the SAC for beryllium, copper and tungsten targets
(each with a nuclear interaction length equivalent to 400 mm Be), as determined with the FLUKA beamline simulation, is shown in \Fig{fig:targ_photons}.
It is seen that the required thicknesses of the converter for beryllium, copper, and tungsten targets of equivalent thickness are 7.3, 5.2, and 3.8 photon conversion lengths, respectively, resulting an effective increase in the $K_L$ flux into the detector of 15\% (28\%) for a copper (tungsten) target relative to a beryllium target. Changing the target material strongly impacts the target design,
as the energy deposit per unit of volume is much higher in a high-$Z$ target than in beryllium (see \Sec{sec:proton_beam}).

Another promising technique to further reduce the photon content in the beam is to use a crystal metal photon converter. If the crystal axis is aligned with the direction of the incoming photons, the coherent effect of the crystal lattice promotes pair production, leading to an effective decrease in the photon conversion length of the material~\cite{Bak:1988bq,Kimball:1985np,Baryshevsky:1989wm}. The effects of coherent interactions increase with photon energy and for decreasing angle of incidence.
A series of exploratory tests was performed with a set of tungsten crystals at the CERN SPS in summer 2018, together with the AXIAL collaboration~\cite{Soldani:2022ekn}. In particular, a commercial quality tungsten crystal of 10-mm thickness was targeted with a tagged photon beam. 
When the $\left<111\right>$ crystal axis was aligned with the beam to within 2.5~mrad, the multiplicity of charged particles was found to be enhanced by a factor 1.6--2.3 for photon energies over the range of 30--100 GeV. Ref.~\cite{Soldani:2022ekn} also reports simulations validated by this result that suggest that a crystal of this type could be used to reduce the thickness of the photon converter by 15--20\% at no cost in effectiveness. Such a solution appears to present little technical difficulty. 



\subsubsection{Neutral kaon beam with extended beamline}
\label{sec:neutral_beam_ext}

From the discussion in \Sec{sec:neutral_beam}, and in particular, from the rates in \Tab{tab:beam_prod}, one can extrapolate that for a total exposure of $5\times10^{19}$ pot, corresponding to 5 years of running, there are about $2\times10^{14}$ $K_L$ mesons entering the KLEVER fiducial volume. While the vast majority of $\Lambda$ baryons decay upstream of the FV, there will also be about $6\times10^8$ 
$\Lambda\to n\pi^0$ decays occurring inside the FV. 
It was initially thought that this background would have been effectively eliminated by cuts on $p_\perp$ and in the $\theta$ vs.\ $p$ plane for the $\pi^0$. As simulations improved however, it became clear that this background is potentially quite dangerous and that the number of $\Lambda$ baryons decaying in the FV must be significantly reduced. 
The following beamline modifications have been studied. They are not mutually exclusive.
\begin{itemize}
\item {\it Lengthening the distance between the target and the start of the fiducial volume} For the 8~mrad production angle, the effective $\Lambda$ decay length is about 17~m in the region of the FV. Lengthening of the beamline by any significant amount would involve new construction, as the planned layout makes use of the almost the entire length of the TCC8+ECN3 complex. Maintaining a beam spot of compact transverse dimensions is important for $K_L\to\pi^0\pi^0$ background rejection, so lengthening the beamline implies using tighter beam collimation, which decreases the $K_L$ flux. 
\item {\it Reducing the beam energy} Reducing the primary proton beam energy from 400~GeV to 300~GeV, for example, softens the secondary momentum spectra, and is particularly effective for the elimination of the high-momentum $\Lambda$s that decay in the FV. Inclusive $K_L$ production is slightly reduced at the lower energy \cite{Rossi:1974if}, but the total flux of $K_L$ in the beam may not change by much, especially after re-optimization of the production angle
\cite{Atherton:1980vj}.
The reduction of the beam energy would have to be agreed upon by all users of the slow-extracted beam. In particular, it would significantly affect high-intensity electron (or positron) and muon beams, as well as any high-momentum negative beams.
For this reason, this option has not yet been investigated seriously.
\item {\it Increasing the targeting angle} Increasing the targeting angle both decreases the ratio of $\Lambda$ to $K_L$ yields (\Fig{fig:beam_prod_ang}) and softens the $\Lambda$ momentum considerably (\ref{fig:beam_mom_ang}). It also decreases the absolute $K_L$ flux and softens the $K_L$ spectrum. For the purposes of the abatement of the $\Lambda$ background, the largest gains are already obtained by the move from 2.4~mrad to 8.0~mrad, as discussed in \Sec{sec:neutral_beam}. Increasing the production angle further provides decreasing returns. 
Any change to the production angle requires re-optimization of the entire layout and revision of the fiducial volume definition, beam opening angle, LAV coverage, energy thresholds (especially on the MEC, for signal events), efficiency requirements, and analysis cuts. On the basis of simulations of a partially re-optimized configuration with $\theta = 20$~mrad, $\Delta\theta = 0.4$~mrad, and the FV moved downstream to $160~{\rm m} < z < 200~{\rm m}$, it appears that the $\Lambda\to n\pi^0$ background could be sufficiently suppressed, but that the sensitivity of the experiment would be significantly decreased, perhaps by a factor of 4 (i.e., to 15 SM $\pi^0\nu\bar{\nu}$ events). These results are highly preliminary and leave space for a smaller change in the production angle to be part of a multi-pronged strategy for dealing with the $\Lambda$ background. 
Increasing the target angle much beyond 8~mrad will require a new target design with larger transverse dimensions, leading to a larger target image and increased beam halo. Alternatively, this could lead to a design with a higher-$Z$ target, with the attendant complications (\Sec{sec:proton_beam}). The feasibility of targeting at large angle (e.g., 20~mrad) needs to be studied from the standpoint of the primary beam steering.
\item {\it Tighter beam collimation} 
The main effect of the beam opening angle on the sensitivity of an experiment like KLEVER is from the change in the quality of the constraint on the decay position in the transverse plane for $\pi^0$ reconstruction, i.e., the beam-spot size. 
The contribution of the beam transverse momentum to the $p_\perp$ of the decay $\pi^0$ is a secondary effect.
In the standard KLEVER configuration, the downstream end of the FV is positioned at least 60~m upstream of the calorimeter to ensure good reconstuction of $p_\perp(\pi^0)$, but a substantial reduction of the beam solid angle (e.g., from $\Delta \theta = 0.4$~mrad to 0.3~mrad) could help to eliminate residual $\Lambda \to n\pi^0$ decays and would definitely improve rejection for $K_L\to\pi^0\pi^0$, at the price of reducing the $K_L$ flux by the ratio of solid angles (e.g., by 0.3$^2$/0.4$^2 = 0.56$). Note that, for extended beamline configurations, the collimation of the neutral beam must be tightened simply to maintain the beam-spot size, with corresponding loss of $K_L$ flux and no significant benefit for additional background rejection. In addition, in the extended beamline configurations, because of the tighter collimation, the beam is more parallel,
but the beam-spot size is larger {\em inside} the FV  than it is in the more compact configurations.
These considerations underscore the importance of the angular constraints from the PSD and/or MEC to improve the $p_\perp(\pi^0)$ determination with decreased reliance on the constraint from beam-spot size.
\end{itemize}
Other experimental approaches to provide additional rejection against $\Lambda\to n\pi^0$ are discussed in \Sec{sec:klever_lambda}.

\begin{figure}
    \centering
    \includegraphics[width=0.60\textwidth]{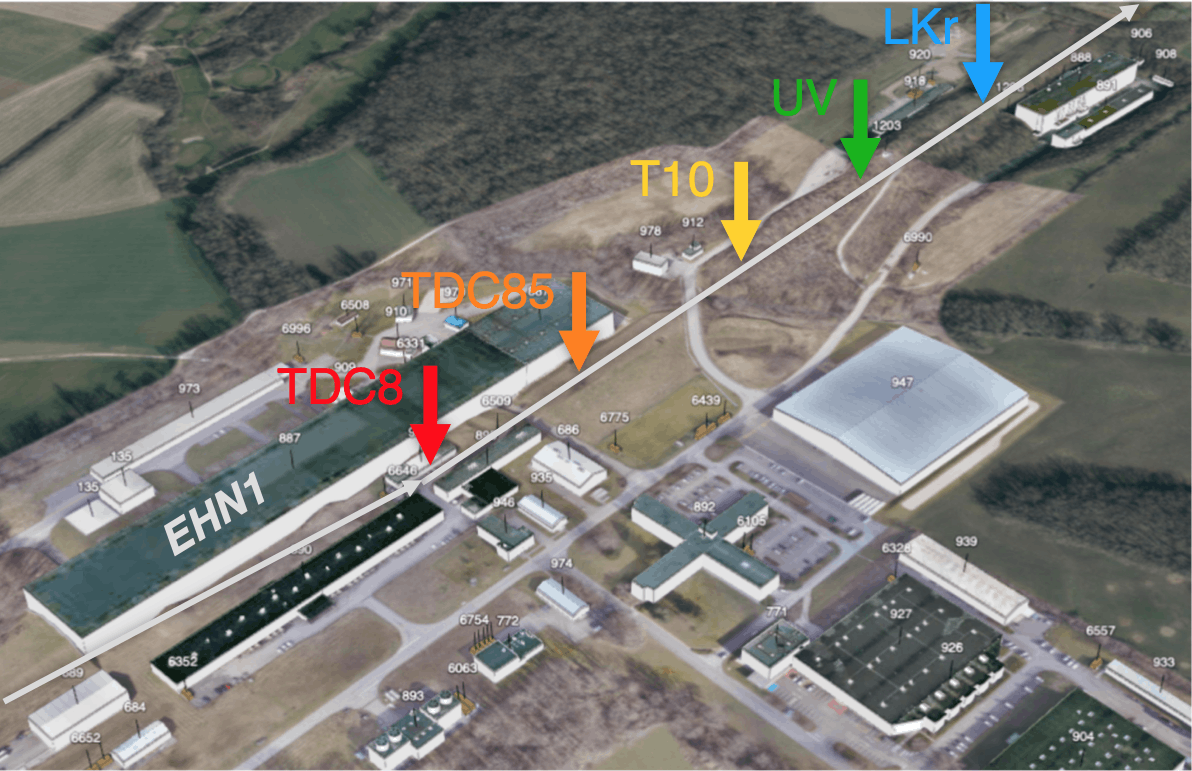} \\[3ex]
    \includegraphics[width=0.85\textwidth]{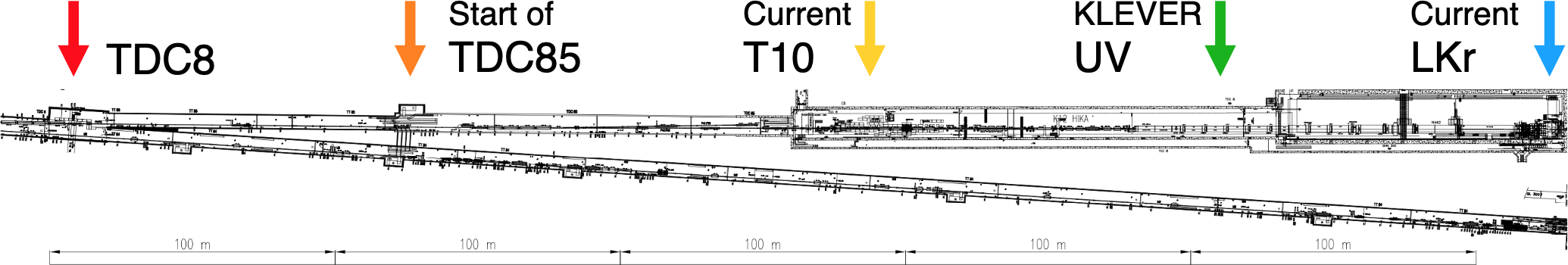}
    \caption{Layout of the P42 and K12 beamlines, with schematic and aerial view of surrounding area, indicating the TDC8 and TDC85 caverns, which are possible target locations for extended beamline configurations, as well as the locations of the current T10 target, the KLEVER UV/AFC, and the current LKr calorimeter for reference.}
    \label{fig:ext_beam}
\end{figure}
\begin{table}
  \centering
  \small
  \begin{tabular}{lcccc}\hline\hline
  Configuration & Standard & L1 & L2 & L3 \\ \hline
  Target pos. [m] & 0 (TCC8) & $-270$ (TDC8) & $-150$ (TDC85) & $-270$ (TDC8) \\
  Final coll. [m] & 120 (ECN3) & 120 (ECN3) & 120 (ECN3) & 30 (TCC8) \\
  Length [m] & 120 & 390 & 270 & 300 \\
  $\Delta \theta$ [mrad] & 0.400 & 0.200 & 0.256 & 0.250 \\ \hline
  \end{tabular}
  \caption{Possible beamline configurations for the KLEVER experiment: the standard configuration and three alternatives with extended length.}
\label{tab:ext_layout}
\end{table}
Of the possibilities mentioned above, at present, one of the most attractive options is to extend the beamline. \Fig{fig:ext_beam} shows the layout of the P42 and K12 beamlines, both as a schematic and as an aerial view of the surface surroundings.
Two possible locations for alternative placement of the target station in the extended beamline scenario are in TDC8, the cavern at which the tunnel for the M2 and P42 beamlines splits into separate tunnels to bring 
the M2 muon beamline to EHN1 (COMPASS) and P42 to TDC85/TCC8/ECN3 (NA62), 
and at the upstream end of the long technical cavern TDC85, which brings the P42 beamline into TCC8. 
In addition to the standard configuration, three alternate configurations with different possible locations for the beam and experiment are listed in \Tab{tab:ext_layout}: 
\begin{itemize}
\item In configuration L1, the target is moved 270~m upstream from its current position into TDC8, while the experiment remains in its current location, and the total beamline length from target to final collimator is 390~m.
Because of the tight collimation required for a beamline of this length, the solid angle and $K_L$ flux are reduced by a factor of four. 
There is space in TDC8 and the cost of infrastructural changes for the installation of the target station alone would be relatively contained.
The tunnel from TDC8 to TCC8 (including TT85 and TDC85) slopes upwards by 9.5~mrad. This places constraints on the possible production angles that can be used, though values from 8--12~mrad can probably be accommodated. The beam emerges into TCC8 and ECN3 with a positive slope, necessitating a complicated 
(although possible, from the standpoint of existing vertical clearance) installation to keep the detectors in the plane of the beam. The target station in TDC8 would also be a source of background for the experiments in EHN1 and the neutrino platform extension, even with extensive additional shielding for the technical galleries and surface structures in the vicinity. Vehicle access to EHN1 via the northeast door would be difficult to maintain because of radiation and shielding concerns. 
\item In configuration L2, the target is moved about 150~m upstream from its current position into TDC8, while the experiment remains in its current location, and the total beamline length from target to final collimator is 270~m.
The solid angle and $K_L$ flux are reduced by a factor of 2.4. 
There is sufficient space in TDC85; the costs would be higher than for configuration L1 because a new target cavern and access shaft would have to be constructed in TDC85. The target could be elevated off of the floor enough to clear the slope of the tunnel, allowing a level installation of the detectors in TCC8/ECN3. From the standpoint of radioprotection, the main concerns are how to protect the street allowing access to the ECN3 surface building and the Lion creek, both of which cross the surface projection of the beamline just downstream of the target station.
\item In configuration L3, the target is located at the same position as in configuration L1, but the experiment is moved upstream into TCC8. This shortens the beamline from 390~m to 300~m, so that a solid angle similar to that for configuration L2 can be used. The engineering and radioprotection considerations are the same as for configuration L1. 
\end{itemize}
Of these configurations, L2 seems to be the most promising. Work at present is focused on fully understanding the radioprotection issues, especially regarding the area around the Lion creek, where the reduced overburden of soil is an additional concern. 
Alternatively, a straightforward plan to extend the beamline would be to lengthen ECN3 in the downstream direction. An extension of ECN3 by at least 100~m would be necessary. The land needed for this extension is on CERN property and is free of structures at present. The primary drawbacks of this solution are the cost and timescale for the civil construction needed, though it is not certain that the civil engineering costs for extension of ECN3 will significantly exceed those for procuring and installing the additional shielding needed for configuration L2. Work on comparative costs and timescale estimates for the two solutions is currently in progress.

\subsection{A high intensity $K^+$ experiment}
\label{sec:NA62x4}

The NA62 experimental techniques has been proven to be successful, as highlighted in the previous sections. Therefore the future high intensity charged kaon program at CERN SPS can benefit from the existing infrastructure of the NA62 experiment, and will be built on its accumulated experience.
The design of the experimental apparatus will follow the same experimental technique of decays in flight, and will need to meet the new physics requirements.
A future charged kaon experiment will face a beam intensity up to a factor four higher than that of NA62, and should allow the study of a broad physics program.
The expansion of the physics program will specifically require a new trigger and data acquisition system.
The intensity increase will drive the hardware choices in the most demanding detectors.

The future experiment aims to make precision measurements and to search for multi-body rare decays with pions and leptons in the final state, along with the measurement of the $K^+\to\pi^+\nu\bar\nu$ decay at a challenging 5\% precision. The basic experimental design (\Fig{fig:NA62}) comprises of a set of beam detectors upstream of the fiducial volume and of detectors to reconstruct the final state, similarly to the NA62 layout~\cite{NA62:2017rwk}. The main beam detectors are: a kaon identification system; a beam tracking device; a device to mark the start of the decay fiducial region. The main final state detectors are: a spectrometer; a particle-identification system for pions, electrons, photons, and muons; calorimeters; a veto system.

In the following, considerations are made regarding each of the sub-detectors, and the extent to which their existing design needs to be modified in order to satisfy the new physics requirements. 

\subsubsection{Beam detectors}
\label{sec:NA62x4_detectors_beam}
The beam detectors will identify the beam particles and measure their positions and momenta. The corresponding CEDAR/KTAG and GTK detectors in NA62 have been proven to allow a substantial reduction of accidental background and excellent kinematic resolution of the reconstructed final state required for rare kaon decay measurements.
In addition, beam detectors allow the investigation of systematic effects relevant for precision measurements, offering an independent method to calibrate the downstream spectrometer for pions, the possibility to study the motion of the beam spot during the spill, and a means to measure the instantaneous intensity. Beam detectors rely on time measurement, whose precision requirement is driven by the beam intensity. A time resolution of $\mathcal{O}(100)$~ps is appropriate at the NA62 intensity, leading to a few percent probability of matching a final state to an accidental beam particle.
This is essential for the NA62 program to keep the non-kaon background of $K^+\to\pi^+\nu\bar\nu$ decays under control, while ensuring high signal acceptance. The foreseen increase of intensity requires an almost linear decrease of time resolution, setting the bar for time resolution at $\mathcal{O}(25)$~ps. A figure of merit justifying the linear scaling is the probability of misidentification of an accidental beam particle as a kaon, measured by NA62. An assumption is made that the beam phase space of the new $K^+$ experiment will remain the same as that of NA62~\cite{NA62:2010xx}.

The CEDAR/KTAG detector used for Cherenkov identification of the beam kaons will operate with $\textrm{H}_2$ gas in the last years of NA62 data taking before LS3, thereby reducing the material along the beam that otherwise could be a source of background at increased beam intensities. The intrinsic potential in terms of CEDAR/KTAG time resolution is suitable for operation at higher intensity. However
the conventional PMTs used at present have a rise time of 200--300~ps, and a four-fold improvement in single-photon time resolution to 75~ps is required to maintain the current performance characteristics during the high-intensity data collection.



MicroChannelPlate photomultipliers are a fast developing technology devices for high-speed single photon counting applications. They combine two fast timing technologies (photo-multiplier tubes and microchannel plates) into a single sensor, resulting in high gain devices with MHz/cm$^2$ rate capabilities. However a current limitation of standard MCP-PMTs is their aging. Damage to the photocathode by accelerated ions produced in the amplification process leads to finite lifetime, followed by a sharp decrease in the quantum efficiency (QE). Ageing effects are quantified by investigating the 
QE as a function of the integrated anode charge. In a highly promising development, MCP-PMT lifetimes have been shown to increase dramatically by treating them with Atomic Layer Deposition coating~\cite{Beaulieu:2009}.
Lifetime corresponding to an integrated anode charge greater than 20~C/cm$^2$ has been reached using improvements of this technique~\cite{Lehmann:2020}. 
These photo-detectors have an excellent timing performance and are therefore suitable for use also in the RICH detector, where the radiation exposure and therefore the aging are not an issue.

Tracking of the $K^+$ beam will require a new beam tracker with time resolution of ${\cal O}(50)$~ps or better, capable of operating
at rates of up to 8~MHz/mm$^2$ (3~GHz over 3$\times$6~cm$^2$), while maintaining a minimum amount of material in the beam ($X/X_0 \approx 0.3-0.5$\% per tracking plane). The use of the temporal information in the pattern recognition makes possible to achieve efficient tracking with a minimum number of planes and associated material but requires a good single hit timing accuracy.

The development of these devices is synergistic with collider requirements since they necessitate ultra-fast detectors, enabling 4D-tracking, to deal with multiple interactions occurring within a bunch crossing. High-resolution position and timing determination and low power consumption are important also for a large fraction of upgrades and new projects, as well as for kaon experiments: the targeted time resolution is about 30~ps for the detectors at LHC and 20--30~ps for the EIC collider, and devices with even more challenging 10~ps timing resolution will be highly desirable for 4D-tracking reconstruction at the foreseen collision pile-up rate of the FCC-hh~\cite{ecfareport:2021}.
While the radiation tolerance of $2.3 \times 10^{15}$ n$_{\rm eq}$/cm$^2$/yr (200~days) does not reach the very challenging level foreseen for hadron colliders, the specific requirement of efficiency above 99\% for beam particle tracking in kaon experiments implies to achieve a 100\% fill factor. One aspect common to high-intensity kaon experiments as well as most future facilities is the requirement for the front-end electronics to transfer off-chip very large data volumes.

At present the two families of sensors that are delivering suitable time performances are thin, planar Low-Gain Avalanche Diodes (LGADs)~\cite{lgad:2014} optimised for timing (also known as Ultra-Fast Silicon Detectors~\cite{lgad:2017}), and 3D-sensors with columns~\cite{3dsensor:2019} or trenches~\cite{3dsensor:2020}. The best performances in terms of temporal precision for these sensors are similar, about 20--30~ps, that are well in line with the requirements of the high-intensity charged kaon experiment. Presently, LGADs are more promising with respect to the low material budget requirement. Further technology advances are under development to reduce the impact of structures between readout pads, such as Trench-isolated LGADs~\cite{tilgad:2020} and Resistive AC-coupled LGAD~\cite{aclgad:2019}.

A guard-ring detector to hardware tag the beginning of the decay region will replace the CHANTI and VetoCounter detectors of NA62.
The CHANTI is a hodoscope-like detector made of scintillator bars, about 3~cm wide, read by SiPMs and arranged in six square stations surrounding the beam just downstream of the final GTK station according to a telescope configuration. The main role of the CHANTI in NA62 is to flag and reject events with charged particles missing the geometrical acceptance of the detectors downstream of the decay region, or not coming from the beam.
These particles originate from hadronic interactions of the beam with the material of the GTK stations, or from muons produced upstream as a result of pion or kaon decays, 
passing through the final collimator, and converging towards the downstream detectors. Historically, the CHANTI has allowed the discovery of
issues related to limited coverage in the final collimator used by NA62 until 2018, leading to its replacement.
The VetoCounter is also made of scintillating bars, and served to detect of early kaon decays upstream of the fiducial region that can represent a background source in specific conditions.

The intensity is the main challenge of such detectors in the future high-intensity $K^+$ experiment.
A possible solution is to build a new detector with $\times 4$ granularity, read with PMs like those already employed by NA62 for KTAG and RICH.
This detector will have a $\mathcal{O}(100)$~ps time resolution (i.e. at least five times better than that of the NA62 CHANTI), which is sufficient to guarantee less than percent loss of signal because of random veto.


\subsubsection{Final state detectors}
\label{sec:NA62x4_detectors_downstream}
The NA62 configuration of the detectors downstream of the decay region that measure the final state particles shows a left-right asymmetry with respect to the beam axis. The 75~GeV/$c$ beam enters the decay region deflected by 1.2~mrad angle in the $X-Z$ plane. 
This compensates the opposite kick 
due to the spectrometer dipole, allowing the beam to reach the electromagnetic calorimeter in the center of the beam pipe.
The size of the beam hole of  spectrometer straw stations and the radius of the beam pipe of the RICH detector are set to minimize 
background to $K^+\to\pi^+\nu\bar\nu$ from final states with particles traveling close to the beam pipe. As a result, the NA62 straw stations are centered on the beam path instead of the beam axis, and the RICH is tilted in the $X-Z$ plane.

A future $K^+$ decay physics program may benefit of a left-right symmetric apparatus, 
while minimal impact on the $K^+\to\pi^+\nu\bar\nu$ analysis must be ensured.
Benefits would include
reduction of systematic effects due to the left-right asymmetry, and
easier switching between charged and neutral beams. These features have been used successfully by NA48 and NA48/2 for CP violation studies~\cite{NA48:2002tmj,NA482:2007ucr}, and by NA62 to test lepton universality in $K_{\ell 2}$ decays using the NA48/2 apparatus~\cite{NA62:2012lny}.
A viable option could be to reduce the dipole field and to increase the size of the beam hole of the spectrometer stations and the radius of the beam pipe. The NA62 experience has also shown that calorimeters around the beam pipe at the end of the apparatus (the HASC detectors) are very effective in detecting particles at low angles.
Eventually the small angle calorimeters can be designed to match the new left-right symmetric geometry.

\subsubsubsection{Spectrometer}


A spectrometer similar to the NA62 straw tracker~\cite{NA62:2017rwk} is planned to reconstruct charged particles in the final state. The straw tubes are expected to stay in the vacuum tank containing the decay region, profiting from the successful technology developed for NA62. Straws with diameter less than 5~mm are necessary to handle the expected particle rates at the higher beam intensity. Straw diameter reduction by a factor $\sim 2$ with respect to the NA62 straws will lead to shorter drift times and improvement in the trailing-edge time resolution from the current ${\sim}30$~ns to ${\sim}6$~ns. Smaller straw diameter also necessitates a change in the geometrical placement of the straws in a single view. Design work based on Monte-Carlo simulations was performed, and the straw layout was optimized taking into account realistic spacing and dimension requirements, which resulted in a choice of eight straw layers per view.

The material used to make new straws using the ultrasonic welding technique will be the same as in the current spectrometer, namely mylar coated with 50~nm of copper and 20~nm of gold on the inside. To reduce the detector material budget, the mylar thickness will be reduced from $36~\upmu$m to either $12~\upmu$m or $19~\upmu$m. The diameter of the gold-plated tungsten anode wires might be reduced from $30~\upmu$m to $20~\upmu$m. The final decision about the thickness and the wire diameter will be made based on mechanical stability tests. Development of small-diameter thin-walled straws has synergies with R\&D work for COMET phase-II at J-PARC~\cite{Nishiguchi:2017gei} and is included in the ECFA detector R\&D roadmap~\cite{ecfareport:2021}.

Based on the design study results, a realistic Geant4-based simulation of the new spectrometer was developed using the same dimensions and positions of the straw chambers, the number and orientation of views in the chamber, the gas composition (Ar+CO$_2$ with 70:30 ratio), and the dipole magnet properties as in the current NA62 layout. A comparison between the two straw detectors is given in Table~\ref{tab:na62_new_straw_comparison}.

The NA62 track reconstruction algorithm was adapted for the new detector, and a preliminary resolution comparison indicates that the new spectrometer could improve the resolution for the reconstructed track angles and momenta by 10--20\% with respect to the existing NA62 spectrometer while maintaining the high track reconstruction efficiency.

\begin{table}[H]
\centering
\caption{Comparison of the NA62 spectrometer and the new straw spectrometer.}
\begin{tabular}{|l|r|r|}
\hline
&Current NA62 spectrometer& New straw spectrometer\\
\hline
Straw diameter & 9.8~mm & 4.8~mm\\
Planes per view & 4 & 8 \\
Straws per plane & 112 & 160\\
\hline
Mylar thickness & $36~\upmu$m  &  (12 or 19)~$\upmu$m \\ 
Anode wire diameter & 30~$\upmu$m & (20 or 30)~$\upmu$m \\
Total material budget & 1.7\% $X_0$ & (1.0 -- 1.5)\% $X_0$ \\
\hline
Maximum drift time &  ${\sim}150~$ns &  ${\sim}80~$ns \\
Hit leading time resolution & (3 -- 4)~ns & (1 -- 4)~ns\\
Hit trailing time resolution & ${\sim}30~$ns & ${\sim}6~$ns \\
Average number of hits hits per view & 2.2 & 3.1\\
\hline
\end{tabular}
\label{tab:na62_new_straw_comparison}
\end{table}




\subsubsubsection{Particle identification and calorimetry}

The $K^+$ decay program can profit from lepton identification with the RICH detector. This is particularly useful for the $K^+\to\pi^+\nu\bar\nu$ measurement, as well as the lepton flavour/number program (Section~\ref{sec:rare_kaon_decays}).
The NA62 neon RICH is well suited for this purpose, and both the vessel and the gas system will be reused in the future experiment.
New mirrors will replace the existing ones which are showing degradation in reflectivity because of ageing. In addition, a design of thinner mirrors than the 2~cm thick glass mirror employed by NA62 is foreseen. This would reduce the material budget in front of the electromagnetic calorimeter, improving the resolution of the electron and photon energy reconstruction. The NA62 RICH reconstructs efficiently positive particles only. For a full usability of the RICH in the charged kaon experiment, the mirror geometry will be adapted to reconstruct efficiently also negative particles.
New photo-detectors will be necessary to guarantee a time resolution in line with KTAG and GTK.
This is essential because the RICH is the main detector to measure the time of the charged particles in the final state, with the upgrade to proceed in parallel with that of the kaon tagging detector.
In addition, a new coupling between the photo-detectors and the radiator will be studied to allow efficient detection also of the UV component of the Cherenkov photons, improving the light yield by about 20\%. The relatively low occupancy of the RICH in a kaon experiment makes the track-based and standalone ring reconstruction used by NA62 adequate also at higher beam intensity.

Planes of fast scintillators located between the RICH and the calorimeter to timestamp the charged particles independently from the RICH are present in NA62 (the CHOD and NewCHOD detectors), and are foreseen in the future experiment. The hodoscopes are used as a trigger of charged particles and are effective as a time reference for RICH ring reconstruction. In addition, these detectors measure the time of charged particles below the RICH Cherenkov threshold (about 15~GeV/$c$ for pions), and provide spatial information for the calorimeter reconstruction. The scintillator slabs of the CHOD detector of NA62 cannot stand the future intensity, as the rate per channel would reach several MHz due to the coarse granularity. The pads of the NewCHOD detector of NA62 are not adequate either, because the time resolution exceeds 1~ns due to the fibers used to extract the signal. The granularity of the new hodoscope must allow the rate of 1~MHz per channel. The time resolution should be in the range of $\mathcal{O}(50)$~ps or better, to match that of RICH, GTK and KTAG. A fast and high-resolution hodoscope will also improve the reconstruction of the direction of charged particles, allowing a more precise track-ring association (useful for multi-track events such as $K^+\to\pi^+\ell^+\ell^-$).
The CHOD/NewCHOD can be replaced with a detector based on modern photo-lithographic technology on flexible and standard printed-circuit board (PCB) supports, which has allowed the invention of novel and robust micro-pattern gaseous
detectors (MPGDs), such as the gaseous electron multiplier (GEM), the thick GEM, and the so-called ``Micromegas''. These detectors exhibit good spatial and time resolution, high rate capability, allow large sensitive areas to be covered with flexible geometry, and guarantee low material budget, good operational stability and radiation hardness.
Two detectors, each made of two $x$-$y$ planes, should be able to reach the required performance in terms of time resolution and will provide fast information for the trigger.

The forward electromagnetic calorimeter is crucial for any kaon program. This detector plays mainly the role of photon veto for the $K^+\to\pi^+\nu\bar\nu$, while it is also essential to reconstruct final states such as $K_{\ell 2}$, $K^+\to\pi^+\ell^+\ell^-$ and $K^+\to\pi^+\gamma\gamma$.
The higher intensity seriously challenges the possibility to reuse the Liquid Krypton electromagnetic calorimeter of NA62.
A synergistic approach between the future charged and neutral kaon program will be adopted, to design and construct a new fast electromagnetic calorimeter to serve both kaon beam configurations (Section~\ref{sec:klever}). A possibility to modify the existing calorimeter readout to stand the particle rates expected in the charged mode is under study as an intermediate step towards a new calorimeter.

A hadron calorimeter made of alternating layers of iron and scintillator similar to that of NA62 is envisaged for the future experiment. The main purpose is to provide pion-muon separation independently from the RICH counter.
This is crucial for the $K^+\to\pi^+\nu\bar\nu$ program, and is also essential for the $K_L$ phase with tracking. The NA62 hadron calorimeter makes use of 60~mm wide scintillator strips, a compromise between particle identification performance, number of channels and expected rate per channel. WLS fibers bring the signal to the PMs mounted to the end of the scintillator strips.
The NA62 calorimeter is split into two modules along the beam axis. The new calorimeter should employ a similar configuration, but with almost a factor two smaller scintillator strips to reduce the rate per channel and to improve particle identification in presence of a significant level of pileup.

The scintillator array technology adopted for the hodoscope will also serve for the muon detector. A configuration with three planes interleaved with iron absorbers will gaurantee the best performance in terms of muon reconstruction and veto detector, similar to that employed by the NA48 experiment~\cite{na48detector:2007}. This detector will serve both online as a veto for the trigger, and offline to complement particle identification. Such a detector must also allow the reconstruction of single and di-muon final states, which are essential for the precision studies of lepton flavor violation.
In order to provide positive identification of muons, the time resolution should be comparable with that of the RICH, hodoscope, GTK and KTAG.
This is crucial to distinguish muons from decays of kaons within the decay region from those from accidental muons coming from the beam line uptream.

Large and small angles veto detectors are essential mainly for the $K^+\to\pi^+\nu\bar\nu$ program. To this purpose the photon detectors studied for the KLEVER experiment (Section~\ref{sec:klever}) are perfectly suited also for a $K^+$ program.
In addition, a cylindrical calorimeter mounted around the beam pipe close to the muon detector will be essential to provide photon and charged particle veto for the $K^+\to\pi^+\nu\bar\nu$ measurement. This detector has to take over the role of the HASC of NA62, and must be adapted to the new left-right symmetric configuration, covering a small angle region larger than that of NA62.

\subsection{$K_L\to\pi^0 \nu \bar{\nu}$: KLEVER}
\label{sec:klever} 

Exploratory work on an experiment to measure ${\rm BR}(K_L\to\pi^0\nu\bar{\nu})$ in a future stage of the ECN3 high-intensity kaon physics program began even before the start of NA62 data taking. As a part of the first Physics Beyond Colliders initiative at CERN, 
the basic design work for the KLEVER experiment 
($K_L$ Experiment for VEry Rare events) was carried out
in preparation for the 2020 update of the European Strategy for Particle Physics~\cite{Ambrosino:2019qvz}.

As in other phases of the extended NA62 program, KLEVER makes use of the 400-GeV primary proton beam from the SPS, slow extracted and transported to the T10 target as described in \Sec{sec:proton_beam} at an intensity of $2\times10^{13}$~ppp, corresponding to about six times the nominal intensity of NA62. The neutral secondary beam for KLEVER is derived at an angle of 8~mrad; the beamline design is described in \Sec{sec:neutral_beam}. 
In practice, the choice of production angle has been optimized together with the limits of the fiducial volume (FV) to maximize the signal sensitivity and minimize backgrounds from $K_L\to\pi^0\pi^0$ and $\Lambda\to n\pi^0$.
For a production angle of 8 mrad, the neutral beam has a mean $K_L$ momentum of 40~GeV, so that 4\% of $K_L$s decay inside an FV extending from 130~m to 170~m downstream of the target. $K_L$s mesons decaying inside this region have a mean momentum of 27~GeV.
Relative to KOTO, the boost from the high-energy beam in KLEVER facilitates the rejection of background channels such as $K_L\to\pi^0\pi^0$ by detection of the additional photons in the final state. On the other hand, the layout poses particular challenges for the design of the small-angle vetoes, which must reject photons from $K_L$ decays escaping through the beam exit amidst an intense background from soft photons and neutrons in the beam. Background from $\Lambda \to n\pi^0$ decays in the beam must also be kept under control.

The KLEVER experiment is designed to achieve a sensitivity of about 60 events for the decay $K_L\to\pi^0\nu\bar{\nu}$ at the SM BR with an $S/B$ ratio of 1. At the SM BR, this would correspond to a relative uncertainty of about 20\%. We would expect to be able to observe a discrepancy with SM predictions with $5\sigma$ significance if the true BR is a bit more than twice or less than one-quarter of the SM BR, or with $3\sigma$ significance if the true BR is less than half of the SM rate. 
As noted in \Sec{sec:neutral_beam}, with a solid angle of $\Delta\theta = 0.4$~mrad,  the yield of $K_L$s in the beam is $2.1\times10^{-5}$ $K_L$ per proton on target (pot).
With a fiducial-volume acceptance of 4\% and a selection efficiency of 5\%, collection of 60 SM events would require a total primary flux of $5\times10^{19}$ pot, corresponding to five years of running at an intensity of $2\times10^{13}$ protons per pulse (ppp) under NA62-like slow-extraction conditions, with
a 16.8~s spill cycle and 100 effective days of running per year.

KLEVER is one part of the comprehensive program for the study of rare kaon decays described in \Sec{sec:sps_program}. 
The timescale for the experiment depends upon the 
point at which the the measurement of $K_L\to\pi^0\nu\bar{\nu}$ becomes the logical next step in 
the North Area kaon physics program.
Two additional considerations also influence the KLEVER scheduling. First, as discussed in \ref{sec:neutral_beam_ext}, in order to provide maximum protection from $\Lambda\to n\pi^0$ background, the KLEVER beamline would need to be lengthened by 150~m. Second, despite the efforts to benchmark the KLEVER neutral beam simulation by comparison with existing data on inclusive particle production by protons on lightweight targets at SPS energies (\Sec{sec:targ_prod}), confidence in the simulations and background estimates would be greatly increased by the possibility to acquire particle production data and measure inclusive rates in the KLEVER beamline setup with a tracking experiment in place. This would be possible in the intermediate experimental phase discussed in \Sec{sec:intermediateKL}, which might run before KLEVER. 
Therefore, although in principle, KLEVER could aim 
to start data taking in LHC Run 4 (for which injector physics is currently foreseen to begin in late 2027 or early 2028), the most natural timescale might envisage KLEVER as the last phase of the comprehensive program, tentatively beginning data taking in LHC Run 5 (for which injector physics is currently estimated to start in 2034).  

\begin{figure}[htb]
    \centering
    \includegraphics[width=0.8\textwidth]{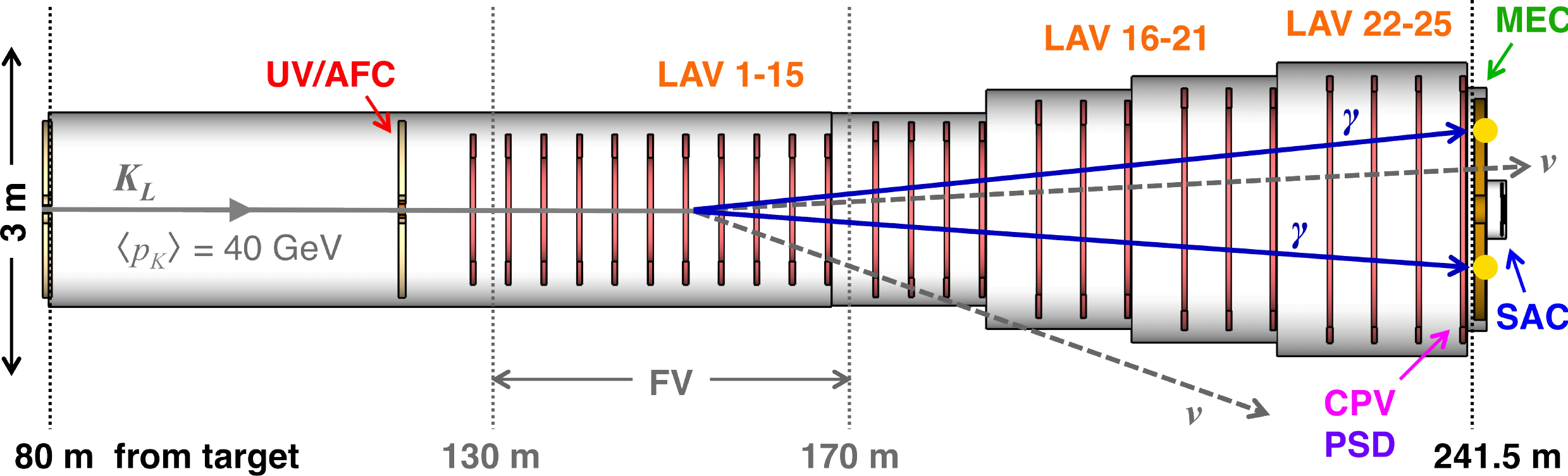}
    \caption{KLEVER experimental apparatus: upstream veto (UV) and active final collimator (AFC), large-angle photon vetoes (LAV), main electromagnetic calorimeter (MEC), small-angle calorimeter (SAC), charged particle veto (CPV), preshower detector (PSD).}
    \label{fig:exp}
\end{figure}
\subsubsection{Layout}
The KLEVER experiment largely consists of a collection of high-efficiency photon detectors arranged around a 160-m-long vacuum volume to guarantee hermetic coverage for photons from $K_L$ decays emitted at polar angles out to 100 mrad and to provide a nearly free path through vacuum up to the main electromagnetic calorimeter (MEC) for photons emitted into a cone of at least 7.5 mrad.
The fiducial region spans about 40 m just downstream of the active final collimator (AFC), but the photon veto coverage extends along the entire length up to the MEC. The layout of the detector elements is schematically illustrated in \Fig{fig:exp}. Note that in the figure, the transverse and longitudinal scales are in the ratio 10:1. The largest elements are about 3~m in diameter. The beginning of the vacuum volume is immediately downstream of the cleaning collimator at $z=80$~m, i.e., 80~m downstream of the T10 target. A 40-m vacuum decay region allows the upstream veto calorimeter (UV) surrounding the AFC to have an unobstructed view for the rejection of $K_L$ decays occurring upstream of the detector volume. The UV and AFC at $z=120$~m define the start of the detector volume, which is lined with 25 large-angle photon and charged-particle veto stations (LAV) in five different sizes, placed at intervals of 4 to 6~m to guarantee hermeticity for decay particles with polar angles out to 100~mrad. The MEC, at the downstream end of the vacuum volume,
replaces the NA48 LKr calorimeter used in NA62: it reconstructs the $\pi^0$ vertex for signal events and helps to reject events with extra photons. A charged-particle veto detector (CPV) in front of the MEC helps to increase the rejection for decays such as $K_{e3}$ and $K_L\to\pi^+\pi^-\pi^0$, and a preshower detector (PSD) allows reconstruction of the angles of incidence for photons, providing additional constraints on signal event candidates even if only one photon converts. 
The small-angle vetoes on the downstream side of the MEC intercept photons from $K_L$ decays that pass through the beam pipe. The small-angle calorimeter (SAC) squarely intercepts the neutral beam—its angular coverage as seen from the target extends to 0.4~mrad. Because of the high rates of neutrons and photons in the beam, the design of this detector is one of the most challenging aspects of the experiment. The intermediate-ring calorimeter (IRC) is a ring-shaped detector between the SAC and MEC and intercepts photons from downstream decays that make it through the calorimeter bore at slightly larger angles.
In addition to the photon vetoes, the experiment makes use of hadronic calorimeters downstream of the MEC (not shown), to reject background from the copious $K_L$ decays into charged particles.
Because of the experimental challenges involved in the measurement of
$K_L\to\pi^0\nu\bar{\nu}$, and in particular, the very high efficiency
required for the photon veto systems, 
most of the subdetector systems for KLEVER will have to be newly constructed. The LAVs, MEC, IRC, and SAC, as well as the hadronic calorimeters, would all be able to be used with the other phases of the experiemental program. The UV, CPV, and PSD have potential reuses as well. 

In the baseline design, the FV covers the region $130~{\rm m}<z<170$~m. The positioning of the FV significantly upstream of the calorimeter, together with the relatively high $K_L$ momentum, is key to obtaining sufficient background rejection against $K_L\to\pi^0\pi^0$ decays with the ring-shaped LAV geometry covering polar angles out to 100~mrad. This comes at a cost in acceptance for signal decays, which increases significantly as the FV is moved closer to the calorimeter. Extending the LAV coverage beyond 100~mrad in the downstream region would allow extension of the FV to increase the signal acceptance; further optimization along these lines is an option under study.

\Fig{fig:exp} shows the experimental configuration before lengthening of the beamline to suppress background from $\Lambda$ decays. 
Lengthening the beamline would change the distance from the cleaning collimator to the AFC, so that the cleaning collimator would be upstream of the start of the 40-m vacuum tank immediately upstream of the AFC, instead of right at its entrance. The detector configuration would be otherwise unchanged. 

\subsubsection{Rates and timing performance}
\label{sec:klever_rates}

\begin{table}[htb]
    \centering
    \begin{tabular}{lc|lc} \hline\hline
        Detector & Event rate (MHz) & Event class & Rate (MHz) \\ \hline
        AFC & 2.3 & Exactly 2 hits on MEC & 4.8  \\
        UV & 7.1 & Exactly 2 photons on MEC & 1.0 \\
        LAV & 14 & 2 hits on MEC with UV, LAV veto & 3.1 \\
        MEC & 18 & 2 hits on MEC, no other hits & 0.007 \\
        IRC & 22 & & \\
        SAC & 95 & & \\ \hline\hline
    \end{tabular}
    \caption{Rates for events with hits on KLEVER detectors, by detector system (left) and for certain event classes (right).}
    \label{tab:klever_rates}
\end{table}
The FLUKA simulation of the beamline (\Sec{sec:neutral_beam}) contains an idealized representation of the experimental setup for the purposes of evaluating rates on the detectors, both from the decays of $K_L$ mesons in the beam and from beam halo. The estimated rates by detector system for events with at least one hit on the system are listed in \Tab{tab:klever_rates}, left. These estimates assume a primary intensity of $2\times10^{13}$ ppp and a 3~s effective spill.
The rates are obtained with a geometrical representation of the detector volumes and assumed probabilities for 
hadrons to register hits. 
For the AFC, UV, LAV, and MEC, the probability for
registering hits is assumed to be 25\% for neutrons and $K_L$ mesons and 100\% for all other particles. For the IRC and SAC, the assumed probability is 10\% for neutral hadrons and 100\% for all other particles. 
The particle tracking threshold is 1~GeV.
Since a particle detected in any of the detectors would effectively constitute a veto (the rates of signal candidates being negligible for these purposes), this provides an estimate of the total veto rate: 134~MHz, of which 104 MHz from events with hits on a single detector and 30 MHz from events with hits on multiple detectors.
The total rate is dominated by the rate of interactions of beam particles in the SAC, consisting of 13 MHz of hits from $K_L$ mesons, 44~MHz of hits from neutrons, and 40~MHz of hits from beamline photons with $E>5$~GeV.
The hadronic interactions can be efficiently recognized offline and we assume only 10\% of the corresponding rate contributes to the inefficiency from accidental coincidence. The remaining veto rate from accidentals is then at most 100 MHz, which we take as a figure of merit. The event time is obtained from the $\pi^0$ candidate reconstructed in the MEC, and the accidental
coincidence rate is dominated by events on the SAC. Assuming a $\pm5\sigma_t$ coincidence window, to limit the inefficiency from accidental coincidence to $<25\%$, the time resolution $\sigma_t$ on the coincidence must be better than 250~ps, and the detector design must minimize any contributions from non-gaussian fluctuations to the timing distribution. These considerations lead to the specification that the nominal time resolution for the most critical detectors (the MEC and the SAC) must be better than 100~ps.

The rates for certain classes of events to be acquired are listed in \Tab{tab:klever_rates}, right: for events with exactly two hits and with exactly two photons on the calorimeter, independently of the other detectors; for events with two hits on the calorimeter without hits on the UV or LAV (presumed to be in online veto); and for events with two hits on the calorimeter and no hits on any of the other detectors. These are estimates of the rates for the physical events---there is no simulation of the detector response apart from the efficiency assumptions outlined above---but these results do suggest that the total rates for event classes that would approximately correspond to level-0 trigger conditions in NA62 are in fact of about the same order of magnitude as in NA62.

\begin{figure}
    \centering
    \includegraphics[width=0.5\textwidth]{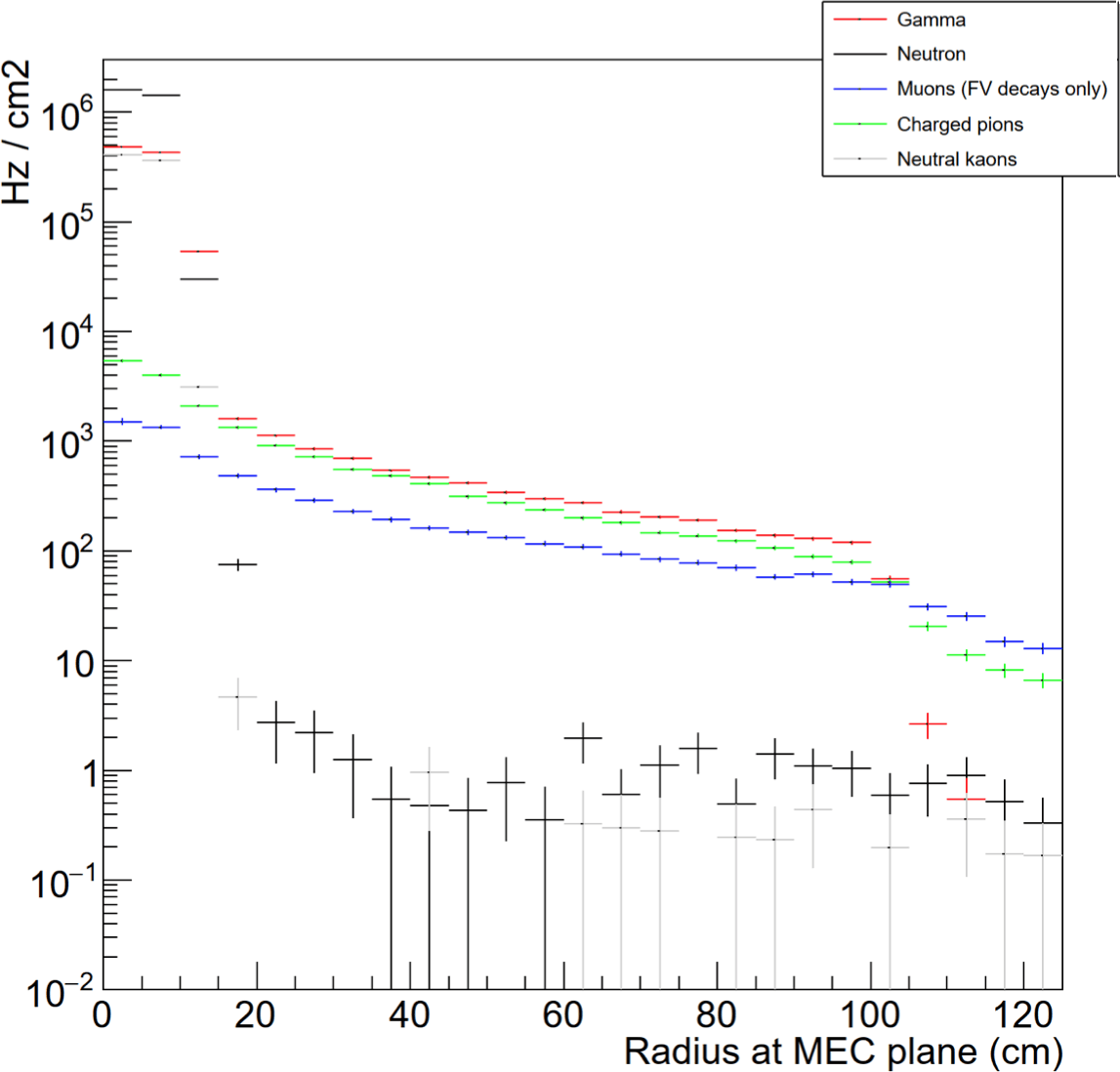}
    \caption{Radial distribution of particle fluxes at the MEC for KLEVER, from beamline simulation with FLUKA.}
    \label{fig:klever_halo}
\end{figure}
Figure~\ref{fig:klever_halo} shows the radial distribution of the particle fluxes at the front face of the MEC, as estimated with the FLUKA simulation of the neutral beam. The beam halo from photons and neutral hadrons drops off rapidly for
$10 < r < 20$~cm, with $r$ the radial distance from the beam axis. For neutrons, the halo-to-core ratio decreases from $2.5\times10^{-2}$ to $8.8\times10^{-5}$ over this interval, correpsonding to a decrease in the total neutron rate in the halo from 12~MHz to 43~kHz. For $r>20$~cm, the neutral hadron fluxes are everywhere on the order of a few Hz/cm$^2$. At $r = 20$~cm, there are 
1~kHz/cm$^2$ of photons and charged pions and a few hundred Hz/cm$^2$ of muons, dropping off more or less linearly with radius. 

\subsubsection{Main electromagnetic calorimeter (MEC)}
\label{sec:klever_mec}

The principal performance requirements for the MEC are excellent intrinsic detection efficiency for high-energy photons, good two-cluster separation for photons, and excellent time resolution.

Early studies indicated that the NA48 liquid-krypton calorimeter
(LKr) \cite{Fanti:2007vi}
currently used in NA62 could be reused as the MEC for reconstruction of
the $\pi^0$ for signal events and rejection of
events with additional photons. 
The energy, position, and time resolution of the LKr calorimeter were measured in NA48 to be
\begin{align}
\frac{\sigma_E}{E} &=0.0042\oplus\frac{0.032}{\sqrt{E {\rm (GeV)}}}\oplus\frac{0.09}{E {\rm (GeV)}}, \\
\sigma_{x, y} & = 0.06~{\rm cm} \oplus \frac{0.42~{\rm cm}}{\sqrt{E {\rm (GeV)}}}, \\
\sigma_t & = \frac{2.5~{\rm ns}}{\sqrt{E {\rm(GeV)}}}.
\end{align}
Indeed, the efficiency
and energy resolution of the
LKr appear to be satisfactory for KLEVER. 
Studies of $K^+\to\pi^+\pi^0$ decays in NA48 data and tests conducted in 2006 with tagged photons from an electron beam confirmed that the LKr has an inefficiency of less than $10^{-5}$ for photons with $E > 10$~GeV, providing the needed rejection for forward photons \cite{NA62+07:M760}.
These studies were fully confirmed in NA62. Notwithstanding the presence of a much larger amount of material upstream of the LKr calorimeter in NA62 than in NA48, a study of single-photon efficiency underpinning the NA62 measurement of ${\rm BR}(K^+\to\pi^+\nu\bar{\nu})$ and used to obtain a limit on ${\rm BR}(\pi^0\to{\rm invisible})$
found an inefficiency of about $10^{-5}$ at $E=20$~GeV, slightly decreasing at higher energies \cite{NA62:2020pwi}.
The LKr time resolution, however, would be a
would be a major liability. The LKr would measure the event time in
KLEVER with $\sim$500 ps resolution, whereas a time resolution of 100~ps is required, as discussed above. 
The LKr time resolution might be somewhat improved via a comprehensive readout
upgrade and/or an increase in the operating voltage,
but an improvement by a factor of 5 is unlikely.
In any case, concerns about the service life of the LKr would remain (it has been in operation continuously since the late 1990s), and the
size of the LKr inner bore would limit the beam solid angle and hence kaon flux.

We are investigating the possibility of replacing the
LKr with a shashlyk-based MEC patterned on the PANDA FS calorimeter, in turn
based on the calorimeter designed for the KOPIO experiment~\cite{Atoian:2007up} (see \Sec{sec:klpnn_status}). This design featured modules $110\times110$~mm$^2$ in cross section made of 
alternating layers of 0.275-mm-thick lead absorber and 1.5-mm-thick injection-molded polystyrene scintillator.
This composition has a radiation length of 3.80 cm and a sampling fraction of 46.7\%. 
The scintillator layers were optically divided into four $55\times55$~mm$^2$ segments; the scintillation light was collected by WLS fibers traversing the stack longitudinally and read out at the back by avalanche photodiodes (APDs). KOPIO was able to obtain an energy and time resolution of 3.3\% and 73 ps at 1 GeV with this design.

For KLEVER, the design would be updated to use silicon photomultipliers (SiPMs) instead of APDs. The final choice of module size and readout granularity has yet to be determined, but on the basis of KLEVER simulations that assume that clusters are resolved if more than 6~cm apart, readout cells of $5\times5$~cm$^2$ seem reasonable. The Moli\`ere radius for the fine-sampling shashlyk design described above is 3.39~cm,
so smaller cells can be used if needed. For comparison, 
the Moli\`ere radius of liquid krypton is 5.86~cm, and the NA48 LKr calorimeter features $2\times2$~cm$^2$ cells.

The MEC would have an inner bore of at least 12~cm to allow the passage of the neutral beam. The bore could be widened to as much as 15~cm to allow the penumbra of beam photons and neutral hadrons to pass through and be intercepted by the IRC. For the sensitivity analysis, we assume that the photons used to 
reconstruct the $\pi^0$ for signal events must have $r > 35$~cm, i.e., any cluster with $r<35$~cm vetoes the event. The sensitive area has an outer radius of 125~cm.
With this design, the radiation dose rate is dominated by photons from $K_L$ decays. The rate is most intense on the innermost layers of the calorimeter, for which it is about 2~kHz/cm$^2$.
Precise dose rate calculations have yet to be performed, but an estimate suggests a dose of 4~kGy/yr to the scintillator for the innermost layers. This estimate would suggest that radiation damage is a concern, but is most likely manageable. Radiation robustness may be a factor in the final choice of scintillator.
Although current information suggests that conventional polystyrene scintillator is sufficiently luminous, fast, and radiation resistant, we are evaluating the advantages that can be obtained with less conventional choices for the light emitter (e.g., perovskite or chalcogenide quantum dots) \cite{Gandini:2020aaa} or matrix material (e.g., polysiloxane) \cite{Acerbi:2020itd}.   

In addition to the basic criteria on energy resolution, efficiency, time resolution, and two-cluster separation, ideally, the MEC would provide information useful for particle identification. For example, identification of pion interactions would provide additional suppression of background for decays with charged particles in the final state, and, as the experience with KOTO suggests, it is crucial to have as much information as possible to assist with $\gamma/n$ discrimination. The fine transverse segmentation of the MEC will play an important role: a simple cut on cluster RMS with the existing LKr can suppress up to 95\% of pion interactions in NA62 data. Fast digitization of the signals from the MEC is expected to provide additional $\gamma/n$ discrimination. The MEC will be backed up with a hadronic veto calorimeter.

\begin{figure}
    \centering
    \includegraphics[width=0.8\textwidth]{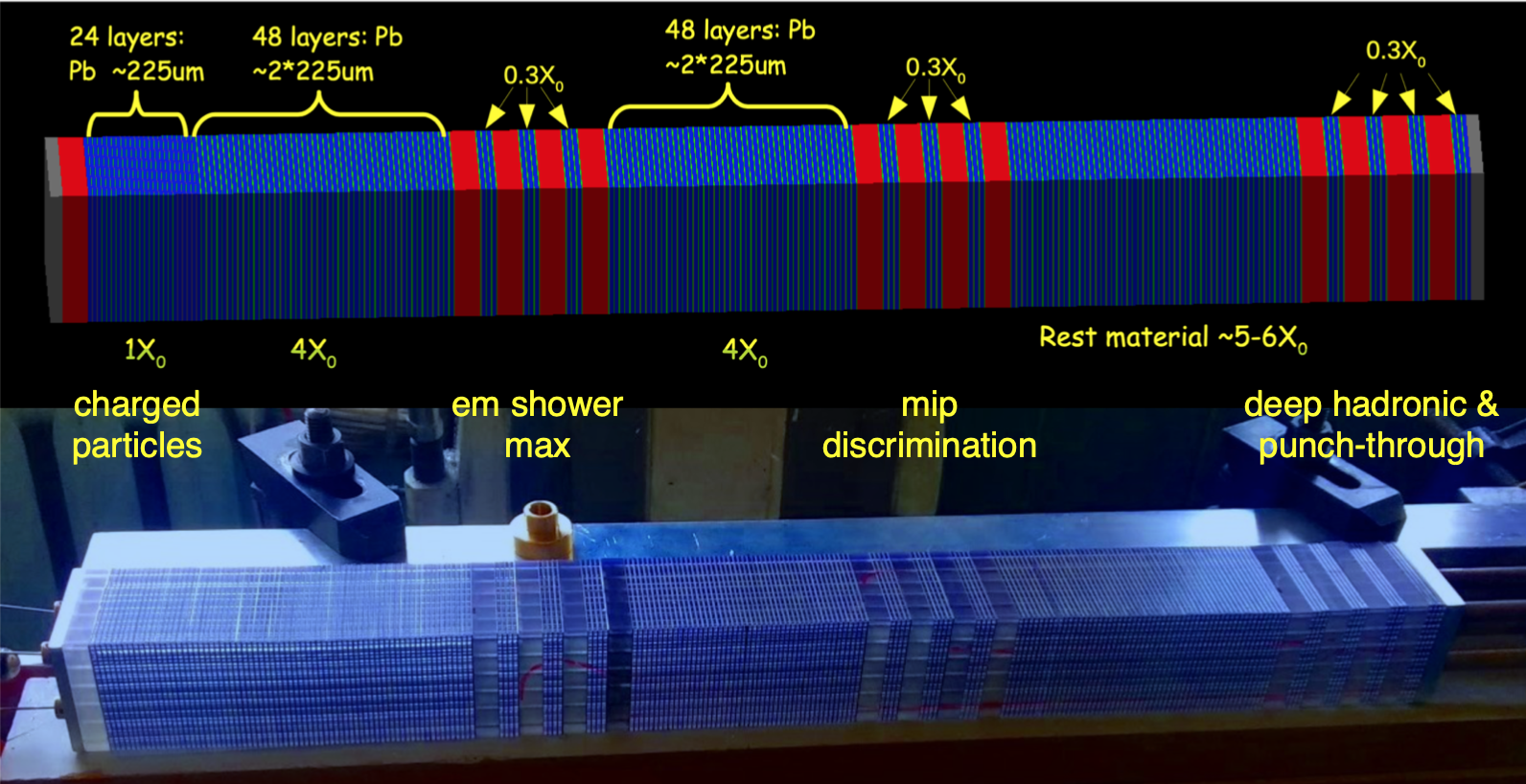}\\ \vspace*{5mm}
    \includegraphics[width=0.8\textwidth]{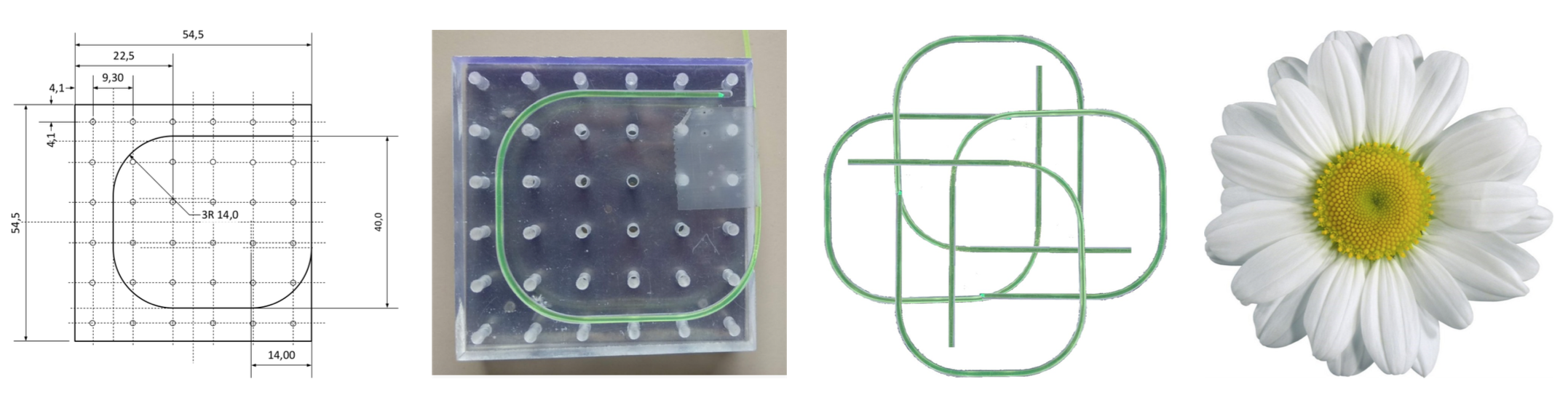}\\
    \caption{Top: Geant4 model of small prototype for {\it romashka} calorimeter, featuring spy tiles placed at key points in the shashlyk stack, together with a photograph of the module constructed at Protvino. Bottom: Fiber routing scheme for independent readout of spy tiles, giving rise to the name {\it romashka} (chamomile).}
    \label{fig:klever_spy_tiles}
\end{figure}
We are also experimenting with concepts to obtain information on the longitudinal shower development from the shashlyk design. One possible design makes use of ``spy tiles'', 10-mm thick scintillator bricks
incorporated into the shashlyk stack but optically isolated from it and
read out by separate WLS fibers. The spy tiles are located at key points
in the longitudinal shower development: near the front of the stack,
near shower maximum, and in the shower tail, as illustrated in  
\Fig{fig:klever_spy_tiles}.
This provides longitudinal sampling of the shower
development, resulting in additional information for $\gamma/n$ separation.
The prototype shown in 
\Fig{fig:klever_spy_tiles} was constructed at Protvino in early 2018. Its basic functionality was tested in the OKA beamline in April 2018, and more comprehensive tests were carried out in September 2019 at DESY in collaboration with LHCb. Simulations suggest (and preliminary test beam data validate, to a certain extent) that the {\it romashka} design with spy tiles can give at least an order of magnitude of additional neutron rejection relative to what can be obtained from the transverse segmentation of the calorimeter alone, providing an overall suppression of up to 99.9\% for neutron interactions. 
The small prototype has a cross sectional area of only $55\times55$~mm$^2$ (one readout cell) and a depth of $14 X_0$ (60\% of the design depth), and both transverse and longitudinal leakage significantly complicated attempts to measure the time resolution. For electrons with $1 < E < 5$~GeV, the measured time resolution was about 200~ps and virtually constant; we expect that this can be significantly improved. The time resolution for hits on the spy tiles (independently of information from the shashlyk stack) was on the order of 500--600~ps, which may be difficult to improve. This is not expected to be a problem, however: the main shashlyk signal establishes the event time and the association of the PID information from the shashlyk tiles is based on the segmentation, with occupancies per cell of at most a few tens of kHz on the innermost layers.  

\subsubsection{Upstream veto (UV) and active final collimator (AFC)}

The upstream veto (UV) rejects $K_L\to\pi^0\pi^0$
decays in the 40~m upstream of the fiducial volume where there are
no large-angle photon vetoes.
The UV is a shashlyk calorimeter with the same basic structure as the MEC
(without the spy tiles). Because the UV does not participate in event reconstruction, the readout granularity can be somewhat coarser than for the MEC. The sensitive area of the UV has inner and outer radii of 10~cm and 100~cm.

The active final collimator (AFC) is inserted into the hole in center
of the UV. The AFC is a LYSO collar counter with angled inner surfaces to
provide the last stage of beam collimation while vetoing photons from $K_L$
that decay in transit through the collimator itself. The collar is made of
24 crystals of trapezoidal cross section, forming a detector with an
inner radius of 60 mm and an outer radius of 100 mm.
The UV and AFC are both 800 mm in depth. The maximum crystal length for
a practical AFC design is about 250 mm, so the detector consists of 3 or
4 longitudinal segments. Each crystal is
read out on the downstream side with two avalanche photodiodes (APDs).
These devices couple well with LYSO and offer high quantum efficiency and
simple signal and HV management. Studies indicate that a light yield
in excess of 4000 p.e./MeV should be easy to achieve.

\subsubsection{Large-angle photon vetoes}

Because of the boost from the high-energy beam, it is sufficient for the
large-angle photon vetoes (LAVs) to cover polar angles out to 100~mrad.
The detectors themselves must have inefficiencies of less than
a few $10^{-4}$ down to at least 100 MeV, so the current NA62 LAVs~\cite{NA62:2017rwk} based
on the OPAL lead glass~\cite{OPAL:1990yff} cannot be reused.
With a photoelectron yield of 0.2 p.e./MeV for electromagnetic showers,
the low-energy efficiency of the OPAL lead glass blocks is
dominated by photon statistics and falls off sharply below 200~MeV. 
A 50~MeV photon gives just 10 photoelectrons, corresponding to
a 5~mV signal, which is the current value of the low threshold for NA62.

\begin{figure}[htbp]
\centering
\includegraphics[width=0.5\textwidth]{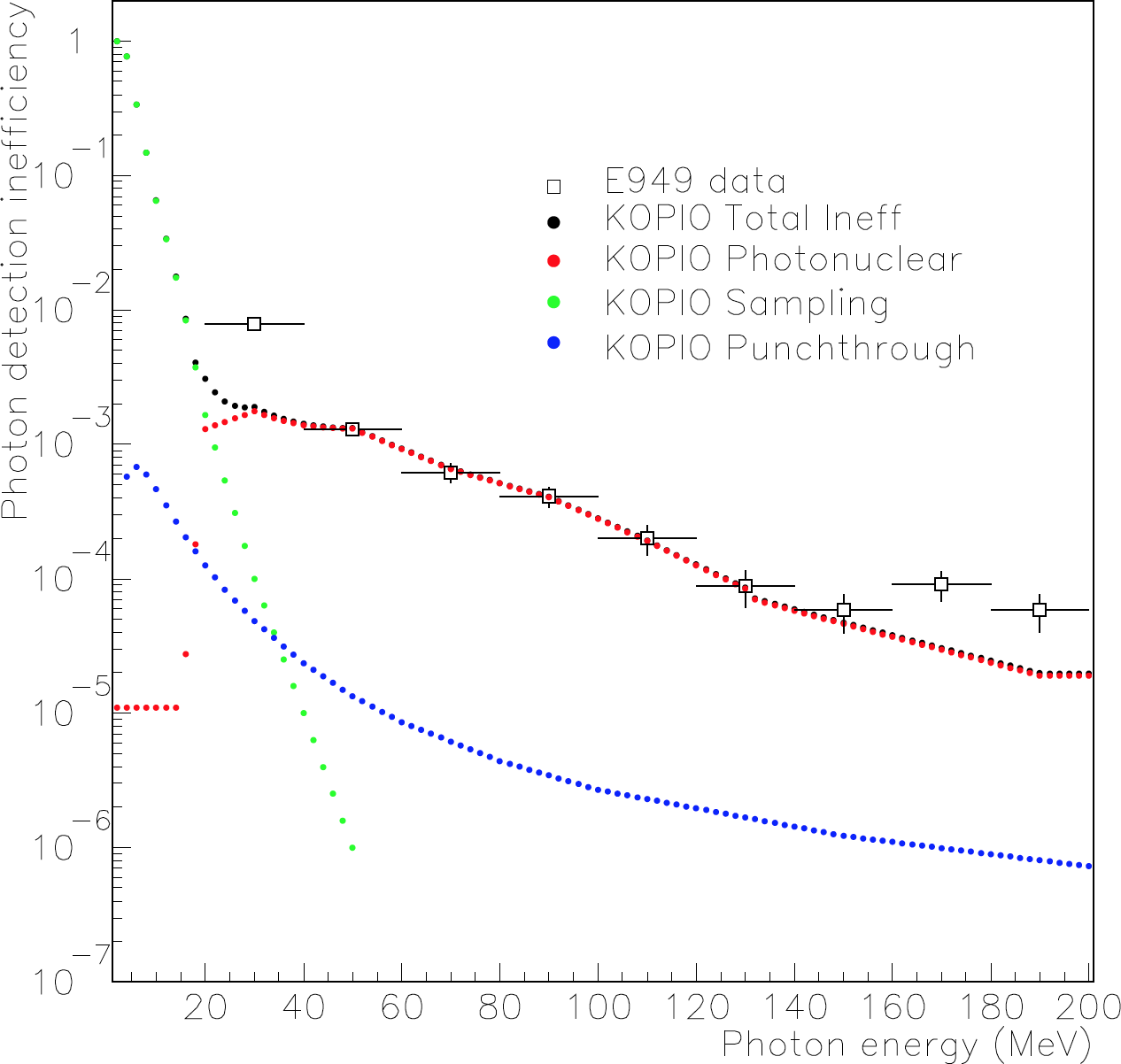}
\caption{Photon detection inefficiency parameterization from
  KOPIO~\cite{KOPIO+05:CDR}, broken down by source. Measured
  inefficiencies for the E949 barrel veto~\cite{Atiya:1992vh} are also plotted.}
\label{fig:kopio_eff}
\end{figure}

As a reference for what low-energy photon detection efficiencies can be
achieved for the new LAVs, \Fig{fig:kopio_eff} shows the inefficiency
parameterization used for the KOPIO proposal~\cite{KOPIO+05:CDR}.
For the energy range
$50~{\rm MeV} < E < 170~{\rm MeV}$,
the overall parameterization (black circles) is based on detection
inefficiencies measured for the
E787/949 barrel photon veto~\cite{Atiya:1992vh} using $K^+\to\pi^+\pi^0$
events; these data are also shown in the figure.
Outside of this range, the parameterization
is guided by FLUKA simulations with different detector designs, and
the overall result is (slightly) adjusted to reflect the segmentation
of the KOPIO shashlyk calorimeter, but for most of the interval
$E < 200$~MeV, the results do not differ much from the E949 measurements.
The contributions to the inefficiency from photonuclear interactions,
sampling fluctuations, and punch through were estimated from
known cross sections, statistical considerations, and mass attenuation 
coefficients.

\begin{figure}[htbp]
\centering
\includegraphics[width=0.6\textwidth]{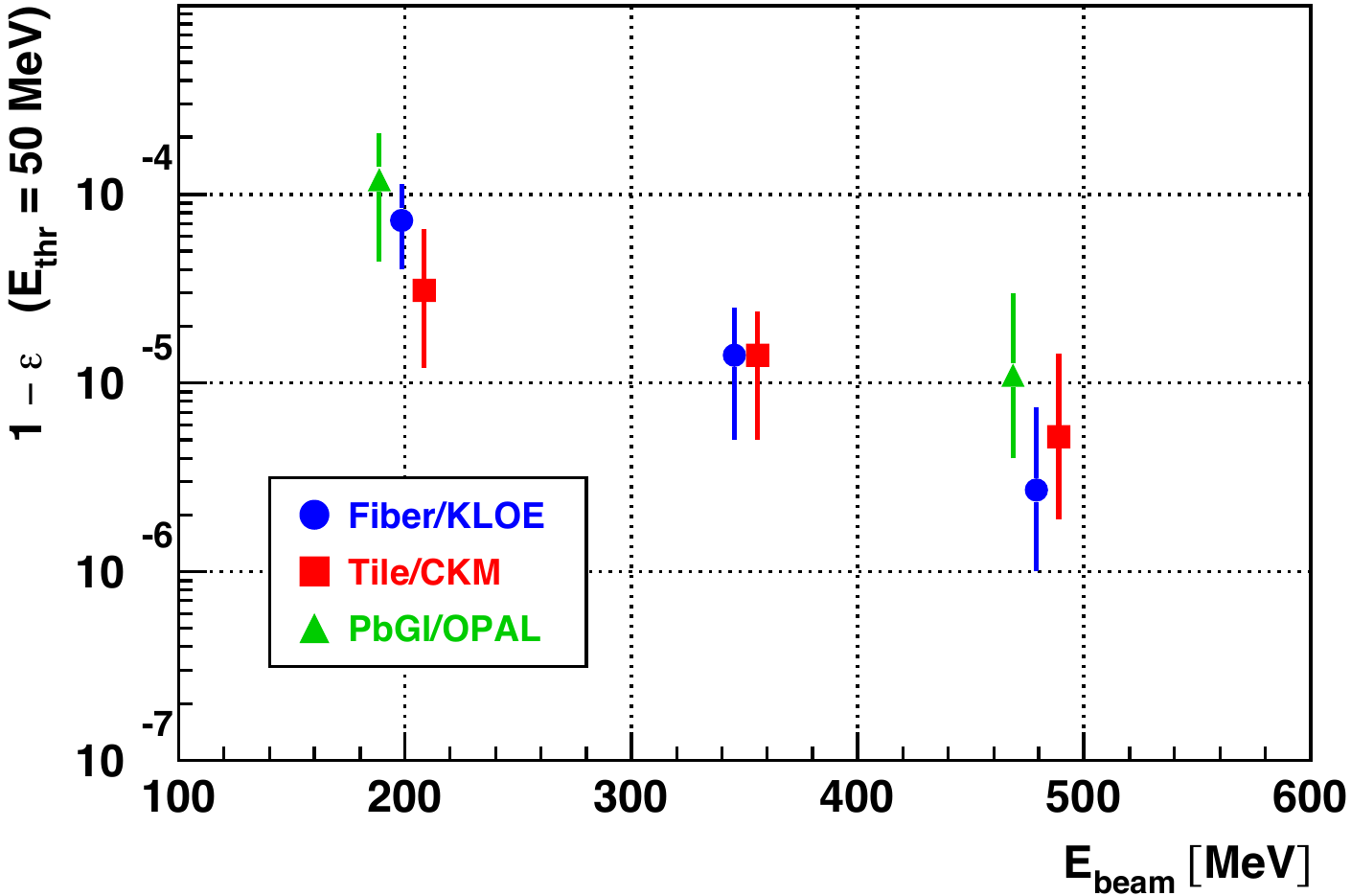}
\caption{Measurements of detection inefficiency for tagged electrons with $E=203$, 350, 483~MeV for three veto prototypes, made at the Frascati BTF in 2007. The red squares are for the CKM VVS prototype.}
\label{fig:hawaii}
\end{figure}

One possible design for the LAVs for the proposed experiment would
be similar to the Vacuum Veto System (VVS) detectors planned for the CKM
experiment at Fermilab~\cite{Frank:2001aa}. The CKM VVS is a lead/scintillator-tile detector with a segmentation of 1~mm Pb + 5~mm scintillator. This segmentation is the same as for the
E787/949 barrel photon veto, so the same low-energy efficiencies
might be expected. The wedge-shaped tiles are stacked into modules and
arranged to form a ring-shaped detector. The scintillation light is
collected and transported by WLS fibers in radial grooves. In the
original VVS design, the fibers brought the light to optical windows for
readout by PMTs outside of the vacuum. Readout by SiPMs
inside the vacuum would make for shorter fibers and would facilitate the
mechanical design---the availability of economical
SiPM bundles with large effective area makes this an attractive option.
In the approximate geometry for the proposed experiment, the LAVs would
consist of 100 layers, for a total thickness of 60~cm, corresponding to
18 $X_0$.  
In the original NA62 proposal~\cite{Anelli:2005xxx},
before the OPAL lead glass became available,
very similar detectors were the baseline solution for the existing LAVs,
and in 2007, the efficiency of the CKM VVS prototype in the energy range 200--500~MeV was measured using a tagged electron beam at the Frascati
Beam-Test Facility (BTF) \cite{Ambrosino:2007ss}.
The results are shown in Fig.~\ref{fig:hawaii}.
Earlier, the efficiency of the same prototype was measured at the
Jefferson National Laboratory with tagged electrons in the interval
500--1200~MeV. An inefficiency of $3\times10^{-6}$ was found at 1200~MeV,
even with a high threshold (80~MeV, or 1 mip)~\cite{Ramberg:2004en}.
The efficiency parameterization for the LAVs used in our simulations is based on the
KOPIO efficiencies up to the point at 129~MeV, and then extrapolated
through the three points measured at the BTF, to $2.5\times10^{-6}$ for photons
with $E > 2.5$~GeV. This is not unreasonable,
considering that the Jefferson Lab measurement shows that nearly this
inefficiency is already obtained at 1.2~GeV.
The tests of this prototype at the BTF were not optimized for the measurement of the time resolution, but indicated a time resolution of better then 250~ps for 500 MeV electrons, which should be sufficient for KLEVER. 

\subsubsection{Small-angle calorimeters}
\label{sec:klever_sac}
The small-angle calorimeter (SAC) sits directly in the neutral beam and must
reject photons from $K_L$ decays that would otherwise escape via the
downstream beam exit. 
The intermediate-ring calorimeter (IRC) is a ring-shaped detector between the SAC
and MEC and intercepts photons from downstream decays that make it through the calorimeter bore
at slightly larger angles.
The same technology could be used for the construction of both detectors, but the performance requirements for the SAC are much more stringent, especially in terms of high-rate operation.

The required SAC photon efficiencies are different for three different intervals in the incident photon energy. 
For photons with $E < 5~{\rm GeV}$, the SAC can be blind,
as the probability is very small for $K_L\to\pi^0\pi^0$ decays in the FV with two photons on the MEC that pass analysis cuts and are not vetoed by the other detectors to have additional photons on the SAC with energies below this threshold.
For photons with $5~{\rm GeV} < E < 30~{\rm GeV}$, the SAC inefficiency must be less than 1\%.
Most photons on the SAC with energies in this range
from $K_L$ decays that are otherwise accepted as signal candidates
are from events in which there are two photons on the SAC, so that the efficiency requirement is relatively relaxed. 
Only for photons with $E > 30~{\rm GeV}$ must the inefficiency be very low ($<10^{-4}$).
For $K_L$ decays passing analysis cuts with a photon on the SAC in this energy range, the other photon is emitted at large angle and has low energy, so the SAC veto is important.

However, even if the SAC efficiency requirements are not intrinsically
daunting, from the simulations described in \Sec{sec:neutral_beam}, there are about 130 MHz of $K_L$ mesons, 440~MHz of neutrons, and 40~MHz of high-energy ($E>5$~GeV) beam photons incident on the SAC, and the required efficiencies must be attained while maintaining insensitivity to the nearly 600~MHz
of neutral hadrons in the beam. 
In order to keep the false veto rate from accidental
coincidence of beam neutrons to an acceptable level, the hadronic component must be reduced to at most a few tens of MHz, so that the total 
accidental rate is dominated by the beam photons and in any case significantly less than the 100~MHz target cited in \Sec{sec:klever_rates}.
These requirements lead to the following considerations:

\begin{itemize}
\item The SAC must be as transparent as possible to the interactions of neutral hadrons. In practice, this means that the nuclear interaction length $\lambda_{\rm int}$ of the SAC in bulk must be much greater than its radiation length $X_0$.
\item The SAC must have good transverse segmentation to provide $\gamma/n$
discrimination. 
\item It would be desirable for the SAC to provide additional information useful for offline $\gamma/n$ discrimination, for example, from longitudinal segmentation, from pulse-shape analysis, or both.
\item The SAC must have a time resolution of 100~ps or less.
\item The SAC must have double-pulse resolution capability at the level of 1~ns.
\end{itemize}
In addition, in five years of operation, the SAC will be exposed to a neutral hadron fluence of about $10^{14}$~$n$/cm$^{2}$, as well as a dose of up to several MGy from photons.

\begin{figure}
    \centering
    \includegraphics[width=0.5\textwidth]{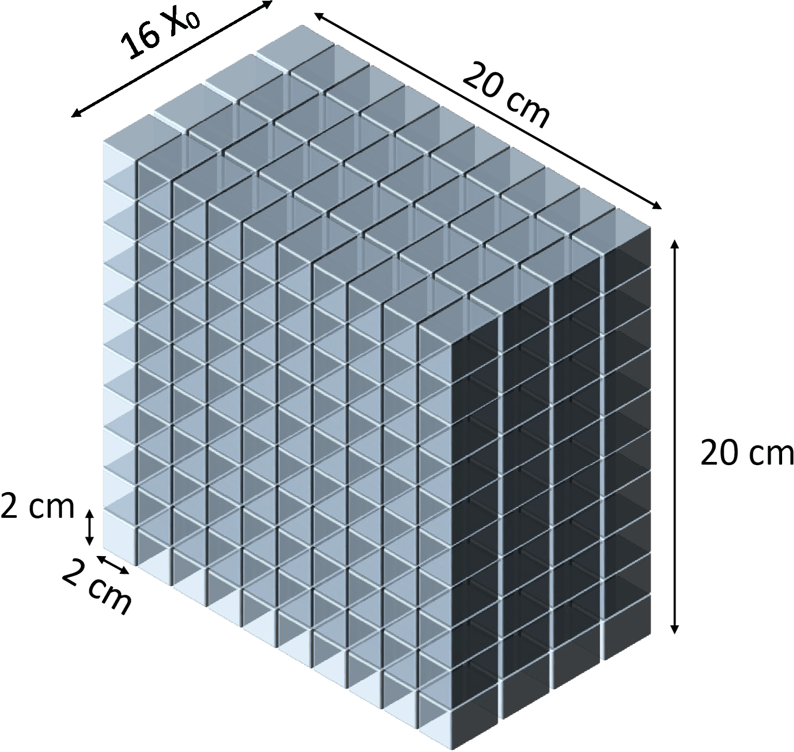}
    \caption{Dimensional sketch of a SAC for KLEVER based on dense, high-$Z$ crystals with both transverse and longitudinal segmentation.}
    \label{fig:klever_sac}
\end{figure}
In principle, all of these requirements could be met with a silicon-tungsten sampling electromagnetic calorimeter. An alternative possibility that is well-matched to the KLEVER requirements would be to use a highly segmented, homogeneous calorimeter with dense, high-$Z$ crystals providing very fast light output. We are currently investigating this latter option. As an example of such a design, the small-angle calorimeter for the PADME experiment used an array of 25 $30\times30\times140~{\rm mm}^3$ crystals of lead fluoride (PbF$_2$). PbF$_2$ is a Cherenkov radiator and provides very fast signals.
For single crystals read out with PMTs, a time resolution of 81~ps and double-pulse separation of 1.8~ns were obtained for 100--400~MeV electrons~\cite{Frankenthal:2018yvf}, satisfying the timing performance requirements for KLEVER. The PADME performance was obtained with fully digitizing waveform readout at 5 GS/s; waveform digitiazion would also be required for KLEVER (see \Sec{sec:readout}).

For KLEVER, a design with longitudinal segmentation is under study, as shown in \Fig{fig:klever_sac}. This design would feature four layers of $20\times20\times40$~mm PbF$_2$ crystals (each $\sim$4$X_0$ in depth). To minimize leakage, the gaps between layers would be kept as small as possible. Compact PMTs such as Hamamatsu's R14755U-100 could fit into a gap of as little as 12~mm. This tube has ideal timing properties for the SAC, including 400~ps rise and fall times and a transit-time spread of 200~ps.
Readout with SiPMs would facilitate a compact SAC design even further, but may require advances in SiPM radiation resistance and timing performance. CRILIN, an electromagnmetic calorimeter under development for the International Muon Collider Collaboration, is also a PbF$_2$-based, highly granular, longitudinally segmented calorimeter with SiPM readout and similar performace requirements~\cite{Cemmi:2021uum}, and development work on the KLEVER SAC is being carried out in collaboration with CRILIN, with particular emphasis on the SiPMs, front-end electronics, and signal readout, as well as on solutions for detector mechanics and SiPM cooling. First KLEVER/CRILIN test beam measurements with individual PbF$_2$ crystals were performed at the Frascati BTF and in the CERN SPS North Area in summer 2021, and tests of a small-scale matrix prototype are scheduled for summer 2022.  

At the doses expected at KLEVER, radiation-induced loss of transparency to Cherenkov light could be significant for PbF$_2$, as suggested by existing studies with ionizing doses of up to O(10~kGy)~\cite{Cemmi:2021uum,Kozma:2002km,Anderson:1989uj}. However, these studies also found significant annealing and dose-rate effects in PbF$_2$, as well as the effectiveness of bleaching with UV light. If the effects of radiation damage to PbF$_2$ prove to be a significant problem, a good, radiation-hard alternative could be recently developed, optimized lead tungstate (PbWO$_4$, PWO), which has a scintillation component with a decay time of 0.8~ns, as well as some slower components with decay times of about 10~ns \cite{PANDA:2011hqx,Auffray:2016xtu,Follin:2021kgn}. 

An intriguing possibility for the construction
of an instrument that is sensitive to photons and relatively insenstive to hadrons is to
exploit the effects of the coherent interactions of high-energy photons in an oriented crystal to induce prompt electromagnetic showering 
(see discussion of the crystal photon converter in the beamline in \Sec{sec:neutral_beam}).
For example, to obtain a compact Si-W sampling calorimeter, crystalline tungsten tiles could be used as the absorber material. Because of the relative ease in producing high-$Z$ optical crystals of high quality, there are good prospects for the construction of compact scintillation or Cherenkov detectors with small radiation length. A decrease by a factor of 5 in the effective radiation length for PWO crystals has been observed for 120~GeV electrons incident to within 1~mrad of the $[001]$ axis~\cite{Bandiera:2018ymh}. The potential gains from orienting some or all of the crystals (e.g., the crystals in the first layer), as well as the procedures and mechanics for aligning the crystals, are under study. It may be possible to eliminate one or more SAC layers by aligning the crystals, for a substantial reduction of the rate of neutral hadron interactions without any loss of photon efficiency. 

\subsubsection{Charged-particle rejection}

For the rejection of charged particles, $K_{e3}$ is a benchmark channel
because of its large BR and because the final state electron can be mistaken
for a photon. Simulations indicate that the needed rejection can be achieved
with two planes of charged-particle veto (CPV) each providing 99.5\%
detection efficiency, supplemented by the $\mu^\pm$ and $\pi^\pm$ recognition
capabilities of the MEC (assumed in this case to be equal to those of the LKr)
and the current NA62 hadronic calorimeters and muon vetoes, which could be
reused in KLEVER. The CPVs are positioned $\sim$3~m upstream of the MEC
and are assumed to be constructed out of thin scintillator tiles.
In thicker scintillation hodoscopes, the detection inefficiency arises
mainly from the gaps between scintillators. For KLEVER, the scintillators
will be only $\sim$5~mm thick ($1.2\%X_0$), and the design will be carefully
optimized to avoid insensitive gaps.

\subsubsection{Preshower detector}

The PSD measures the directions for photons incident on the MEC.
Without the PSD, the $z$-position of the $\pi^0$ decay vertex can only be
reconstructed by assuming that two clusters on the MEC are indeed
photons from the decay of a single $\pi^0$. With the PSD, a vertex can be
reconstructed by projecting the photon trajectories to the beamline.
The invariant mass is then an independent quantity, and
$K_L\to\pi^0\pi^0$ decays with mispaired photons can be efficiently
rejected.
The vertex can be reconstructed using a single photon and the constraint
from the nominal beam axis. 
Simulations show that with $0.5 X_0$ of converter
(corresponding to a probability of at least one conversion of 50\%)
and two tracking planes with a spatial resolution of 100~$\upmu$m,
placed 50~cm apart, the mass resolution is about 20~MeV and the
vertex position resolution is about 10~m. The tracking detectors
must cover a surface of about 5 m$^2$ with minimal material.
Micropattern gas detector (MPGD) technology seems perfectly suited for
the PSD. Information from the PSD will be used for bifurcation
studies of the background and for the selection of control samples,
as well as in signal selection.

\subsubsection{Expected performance for $K_L\to\pi^0\nu\bar{\nu}$}
\label{sec:klever_sens}
\label{sec:klever_lambda}

Simulations of the experiment in the standard configuration 
carried out with fast-simulation techniques (idealized geometry, parameterized detector response, etc.) suggest that the target sensitivity is achievable (60 SM events with $S/B = 1$). 
Background channels considered at high simulation statistics include $K_L\to\pi^0\pi^0$
(including events with
reconstructed photons from different $\pi^0$s and events with overlapping
photons on the MEC), $K_L\to 3\pi^0$, and $K_L\to\gamma\gamma$.

\begin{figure}[htb]
  \centering
  \includegraphics[width=0.8\textwidth]{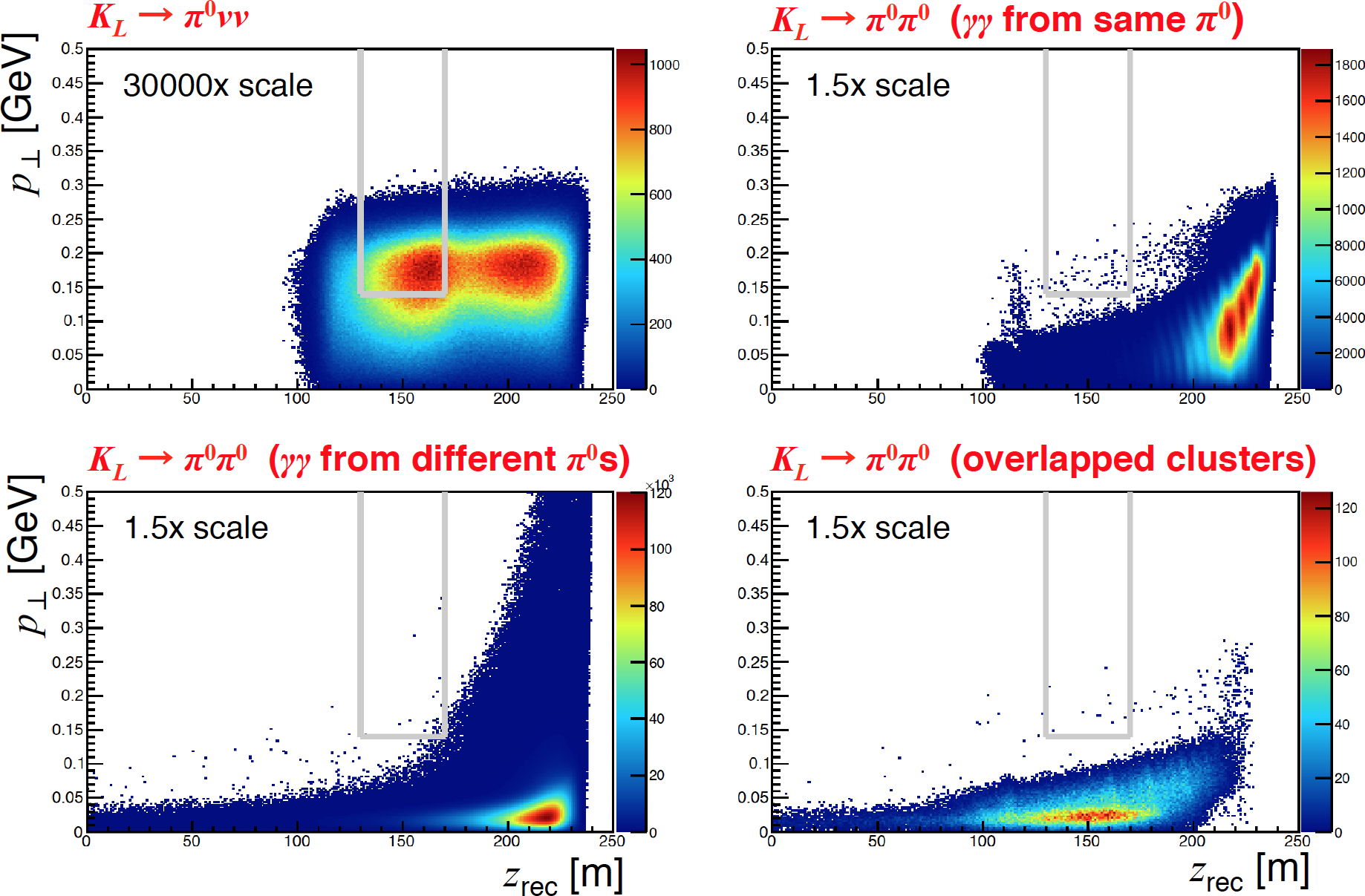}
  \caption{Distributions of events in plane of $(z_{\rm rec}, p_\perp)$
    after basic event selection cuts, from fast MC simulation, for 
    $K_L\to\pi^0\nu\bar{\nu}$ events (top left) and for
    $K_L\to\pi^0\pi^0$ events with two photons from the same
    $\pi^0$ (top right),
    two photons from different $\pi^0$s (bottom left), and
    with two or more indistinguishable overlapping photon clusters
    (bottom right).}
  \label{fig:sel}
\end{figure}
\paragraph{Signal selection and rejection of $K_L$ background}
Fig.~\ref{fig:sel} illustrates a possible scheme for differentiating signal
events from $K_L\to\pi^0\pi^0$ background. Events with exactly two photons
on the MEC and no other activity in the detector are selected. The clusters
on the MEC from both photons must also be more than 35~cm from the beam axis
(this helps to increase the rejection for events with overlapping clusters).
If one or both photons convert in the PSD, the reconstructed vertex
position $z_{\rm PSD}$ must be inside the fiducial volume. The plots show the
distributions of the events satisfying these minimal criteria in the
plane of $p_\perp$ vs.\ $z_{\rm rec}$ for the $\pi^0$, where $z_{\rm rec}$ is obtained from the transverse
separation of the two photon clusters on the MEC, assuming that they come from a
$\pi^0$ ($M_{\gamma\gamma} = m_{\pi^0}$). Signal candidates must have $z_{\rm rec}$ inside the FV and $p_\perp > 140$~MeV. 
This scheme is far from final, but it does
demonstrate that it should be possible to obtain $S/B\sim1$ 
with respect to other $K_L$ decays. 
We are continuing to investigate potential improvements to both the experiment and the analysis to increase the sensitivity, reduce the background, and increase redundancy, including the following:
\begin{itemize}
    \item Increased reliance on the PSD, with the requirement that at least one photon for signal candidates must convert and be tracked. This is particularly helpful in the downstream region, where the angles of incidence are greatest and the $p_\perp$ reconstruction is poorest.
    \item Requiring at least one photon conversion in the PSD halves the signal acceptance. This may be recovered by extending the FV in the downstream direction. This may be possible with the increased rejection from the PSD for $K_L\to\pi^0\pi^0$ events with candidate photons from different $\pi^0$s, and may also require extension of the LAV coverage to larger polar angles for the downstream stations.
    \item Use of multivariate analysis techniques to efficiently combine the above criteria. While efforts have been made in this direction, current attention is on development of the full simulation, which is prerequisite to establishing a definitive search strategy. 
\end{itemize}

\paragraph{$\Lambda\to n\pi^0$ background and beamline extension}
As noted in \Sec{sec:neutral_beam_ext}, additional measures must be taken to ensure sufficient suppression of background from $\Lambda\to n\pi^0$ decays.
The most attractive option is to maintain the 8~mrad production angle and increase the length of the beamline from target to AFC. Fast simulations of the extended-beamline configurations L1 and L2 (\Tab{tab:ext_layout}) have been performed and suggest that increasing the beamline length by 150~m as in L2 is sufficient to reduce the $\Lambda$ background to an acceptable level (L3 has not yet been simulated, but the total beamline length is similar to that for L2). In particular, the number of $\Lambda$s decaying in the FV is reduced by a factor of 6400 to about 100\,000, of which fewer than 200 pass the standard analysis cuts. Most of these remaining $\Lambda$ events can be eliminated using a hadronic calorimeter at small angle backing up the SAC to veto the high-energy neutron from the $\Lambda\to n\pi^0$ decay. Because of the mass asymmetry in the decay, the neutrons from decays in the FV that pass analysis cuts are {\em always} emitted into the SAC acceptance and have an energy distribution with a mean of 220~GeV; 98\% of these neutrons have $E>150$~GeV. Relatively few of the beam neutrons have energies this high (see \Fig{fig:beam_flux}); a hadronic calorimeter with a threshold at $E=150$~GeV would add less than 20~MHz to the total 140~MHz veto rate. The concept requires full simulation, but it seems reasonable to assume that the residual $\Lambda$ background could be reduced by an order of magnitude or more.
Because of the loss of solid angle due to the beamline extension, the $K_L$ flux is reduced by a factor of 2.4 with respect to the standard configuration. With the performance gains expected from reoptimization of the analysis, with particular attention to extension of the FV, better use of the information from the PSD, and, if needed, an increase of the angular coverage of the most downstream LAVs, this loss can be largely recovered, so that the original sensitivity goal remains within reach.

For the purposes of this discussion, the beam flux has simply been scaled by the solid-angle ratio (0.41). In reality, all of the rates in \Tab{tab:beam_rates} would be expected to decrease by a similar factor, as well as the total expected veto rate,
so one beneficial effect of the beamline extension would be the loosening of the timing and rate requirements. 
Once the extended beamline layout is finalized, it will be necessary to re-run the FLUKA-based simulation to obtain more reliable estimates of the beam flux, halo, and veto rates in the detectors.  

\paragraph {Beam-gas interactions}
The background from single $\pi^0$ production in interactions of beam neutrons on residual gas has been estimated with the FLUKA-based simulation of \Sec{sec:neutral_beam} to be a most a few percent of the expected signal for a residual gas pressure of $10^{-7}$ mbar.

\paragraph{Outlook} An effort is underway to develop a comprehensive simulation and to use it to validate the results obtained so far. This work is being carried out within the framework of the flexible Monte Carlo platform for the integrated program, allowing new configurations of the existing and proposed detectors to be simulated with the full NA62 Monte Carlo.
Of particular note,
backgrounds from radiative $K_L$ decays and cascading hyperon decays
remain to be studied, and the neutral-beam halo from our detailed
FLUKA beamline simulation needs to be fully incorporated into
the simulation of the experiment.
While mitigation of potential background contributions from one or more
of these sources might ultimately require specific modifications to the
experimental setup, we expect this task to be less complicated than
dealing with the primary challenges from $K_L\to\pi^0\pi^0$ and
$\Lambda\to n\pi^0$.

\subsection{An intermediate experimental phase targeting $K_L\to\pi^0 \ell^+\ell^-$ decays}
\label{sec:intermediateKL}
An intermediate experimental phase, 
exposing most of the final state detectors (Section~\ref{sec:NA62x4_detectors_downstream}) required for the high-statistics charged kaon experiment to a neutral beam (Section~\ref{sec:neutral_beam}) with similar specifications to the one needed for the KLEVER experiment, could be well suited to measure the decay properties of the $K_L \to (\pi^0)\ell^+\ell^-$ modes.
Minor adjustments to the detector layout might be needed as, for instance, the re-alignment of the straw chambers to the neutral beam axis.

The characteristics of the experimental apparatus aiming at $K_L\to \pi^0\ell^+\ell^-$ measurements, including precision tracking, redundant particle identification, high performance calorimetry, hermetic photon vetoes, high rate capabilities, and the required statistics of ${\cal O}(10^{13})$ kaon decays, allow the exploration of many rare and forbidden $K_L$ and hyperon decays.

Besides the valuable physics reach achievable, this intermediate phase would allow to characterize and benchmark the neutral beam and its halo, exploiting the advantages of the tracking detectors,
providing essential input to finalize the design of the KLEVER layout.
\subsection{Trigger and data acquisition}
\label{sec:readout}

Trigger and data acquisition will be designed together for the whole of the kaon programme.

Preliminary studies indicate that
the hit and event rates on most of the detectors are on the order of
a few tens of MHz, a few times larger than in NA62, with the notable
exception of the SAC, which will require an innovative readout solution
to handle rates of 100 MHz.
Digitization of the signals from the KLEVER detectors with FADCs at
high frequency (up to 1 GHz) would help to
efficiently veto background events without inducing dead time.
For the SAC, this is strictly necessary, since
the signal duration will be long compared to the mean interval between
events on a single channel. Detailed signal analysis may also assist with
particle identification and discrimination of uncorrelated background,
for example, from neutron interactions in the MEC or muon halo in the LAVs.

A streaming readout system that transfers all data off-detector has various advantages. Latency constraints can be relaxed since data are not buffered in the limited front-end memories and not susceptible to radiation-induced corruption. There is no need for a fast, synchronous trigger: online data reduction can be achieved through more sophisticated and flexible algorithms on devices with substantial computing power, such as high-end servers optionally hosting accelerators (GPUs or FPGAs). In principle, offline-like criteria may be used to implement the trigger.

On the other hand, a free-running, continuously digitizing readout implies very challenging data rates. Assuming 100~MHz of interactions in the SAC and current hit multiplicity and event size estimates, the data flow from the SAC alone would be 100~GB/s. A possible scheme for the readout system includes a low-level front-end layer (L0) in which signals are digitized and timestamped using TDCs with 100 ps precision, and then transferred, for example, via optical GBT link, to a readout board such as the PCIe40 developed for LHCb or the FELIX developed by ATLAS. A first layer of data processing may be implemented on this board, e.g., methods for $\gamma/n$ discrimination in the SAC or cluster-finding algorithms for the MEC may be implemented. The extracted features would be transferred, together with the data, to the host memory and dispatched over a high-speed network for further data aggregation and filtering. 

The specific solutions discussed here are intended as examples to demonstrate that a readout system with the needed requirements is within reach. Further evolution (including R\&D for this program) is expected before actual solutions are chosen.

\section{Experiments at the CERN LHC}
\subsection{LHCb in Run-I and II}
\label{sec:lhcb}

The LHCb detector~\cite{LHCb:2008vvz} is a single-arm forward spectrometer, covering the pseudorapidity
range $2 < \eta < 5$, which collected data in proton-proton collisions at the Large Hadron Collider at CERN in runs I and II. It was composed of a silicon-strip vertex detector surrounding the pp interaction
region (VELO), with a length of about 1 meter from the interaction point, a large-area
silicon-strip detector (TT) located upstream of a dipole magnet and three tracking stations
of silicon-strip detectors and straw drift tubes placed downstream of the magnet. Particle
identification was provided by two ring-imaging Cherenkov detectors, an electromagnetic
and a hadronic calorimeter, and a muon system composed of alternating layers of iron
and multiwire proportional chambers. The trigger system consisted in two stages: a hardware stage called Level-0 (L0), which identifies a high transverse momentum object in the event, and a software High Level Trigger (HLT) based on particle reconstruction and dedicated selections. The HLT was divided in two consecutive stages: HLT1 and HLT2.
LHCb has collected an integrated luminosity
of about 9 fb$^{-1}$ and is being upgraded for the next run of the LHC. The combined $K^{\pm}$ and $K^0$ production cross-sections at the LHC exceed 1 barn~\cite{Borsato:2018tcz,AlvesJunior:2018ldo}. Approximately $20\%$ of the kaons produced at LHCb interaction point are within the geometrical acceptance of the spectrometer.
The efficiency of detecting strange-hadron decays is, however, not be the same as for
heavy flavour for several reasons. The detector layout, which is optimised for beauty decays,
implies a relatively lower acceptance for $K_S^0$, with $K_L^0$ and $K^+$ efficiencies diminished even further. This is due to the differing flight lengths of the different mesons. The typical decay length of a beauty hadron is $\sim 1$ cm, $K_S^0$ can fly a distance of nearly one meter (hence roughly half of them escape the VELO), while
$K^{\pm}$ and $K_L^0$ traverse distances longer than the full LHCb detector length on average.
Another difficulty to reconstruct kaon decays at LHCb is the relatively low transverse momentum of the daughter particles, of the order of $100$ MeV, which is one order of magnitude smaller than those from beauty decays. As a reference, approximately $1\%$ of the $K_S^0\rightarrow\mu^+\mu^-$ decays produced at the interaction point could be reconstructed by the detector in the absence of trigger constraints.
During the first run of the LHC, no dedicated kaon triggers were enabled by LHCb, with a total trigger efficiency at the order of $1\%$ for the $K_S^0\rightarrow\mu^+\mu^-$ decays that would have otherwise been reconstructed. Despite of a total reconstruction and trigger efficiency of $\sim 10^{-4}$ the first run of the LHC provided two world best results on rare strange decays: an upper limit on the $K_S^0\rightarrow\mu^+\mu^-$ branching fraction thirty times more precise than the previous value, and an evidence for $\Sigma\to p\mu^+\mu^-$~\cite{LHCb:2017rdd} with a branching fraction compatible with the HyperCP evidence~\cite{HyperCP:2005mvo}, although with no sign of an anomalous dimuon invariant mass.
In 2015, changes made in HLT2 allowed to double the HLT efficiency. For the rest of the second run of the LHC, a dedicated reconstruction for low transverse momentum muons at the HLT1 was developed, as well as corresponding dimuon selection lines. This increased the trigger efficiency from $\sim 2\%$ to $\sim 20\%$~\cite{Dettori:2017ycr}, then limited by the L0. 
The data taken during the second run allowed to further improve the upper limit on ${\rm BR}(K_S^0\rightarrow\mu^+\mu^-)$ by yet another factor of four~\cite{LHCb:2020ycd}, yielding a final result:
\begin{equation}
{\rm BR} (K_S^0\rightarrow\mu^+\mu^-) < 2.1\times 10^{-10}
\end{equation}
\noindent at $90\%$ confidence level. An update of the $\Sigma\rightarrow p\mu^+\mu^-$~\cite{LHCb:2017rdd} is ongoing. The recorded data could also allow for more measurements of kaon and hyperon decays. 

The upgraded LHCb detector will have no hardware trigger, and instead a GPU-based software trigger as will discussed in Section~\ref{sec:lhcb_upgrade}.

\subsection{Future results of LHCb and its Upgraded detectors}
\label{sec:lhcb_upgrade}
The LHCb experiment will resume data taking in 2022 with an upgraded detector and new trigger system. The upgraded detector has a fully renewed tracking system and improved electronics to fully acquire data at 40 MHz. 
The hardware stage of the trigger, L0, is absent in the upgrade. Instead, a first software-based HLT will be running on a farm of graphics processing units (GPUs) ~\cite{Aaij:2019zbu,LHCbCollaboration:2717938}. This will provide efficiencies almost one order of magnitude bigger than those in Run-II. The LHCb upgrade plans to collect an integrated luminosity of $50 fb^{-1}$, which, together with the improved trigger efficiency, will provide strangeness sample sizes almost 100 times bigger than the ones achieved in Run-II. The expected time period for the LHCb Upgrade data taking is 2022-2030.
Furthermore , a Phase-II upgrade has been proposed ~\cite{LHCb:2018roe,Aaij:2244311}, which aims at collecting 300 $fb^{-1}$ of integrated luminosity in the period 2031-2035. 
The future capabilities of the LHCb upgrade for strangeness physics have been documented in ~\cite{AlvesJunior:2018ldo,Cerri:2018ypt}. In particular:
\begin{itemize}
    \item The LHCb Upgrade and its Phase-II could reach a sensitivity in the branching fraction of $K_S^0\rightarrow\mu^+\mu^-$ close to the SM prediction.
    \item The LHCb upgrade could quickly overtake NA48's results on the the branching fraction of $K_S^0\rightarrow\pi^0\mu^+\mu^-$ and read a statistical precision of $0.25\times10^{-9}$ with 50 $fb^{-1}$. It would also be able to measure its differential decay rate, allowing for $25\%$ ($10\%$) precision measurement of $|a_S|$ without the need of any theoretical input on $b_S$ with 50 (300) $fb^{-1}$
    \item World best measurements in rare $K_S^0$ rare like $K_S^0\rightarrow4l$,  $K_S^0\rightarrow X^0\mu\mu$, or $K_S^0\rightarrow X^0\pi\mu$.
    
\end{itemize}
The LHCb and its upgrade can also be a world leading experiment in hyperon physics. 
\begin{itemize}
\item First and foremost, aside from observing  $\Sigma^+\to p \mu^+ \mu^-$, which could happen already in Run~2, precision measurements of the differential branching fraction, CP violation and forward-backward asymmetries will be performed~\cite{He:2018yzu}. The $\Sigma^+\to p e^+ e^-$ will also be in reach, giving the possibility to perform also lepton flavour universality tests. 
\item CP violation measurements in neutral hyperons will be accessible both directly (e.g. for $\Lambda \to p \pi^- e^+ e^-$  decays), or through cascade decays such as $\Xi^- \to \Lambda \pi^-$. \item Tests of Lepton Flavor Universality in Semileptonic Hyperon Decays: while the branching fractions of the electron modes are known very precisely, the muonic modes have currently very large experimental uncertainties that LHCb could in principle significantly improve. In particular, a measurement of $B(\Lambda^0\to p\mu\bar{\nu})$ is ongoing using Run-II data~\cite{BreaRodriguez:2020jlg}.
\item Searches for Dark Baryons in hyperon decays~\cite{Alonso-Alvarez:2021oaj}, with sensitivities down to $O(2\times 10^{-6}$) level for $\Xi^0\to\pi^+\pi^-\psi_{DS}(940)$ and $O(7\times10^{-11})$ for $\Xi^0\to\mu^+\mu^-\psi_{DS}(940)$, according to fast simulation studies.
    \item In addition, a large number of lepton and/or baryon number violating decays will be accessible with unprecedented statistics, leading to world best limits. 
\end{itemize}

\section{Outlook and Opportunities for US Collaborators}
A number of US groups are participating in the current program of experiments. For KOTO, Arizona State University, University of Michigan, and the University of Chicago joined the experiment since 2007. Contributions include the KTeV CsI crystal calorimeter (transferred from Fermilab to J-PARC), and the DAQ system including front-end digitization, digital pipelined trigger, and readout. Beyond the hardware, the US groups contribute to all aspect of the data taking and analysis, particularly the introduction of AI neural-net techniques to further eliminate backgrounds. One of the PIs has been elected as co-spokesperson and served since the beginning. In addition to several US graduate students who did their dissertation research on KOTO, many Michigan undergraduates participated in KOTO via the National Science Foundation REU program.  In a 2018 DOE review, the committee's report concluded that moving towards KOTO Step-2 is well justified and that effort to recruit new collaborators is encouraged. 

The HEP group at George Mason University (GMU)
is the sole US member of NA62, and consists of a PI, two senior consultants, and both graduate and undergraduate students. GMU has been an
institutional member of the NA62 collaboration since 2006. In that time,
six graduate students and approximately thirty undergraduate students
have worked on the experiment. The group has contributed to detector construction,
commissioning, and maintenance; development of software trigger
algorithms; development and maintenance of on-line utilities and
detector control software; development of simulation and data analysis
software; and data analyses.  The PI has served as run coordinator, and
he and one of the senior consultants are members of the collaboration's
editorial board.  In addition to its ongoing involvement in NA62, the
group is working on simulation tools for design of future kaon physics
experiments at CERN. 

Groups at Massachusetts Institute of Technology, University of Cincinnati, University of Maryland, Los Alamos National Laboratory, and Syracuse University are members of the LHCb collaboration. These groups play major roles in physics analysis, hardware development, and computing at LHCb. MIT is the host for the US LHCb Tier 2 computing center and US groups are involved in online reconstruction and selection. Syracuse University was one of the contributing institutions to the LHCb silicon vertex detector (VELO), US groups are leading the Upstream Tracker (UT) Phase-I upgrade for LHC Run III, which is expected to improve triggering speed substantially, and US groups are currently designing a Si-W calorimeter that is an option being considered for the LHCb Phase-II upgrade. 


Upgrades and/or next-generation experiments at JPARC, CERN's SPS, and LHCb are proposed, as described in this document. All of these proposals represent significant increases in complexity and require that the collaborations grow, which opens the door for expanded US participation. 

One potential US contribution would be development and implementation of data management frameworks necessary for the large data sets and international collaborations envisioned in next-generation experiments. As a potential model, Brookhaven National Lab (BNL) is responsible for data management at Belle-II~\cite{refId0}, which is a Japan-based experiment with similar scale and data handling challenges as next-generation kaon experiments. One of scientists involved in this effort was also responsible for developing the NA62 offline computing architecture, building on solutions developed within ATLAS\cite{Laycock:2019vxl}.  BNL is exploring the possibility of implementing a Rucio~\cite{Barisits:2019fyl} service for KOTO, to facilitate data transfer between the host laboratory and international collaborators and to better optimize use of resources at the host lab. Exploratory work on this possibility as well as the possibility of producing large Monte Carlo simulations at US computing facilities has been supported by the U.S.-Japan Science and Technology Cooperation Program in High Energy Physics (``US-Japan'') and would be even more compelling for the next-generation experiments at JPARC and CERN SPS described in this document. US groups already have significant computing responsibility in LHCb, as mentioned above.

As described above, US institutions have made significant contributions to KOTO's trigger and DAQ, with the 2022 upgrade described in Section~\ref{sect:kotodaqupgrade} being supported in large part by US-Japan funding. KOTO Step 2 will have higher rates and more detector elements, so advancements in trigger and DAQ technology will be required; given the expertise of current US KOTO collaborators, these would be natural places for expanded US involvement. But, with the substantial size increases of the KOTO Step 2 detector, in particular the forward calorimeter and the barrel veto, there are numerous other opportunities to make significant hardware contributions. For example, the barrel veto needs to be segmented longitudinally to reduce rates and control splash-back, and perhaps photon angle measurement which could be significant to the physics signatures. 

Similarly,  the whole complement of NA62 detectors will have to be rebuilt for the future program, and in the short term there are significant and challenging R\&D initiatives in GHz rate silicon pixel tracking, high-rate straw tracking, and high performance calorimetry and photon detection with innovative materials. Several options are under investigation for high-performance calorimetry for NA62/KLEVER, including Si-W sampling calorimeters, shashlyk calorimeters featuring innovative, radiation hard materials, and compact crystal calorimeters. The effort to develop these concepts has clear synergies with calorimetry proposals for the LHCb Phase-II upgrade, including some R\&D already being pursued by US LHCb groups.



Participation in kaon experiments at JPARC and/or CERN is currently the only opportunity for US physicists to contribute to this vital area of research, would add breadth to the US research program, and would provide excellent opportunities for students to make major hardware and analysis contributions in moderately sized collaborations. As part of the Snowmass process, the US HEP community is exploring possible expansions to our physics program that could be
achieved with future upgrades to the Fermilab proton accelerator complex~\cite{Ainsworth:2021ahm,Arrington:2022pon}, envisioned to take place in the 2030s. This could include additional (next-next generation) experiments exploring the physics accessible with high-intensity kaon beams. The best way for the US HEP community to be well-positioned to take advantage of this opportunity would be to explore this physics and develop expertise in modern kaon experiments by participating in the international programs described in this document.

\bibliographystyle{elsarticle-num-mod}
\bibliography{main}

\begin{thebibliography}{100}
\expandafter\ifx\csname url\endcsname\relax
  \def\url#1{\texttt{#1}}\fi
\expandafter\ifx\csname urlprefix\endcsname\relax\def\urlprefix{URL }\fi
\expandafter\ifx\csname href\endcsname\relax
  \def\href#1#2{#2} \def\path#1{#1}\fi

\bibitem{Rochester:1947}
G.~Rochester, C.~C. Butler, Nature 160 (1947) 855--857.

\bibitem{Brown:1949}
R.~Brown, et~al., Nature 163 (1949) 82.

\bibitem{Cronin:2013}
J.~Cronin, $\it{Fermi}$ $\it{Remembered}$, University of Chicago Press ISBN
  10:022610088X (2013) p.XX.

\bibitem{Pais:1955}
A.~H. Pais, M.~Gell-Mann, Phys. Rev. 97 (1955) 1387.

\bibitem{Lederman:1956}
K.~Lande, et~al., Phys. Rev. 103 (1956) 1901.

\bibitem{Wu:1957}
C.~S. Wu, et~al., Phys. Rev. 105 (1957) 1413.

\bibitem{Christenson:1964}
J.~H. Christenson, et~al., Phys. Rev. Lett. 13 (1964) 138.

\bibitem{Cirigliano:2011ny}
V.~Cirigliano, G.~Ecker, H.~Neufeld, A.~Pich, J.~Portoles, Rev. Mod. Phys. 84
  (2012) 399.
\newblock \href {http://arxiv.org/abs/1107.6001} {\path{arXiv:1107.6001}},
  \href {https://doi.org/10.1103/RevModPhys.84.399}
  {\path{doi:10.1103/RevModPhys.84.399}}.

\bibitem{Buras:2015qea}
A.~J. Buras, D.~Buttazzo, J.~Girrbach-Noe, R.~Knegjens, JHEP 11 (2015) 033.
\newblock \href {http://arxiv.org/abs/1503.02693} {\path{arXiv:1503.02693}},
  \href {https://doi.org/10.1007/JHEP11(2015)033}
  {\path{doi:10.1007/JHEP11(2015)033}}.

\bibitem{Brod:2021hsj}
J.~Brod, M.~Gorbahn, E.~Stamou, PoS BEAUTY2020 (2021) 056.
\newblock \href {http://arxiv.org/abs/2105.02868} {\path{arXiv:2105.02868}},
  \href {https://doi.org/10.22323/1.391.0056} {\path{doi:10.22323/1.391.0056}}.

\bibitem{Buras:2021nns}
A.~J. Buras, E.~Venturini (9 2021).
\newblock \href {http://arxiv.org/abs/2109.11032} {\path{arXiv:2109.11032}}.

\bibitem{Buras:2014zga}
A.~J. Buras, D.~Buttazzo, J.~Girrbach-Noe, R.~Knegjens, JHEP 11 (2014) 121.
\newblock \href {http://arxiv.org/abs/1408.0728} {\path{arXiv:1408.0728}},
  \href {https://doi.org/10.1007/JHEP11(2014)121}
  {\path{doi:10.1007/JHEP11(2014)121}}.

\bibitem{Buras:2015yca}
A.~J. Buras, D.~Buttazzo, R.~Knegjens, JHEP 11 (2015) 166.
\newblock \href {http://arxiv.org/abs/1507.08672} {\path{arXiv:1507.08672}},
  \href {https://doi.org/10.1007/JHEP11(2015)166}
  {\path{doi:10.1007/JHEP11(2015)166}}.

\bibitem{Isidori:2006qy}
G.~Isidori, F.~Mescia, P.~Paradisi, C.~Smith, S.~Trine, JHEP 08 (2006) 064.
\newblock \href {http://arxiv.org/abs/hep-ph/0604074}
  {\path{arXiv:hep-ph/0604074}}, \href
  {https://doi.org/10.1088/1126-6708/2006/08/064}
  {\path{doi:10.1088/1126-6708/2006/08/064}}.

\bibitem{Blanke:2009pq}
M.~Blanke, Acta Phys. Polon. B41 (2010) 127.
\newblock \href {http://arxiv.org/abs/0904.2528} {\path{arXiv:0904.2528}}.

\bibitem{Blanke:2009am}
M.~Blanke, A.~J. Buras, B.~Duling, S.~Recksiegel, C.~Tarantino, Acta Phys.
  Polon. B41 (2010) 657--683.
\newblock \href {http://arxiv.org/abs/0906.5454} {\path{arXiv:0906.5454}}.

\bibitem{Buras:2012jb}
A.~J. Buras, F.~De~Fazio, J.~Girrbach, JHEP 02 (2013) 116.
\newblock \href {http://arxiv.org/abs/1211.1896} {\path{arXiv:1211.1896}},
  \href {https://doi.org/10.1007/JHEP02(2013)116}
  {\path{doi:10.1007/JHEP02(2013)116}}.

\bibitem{Blanke:2008yr}
M.~Blanke, A.~J. Buras, B.~Duling, K.~Gemmler, S.~Gori, JHEP 03 (2009) 108.
\newblock \href {http://arxiv.org/abs/0812.3803} {\path{arXiv:0812.3803}},
  \href {https://doi.org/10.1088/1126-6708/2009/03/108}
  {\path{doi:10.1088/1126-6708/2009/03/108}}.

\bibitem{Grossman:1997sk}
Y.~Grossman, Y.~Nir, Phys. Lett. B398 (1997) 163--168.
\newblock \href {http://arxiv.org/abs/hep-ph/9701313}
  {\path{arXiv:hep-ph/9701313}}, \href
  {https://doi.org/10.1016/S0370-2693(97)00210-4}
  {\path{doi:10.1016/S0370-2693(97)00210-4}}.

\bibitem{Fuyuto:2014cya}
K.~Fuyuto, W.-S. Hou, M.~Kohda, Phys. Rev. Lett. 114 (2015) 171802.
\newblock \href {http://arxiv.org/abs/1412.4397} {\path{arXiv:1412.4397}},
  \href {https://doi.org/10.1103/PhysRevLett.114.171802}
  {\path{doi:10.1103/PhysRevLett.114.171802}}.

\bibitem{Kitahara:2019lws}
T.~Kitahara, T.~Okui, G.~Perez, Y.~Soreq, K.~Tobioka, Phys. Rev. Lett. 124
  (2020) 071801.
\newblock \href {http://arxiv.org/abs/1909.11111} {\path{arXiv:1909.11111}},
  \href {https://doi.org/10.1103/PhysRevLett.124.071801}
  {\path{doi:10.1103/PhysRevLett.124.071801}}.

\bibitem{Egana-Ugrinovic:2019wzj}
D.~Egana-Ugrinovic, S.~Homiller, P.~Meade, Phys. Rev. Lett. 124 (2020) 191801.
\newblock \href {http://arxiv.org/abs/1911.10203} {\path{arXiv:1911.10203}},
  \href {https://doi.org/10.1103/PhysRevLett.124.191801}
  {\path{doi:10.1103/PhysRevLett.124.191801}}.

\bibitem{Ziegler:2020ize}
R.~Ziegler, J.~Zupan, R.~Zwicky, JHEP 07 (2020) 229.
\newblock \href {http://arxiv.org/abs/2005.00451} {\path{arXiv:2005.00451}},
  \href {https://doi.org/10.1007/JHEP07(2020)229}
  {\path{doi:10.1007/JHEP07(2020)229}}.

\bibitem{Batley:2002gn}
J.~R. Batley, et~al., Phys. Lett. B544 (2002) 97--112.
\newblock \href {http://arxiv.org/abs/hep-ex/0208009}
  {\path{arXiv:hep-ex/0208009}}, \href
  {https://doi.org/10.1016/S0370-2693(02)02476-0}
  {\path{doi:10.1016/S0370-2693(02)02476-0}}.

\bibitem{Abouzaid:2010ny}
E.~Abouzaid, et~al., Phys. Rev. D83 (2011) 092001.
\newblock \href {http://arxiv.org/abs/1011.0127} {\path{arXiv:1011.0127}},
  \href {https://doi.org/10.1103/PhysRevD.83.092001}
  {\path{doi:10.1103/PhysRevD.83.092001}}.

\bibitem{ParticleDataGroup:2020ssz}
P.~A. Zyla, et~al., PTEP 2020 (2020) 083C01.
\newblock \href {https://doi.org/10.1093/ptep/ptaa104}
  {\path{doi:10.1093/ptep/ptaa104}}.

\bibitem{Bai:2015nea}
Z.~Bai, et~al., Phys. Rev. Lett. 115 (2015) 212001.
\newblock \href {http://arxiv.org/abs/1505.07863} {\path{arXiv:1505.07863}},
  \href {https://doi.org/10.1103/PhysRevLett.115.212001}
  {\path{doi:10.1103/PhysRevLett.115.212001}}.

\bibitem{RBC:2020kdj}
R.~Abbott, et~al., Phys. Rev. D 102 (2020) 054509.
\newblock \href {http://arxiv.org/abs/2004.09440} {\path{arXiv:2004.09440}},
  \href {https://doi.org/10.1103/PhysRevD.102.054509}
  {\path{doi:10.1103/PhysRevD.102.054509}}.

\bibitem{Aebischer:2020jto}
J.~Aebischer, C.~Bobeth, A.~J. Buras, Eur. Phys. J. C 80 (2020) 705.
\newblock \href {http://arxiv.org/abs/2005.05978} {\path{arXiv:2005.05978}},
  \href {https://doi.org/10.1140/epjc/s10052-020-8267-1}
  {\path{doi:10.1140/epjc/s10052-020-8267-1}}.

\bibitem{Cirigliano:2019cpi}
V.~Cirigliano, H.~Gisbert, A.~Pich, A.~Rodr\'\i{}guez-S\'anchez, JHEP 02 (2020)
  032.
\newblock \href {http://arxiv.org/abs/1911.01359} {\path{arXiv:1911.01359}},
  \href {https://doi.org/10.1007/JHEP02(2020)032}
  {\path{doi:10.1007/JHEP02(2020)032}}.

\bibitem{Aebischer:2020mkv}
J.~Aebischer, A.~J. Buras, J.~Kumar, JHEP 12 (2020) 097.
\newblock \href {http://arxiv.org/abs/2006.01138} {\path{arXiv:2006.01138}},
  \href {https://doi.org/10.1007/JHEP12(2020)097}
  {\path{doi:10.1007/JHEP12(2020)097}}.

\bibitem{Aebischer:2022vky}
J.~Aebischer, A.~J. Buras, J.~Kumar, {On the Importance of Rare Kaon Decays: A
  Snowmass 2021 White Paper}, in: {2022 Snowmass Summer Study}, 2022.
\newblock \href {http://arxiv.org/abs/2203.09524} {\path{arXiv:2203.09524}}.

\bibitem{Bordone:2017lsy}
M.~Bordone, D.~Buttazzo, G.~Isidori, J.~Monnard, Eur. Phys. J. C77 (2017) 618.
\newblock \href {http://arxiv.org/abs/1705.10729} {\path{arXiv:1705.10729}},
  \href {https://doi.org/10.1140/epjc/s10052-017-5202-1}
  {\path{doi:10.1140/epjc/s10052-017-5202-1}}.

\bibitem{Buttazzo:2017ixm}
D.~Buttazzo, A.~Greljo, G.~Isidori, D.~Marzocca, JHEP 11 (2017) 044.
\newblock \href {http://arxiv.org/abs/1706.07808} {\path{arXiv:1706.07808}},
  \href {https://doi.org/10.1007/JHEP11(2017)044}
  {\path{doi:10.1007/JHEP11(2017)044}}.

\bibitem{Fajfer:2018bfj}
S.~Fajfer, N.~Košnik, L.~Vale~Silva, Eur. Phys. J. C78 (2018) 275.
\newblock \href {http://arxiv.org/abs/1802.00786} {\path{arXiv:1802.00786}},
  \href {https://doi.org/10.1140/epjc/s10052-018-5757-5}
  {\path{doi:10.1140/epjc/s10052-018-5757-5}}.

\bibitem{Marzocca:2021miv}
D.~Marzocca, S.~Trifinopoulos, E.~Venturini (6 2021).
\newblock \href {http://arxiv.org/abs/2106.15630} {\path{arXiv:2106.15630}}.

\bibitem{Adler:2001xv}
S.~Adler, et~al., Phys. Rev. Lett. 88 (2002) 041803.
\newblock \href {http://arxiv.org/abs/hep-ex/0111091}
  {\path{arXiv:hep-ex/0111091}}, \href
  {https://doi.org/10.1103/PhysRevLett.88.041803}
  {\path{doi:10.1103/PhysRevLett.88.041803}}.

\bibitem{Artamonov:2009sz}
A.~V. Artamonov, et~al., Phys. Rev. D79 (2009) 092004.
\newblock \href {http://arxiv.org/abs/0903.0030} {\path{arXiv:0903.0030}},
  \href {https://doi.org/10.1103/PhysRevD.79.092004}
  {\path{doi:10.1103/PhysRevD.79.092004}}.

\bibitem{Comfort:2011zz}
J.~Comfort, et~al., {ORKA:} measurement of the $k^ \to \pi^+ \nu \bar{\nu}$
  decay at {Fermilab}, fERMILAB-PROPOSAL-1021 (2011).
\newblock \href {https://doi.org/10.2172/1041571} {\path{doi:10.2172/1041571}}.

\bibitem{Frank:2001aa}
J.~Frank, et~al., {Charged Kaons at the Main injector (CKM)}: A proposal for a
  precision measurement of the decay {$K^+\to\pi^+\nu\bar{\nu}$} and other rare
  {$K^+$} processes at {Fermilab} using the {Main Injector} (2001).
\newblock \href {https://doi.org/10.2172/878912} {\path{doi:10.2172/878912}}.

\bibitem{Cooper:2005zz}
P.~S. Cooper, {Kplus} - {A} cost effective and competitive precision
  measurement of the decay {$K^+\to\pi^+\nu\bar{\nu}$} (2005).

\bibitem{Anelli:2005xxx}
G.~Anelli, et~al., \href{https://cds.cern.ch/record/832885}{Proposal to measure
  the rare decay $k^+\to\pi^+\nu\bar{nu}$ at the cern sps},
  cERN-SPSC-2005-013/SPSC-P-326 (2005).
\newline\urlprefix\url{https://cds.cern.ch/record/832885}

\bibitem{NA62:2010xx}
F.~Hahn~(ed.), et~al., \href{http://cds.cern.ch/record/1404985}{{NA62}
  technical design document}, {NA62-10-07} (2010).
\newline\urlprefix\url{http://cds.cern.ch/record/1404985}

\bibitem{NA62:2017rwk}
E.~Cortina~Gil, et~al., JINST 12 (2017) P05025.
\newblock \href {http://arxiv.org/abs/1703.08501} {\path{arXiv:1703.08501}},
  \href {https://doi.org/10.1088/1748-0221/12/05/P05025}
  {\path{doi:10.1088/1748-0221/12/05/P05025}}.

\bibitem{NA62:2018ctf}
E.~Cortina~Gil, et~al., Phys. Lett. B 791 (2019) 156--166.
\newblock \href {http://arxiv.org/abs/1811.08508} {\path{arXiv:1811.08508}},
  \href {https://doi.org/10.1016/j.physletb.2019.01.067}
  {\path{doi:10.1016/j.physletb.2019.01.067}}.

\bibitem{NA62:2020fhy}
E.~Cortina~Gil, et~al., JHEP 11 (2020) 042.
\newblock \href {http://arxiv.org/abs/2007.08218} {\path{arXiv:2007.08218}},
  \href {https://doi.org/10.1007/JHEP11(2020)042}
  {\path{doi:10.1007/JHEP11(2020)042}}.

\bibitem{NA62:2021zjw}
E.~Cortina~Gil, et~al., JHEP 06 (2021) 093.
\newblock \href {http://arxiv.org/abs/2103.15389} {\path{arXiv:2103.15389}},
  \href {https://doi.org/10.1007/JHEP06(2021)093}
  {\path{doi:10.1007/JHEP06(2021)093}}.

\bibitem{NA62:2020pwi}
E.~Cortina~Gil, et~al., JHEP 02 (2021) 201.
\newblock \href {http://arxiv.org/abs/2010.07644} {\path{arXiv:2010.07644}},
  \href {https://doi.org/10.1007/JHEP02(2021)201}
  {\path{doi:10.1007/JHEP02(2021)201}}.

\bibitem{NA62:2019xxx}
{NA62 Collaboration}, \href{https://cds.cern.ch/record/2691873}{Addendum i to
  p326: Continuation of the physics programme of the na62 experiment},
  cERN-SPSC-2019-039/SPSC-P326-ADD-1 (2019).
\newline\urlprefix\url{https://cds.cern.ch/record/2691873}

\bibitem{NA62:2021xxx}
{NA62 Collaboration}, \href{https://cds.cern.ch/record/2759577}{2021 {NA62}
  status report to the {CERN} {SPSC}}, cERN-SPSC-2021-009/SPSC-SR-286 (2021).
\newline\urlprefix\url{https://cds.cern.ch/record/2759577}

\bibitem{Littenberg:1989ix}
L.~S. Littenberg, Phys. Rev. D 39 (1989) 3322--3324.
\newblock \href {https://doi.org/10.1103/PhysRevD.39.3322}
  {\path{doi:10.1103/PhysRevD.39.3322}}.

\bibitem{Graham:1992pk}
G.~E. Graham, et~al., Phys. Lett. B 295 (1992) 169--173.
\newblock \href {https://doi.org/10.1016/0370-2693(92)90107-F}
  {\path{doi:10.1016/0370-2693(92)90107-F}}.

\bibitem{E779:1994amx}
M.~Weaver, et~al., Phys. Rev. Lett. 72 (1994) 3758--3761.
\newblock \href {https://doi.org/10.1103/PhysRevLett.72.3758}
  {\path{doi:10.1103/PhysRevLett.72.3758}}.

\bibitem{KTeV:1998taf}
J.~Adams, et~al., Phys. Lett. B 447 (1999) 240--245.
\newblock \href {http://arxiv.org/abs/hep-ex/9806007}
  {\path{arXiv:hep-ex/9806007}}, \href
  {https://doi.org/10.1016/S0370-2693(98)01593-7}
  {\path{doi:10.1016/S0370-2693(98)01593-7}}.

\bibitem{E799-IIKTeV:1999iym}
A.~Alavi-Harati, et~al., Phys. Rev. D 61 (2000) 072006.
\newblock \href {http://arxiv.org/abs/hep-ex/9907014}
  {\path{arXiv:hep-ex/9907014}}, \href
  {https://doi.org/10.1103/PhysRevD.61.072006}
  {\path{doi:10.1103/PhysRevD.61.072006}}.

\bibitem{E391a:2006fxm}
J.~K. Ahn, et~al., Phys. Rev. D 74 (2006) 051105, [Erratum: Phys.Rev.D 74,
  079901 (2006)].
\newblock \href {http://arxiv.org/abs/hep-ex/0607016}
  {\path{arXiv:hep-ex/0607016}}, \href
  {https://doi.org/10.1103/PhysRevD.74.051105}
  {\path{doi:10.1103/PhysRevD.74.051105}}.

\bibitem{Ahn:2007cd}
J.~K. Ahn, et~al., Phys. Rev. Lett. 100 (2008) 201802.
\newblock \href {http://arxiv.org/abs/0712.4164} {\path{arXiv:0712.4164}},
  \href {https://doi.org/10.1103/PhysRevLett.100.201802}
  {\path{doi:10.1103/PhysRevLett.100.201802}}.

\bibitem{E391a:2009jdb}
J.~K. Ahn, et~al., Phys. Rev. D 81 (2010) 072004.
\newblock \href {http://arxiv.org/abs/0911.4789} {\path{arXiv:0911.4789}},
  \href {https://doi.org/10.1103/PhysRevD.81.072004}
  {\path{doi:10.1103/PhysRevD.81.072004}}.

\bibitem{KOTO:2016vwr}
J.~K. Ahn, et~al., PTEP 2017 (2017) 021C01.
\newblock \href {http://arxiv.org/abs/1609.03637} {\path{arXiv:1609.03637}},
  \href {https://doi.org/10.1093/ptep/ptx001} {\path{doi:10.1093/ptep/ptx001}}.

\bibitem{Ahn:2018mvc}
J.~K. Ahn, et~al., Phys. Rev. Lett. 122 (2019) 021802.
\newblock \href {http://arxiv.org/abs/1810.09655} {\path{arXiv:1810.09655}},
  \href {https://doi.org/10.1103/PhysRevLett.122.021802}
  {\path{doi:10.1103/PhysRevLett.122.021802}}.

\bibitem{KOTO:2020prk}
J.~K. Ahn, et~al., Phys. Rev. Lett. 126 (2021) 121801.
\newblock \href {http://arxiv.org/abs/2012.07571} {\path{arXiv:2012.07571}},
  \href {https://doi.org/10.1103/PhysRevLett.126.121801}
  {\path{doi:10.1103/PhysRevLett.126.121801}}.

\bibitem{Comfort:2015xx}
J.~Comfort, et~al., {KOPIO} project: Conceptual design report (2005).

\bibitem{KAMI:2001cyd}
T.~Alexopoulos, et~al. (4 2001).

\bibitem{Ecker:1991ru}
G.~Ecker, A.~Pich, Nucl. Phys. B366 (1991) 189--205.
\newblock \href {https://doi.org/10.1016/0550-3213(91)90056-4}
  {\path{doi:10.1016/0550-3213(91)90056-4}}.

\bibitem{Isidori:2003ts}
G.~Isidori, R.~Unterdorfer, JHEP 01 (2004) 009.
\newblock \href {http://arxiv.org/abs/hep-ph/0311084}
  {\path{arXiv:hep-ph/0311084}}, \href
  {https://doi.org/10.1088/1126-6708/2004/01/009}
  {\path{doi:10.1088/1126-6708/2004/01/009}}.

\bibitem{DAmbrosio:2017klp}
G.~D'Ambrosio, T.~Kitahara, Phys. Rev. Lett. 119 (2017) 201802.
\newblock \href {http://arxiv.org/abs/1707.06999} {\path{arXiv:1707.06999}},
  \href {https://doi.org/10.1103/PhysRevLett.119.201802}
  {\path{doi:10.1103/PhysRevLett.119.201802}}.

\bibitem{KLMuMu_theory}
G.~D'Ambrosio, G.~Ecker, G.~Isidori, H.~Neufeld,
  \href{http://preprints.cern.ch/cgi-bin/setlink?base=preprint&categ=cern&id=th-7503-94}{{Radiative
  non-leptonic kaon decays}}, in: {2nd DAPHNE Physics Handbook:265-313}, 1994,
  pp. 265--313.
\newblock \href {http://arxiv.org/abs/hep-ph/9411439}
  {\path{arXiv:hep-ph/9411439}}.
\newline\urlprefix\url{http://preprints.cern.ch/cgi-bin/setlink?base=preprint&categ=cern&id=th-7503-94}

\bibitem{Dorsner:2011ai}
I.~Dorsner, J.~Drobnak, S.~Fajfer, J.~F. Kamenik, N.~Kosnik, JHEP 11 (2011)
  002.
\newblock \href {http://arxiv.org/abs/1107.5393} {\path{arXiv:1107.5393}},
  \href {https://doi.org/10.1007/JHEP11(2011)002}
  {\path{doi:10.1007/JHEP11(2011)002}}.

\bibitem{Mandal:2019gff}
R.~Mandal, A.~Pich, JHEP 12 (2019) 089.
\newblock \href {http://arxiv.org/abs/1908.11155} {\path{arXiv:1908.11155}},
  \href {https://doi.org/10.1007/JHEP12(2019)089}
  {\path{doi:10.1007/JHEP12(2019)089}}.

\bibitem{Bobeth:2017ecx}
C.~Bobeth, A.~J. Buras, JHEP 02 (2018) 101.
\newblock \href {http://arxiv.org/abs/1712.01295} {\path{arXiv:1712.01295}},
  \href {https://doi.org/10.1007/JHEP02(2018)101}
  {\path{doi:10.1007/JHEP02(2018)101}}.

\bibitem{Chobanova:2017rkj}
V.~Chobanova, G.~D'Ambrosio, T.~Kitahara, M.~Lucio~Martinez,
  D.~Martinez~Santos, I.~S. Fernandez, K.~Yamamoto, JHEP 05 (2018) 024.
\newblock \href {http://arxiv.org/abs/1711.11030} {\path{arXiv:1711.11030}},
  \href {https://doi.org/10.1007/JHEP05(2018)024}
  {\path{doi:10.1007/JHEP05(2018)024}}.

\bibitem{E871:2000wvm}
D.~Ambrose, et~al., Phys. Rev. Lett. 84 (2000) 1389--1392.
\newblock \href {https://doi.org/10.1103/PhysRevLett.84.1389}
  {\path{doi:10.1103/PhysRevLett.84.1389}}.

\bibitem{Akagi:1994bb}
T.~Akagi, et~al., Phys. Rev. D 51 (1995) 2061--2089.
\newblock \href {https://doi.org/10.1103/PhysRevD.51.2061}
  {\path{doi:10.1103/PhysRevD.51.2061}}.

\bibitem{E791:1994xxb}
A.~Heinson, et~al., Phys. Rev. D 51 (1995) 985--1013.
\newblock \href {https://doi.org/10.1103/PhysRevD.51.985}
  {\path{doi:10.1103/PhysRevD.51.985}}.

\bibitem{LHCb:2020ycd}
R.~Aaij, et~al., Phys. Rev. Lett. 125 (2020) 231801.
\newblock \href {http://arxiv.org/abs/2001.10354} {\path{arXiv:2001.10354}},
  \href {https://doi.org/10.1103/PhysRevLett.125.231801}
  {\path{doi:10.1103/PhysRevLett.125.231801}}.

\bibitem{Dery:2021mct}
A.~Dery, M.~Ghosh, Y.~Grossman, S.~Schacht, JHEP 07 (2021) 103.
\newblock \href {http://arxiv.org/abs/2104.06427} {\path{arXiv:2104.06427}},
  \href {https://doi.org/10.1007/JHEP07(2021)103}
  {\path{doi:10.1007/JHEP07(2021)103}}.

\bibitem{Smith:2014mla}
C.~Smith, {Rare K decays: Challenges and Perspectives} (2014).
\newblock \href {http://arxiv.org/abs/1409.6162} {\path{arXiv:1409.6162}}.

\bibitem{Mertens:2011ts}
P.~Mertens, C.~Smith, JHEP 08 (2011) 069.
\newblock \href {http://arxiv.org/abs/1103.5992} {\path{arXiv:1103.5992}},
  \href {https://doi.org/10.1007/JHEP08(2011)069}
  {\path{doi:10.1007/JHEP08(2011)069}}.

\bibitem{AlaviHarati:2003mr}
A.~Alavi-Harati, et~al., Phys. Rev. Lett. 93 (2004) 021805.
\newblock \href {http://arxiv.org/abs/hep-ex/0309072}
  {\path{arXiv:hep-ex/0309072}}, \href
  {https://doi.org/10.1103/PhysRevLett.93.021805}
  {\path{doi:10.1103/PhysRevLett.93.021805}}.

\bibitem{AlaviHarati:2000hs}
A.~Alavi-Harati, et~al., Phys. Rev. Lett. 84 (2000) 5279--5282.
\newblock \href {http://arxiv.org/abs/hep-ex/0001006}
  {\path{arXiv:hep-ex/0001006}}, \href
  {https://doi.org/10.1103/PhysRevLett.84.5279}
  {\path{doi:10.1103/PhysRevLett.84.5279}}.

\bibitem{Batley:2003mu}
J.~R. Batley, et~al., Phys. Lett. B576 (2003) 43--54.
\newblock \href {http://arxiv.org/abs/hep-ex/0309075}
  {\path{arXiv:hep-ex/0309075}}, \href
  {https://doi.org/10.1016/j.physletb.2003.10.001}
  {\path{doi:10.1016/j.physletb.2003.10.001}}.

\bibitem{Batley:2004wg}
J.~R. Batley, et~al., Phys. Lett. B599 (2004) 197--211.
\newblock \href {http://arxiv.org/abs/hep-ex/0409011}
  {\path{arXiv:hep-ex/0409011}}, \href
  {https://doi.org/10.1016/j.physletb.2004.08.058}
  {\path{doi:10.1016/j.physletb.2004.08.058}}.

\bibitem{Goudzovski:2022vbt}
E.~Goudzovski, et~al. (1 2022).
\newblock \href {http://arxiv.org/abs/2201.07805} {\path{arXiv:2201.07805}}.

\bibitem{Perrin-Terrin:2021jtl}
M.~Perrin-Terrin (12 2021).
\newblock \href {http://arxiv.org/abs/2112.12848} {\path{arXiv:2112.12848}}.

\bibitem{Muong-2:2021ojo}
B.~Abi, et~al., Phys. Rev. Lett. 126 (2021) 141801.
\newblock \href {http://arxiv.org/abs/2104.03281} {\path{arXiv:2104.03281}},
  \href {https://doi.org/10.1103/PhysRevLett.126.141801}
  {\path{doi:10.1103/PhysRevLett.126.141801}}.

\bibitem{Hurth:2021nsi}
T.~Hurth, F.~Mahmoudi, D.~M. Santos, S.~Neshatpour, Phys. Lett. B 824 (2022)
  136838.
\newblock \href {http://arxiv.org/abs/2104.10058} {\path{arXiv:2104.10058}},
  \href {https://doi.org/10.1016/j.physletb.2021.136838}
  {\path{doi:10.1016/j.physletb.2021.136838}}.

\bibitem{Isidori:2021vtc}
G.~Isidori, D.~Lancierini, P.~Owen, N.~Serra, Phys. Lett. B 822 (2021) 136644.
\newblock \href {http://arxiv.org/abs/2104.05631} {\path{arXiv:2104.05631}},
  \href {https://doi.org/10.1016/j.physletb.2021.136644}
  {\path{doi:10.1016/j.physletb.2021.136644}}.

\bibitem{Heeck:2022znj}
J.~Heeck, A.~Thapa (2 2022).
\newblock \href {http://arxiv.org/abs/2202.08854} {\path{arXiv:2202.08854}}.

\bibitem{Borsato:2018tcz}
M.~Borsato, V.~V. Gligorov, D.~Guadagnoli, D.~Martinez~Santos, O.~Sumensari,
  Phys. Rev. D 99 (2019) 055017.
\newblock \href {http://arxiv.org/abs/1808.02006} {\path{arXiv:1808.02006}},
  \href {https://doi.org/10.1103/PhysRevD.99.055017}
  {\path{doi:10.1103/PhysRevD.99.055017}}.

\bibitem{Cirigliano:2007xi}
V.~Cirigliano, I.~Rosell, Phys. Rev. Lett. 99 (2007) 231801.
\newblock \href {http://arxiv.org/abs/0707.3439} {\path{arXiv:0707.3439}},
  \href {https://doi.org/10.1103/PhysRevLett.99.231801}
  {\path{doi:10.1103/PhysRevLett.99.231801}}.

\bibitem{PhysRevD.74.011701}
A.~Masiero, P.~Paradisi, R.~Petronzio, Phys. Rev. D 74 (2006) 011701.
\newblock \href {https://doi.org/10.1103/PhysRevD.74.011701}
  {\path{doi:10.1103/PhysRevD.74.011701}},
  \href{https://link.aps.org/doi/10.1103/PhysRevD.74.011701}{[link]}.
\newline\urlprefix\url{https://link.aps.org/doi/10.1103/PhysRevD.74.011701}

\bibitem{BUCHMULLER1987442}
W.~Buchmüller, R.~Rückl, D.~Wyler, Physics Letters B 191 (1987) 442--448.
\newblock \href {https://doi.org/https://doi.org/10.1016/0370-2693(87)90637-X}
  {\path{doi:https://doi.org/10.1016/0370-2693(87)90637-X}},
  \href{https://www.sciencedirect.com/science/article/pii/037026938790637X}{[link]}.
\newline\urlprefix\url{https://www.sciencedirect.com/science/article/pii/037026938790637X}

\bibitem{Davidson:1993qk}
S.~Davidson, D.~C. Bailey, B.~A. Campbell, Z. Phys. C 61 (1994) 613--644.
\newblock \href {http://arxiv.org/abs/hep-ph/9309310}
  {\path{arXiv:hep-ph/9309310}}, \href {https://doi.org/10.1007/BF01552629}
  {\path{doi:10.1007/BF01552629}}.

\bibitem{lacker2010simultaneous}
H.~Lacker, A.~Menzel, Journal of High Energy Physics 2010 (2010) 1--20.

\bibitem{Abada:2012mc}
A.~Abada, D.~Das, A.~M. Teixeira, A.~Vicente, C.~Weiland, JHEP 02 (2013) 048.
\newblock \href {http://arxiv.org/abs/1211.3052} {\path{arXiv:1211.3052}},
  \href {https://doi.org/10.1007/JHEP02(2013)048}
  {\path{doi:10.1007/JHEP02(2013)048}}.

\bibitem{PhysRevLett.29.1274}
A.~R. Clark, B.~Cork, T.~Elioff, L.~T. Kerth, J.~F. McReynolds, D.~Newton,
  W.~A. Wenzel, Phys. Rev. Lett. 29 (1972) 1274--1277.
\newblock \href {https://doi.org/10.1103/PhysRevLett.29.1274}
  {\path{doi:10.1103/PhysRevLett.29.1274}},
  \href{https://link.aps.org/doi/10.1103/PhysRevLett.29.1274}{[link]}.
\newline\urlprefix\url{https://link.aps.org/doi/10.1103/PhysRevLett.29.1274}

\bibitem{HEARD1975327}
K.~Heard, J.~Heintze, G.~Heinzelmann, P.~Igo-Kemenes, W.~Kalbreier, E.~Mittag,
  H.~Rieseberg, B.~Schürlein, H.~Siebert, V.~Soergel, K.~Streit, A.~Wagner,
  A.~Walenta, Physics Letters B 55 (1975) 327--330.
\newblock \href {https://doi.org/https://doi.org/10.1016/0370-2693(75)90613-9}
  {\path{doi:https://doi.org/10.1016/0370-2693(75)90613-9}},
  \href{https://www.sciencedirect.com/science/article/pii/0370269375906139}{[link]}.
\newline\urlprefix\url{https://www.sciencedirect.com/science/article/pii/0370269375906139}

\bibitem{HEINTZE1976302}
J.~Heintze, G.~Heinzelmann, P.~Igo-Kemenes, R.~Mundhenke, H.~Rieseberg,
  B.~Schürlein, H.~Siebert, V.~Soergel, H.~Stelzer, K.~Streit, A.~Walenta,
  Physics Letters B 60 (1976) 302--304.
\newblock \href {https://doi.org/https://doi.org/10.1016/0370-2693(76)90306-3}
  {\path{doi:https://doi.org/10.1016/0370-2693(76)90306-3}},
  \href{https://www.sciencedirect.com/science/article/pii/0370269376903063}{[link]}.
\newline\urlprefix\url{https://www.sciencedirect.com/science/article/pii/0370269376903063}

\bibitem{NA62:2012lny}
C.~Lazzeroni, et~al., Phys. Lett. B 719 (2013) 326--336.
\newblock \href {http://arxiv.org/abs/1212.4012} {\path{arXiv:1212.4012}},
  \href {https://doi.org/10.1016/j.physletb.2013.01.037}
  {\path{doi:10.1016/j.physletb.2013.01.037}}.

\bibitem{Crivellin:2016vjc}
A.~Crivellin, G.~D'Ambrosio, M.~Hoferichter, L.~C. Tunstall, Phys. Rev. D 93
  (2016) 074038.
\newblock \href {http://arxiv.org/abs/1601.00970} {\path{arXiv:1601.00970}},
  \href {https://doi.org/10.1103/PhysRevD.93.074038}
  {\path{doi:10.1103/PhysRevD.93.074038}}.

\bibitem{Ecker:1987}
G.~Ecker, A.~Pich, E.~de~Rafael, Nucl. Phys. B291 (1987) 692.
\newblock \href {https://doi.org/10.1016/0550-3213(87)90491-3}
  {\path{doi:10.1016/0550-3213(87)90491-3}}.

\bibitem{DAmbrosio:1998gur}
G.~D'Ambrosio, G.~Ecker, G.~Isidori, J.~Portoles, JHEP 08 (1998) 004.
\newblock \href {http://arxiv.org/abs/hep-ph/9808289}
  {\path{arXiv:hep-ph/9808289}}, \href
  {https://doi.org/10.1088/1126-6708/1998/08/004}
  {\path{doi:10.1088/1126-6708/1998/08/004}}.

\bibitem{DAmbrosio:2018ytt}
G.~D'Ambrosio, D.~Greynat, M.~Knecht, JHEP 02 (2019) 049.
\newblock \href {http://arxiv.org/abs/1812.00735} {\path{arXiv:1812.00735}},
  \href {https://doi.org/10.1007/JHEP02(2019)049}
  {\path{doi:10.1007/JHEP02(2019)049}}.

\bibitem{E865:1999ker}
R.~Appel, et~al., Phys. Rev. Lett. 83 (1999) 4482--4485.
\newblock \href {http://arxiv.org/abs/hep-ex/9907045}
  {\path{arXiv:hep-ex/9907045}}, \href
  {https://doi.org/10.1103/PhysRevLett.83.4482}
  {\path{doi:10.1103/PhysRevLett.83.4482}}.

\bibitem{NA482:2009pfe}
J.~R. Batley, et~al., Phys. Lett. B 677 (2009) 246--254.
\newblock \href {http://arxiv.org/abs/0903.3130} {\path{arXiv:0903.3130}},
  \href {https://doi.org/10.1016/j.physletb.2009.05.040}
  {\path{doi:10.1016/j.physletb.2009.05.040}}.

\bibitem{NA482:2010zrc}
J.~R. Batley, et~al., Phys. Lett. B 697 (2011) 107--115.
\newblock \href {http://arxiv.org/abs/1011.4817} {\path{arXiv:1011.4817}},
  \href {https://doi.org/10.1016/j.physletb.2011.01.042}
  {\path{doi:10.1016/j.physletb.2011.01.042}}.

\bibitem{Mohapatra:1979ia}
R.~N. Mohapatra, G.~Senjanovic, Phys. Rev. Lett. 44 (1980) 912.
\newblock \href {https://doi.org/10.1103/PhysRevLett.44.912}
  {\path{doi:10.1103/PhysRevLett.44.912}}.

\bibitem{NA62:2019eax}
E.~Cortina~Gil, et~al., Phys. Lett. B 797 (2019) 134794.
\newblock \href {http://arxiv.org/abs/1905.07770} {\path{arXiv:1905.07770}},
  \href {https://doi.org/10.1016/j.physletb.2019.07.041}
  {\path{doi:10.1016/j.physletb.2019.07.041}}.

\bibitem{NA62:2021zxl}
E.~Cortina~Gil, et~al., Phys. Rev. Lett. 127 (2021) 131802.
\newblock \href {http://arxiv.org/abs/2105.06759} {\path{arXiv:2105.06759}},
  \href {https://doi.org/10.1103/PhysRevLett.127.131802}
  {\path{doi:10.1103/PhysRevLett.127.131802}}.

\bibitem{NA62:2022tte}
E.~Cortina~Gil, et~al. (2 2022).
\newblock \href {http://arxiv.org/abs/2202.00331} {\path{arXiv:2202.00331}}.

\bibitem{Bijnens:1992en}
J.~Bijnens, G.~Ecker, J.~Gasser, Nucl. Phys. B 396 (1993) 81--118.
\newblock \href {http://arxiv.org/abs/hep-ph/9209261}
  {\path{arXiv:hep-ph/9209261}}, \href
  {https://doi.org/10.1016/0550-3213(93)90259-R}
  {\path{doi:10.1016/0550-3213(93)90259-R}}.

\bibitem{Ecker:1987hd}
G.~Ecker, A.~Pich, E.~de~Rafael, Nucl. Phys. B 303 (1988) 665--702.
\newblock \href {https://doi.org/10.1016/0550-3213(88)90425-7}
  {\path{doi:10.1016/0550-3213(88)90425-7}}.

\bibitem{Donoghue:1997rr}
J.~F. Donoghue, F.~Gabbiani, Phys. Rev. D 56 (1997) 1605--1611.
\newblock \href {http://arxiv.org/abs/hep-ph/9702278}
  {\path{arXiv:hep-ph/9702278}}, \href
  {https://doi.org/10.1103/PhysRevD.56.1605}
  {\path{doi:10.1103/PhysRevD.56.1605}}.

\bibitem{Gabbiani:1998tj}
F.~Gabbiani, Phys. Rev. D 59 (1999) 094022.
\newblock \href {http://arxiv.org/abs/hep-ph/9812419}
  {\path{arXiv:hep-ph/9812419}}, \href
  {https://doi.org/10.1103/PhysRevD.59.094022}
  {\path{doi:10.1103/PhysRevD.59.094022}}.

\bibitem{Bijnens:1989mr}
J.~Bijnens, Nucl. Phys. B 337 (1990) 635.
\newblock \href {https://doi.org/10.1016/0550-3213(90)90509-C}
  {\path{doi:10.1016/0550-3213(90)90509-C}}.

\bibitem{DAmbrosio:1996jmq}
G.~D'Ambrosio, G.~Ecker, G.~Isidori, H.~Neufeld, Z. Phys. C 76 (1997) 301--310.
\newblock \href {http://arxiv.org/abs/hep-ph/9612412}
  {\path{arXiv:hep-ph/9612412}}, \href {https://doi.org/10.1007/s002880050554}
  {\path{doi:10.1007/s002880050554}}.

\bibitem{Aoyama:2020ynm}
T.~Aoyama, et~al., Phys. Rept. 887 (2020) 1--166.
\newblock \href {http://arxiv.org/abs/2006.04822} {\path{arXiv:2006.04822}},
  \href {https://doi.org/10.1016/j.physrep.2020.07.006}
  {\path{doi:10.1016/j.physrep.2020.07.006}}.

\bibitem{KTeV:2008pev}
E.~Abouzaid, et~al., Phys. Rev. Lett. 100 (2008) 182001.
\newblock \href {http://arxiv.org/abs/0802.2064} {\path{arXiv:0802.2064}},
  \href {https://doi.org/10.1103/PhysRevLett.100.182001}
  {\path{doi:10.1103/PhysRevLett.100.182001}}.

\bibitem{KTeV:2006pwx}
E.~Abouzaid, et~al., Phys. Rev. D 75 (2007) 012004.
\newblock \href {http://arxiv.org/abs/hep-ex/0610072}
  {\path{arXiv:hep-ex/0610072}}, \href
  {https://doi.org/10.1103/PhysRevD.75.012004}
  {\path{doi:10.1103/PhysRevD.75.012004}}.

\bibitem{Husek:2014tna}
T.~Husek, K.~Kampf, J.~Novotn\'y, Eur. Phys. J. C 74 (2014) 3010.
\newblock \href {http://arxiv.org/abs/1405.6927} {\path{arXiv:1405.6927}},
  \href {https://doi.org/10.1140/epjc/s10052-014-3010-4}
  {\path{doi:10.1140/epjc/s10052-014-3010-4}}.

\bibitem{NA62:2020xlg}
E.~Cortina~Gil, et~al., JHEP 03 (2021) 058.
\newblock \href {http://arxiv.org/abs/2011.11329} {\path{arXiv:2011.11329}},
  \href {https://doi.org/10.1007/JHEP03(2021)058}
  {\path{doi:10.1007/JHEP03(2021)058}}.

\bibitem{NA62:2017qcd}
E.~Cortina~Gil, et~al., Phys. Lett. B 778 (2018) 137--145.
\newblock \href {http://arxiv.org/abs/1712.00297} {\path{arXiv:1712.00297}},
  \href {https://doi.org/10.1016/j.physletb.2018.01.031}
  {\path{doi:10.1016/j.physletb.2018.01.031}}.

\bibitem{NA62:2020mcv}
E.~Cortina~Gil, et~al., Phys. Lett. B 807 (2020) 135599.
\newblock \href {http://arxiv.org/abs/2005.09575} {\path{arXiv:2005.09575}},
  \href {https://doi.org/10.1016/j.physletb.2020.135599}
  {\path{doi:10.1016/j.physletb.2020.135599}}.

\bibitem{NA62:2021bji}
E.~Cortina~Gil, et~al., Phys. Lett. B 816 (2021) 136259.
\newblock \href {http://arxiv.org/abs/2101.12304} {\path{arXiv:2101.12304}},
  \href {https://doi.org/10.1016/j.physletb.2021.136259}
  {\path{doi:10.1016/j.physletb.2021.136259}}.

\bibitem{NA62:2019meo}
E.~Cortina~Gil, et~al., JHEP 05 (2019) 182.
\newblock \href {http://arxiv.org/abs/1903.08767} {\path{arXiv:1903.08767}},
  \href {https://doi.org/10.1007/JHEP05(2019)182}
  {\path{doi:10.1007/JHEP05(2019)182}}.

\bibitem{HyperCP:2004kbv}
R.~A. Burnstein, et~al., Nucl. Instrum. Meth. A 541 (2005) 516--565.
\newblock \href {http://arxiv.org/abs/hep-ex/0405034}
  {\path{arXiv:hep-ex/0405034}}, \href
  {https://doi.org/10.1016/j.nima.2004.12.031}
  {\path{doi:10.1016/j.nima.2004.12.031}}.

\bibitem{BESIII:2018cnd}
M.~Ablikim, et~al., Nature Phys. 15 (2019) 631--634.
\newblock \href {http://arxiv.org/abs/1808.08917} {\path{arXiv:1808.08917}},
  \href {https://doi.org/10.1038/s41567-019-0494-8}
  {\path{doi:10.1038/s41567-019-0494-8}}.

\bibitem{Ireland:2019uja}
D.~G. Ireland, M.~D\"oring, D.~I. Glazier, J.~Haidenbauer, M.~Mai,
  R.~Murray-Smith, D.~R\"onchen, Phys. Rev. Lett. 123 (2019) 182301.
\newblock \href {http://arxiv.org/abs/1904.07616} {\path{arXiv:1904.07616}},
  \href {https://doi.org/10.1103/PhysRevLett.123.182301}
  {\path{doi:10.1103/PhysRevLett.123.182301}}.

\bibitem{Geng:2021fog}
L.-S. Geng, J.~M. Camalich, R.-X. Shi (12 2021).
\newblock \href {http://arxiv.org/abs/2112.11979} {\path{arXiv:2112.11979}}.

\bibitem{Lyagin:1962}
I.~Lyagin, E.~Ginzburg, Sov. Phys. JETP 14 (1962) 653.

\bibitem{HyperCP:2005mvo}
H.~Park, et~al., Phys. Rev. Lett. 94 (2005) 021801.
\newblock \href {http://arxiv.org/abs/hep-ex/0501014}
  {\path{arXiv:hep-ex/0501014}}, \href
  {https://doi.org/10.1103/PhysRevLett.94.021801}
  {\path{doi:10.1103/PhysRevLett.94.021801}}.

\bibitem{LHCb:2017rdd}
R.~Aaij, et~al., Phys. Rev. Lett. 120 (2018) 221803.
\newblock \href {http://arxiv.org/abs/1712.08606} {\path{arXiv:1712.08606}},
  \href {https://doi.org/10.1103/PhysRevLett.120.221803}
  {\path{doi:10.1103/PhysRevLett.120.221803}}.

\bibitem{KOTOproposal}
J.~Comfort, et~al., {Proposal for $K_L \rightarrow \pi^0 \nu \overline{\nu}$
  Experiment at J-PARC},
  \url{https://j-parc.jp/researcher/Hadron/en/pac_0606/pdf/p14-Yamanaka.pdf}
  (2006).

\bibitem{ref:kaon2019}
T.~Nomura, J. Phys. Conf. Ser. 1526 (2020) 012027.
\newblock \href {https://doi.org/10.1088/1742-6596/1526/1/012027}
  {\path{doi:10.1088/1742-6596/1526/1/012027}}.

\bibitem{Kamiji:2017deh}
I.~Kamiji, K.~Nakagiri, J. Phys. Conf. Ser. 800 (2017) 012041.
\newblock \href {https://doi.org/10.1088/1742-6596/800/1/012041}
  {\path{doi:10.1088/1742-6596/800/1/012041}}.

\bibitem{Maeda:2014pga}
Y.~Maeda, et~al., PTEP 2015 (2015) 063H01.
\newblock \href {https://doi.org/10.1093/ptep/ptv074}
  {\path{doi:10.1093/ptep/ptv074}}.

\bibitem{LHCcomm:2022}
\href{https://lhc-commissioning.web.cern.ch/schedule/LHC-long-term.htm}{[link]}.
\newline\urlprefix\url{https://lhc-commissioning.web.cern.ch/schedule/LHC-long-term.htm}

\bibitem{Bartosik:2018xxx}
H.~Bartosik, et~al., \href{https://cds.cern.ch/record/2650722}{{SPS} operation
  and future proton sharing scenarios for the {SHiP} experiment at the {BDF}
  facility}, Tech. Rep. CERN-ACC-NOTE-2018-002, CERN (2018).
\newline\urlprefix\url{https://cds.cern.ch/record/2650722}

\bibitem{Banerjee:2018xxx}
D.~Banerjee, et~al., \href{http://cds.cern.ch/record/2650193/}{The report of
  the {Conventional Beams Working Group} to the {Physics Beyond Colliders}
  study and to the {European Strategy for Particle Physics}}, Tech. Rep.
  CERN-PBC-NOTES-2018-005, CERN (2018).
\newline\urlprefix\url{http://cds.cern.ch/record/2650193/}

\bibitem{Gatignon:2018xxx}
L.~Gatignon, et~al., \href{https://cds.cern.ch/record/2650989}{Report from the
  {Conventional Beams Working Group} to the {Physics Beyond Colliders} study
  and to the {European Strategy for Particle Physics}}, Tech. Rep.
  CERN-PBC-REPORT-2018-002, CERN (2018).
\newline\urlprefix\url{https://cds.cern.ch/record/2650989}

\bibitem{Ambrosino:2019qvz}
F.~Ambrosino, et~al. (1 2019).
\newblock \href {http://arxiv.org/abs/1901.03099} {\path{arXiv:1901.03099}}.

\bibitem{Beacham:2019nyx}
J.~Beacham, et~al., J. Phys. G 47 (2020) 010501.
\newblock \href {http://arxiv.org/abs/1901.09966} {\path{arXiv:1901.09966}},
  \href {https://doi.org/10.1088/1361-6471/ab4cd2}
  {\path{doi:10.1088/1361-6471/ab4cd2}}.

\bibitem{Solieri:2020tar}
N.~Solieri, A.~Ciccotelli, Analysis of the survival of the {NA62} {T10} target
  for use in the proposed {KLEVER} experiment, Tech. rep., CERN, Geneva
  (January 2020).

\bibitem{Bruno:2006xxx}
L.~Bruno, The {CNGS} target: {E}xpectation vs.\ experience, Tech. rep., CERN,
  Geneva (September 2006).

\bibitem{Takahashi:2015xxx}
H.~Takahashi, et~al., J.\ Radioanal.\ Nucl.\ Chem. 305 (2015) 803.
\newblock \href {https://doi.org/10.1007/s10967-015-3940-9}
  {\path{doi:10.1007/s10967-015-3940-9}}.

\bibitem{Solieri:2020tax}
N.~Solieri, A.~Ciccotelli, {FEM} analysis of the survival of the {K12} {TAX}
  for use in the proposed {KLEVER} and {NA62-4x} experiments, Tech. rep., CERN,
  Geneva (January 2020).

\bibitem{Atherton:1980vj}
H.~W. Atherton, C.~Bovet, N.~Doble, G.~von Holtey, L.~Piemontese, A.~Placci,
  M.~Placidi, D.~E. Plane, M.~Reinharz, E.~Rossa, {Precise Measurements of
  Particle Production by 400-{GeV}/$c$ Protons on Beryllium Targets} (1980).

\bibitem{Lundberg:1984pj}
B.~G. Lundberg, {Neutral Strange Particle Production and Polarization at Large
  p(t)}, Ph.D. thesis, Wisconsin U., Madison (1984).
\newblock \href {https://doi.org/10.2172/1433363} {\path{doi:10.2172/1433363}}.

\bibitem{Beretvas:1986km}
A.~Beretvas, et~al., Phys. Rev. D 34 (1986) 53--74.
\newblock \href {https://doi.org/10.1103/PhysRevD.34.53}
  {\path{doi:10.1103/PhysRevD.34.53}}.

\bibitem{Jones:1980vp}
L.~W. Jones, M.~R. Whalley, H.~R. Gustafson, K.~J. Heller, M.~J. Longo, T.~J.
  Roberts, {Inclusive neutron production by 400-GeV protons}, in: {X
  International Symposium on Multiparticle Dynamics}, 1980.

\bibitem{vanDijk:2018aaa}
M.~van Dijk, M.~Rosenthal, Target studies for the proposed {KLEVER} experiment,
  cERN-ACC-NOTE-2018-0066, CERN-PBC-Notes-2018-002, KLEVER-PUB-18-01 (2018).

\bibitem{Malensek:1981em}
A.~J. Malensek (10 1981).

\bibitem{VanDijk:2019oml}
M.~Van~Dijk, et~al. (2019) THPGW061\href
  {https://doi.org/10.18429/JACoW-IPAC2019-THPGW061}
  {\path{doi:10.18429/JACoW-IPAC2019-THPGW061}}.

\bibitem{Bak:1988bq}
J.~F. Bak, et~al., Phys. Lett. B202 (1988) 615--619.
\newblock \href {https://doi.org/10.1016/0370-2693(88)91874-6}
  {\path{doi:10.1016/0370-2693(88)91874-6}}.

\bibitem{Kimball:1985np}
J.~C. Kimball, N.~Cue, Phys. Rept. 125 (1985) 69--101.
\newblock \href {https://doi.org/10.1016/0370-1573(85)90021-3}
  {\path{doi:10.1016/0370-1573(85)90021-3}}.

\bibitem{Baryshevsky:1989wm}
V.~G. Baryshevsky, V.~V. Tikhomirov, Sov. Phys. Usp. 32 (1989) 1013--1032,
  [Usp. Fiz. Nauk159,529(1989)].
\newblock \href {https://doi.org/10.1070/PU1989v032n11ABEH002778}
  {\path{doi:10.1070/PU1989v032n11ABEH002778}}.

\bibitem{Soldani:2022ekn}
M.~Soldani, et~al. (3 2022).
\newblock \href {http://arxiv.org/abs/2203.07163} {\path{arXiv:2203.07163}}.

\bibitem{Rossi:1974if}
A.~M. Rossi, G.~Vannini, A.~Bussiere, E.~Albini, D.~D'Alessandro,
  G.~Giacomelli, Nucl. Phys. B 84 (1975) 269--305.
\newblock \href {https://doi.org/10.1016/0550-3213(75)90307-7}
  {\path{doi:10.1016/0550-3213(75)90307-7}}.

\bibitem{Beaulieu:2009}
D.~Beaulieu, et~al., Nucl. Instrum. Methods A 607 (2009) 81.

\bibitem{Lehmann:2020}
A.~Lehmann, et~al., Nucl. Instrum. Methods A 952 (2020) 161821.

\bibitem{ecfareport:2021}
ECFA (2021).
\newblock \href {https://doi.org/10.17181/CERN.XDPL.W2EX}
  {\path{doi:10.17181/CERN.XDPL.W2EX}}.

\bibitem{lgad:2014}
G.~Pellegrini, et~al., Nucl. Instrum. Methods A 765 (2014) 12--16.

\bibitem{lgad:2017}
H.~Sadrozinski, A.~Seiden, N.~Cartiglia, Rep. Prog. Phys. 026101 (2017).

\bibitem{3dsensor:2019}
G.~Kramberger, et~al., Nucl. Instrum. Methods A 934 (2019) 26--32.

\bibitem{3dsensor:2020}
A.~Lai, et~al., Nucl. Instrum. Methods A 981 (2020) 164491.

\bibitem{tilgad:2020}
G.~Paternoster, et~al., IEEE Electr. Device L. 41 (2020) 884.

\bibitem{aclgad:2019}
M.~Mandurrino, et~al., IEEE Electr. Device L. 40 (2019) 1780.

\bibitem{NA48:2002tmj}
J.~R. Batley, et~al., Phys. Lett. B 544 (2002) 97--112.
\newblock \href {http://arxiv.org/abs/hep-ex/0208009}
  {\path{arXiv:hep-ex/0208009}}, \href
  {https://doi.org/10.1016/S0370-2693(02)02476-0}
  {\path{doi:10.1016/S0370-2693(02)02476-0}}.

\bibitem{NA482:2007ucr}
J.~R. Batley, et~al., Eur. Phys. J. C 52 (2007) 875--891.
\newblock \href {http://arxiv.org/abs/0707.0697} {\path{arXiv:0707.0697}},
  \href {https://doi.org/10.1140/epjc/s10052-007-0456-7}
  {\path{doi:10.1140/epjc/s10052-007-0456-7}}.

\bibitem{Nishiguchi:2017gei}
H.~Nishiguchi, et~al., Nucl. Instrum. Methods A 958 (2020) 162800.

\bibitem{na48detector:2007}
V.~Fanti, et~al., Nucl. Instrum. Methods A 574 (2007) 433.

\bibitem{Fanti:2007vi}
V.~Fanti, et~al., Nucl. Instrum. Meth. A574 (2007) 433--471.
\newblock \href {https://doi.org/10.1016/j.nima.2007.01.178}
  {\path{doi:10.1016/j.nima.2007.01.178}}.

\bibitem{NA62+07:M760}
{NA62 Collaboration}, Tech. Rep. SPSC-2007-035 (M-760), CERN (2007).

\bibitem{Atoian:2007up}
G.~S. Atoian, et~al., Nucl. Instrum. Meth. A584 (2008) 291--303.
\newblock \href {http://arxiv.org/abs/0709.4514} {\path{arXiv:0709.4514}},
  \href {https://doi.org/10.1016/j.nima.2007.10.022}
  {\path{doi:10.1016/j.nima.2007.10.022}}.

\bibitem{Gandini:2020aaa}
M.~Gandini, et~al., Nat. Nanotechnol. 15 (2020) 462--468.
\newblock \href {https://doi.org/https://doi.org/10.1038/s41565-020-0683-8}
  {\path{doi:https://doi.org/10.1038/s41565-020-0683-8}}.

\bibitem{Acerbi:2020itd}
F.~Acerbi, et~al., Nucl. Instrum. Meth. A 956 (2020) 163379.
\newblock \href {http://arxiv.org/abs/2001.03130} {\path{arXiv:2001.03130}},
  \href {https://doi.org/10.1016/j.nima.2019.163379}
  {\path{doi:10.1016/j.nima.2019.163379}}.

\bibitem{OPAL:1990yff}
K.~Ahmet, et~al., Nucl. Instrum. Meth. A 305 (1991) 275--319.
\newblock \href {https://doi.org/10.1016/0168-9002(91)90547-4}
  {\path{doi:10.1016/0168-9002(91)90547-4}}.

\bibitem{KOPIO+05:CDR}
J.~Comfort, et~al., {KOPIO} project: Conceptual design report (2005).

\bibitem{Atiya:1992vh}
M.~S. Atiya, et~al., Nucl. Instrum. Meth. A321 (1992) 129--151.
\newblock \href {https://doi.org/10.1016/0168-9002(92)90382-E}
  {\path{doi:10.1016/0168-9002(92)90382-E}}.

\bibitem{Ambrosino:2007ss}
F.~Ambrosino, et~al., {A Prototype large-angle photon veto detector for the
  P326 experiment at CERN}, in: {Proceedings, 2007 IEEE Nuclear Science
  Symposium and Medical Imaging Conference (NSS/MIC 2007): Honolulu, Hawaii,
  October 28-November 3, 2007}, Vol.~1, 2007, pp. 57--64.
\newblock \href {http://arxiv.org/abs/0711.3398} {\path{arXiv:0711.3398}},
  \href {https://doi.org/10.1109/NSSMIC.2007.4436288}
  {\path{doi:10.1109/NSSMIC.2007.4436288}}.

\bibitem{Ramberg:2004en}
E.~Ramberg, P.~Cooper, R.~Tschirhart, IEEE Trans. Nucl. Sci. 51 (2004)
  2201--2204.
\newblock \href {https://doi.org/10.1109/TNS.2004.836738}
  {\path{doi:10.1109/TNS.2004.836738}}.

\bibitem{Frankenthal:2018yvf}
A.~Frankenthal, et~al., Nucl. Instrum. Meth. A 919 (2019) 89--97.
\newblock \href {http://arxiv.org/abs/1809.10840} {\path{arXiv:1809.10840}},
  \href {https://doi.org/10.1016/j.nima.2018.12.035}
  {\path{doi:10.1016/j.nima.2018.12.035}}.

\bibitem{Cemmi:2021uum}
A.~Cemmi, et~al. (7 2021).
\newblock \href {http://arxiv.org/abs/2107.12307} {\path{arXiv:2107.12307}}.

\bibitem{Kozma:2002km}
P.~Kozma, P.~Kozma~Jr., R.~Bajgar, Nucl. Instrum. Meth. A 484 (2002) 149--152.
\newblock \href {https://doi.org/10.1016/S0168-9002(01)02011-3}
  {\path{doi:10.1016/S0168-9002(01)02011-3}}.

\bibitem{Anderson:1989uj}
D.~F. Anderson, M.~Kobayashi, Y.~Yoshimura, C.~L. Woody, Nucl. Instrum. Meth. A
  290 (1990) 385--389.
\newblock \href {https://doi.org/10.1016/0168-9002(90)90553-I}
  {\path{doi:10.1016/0168-9002(90)90553-I}}.

\bibitem{PANDA:2011hqx}
M.~Kavatsyuk, et~al., Nucl. Instrum. Meth. A 648 (2011) 77--91.
\newblock \href {https://doi.org/10.1016/j.nima.2011.06.044}
  {\path{doi:10.1016/j.nima.2011.06.044}}.

\bibitem{Auffray:2016xtu}
E.~Auffray, et~al., J. Lumin. 178 (2016) 54--60.
\newblock \href {https://doi.org/10.1016/j.jlumin.2016.05.015}
  {\path{doi:10.1016/j.jlumin.2016.05.015}}.

\bibitem{Follin:2021kgn}
M.~Follin, V.~Sharyy, J.-P. Bard, M.~Korzhik, D.~Yvon, JINST 16 (2021) P08040.
\newblock \href {http://arxiv.org/abs/2103.13106} {\path{arXiv:2103.13106}},
  \href {https://doi.org/10.1088/1748-0221/16/08/P08040}
  {\path{doi:10.1088/1748-0221/16/08/P08040}}.

\bibitem{Bandiera:2018ymh}
L.~Bandiera, et~al., Phys. Rev. Lett. 121 (2018) 021603.
\newblock \href {http://arxiv.org/abs/1803.10005} {\path{arXiv:1803.10005}},
  \href {https://doi.org/10.1103/PhysRevLett.121.021603}
  {\path{doi:10.1103/PhysRevLett.121.021603}}.

\bibitem{LHCb:2008vvz}
A.~A. Alves, Jr., et~al., JINST 3 (2008) S08005.
\newblock \href {https://doi.org/10.1088/1748-0221/3/08/S08005}
  {\path{doi:10.1088/1748-0221/3/08/S08005}}.

\bibitem{AlvesJunior:2018ldo}
A.~A. Alves~Junior, et~al., JHEP 05 (2019) 048.
\newblock \href {http://arxiv.org/abs/1808.03477} {\path{arXiv:1808.03477}},
  \href {https://doi.org/10.1007/JHEP05(2019)048}
  {\path{doi:10.1007/JHEP05(2019)048}}.

\bibitem{Dettori:2017ycr}
F.~Dettori, D.~Martinez~Santos, J.~Prisciandaro (2017).

\bibitem{Aaij:2019zbu}
R.~Aaij, et~al., Comput. Softw. Big Sci. 4 (2020) 7.
\newblock \href {http://arxiv.org/abs/1912.09161} {\path{arXiv:1912.09161}},
  \href {https://doi.org/10.1007/s41781-020-00039-7}
  {\path{doi:10.1007/s41781-020-00039-7}}.

\bibitem{LHCbCollaboration:2717938}
C.~M. LHCb~Collaboration, \href{https://cds.cern.ch/record/2717938}{{LHCb
  Upgrade GPU High Level Trigger Technical Design Report}}, Tech. rep., CERN,
  Geneva (May 2020).
\newline\urlprefix\url{https://cds.cern.ch/record/2717938}

\bibitem{LHCb:2018roe}
R.~Aaij, et~al. (8 2018).
\newblock \href {http://arxiv.org/abs/1808.08865} {\path{arXiv:1808.08865}}.

\bibitem{Aaij:2244311}
R.~Aaij, et~al., \href{https://cds.cern.ch/record/2244311}{{Expression of
  Interest for a Phase-II LHCb Upgrade: Opportunities in flavour physics, and
  beyond, in the HL-LHC era}}, Tech. rep., CERN, Geneva (Feb 2017).
\newline\urlprefix\url{https://cds.cern.ch/record/2244311}

\bibitem{Cerri:2018ypt}
A.~Cerri, et~al., CERN Yellow Rep. Monogr. 7 (2019) 867--1158.
\newblock \href {http://arxiv.org/abs/1812.07638} {\path{arXiv:1812.07638}},
  \href {https://doi.org/10.23731/CYRM-2019-007.867}
  {\path{doi:10.23731/CYRM-2019-007.867}}.

\bibitem{He:2018yzu}
X.-G. He, J.~Tandean, G.~Valencia, JHEP 10 (2018) 040.
\newblock \href {http://arxiv.org/abs/1806.08350} {\path{arXiv:1806.08350}},
  \href {https://doi.org/10.1007/JHEP10(2018)040}
  {\path{doi:10.1007/JHEP10(2018)040}}.

\bibitem{BreaRodriguez:2020jlg}
A.~Brea~Rodr\'\i{}guez, J. Phys. Conf. Ser. 1526 (2020) 012022.
\newblock \href {https://doi.org/10.1088/1742-6596/1526/1/012022}
  {\path{doi:10.1088/1742-6596/1526/1/012022}}.

\bibitem{Alonso-Alvarez:2021oaj}
G.~Alonso-\'Alvarez, G.~Elor, M.~Escudero, B.~Fornal, B.~Grinstein, J.~M.
  Camalich (11 2021).
\newblock \href {http://arxiv.org/abs/2111.12712} {\path{arXiv:2111.12712}}.

\bibitem{refId0}
{Serfon, C\'edric}, {Mashinistov, Ruslan}, {De Stefano, John Steven},
  {Hern\'andez Villanueva, Michel}, {Ito, Hironori}, {Kato, Yuji}, {Laycock,
  Paul}, {Miyake, Hideki}, {Ueda, Ikuo}, EPJ Web Conf. 251 (2021) 02057.
\newblock \href {https://doi.org/10.1051/epjconf/202125102057}
  {\path{doi:10.1051/epjconf/202125102057}},
  \href{https://doi.org/10.1051/epjconf/202125102057}{[link]}.
\newline\urlprefix\url{https://doi.org/10.1051/epjconf/202125102057}

\bibitem{Laycock:2019vxl}
P.~Laycock, EPJ Web Conf. 214 (2019) 02017.
\newblock \href {https://doi.org/10.1051/epjconf/201921402017}
  {\path{doi:10.1051/epjconf/201921402017}}.

\bibitem{Barisits:2019fyl}
M.~Barisits, et~al., Comput. Softw. Big Sci. 3 (2019) 11.
\newblock \href {http://arxiv.org/abs/1902.09857} {\path{arXiv:1902.09857}},
  \href {https://doi.org/10.1007/s41781-019-0026-3}
  {\path{doi:10.1007/s41781-019-0026-3}}.

\bibitem{Ainsworth:2021ahm}
R.~Ainsworth, et~al. (6 2021).
\newblock \href {http://arxiv.org/abs/2106.02133} {\path{arXiv:2106.02133}}.

\bibitem{Arrington:2022pon}
J.~Arrington, et~al. (3 2022).
\newblock \href {http://arxiv.org/abs/2203.03925} {\path{arXiv:2203.03925}}.

\end{thebibliography}

\end{document}